\documentclass[a4paper, 10pt, twocolumn, openany]{book}

\usepackage[UKenglish]{babel} 
\usepackage[T1]{fontenc}
\usepackage[ansinew]{inputenc}
\usepackage{color}

\usepackage{lmodern} %Type1-font for non-english texts and characters

\usepackage{graphicx} %%For loading graphic files

\usepackage{amsmath}
\usepackage{amsthm}
\usepackage{amsfonts}
\usepackage{amssymb}
\usepackage{epsfig}
\usepackage{upgreek}
\usepackage{balance}

\usepackage{cite}
\usepackage{xspace}

\usepackage{fixltx2e}
\usepackage{fancyhdr}
\usepackage{titlesec}

\usepackage{pgfgantt}

\usepackage[toc]{multitoc}

\usepackage{booktabs}		%% nice layout of tables
\usepackage{siunitx}		%% correct typesetting of units
\DeclareSIUnit[number-unit-product = {}]\c{$c$}
\usepackage{caption}		%% customize captions

\newcommand{\Michel}{$\mu \rightarrow e\nu\nu$\xspace}
\newcommand{\meg}{$\mu \rightarrow e\gamma$\xspace}
\newcommand{\mte}{$\mu \rightarrow eee$\xspace}
\newcommand{\mteee}{$\mu \rightarrow eee$\xspace}
\newcommand{\mtenunu}{$\mu \rightarrow eee\nu\nu$\xspace}

\newcommand{\ord}{{\cal O}}

\titleformat{\chapter}[display]{\sc}{Chapter \thechapter}{20.0pt}{\includegraphics[width=0.2\textwidth]{common/figures/logo_drawing} \Huge}

\titleformat{\part}[display]{\sc \begin{center}}{\Large Part \thepart}{35.0pt}{\includegraphics[width=0.3\textwidth]{common/figures/logo_drawing}\\ \vspace{1cm} \Huge}[\end{center}]

\titleformat{\section}{}{\Large \thesection~}{0cm}{\Large}[]

\titleformat{\subsection}{\sc}{\thesubsection~}{0cm}{}[]

\begin{document}

\onecolumn
\pagestyle{empty} %No headings for the first pages.

%% Title Page %%%%%%%%%%%%%%%%%%%%%%%%%%%%%%%%%%%%%%%%%%%%%%%
\vspace{-3cm}
\hspace{-3cm}

\includegraphics[width=0.30\textwidth]{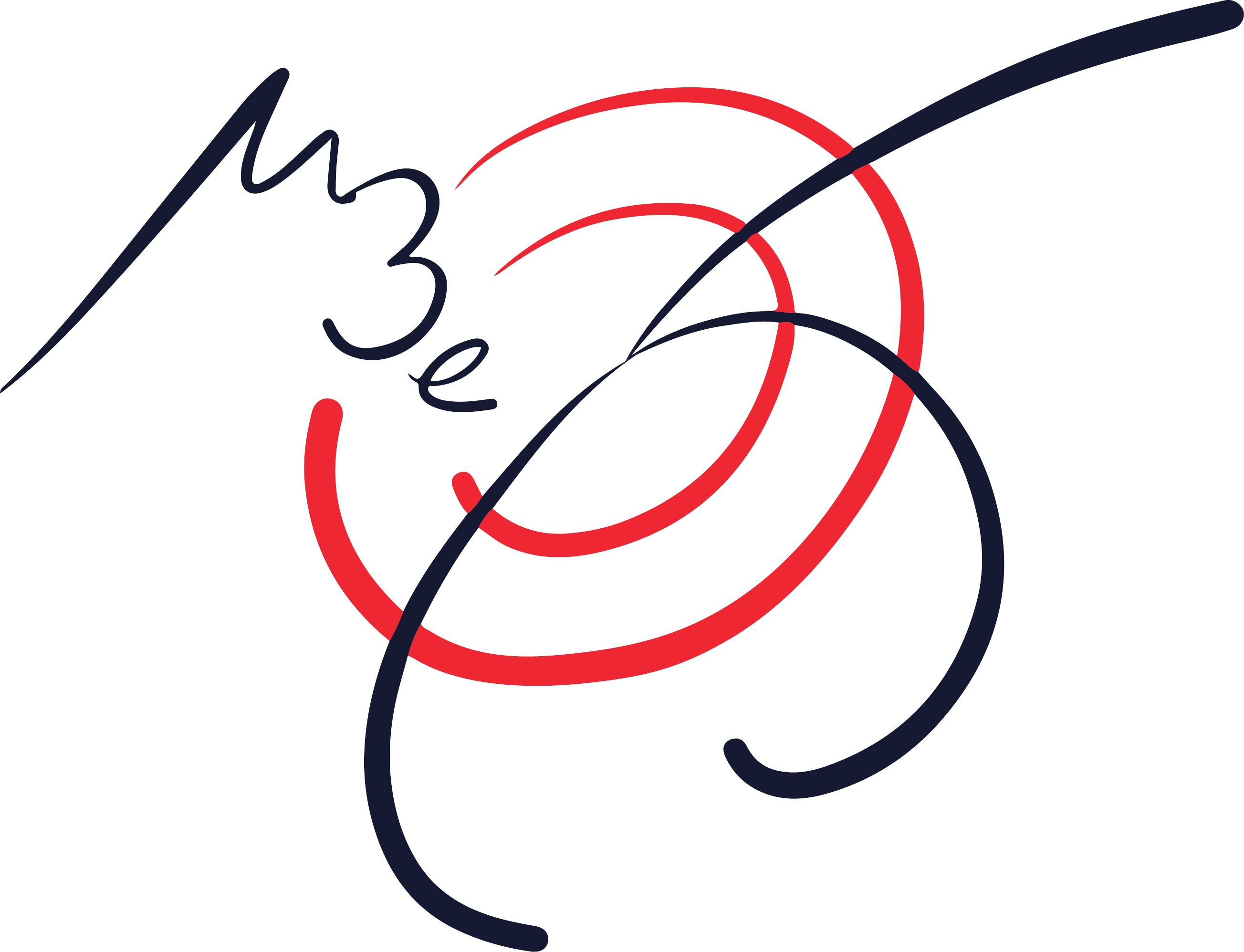} 
\vspace{0.2cm}

{
\centering
{\Huge
Research Proposal for an Experiment to Search for the Decay \mte
}
\vspace{2cm}
%\author{Author and Institute List}

%\author{
A.~Blondel, A.~Bravar, M.~Pohl\\ \emph{D\'epartement de physique nucl\'eaire et
  corpusculaire,} \\ \emph{Universit\'e de Gen\`eve, Gen\`eve}\\ \vspace{0.7cm}
S.~Bachmann, N.~Berger, M.~Kiehn,  
A.~Sch\"oning, D.~Wiedner, B.~Windelband\\\emph{Physikalisches Institut, Universit\"at
  Heidelberg, Heidelberg}\\ \vspace{0.7cm}
 P.~Eckert, H.-C.~Schultz-Coulon, W.~Shen\\\emph{Kirchoff Institut f\"ur Physik, Universit\"at
  Heidelberg, Heidelberg}\\ \vspace{0.7cm} 
P.~Fischer, I.~Peri\'c\\ \emph{Zentralinstitut f\"ur Informatik, Universit\"at
  Heidelberg, Mannheim} \\ \vspace{0.7cm}
M.~Hildebrandt, P.-R.~Kettle, A.~Papa, S.~Ritt, A.~Stoykov\\ \emph{Paul Scherrer Institut, Villigen} \\ \vspace{0.7cm}
G.~Dissertori, C.~Grab, R.~Wallny\\ \emph{Eidgen\"ossiche Technische Hochschule
  Z\"urich, Z\"urich} \\ \vspace{0.7cm}
R.~Gredig, P.~Robmann, U.~Straumann\\ \emph{Universit\"at Z\"urich, Z\"urich}\\
\vspace{1.4cm}
December 10$^{th}$, 2012\\
}

\twocolumn
\setcounter{tocdepth}{1}
\tableofcontents{}

\newpage

\newpage
\balance
\chapter*{Executive Summary}
\addcontentsline{toc}{chapter}{Executive Summary}

We propose an experiment (\emph{Mu3e}) to search for the lepton flavour violating (LFV) decay
$\mu^{+} \rightarrow e^{+}e^{-}e^{+}$. 
We aim for an ultimate sensitivity of one in $\num{e16}$ $\mu$-decays, four orders of
magnitude better than previous searches. 
This sensitivity is made possible by exploiting modern silicon pixel detectors providing high spatial resolution
and hodoscopes using scintillating fibres and tiles providing precise timing
information at high particle rates.

Existing beamlines available at PSI providing rates of order $10^8$ muons per
second allow to test for the decay $\mu^{+}
\rightarrow e^{+}e^{-}e^{+}$ in one of $\num{e15}$ muon decays. In a first phase of the experiment, we plan to make use of this and establish the experimental technique whilst at the same time pushing the sensitivity by three orders of magnitude.

The installation of a new muon beamline at the spallation neutron source is
currently under discussion at PSI. Such a \emph{High Intensity Muon Beam}
(HiMB) will provide intensities in excess of $\num{e9}$ muons per second, which in turn are required to reach the aimed sensitivity of 
$\textrm{B}(\mu^{+} \rightarrow e^{+}e^{-}e^{+}) \sim \num{e-16}$.

The proposed experiment is highly complementary to other LFV
searches for physics beyond the standard model, i.e.~direct searches 
performed at the Large Hadron Collider (LHC)
and indirected searches in the decay of taus and muons, such as the decay
 $\mu^{+} \rightarrow e^{+} \gamma$, which is the subject of the \emph{MEG} experiment currently in operation at PSI. 
The proposed experiment for the search $\mu^{+} \rightarrow e^{+}e^{-}e^{+}$
will test lepton flavour violating models of physics beyond the Standard Model 
with unprecedented sensitivity. 

This sensitivity is experimentally achieved by a novel experimental design
exploiting silicon pixel detectors based on High Voltage Monolithic Active Pixel Sensors (HV-MAPS). This technology provides high granularity, important for precision tracking and vertexing, and allows one to significantly reduce the material budget by thinning down the sensors and by integrating 
the hit digitisation and readout circuitry in the sensor itself.
The detector geometry is optimized to reach the highest possible momentum
resolution in a multiple Coulomb scattering environment, which is needed to
suppress the dominating background from the radiative muon decay with internal conversion, $\mu \rightarrow eee\nu\bar{\nu}$.
The time information of the decay electrons\footnote{Here and in the following, the term “electron” denotes generically both decay electrons and positrons.}, obtained from the pixel detector is further improved by a 
time-of-flight system consisting of a scintillating fiber hodoscope  and  tiles with Silicon Photo-Multipliers (SiPM) for light detection.
By combining both detector systems accidental background can be reduced below the aimed sensitivity of  $\textrm{B}(\mu^{+} \rightarrow e^{+}e^{-}e^{+}) \sim \num{e-16}$.

We will complete the sensor development and start constructing the detector in
2013, in order to be ready for first exploratory data taking at an existing
beam line with a first minimal detector setup in 2015. 
A detector capable of taking data rates of order $10^8$ muons per second and capable of reaching a sensitivity of 
$\textrm{B}(\mu^{+} \rightarrow e^{+}e^{-}e^{+}) \sim \num{e-15}$ will be available in 2016. This \emph{Phase I} detector is the main focus of this proposal.

In \emph{Phase II}, beyond 2017, the experiment will reach the ultimate sensitivity by exploiting a possible new high intensity muon beamline with an intensity of $> \num{2e9}$ muons per second. In the absence of a signal, LFV muon decays can then be excluded for $\textrm{B}(\mu^{+} \rightarrow e^{+}e^{-}e^{+}) \num{<e-16}$ at $\SI{90}{\percent}$ confidence level.

\part{Introduction}

\pagestyle{fancy}

\chapter{Motivation}
\label{sec:Motivation}

\nobalance
\pagestyle{fancy}

In the Standard Model (SM) of elementary particle physics, the number of
leptons of each family (lepton flavour) is conserved  at tree level. 
In the neutrino sector,
lepton flavour violation (LFV) has however been observed in the form of
neutrino mixing by the Super-Kamiokande \cite{Fukuda:1998mi}, SNO
\cite{Ahmad:2001an}, KamLAND \cite{Eguchi:2002dm} and subsequent
experiments. Consequently, lepton flavour is a broken symmetry, the
standard model has to be adapted to incorporate massive neutrinos and lepton
flavour violation is also expected in the charged lepton sector. The exact
mechanism and size of LFV being unknown, its study is of large interest, as it
is linked to neutrino mass generation, CP violation and new physics beyond the
SM (BSM).

The non-observation of LFV of charged leptons in past and present experiments
might at a first glance be surprising, as the mixing angles in the neutrino matrix 
have been measured to be large (maximal).
This huge suppression of LFV effects is however accidental
and due to the fact that (a) neutrinos are so much lighter than charged
leptons and (b) the mass differences between neutrinos (more precisely the square of the
mass differences) are very small compared to the W-boson mass.

The situation completely changes if new particles beyond the SM are introduced.
If e.g. SUSY is realized at the electroweak scale, the scalar partners of
the charged leptons (sleptons) will have large masses, and if not fully
degenerate, induce LFV interactions via loop corrections.
These LFV effects from new particles at the TeV scale are naturally
generated in many models and are therefore considered to be a prominent
signature for new physics. 
%This argument applies for many models where the existence of new particles 
%coupling to leptons are predicted. 
%Many of these models have als the feature to also explain the smallness of the
%neutrino masses.

In many extensions of the SM, such as grand unified models
\cite{Pati:1974yy,Georgi:1974sy,Langacker1981}, supersymmetric models
\cite{Haber:1984rc} (see Figure~\ref{fig:m3e_feyn2}), left-right symmetric
models \cite{Mohapatra:1974hk, Mohapatra:1974gc,Senjanovic:1975rk}, models
with an extended Higgs sector \cite{Kakizaki:2003jk} and models where
electroweak symmetry is broken dynamically \cite{Hill:2002ap}, an
experimentally accessible amount of LFV is predicted in a large
region of the parameter space.

\begin{table*}
	\centering
		\begin{tabular}{llll}
		% \hline
		\toprule
		\sc Decay channel & \sc Experiment & \sc Branching ratio limit & \sc Reference \\
		% \hline \hline
		\midrule
			$\mu \rightarrow e\gamma$ & \emph{MEGA} & $\num{< 1.2e-11}$ & \cite{Brooks:1999pu} \\
			                          & \emph{MEG}  & $\num{< 2.4e-12}$ & \cite{Adam:2011ch} \\
			% \hline                          
	    $\mu \;\rightarrow eee$     & \emph{SINDRUM} & $\num{< 1.0e-12}$ & \cite{Bellgardt:1987du}\\
			% \hline                          
	    $\mu \; Au \rightarrow e Au$     & \emph{SINDRUM II} & $\num{< 7e-13}$ & \cite{Bertl:2006up}\\
		% \hline
		\bottomrule
		\end{tabular}
	\caption{Experimental limits on LFV muon decays}
	\label{tab:LimitsOnLFVMuonDecays}
\end{table*}

Seesaw and Left-Right symmetric (supersymmetric) models are good candidates 
for realising grand unification, which also unify 
quark and lepton mass matrices. 
Moreover, it has been shown that LFV effects
in the low energy limit can be related to mixing parameters at the GUT scale
or to heavy Majorana masses in these models
\cite{Calibbi:2006nq,Antusch:2006vw}.
Seesaw models are therefore very attractive in the context of LFV as they are also able to naturally explain
the smallness of the masses of the left handed neutrinos.
In this context the recent results from neutrino oscillation
experiments are very interesting, as they measured a large mixing angle
$\theta_{13}$, which enhances the LFV-muon decays in most models which try to
explain the small neutrino masses and the large mixing.

Currently the most accurate measurement is provided by the \emph{Daya Bay}
reactor neutrino experiment \cite{DayaBay2}
yielding $\sin^2({2 \theta_{13}})=0.089 \pm 0.010 \textrm{(stat)} \pm 0.005 \textrm{(syst)}$,
excluding the no-oscillation hypothesis at $7.7$ standard deviantions. The Daya Bay 
measurement is in good agreement with measurements by the \emph{RENO} \cite{Ahn:2012nd}, 
 \emph{Double Chooz} \cite{Abe:2012tg} and \emph{T2K} \cite{Abe:2011sj} experiments.
These results are very encouraging, as large values of $\sin^2{(2 \theta_{13})}$
lead to large LFV effects in many BSM models.

The observation of LFV in the charged lepton sector would be a sign for new physics, possibly at scales far beyond the reach of direct observation at the large hadron collider (LHC).
Several experiments have been performed or are in operation to search for LFV in
the decays of muons or taus. Most prominent are the search for the radiative
muon decay \meg \cite{Brooks:1999pu,Nicolo:2003at,Adam:2009ci,Adam:2011ch},
the decay \mte \cite{Bellgardt:1987du}, the conversion of captured muons to
electrons \cite{Bertl:2006up} and LFV tau decays
\cite{Hayasaka2011,Hayasaka:2007vc,Aubert:2009ag,Lees:2010ez,Aubert:2009tk,
Aubert:2009my,Aubert:2009ys,Aubert:2007kx,Aubert:2006cz,Hayasaka:2010np,Miyazaki:2010qb,
Miyazaki:2009wc,Miyazaki:2008mw,Miyazaki:2007zw,Nishio:2008zx,Miyazaki:2007jp,Miyazaki:2006sx,Aushev:2010bq,Abe:2010sj}.

\balance

The recent search performed by the \emph{MEG}-Collaboration yields
 ${\textrm{B}}(\mu \rightarrow e \gamma)\num{ < 2.4e-12}$ \cite{Adam:2011ch}
and sets currently the most stringent limit on many LFV models. The \emph{MEG} collaboration
plans to continue operation into 2013 in order to increase the
number of stopped muons and to reach a sensitivity of a few times
$\num{e-13}$. Plans to upgrade the experiment to further improve the sensitivity are currently under discussion.

In the near future the \emph{DeeMe} experiment 
\cite{Aoki:2012zza} 
at J-PARC plans to improve the
current muon-to-electron conversion limit of ${\textrm{B}}(\mu \; Au \rightarrow
e \; Au)\num{<7e-13}$ \cite{Bertl:2006up}
by almost two orders of magnitude. By the end of the decade this limit could be
improved by even four orders of magnitude by \emph{COMET} at J-PARC \cite{Akhmetshin2012} and \emph{Mu2e} at Fermilab \cite{Bartoszek2012,Brown:2012zzd}.

Selected limits for lepton flavour violating muon decays and muon-to-electron
conversion experiments, which are of high relevance for the proposed
experiment, are shown in Table \ref{tab:LimitsOnLFVMuonDecays}.
A search for the LFV decay \mte with an unprecedented  sensitivity of
$\num{<e-16}$ as proposed here would provide a unique opportunity for discoveries
of physics beyond the SM in the coming years.

%come from the MEGA ($BR(\mu \rightarrow e\gamma) < 1.2 \cdot 10^{-11}$, \cite{Brooks:1999pu}) and MEG (Published: $BR(\mu \rightarrow e\gamma) < 2.8 \cdot 10^{-11}$ \cite{Adam:2009ci}, recent preliminary result: $BR(\mu \rightarrow e\gamma) < 1.5 \cdot 10^{-11}$ \cite{Signorelli2011}) as well as the SINDRUM ($BR(\mu \rightarrow eee) < 1.0 \cdot 10^{-12}$ \cite{Bellgardt:1987du}) experiments. 

\chapter{Theory}
\label{sec:Theory}

\nobalance

\begin{figure}[b!]
	\centering
		\includegraphics[width=0.49\textwidth]{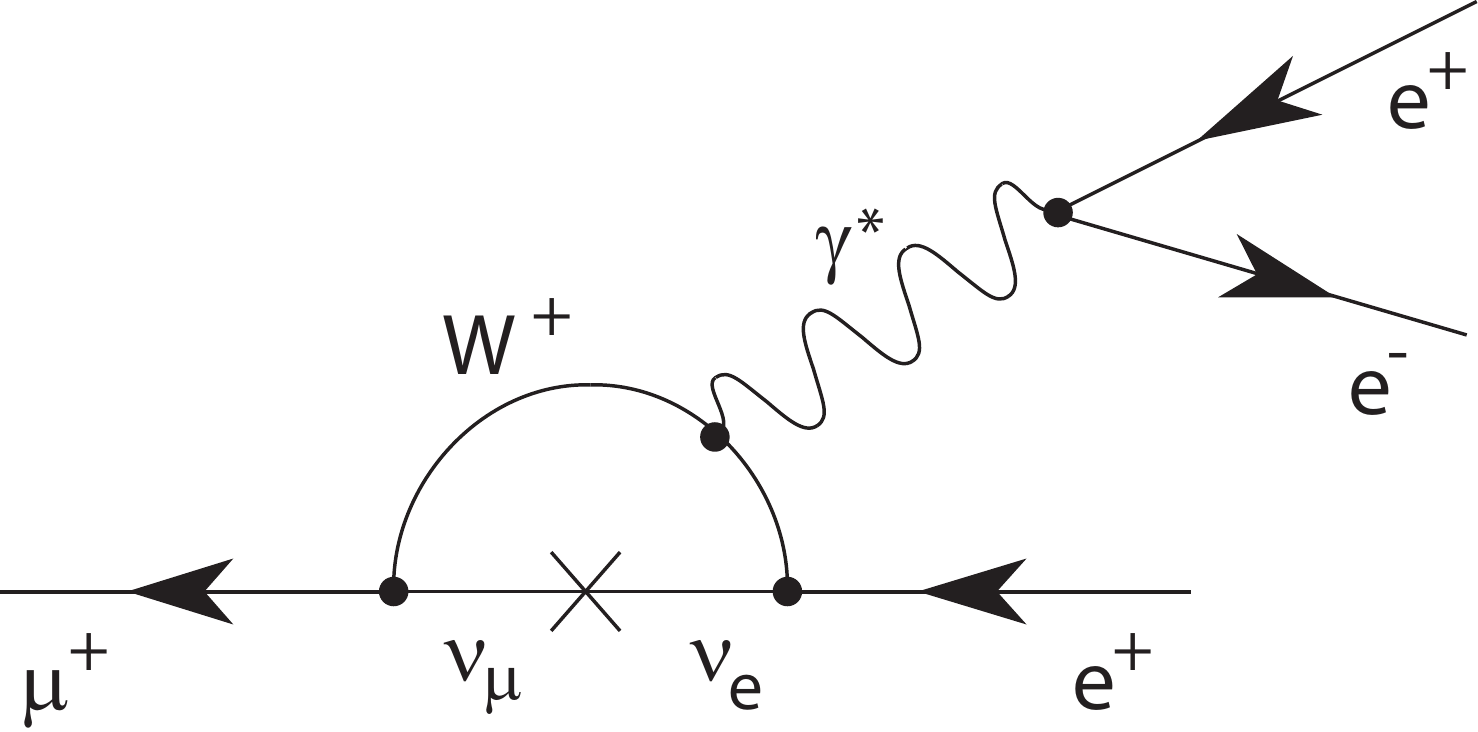} %\hspace{0.3cm}
	\caption{Feynman diagram for the \mte process via neutrino mixing
          (indicated by the cross). }
	\label{fig:m3e_feyn1}
\end{figure}

\begin{figure}[t!]
	\centering
		\includegraphics[width=0.49\textwidth]{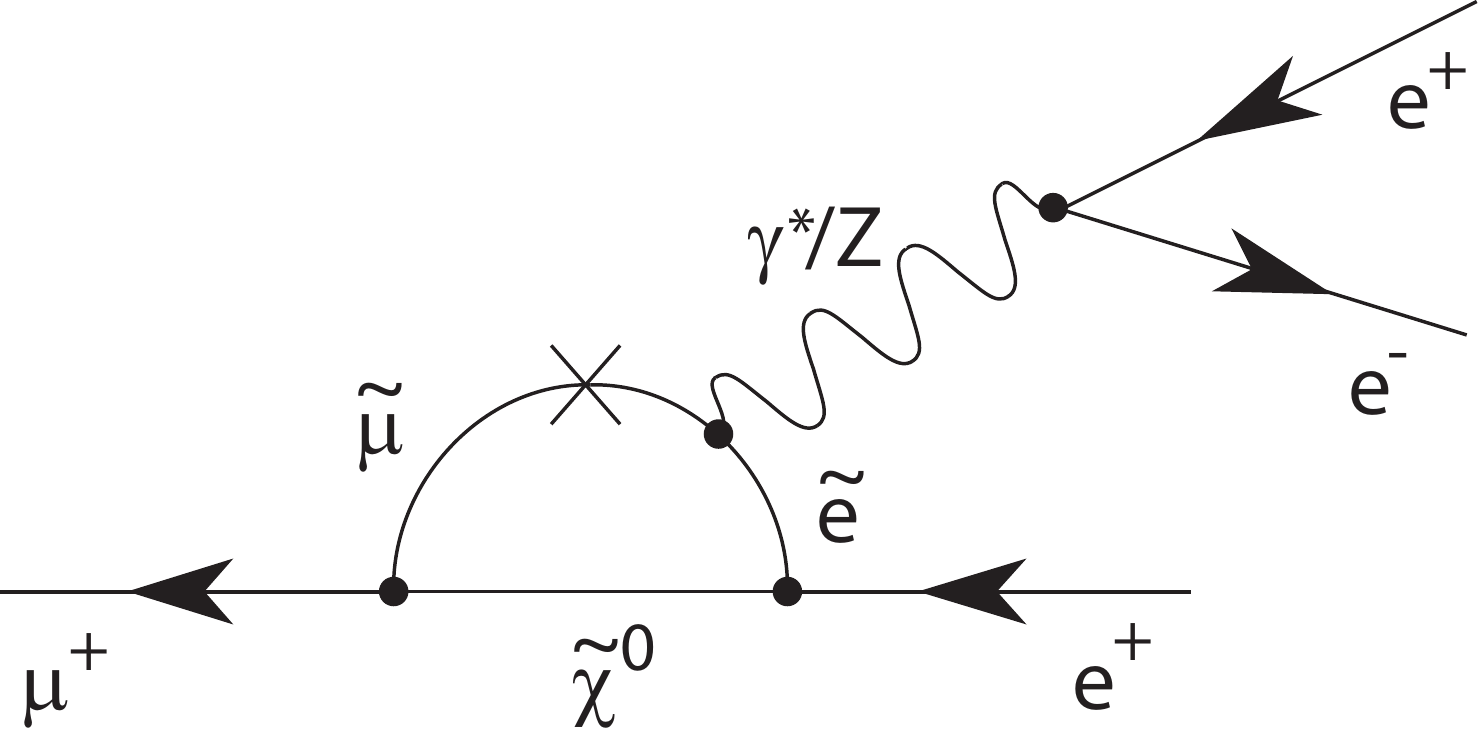}%\hspace{0.3cm}
\caption{Diagram for lepton flavour violation involving supersymmetric particles.}
	\label{fig:m3e_feyn2}
\end{figure}

\begin{figure}[b!]
	\centering
		\includegraphics[width=0.49\textwidth]{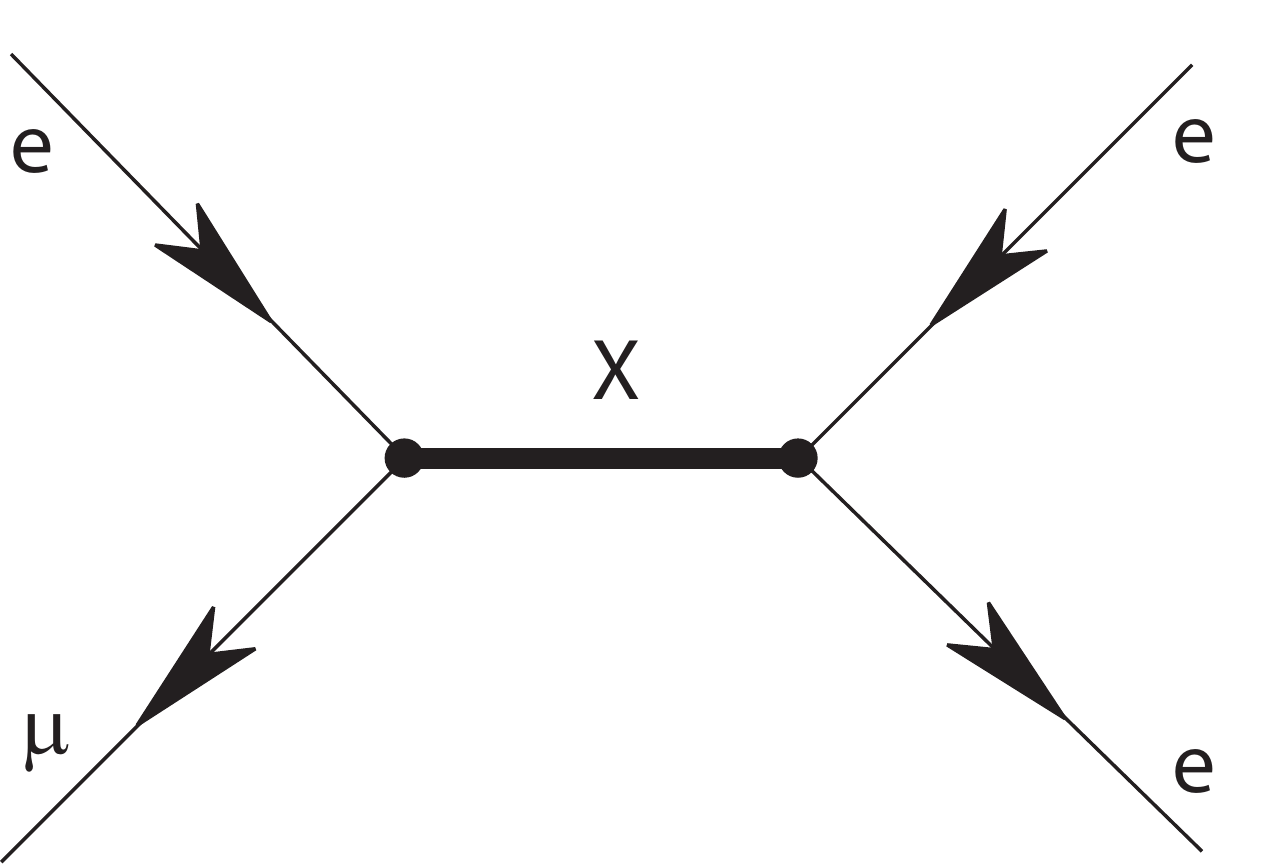}
	\caption{Diagram for lepton flavour violation at tree level.}
	\label{fig:m3e_feyn3}
\end{figure}

In the SM, charged lepton flavour violating reactions are forbidden at tree
level and can only be induced by lepton mixing through higher order loop
diagrams. 
However, the dominant neutrino mixing loop diagram, see Figure~\ref{fig:m3e_feyn1}, is
strongly suppressed in the SM with  ${\textrm{B}}\;\num{ << e-50}$  
and thus giving potentially high sensitivity to LFV reactions in models beyond the
Standard Model (BSM). 

Such an example is shown in
Figure~\ref{fig:m3e_feyn2}, where a $\gamma$/$Z$-penguin diagram is shown
with new supersymmetric (SUSY) particles running in a loop.
These loop contributions are important basically for all models, 
where new particle couplings to electrons and muons are introduced.
Lepton flavor violation can also be mediated by tree couplings
as shown in Figure~\ref{fig:m3e_feyn3}. These couplings could be mediated by
new particles, like Higgs particles or doubly charged Higgs particles, R-parity
violating scalar neutrinos or new heavy vector bosons,
the latter being motivated by models with extra dimensions~\cite{Randall:1999ee,ArkaniHamed:1999dc}. 
%In contrast to the purely leptonic LFV processes, which are best tested in muon or tau decays, 
These models usually also predict semihadronic decays of tau leptons or the muon conversion process $\mu q \rightarrow e q^\prime$, which is experimentally best tested in muon capture experiments.

The lepton flavor violating three electron decay of the muon can be mediated,
depending on the model, via virtual loop (Figure~\ref{fig:m3e_feyn2}) and box
diagrams or via tree diagrams (Figure~\ref{fig:m3e_feyn3}). 
The most general Lagrangian for this decay can be parameterized as \cite{Kuno:1999jp}
\footnote{A representation of Lagrangian containing explicitly the contributions
from the loop and box diagrams can be found in \cite{Hisano:1995cp}.}:
\begin{equation} \label{eq:lagrangian}
\begin{split}
L_{\mu \rightarrow eee} \;  = & \; \; 
\frac{4G_F}{2}   \left[ 
  m_\mu A_R \; \overline{\mu_R} \sigma^{\mu \nu} e_L F_{\mu \nu} \right.\\
& + m_\mu A_L \; \overline{\mu_L} \sigma^{\mu \nu} e_R F_{\mu \nu}  \\
 & +  \;g_1 \; (\overline{\mu_R} e_L) \; (\overline{e_R} e_L) \\
 & + g_2 \; (\overline{\mu_L} e_R)\; (\overline{e_L}e_R)  \\
 & +  \;g_3 \;  (\overline{\mu_R} \gamma^\mu e_R) \; (\overline{e_R} \gamma_\mu e_R) \\
 & + g_4 \; (\overline{\mu_L} \gamma^\mu e_L) \; (\overline{e_L} \gamma_\mu e_L)  \\
 & + \; g_5 \;  (\overline{\mu_R} \gamma^\mu e_R) \; (\overline{e_L} \gamma_\mu e_L) \\
 & + g_6 \;\left. (\overline{\mu_L} \gamma^\mu e_L) \; (\overline{e_R} \gamma_\mu e_R) \; + \; H.c.
\; \right] 
\end{split}
\end{equation}
The form factors $A_{R,L}$ describe tensor type (dipole) couplings, mostly acquiring 
contributions from the photon penguin diagram, whereas the scalar-type ($g_{1,2}$) and
vector-type ($g_{3}-g_{6}$) form factors can be regarded as four fermion contact
interactions, to which the tree diagram contributes in leading order.
In addition also off shell form factors from the penguin diagrams, which are not testable in the \meg decay 
contribute to $g_{1}-g_{6}$ \cite{Hisano:1995cp}.
In case of non-zero dipole  
and four-fermion couplings also interference effects have to be considered,
which can be exploited to investigate violation of time reversal ($T$-invariance).

%The couplings of the radiative process, see penguin diagrams in Figure~\ref{fig:m3e_feyn1} and~Figure~\ref{fig:m3e_feyn2},
%are of tensor type (dipole couplings) and are described by the $A_{R,L}$ terms.
%Using the contact interaction formalism, 
%the four fermion couplings of the tree diagram processes are described by
%scalar-type ($g_{1,2}$) and vector-type ($g_{3}-g_{6}$) couplings. 
%This ansatz is motivated by the high mass of the intermediate particle
%(see Figure~\ref{fig:m3e_feyn3}). 

%Off shell photon form factors, which can not be tested in 
%$\mu \rightarrow e \gamma$ decays,
%also contribute to the terms $g_3$ - $g_6$.
%The contributions to the different LFV couplings depend on the model under
%consideration. 

By neglecting higher order terms in $m_e$, the total branching ratio of the decay can be expressed by:
\begin{equation}
\begin{split}
\textrm{B}(\mu \rightarrow eee) = &
 \frac{g_1^2+g_2^2}{8} + 2\; (g_3^2+g_4^2) \; + \; g_5^2+g_6^2 \; \\
 + &\; 32 \; e A^2 \; (\ln{\frac{m^2_\mu}{m^2_e}} - 11/4 ) \\
 + & 16 \; \eta \;e  A \; \sqrt{g_3^2+g_4^2} \\ 
 + & 8 \;\eta^\prime \; e  A \; \sqrt{g_5^2+g_5^2}  \quad ,
\end{split}
\end{equation}
where the definition $A^2=A_L^2+A_R^2$ is used.
The term proportional to $A^2$ is logarithmically enhanced and can be assigned
to the photon penguin diagram. 
The constants $\eta$ and $\eta^\prime$ are $T$-violating mixing parameters.
In case of a signal, the different terms can be measured from the angular
distribution of \mte decay particles using a polarized muon beam.

\section{Comparison \mte versus \meg}
In the decay \meg physics beyond the SM is only tested by 
photon penguin diagrams, in contrast to \mteee where also tree, $Z$-penguin and
box diagrams contribute.
To compare the new physics mass scale reach between the processes \mteee and
\meg a simplified model is chosen; it is assumed that the photon penguin
diagram Figure~\ref{fig:m3e_feyn2} and the tree diagram
Figure~\ref{fig:m3e_feyn3} are the only relevant contributions. The 
Lagrangian then simplifies to\footnote{A similar study was presented in \cite{Gouvea2009}}:
\begin{equation}
\begin{split}
L_{LFV} \;  = \; &  {\left[ \frac{m_\mu}{(\kappa+1) \Lambda^2} \ \overline{\mu_R}
\sigma^{\mu \nu} e_L F_{\mu \nu} \right]}_{\gamma-\textrm{penguin}}  \\ 
& + {\left[
\frac{\kappa}{(\kappa+1) \Lambda^2} \ (\overline{\mu_L} \gamma^\mu e_L) \;
(\overline{e_L} \gamma_\mu e_L) \right]}_{\textrm{tree}}   \label{eq:kappa}
\end{split}
\end{equation}
where for the contact interaction (``tree'') term  exemplarily a left-left
vector coupling is chosen.
In this definition a common mass scale $\Lambda$ is introduced and the
parameter $\kappa$ describes the ratio of the amplitudes
of the vector-type (tree) term  over the tensor ($\gamma-\textrm{penguin}$) term.
%, which is the only relevant one for the \meg process. 

\begin{figure}[tb!]
	\centering
		\includegraphics[width=0.49\textwidth]{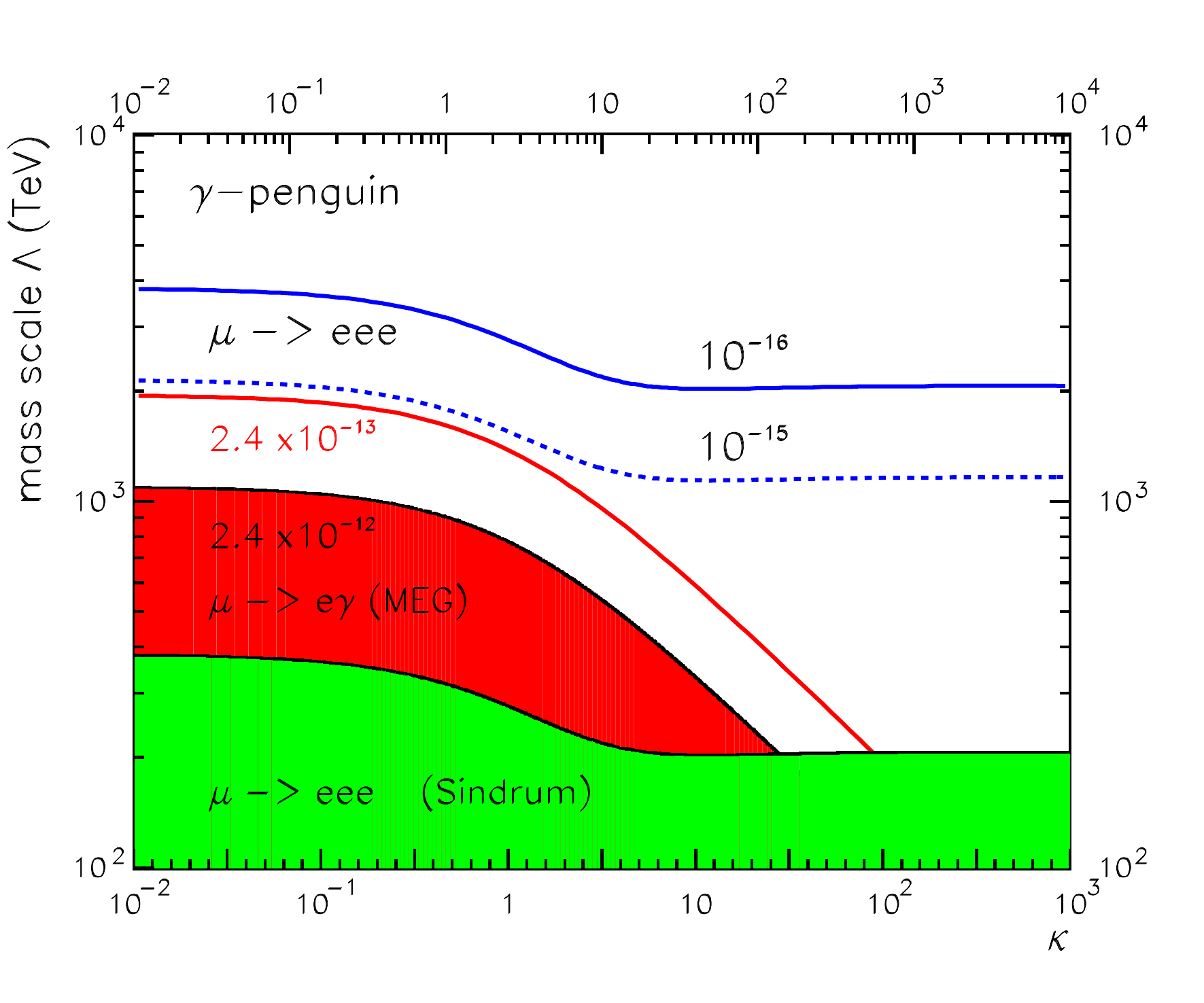}
	\caption{Experimental limits and projected limits on the LFV mass scale $\Lambda$ as a function of the parameter $\kappa$ (see equation \ref{eq:kappa}) assuming negligible contributions from $Z^0$ penguins; based on \cite{Gouvea2009}.}
	\label{fig:mu3ekappa}
\end{figure}

Limits on the common mass scale $\Lambda$ as obtained from the experimental bounds on 
$\textrm{B}(\mu \rightarrow e \gamma)\num{<2.4e-12}$ ($\SI{90}{\percent}$ CL \emph{MEG} 2011)
and $\textrm{B}(\mu \rightarrow eee)\num{<1.0e-12}$ ($\SI{90}{\percent}$ CL \emph{SINDRUM})
are shown in Figure~\ref{fig:mu3ekappa} as function of the parameter
$\kappa$. 
Experimentally, for small values of $\kappa$ (dipole coupling) the mass scale $\Lambda$ is best constrained by
the \emph{MEG} experiment whereas the four fermion contact
interaction region with $\kappa \gtrsim 10$ is
best constrained by the \emph{SINDRUM} experiment.

%The process $\mu \rightarrow e \gamma$ constrains the mass scale at
%low values of $\kappa$ (dipole coupling) whereas for $\kappa \gtrsim 10$ the
%$\mu \rightarrow eee$ is constraining the four fermion contact interaction region. 
For comparison also a hypothetical ten times improved limit is shown for the
\emph{MEG} experiment (post-upgrade)
and compared to the sensitivities of the
proposed $\mu \rightarrow eee$ experiment of
$\num{e-15}$ (phase~I) and $\num{e-16}$ (phase~II). 
It can be seen 
that in this simple model comparison high mass scales $\Lambda$
%for $\textrm{B}(\mu \rightarrow eee) \lesssim \num{e-15}$ LFV processes
will be best constrained by the proposed $\mu \rightarrow eee$ experiment for all 
values of $\kappa$ already in phase~I.

In case of dominating tensor couplings ($A \ne 0$, $\kappa \to 0$) a quasi model independent relation between the \mte decay rate and the 
$\mu \rightarrow e \gamma$ decay rate can be derived:
\begin{equation}
\frac{\textrm{B}(\mu \rightarrow eee)}{\textrm{B}(\mu \rightarrow e \gamma)} \approx 0.006  \label{eq:ratio}
\end{equation}
This ratio applies for many supersymmetric models, where LFV effects are 
predominantly mediated by gauge bosons and where the masses of the
scalar leptons or gauginos are of electroweak scale. 
In these models, which are already heavily constrained or even excluded by the
recent LHC results, the sensitivity of the proposed   \emph{Mu3e} experiment
in terms of branching ratio has to be two orders of magnitude higher 
than that of the \emph{MEG} experiment in order to be competitive.

%In order to set competitive constraints on LFV dipole couplings, a limit on 
%the branching ratio of the decay $\mu \rightarrow eee$ needs to be about
%two orders of magnitude smaller than for the decay $\mu \rightarrow e \gamma$. 

\begin{figure}[tb!]
	\centering
		\includegraphics[width=0.49\textwidth]{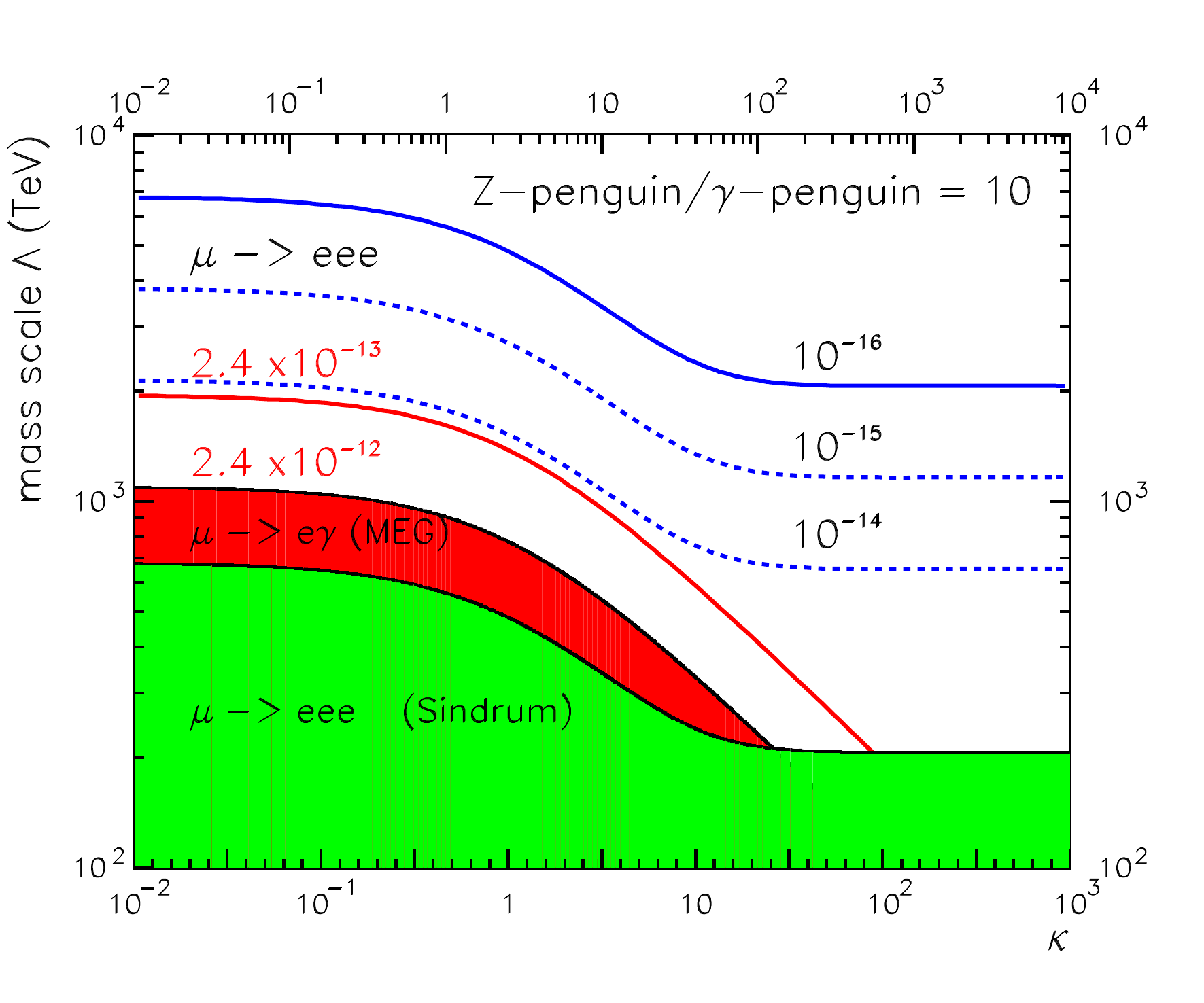}
	\caption{Experimental limits and projected limits on the LFV mass scale $\Lambda$ as a function of the parameter $\kappa$ (see equation \ref{eq:kappa}) assuming contributions from $Z^0$ penguins ten times larger than the photon contribution.}
	\label{fig:mu3ekappaZ}
\end{figure}

%%%
\subsection{$Z$-penguin Contribution}
However, besides the tree and $\gamma$-penguin diagrams also the $Z$-penguin diagram can
significantly contribute to the process \mteee. 
The $Z$-penguin diagram is of particular importance if the new physics scale
is higher than the electromagnetic scale, as can be easily derived from a
dimensional analysis.
The enhancement of the $Z$-penguin contribution over the $\gamma$-penguin
contribution and its non-decoupling behaviour when going to high mass scales
was discussed for Little Higgs models~\cite{delAguila:2008zu,delAguila:2011wk} as well as for several SUSY
models~\cite{Arganda:2005ji,Hirsch:2012ax,Dreiner:2012mx,Abada:2012cq,Hirsch:2012kv}.
SUSY models with $R$-parity violation and right handed neutrinos received
recently quite some attention in this context, as approximate cancellations of
different $Z$-diagram contributions are not present in extended Minimum
Super-Symmetric Standard Models (MSSM).

The effect of such an enhanced Z-penguin coupling, where the LFV contribution 
to the \mte process is exemplarily enhanced by a factor of ten relative to the photon-penguin
contribution, is shown in Figure~\ref{fig:mu3ekappaZ}.  
It can be seen that the sensitivity of the \mte process to new physics is
significantly enhanced at small values of $\kappa$
and that in such a case a sensitivity of \num{e-14} is already sufficient to be competitive
with the \meg process with a sensitivity of a few times \num{e-13}.

%\subsection{LFV Models}

%\begin{figure*}[tb!]
%	\centering
%              \includegraphics[width=1.00\textwidth]{Theory/figures/mu3eSUSY_GUT2_again22}
%%              \includegraphics[width=1.00\textwidth]{figures/mu3eSUSY_GUT2_nomeg2}
%%              \includegraphics[width=1.00\textwidth]{figures/mu3eSUSY_GUT2}
%%		\includegraphics[width=1.00\textwidth]{figures/masiero}
%	\caption{Radiative branching ratios $\mu \rightarrow e \gamma$ vs. 
%$\tau \rightarrow \mu \gamma$
%from a SUSY-GUT model parameter scan for 
%$\tan{\beta=40}$ 
%\protect{\cite{Calibbi:2006nq}}. 
%The different colors represent different flavor mass matrices at the GUT
%scale. The corresponding  $\mu \rightarrow eee$ branching ratio
%is given by $0.006 \times \textrm{B}(\mu \rightarrow e \gamma)$. 
%The envisaged sensitivity of $10^{-16}$ for B$(\mu \rightarrow eee$) 
%corresponds here to $\sim \num{1.5e-14}$ for B$(\mu \rightarrow e\gamma$).
%In addition, sensitivities of current and future B-factories for the complementary
%LFV tau decay $\tau \rightarrow \mu \gamma$ are shown.
%}
%	\label{fig:masiero}
%\end{figure*}

\section{Discussion of Specific Models}
%New heavy particles as predicted by Higgs Triplet
%models, models with additional $\textrm{U}'(1)$ gauge symmetries or models with extra dimensions can also induce sizable LFV effects.
In the following, selected models are discussed in more detail in the context of the proposed experiment.

\subsection{Inverse Seesaw SUSY Model}

\begin{figure}
	\centering
       \includegraphics[width=0.48\textwidth]{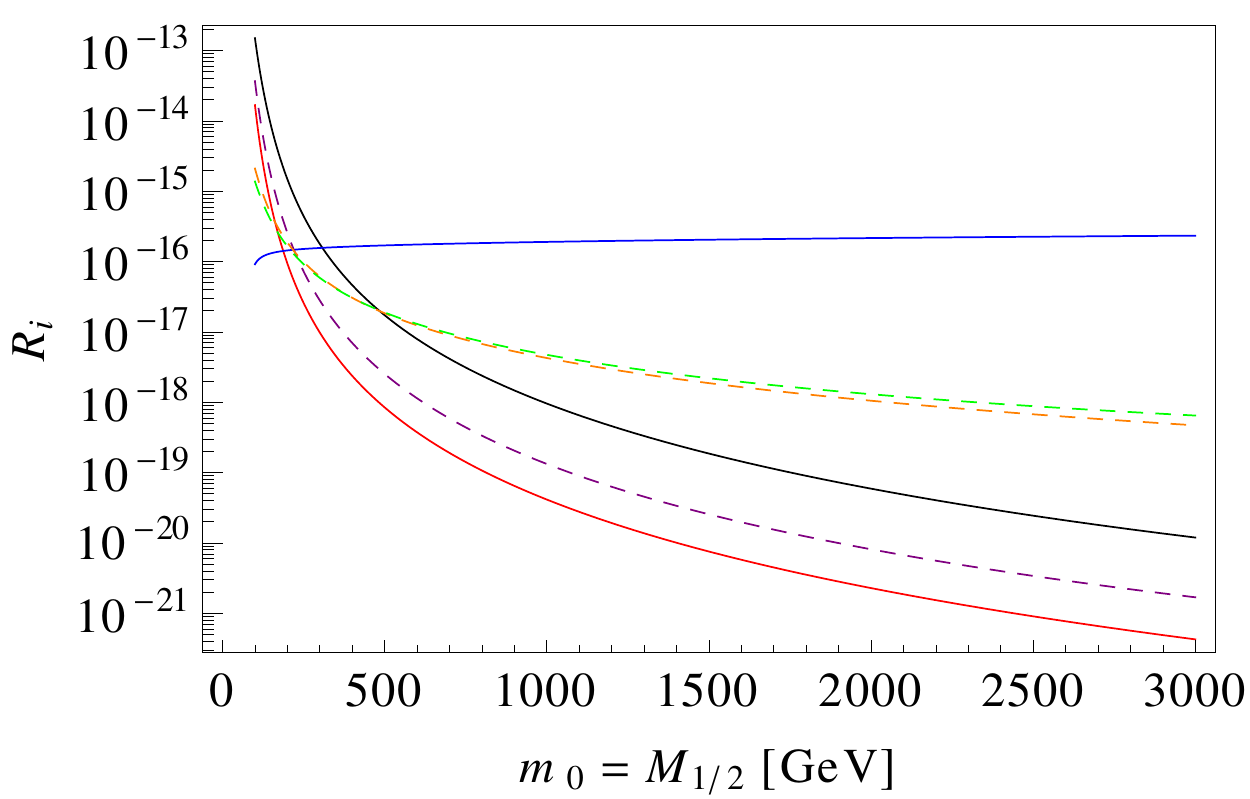}
	\caption{Inverse Seesaw SUSY Model: Contributions to $B(\mu \rightarrow eee)$ as a function of
          $m_0 = M_ {1/2}$ for a degenerate singlet spectrum with $\hat{M}_R = \SI{10}{TeV}$ and $M = \SI{e11}{GeV}$. The rest of the cMSSM parameters are set to $A_0 = \SI{-300}{GeV}$, $B_0 = 0$,
$\tan \beta = 10$ and sign($\mu$) $= +$. Solid lines represent individual contributions,
$\gamma$ (black), $Z$ (blue) and $h$ (red) whereas the dashed lines represent interference terms, $\gamma - Z$ (green), $\gamma -h$ (purple) and $Z -h$ (orange). Note that in this case $h$ includes both Higgs and
box contributions. Taken from \protect{\cite{Abada:2012cq}}.}
	\label{fig:abada}
\end{figure}

Despite the fact that the most simple supersymmetric models with light squarks
and gluinos were
recently excluded by LHC experiments
\cite{ATLAS:2012ds, ATLAS:2012dp,ATLAS:2012kr,ATLAS:2012jp,ATLAS:2012ht,ATLAS:2012uu,ATLAS:2012yr,ATLAS:2012ae,ATLAS:2012ms, ATLAS:2012tx,ATLAS:2012ku,ATLAS:2012gg,ATLAS:2012ar,ATLAS:2012rz,Aad:2012pq,Aad:2012hm, CMS:2012un, CMS:2012ew,Chatrchyan:2012jx,CMS:2012mfa,Chatrchyan:2012te,CMS:2012th,Chatrchyan:2012sa,Chatrchyan:2012mea,Chatrchyan:2012qka}
% the following is old... 
%ATLAS:2011cwa,ATLAS:2011ad,ATLAS:2011ib,CMS:2011zy}, 
supersymmetry can still
exist in nature, just at higher mass scales or in more complex realisations.
In many of these realisations with a non-minimal particle content 
the $Z$-penguin contribution discussed above gets significantly enhanced.

As a first example results obtained by a supersymmetric model with an inverse seesaw 
mechanism \cite{Abada:2012cq} are discussed here.
The inverse seesaw model \cite{Mohapatra:1986bd} constitutes a very appealing
alternative to the standard seesaw realization and can be embedded in a minimal
extension of the MSSM by the addition 
of two extra gauge singlet superfields, with opposite lepton numbers.
Similar to other models, e.g.~flavour violating Higgs decays in the MSSM,
the $Z$-penguin exhibits here a non-decoupling behaviour, which is shown in
Figure~\ref{fig:abada} for an effective right-handed neutrino mass of $M =
\SI{e11}{GeV}$ and degenerate sterile neutrino masses of $\hat{M}_R = \SI{10}{TeV}$.
At small mass scales $m_0=M_{1/2}$ of the constrained MSSM (cMSSM) the
photon-mediated penguin contribution clearly dominates over the other
contributions from Higgs-mediated penguin and Z-mediated penguin diagrams.
This picture completely changes at higher mass scales above $200$-$300$~GeV,
where the $Z$-mediated penguin diagram becomes dominant. 
The non-decoupling behaviour of the $Z$-penguin is clearly
visible which will allow to test this model at any SUSY mass scale for the
seesaw parameters given in this example at phase~II of the proposed experiment.

%Lepton Flavor Violation in the framework of SUSY-GUT models was
%discussed in  \cite{Calibbi:2006nq} for two different 
%realisations of the flavor matrix at the GUT scale.
%The ``CKM-type'' parameterisation has CKM-like fermion mixing angles
%at the GUT scale whereas
%the ``PMNS-type'' parameterisation has neutrino like mixing angles 
%(bi-maximal mixing) with $\theta_{13}$ being a free parameter.
%In Figure~\ref{fig:masiero} the branching ratios for the LFV radiative
%decays of the tau and the muon is shown for a Higgs Doublet expectation
%value ratio of $\tan{\beta}=40$.
%It can be seen that the PMNS scenario can be almost completely
%ruled out 
%independent of the value of $\theta_{13}$ for $\textrm{B}(\mu \rightarrow e
%\gamma)\num{<e-14}$, 
%which corresponds to the aimed sensitivity of the 
%$\mu \rightarrow eee$ search with  $\textrm{B}(\mu \rightarrow eee) \num{<e-16}$
%using equation~\ref{eq:ratio}. 
%The figure also shows the complementarity to LFV searches using taus in 
%the process $\tau \rightarrow \mu \gamma$. All parameter sets which might be
%tested in future B-factories might be completely ruled out by the proposed 
%experiment.

\subsection{Supersymmetric $SU(3)_c \times SU(2)_L \times U(1)_{B-L} \times
  U(1)_R$ Model}

\begin{figure}
	\centering
   \includegraphics[width=0.48\textwidth]{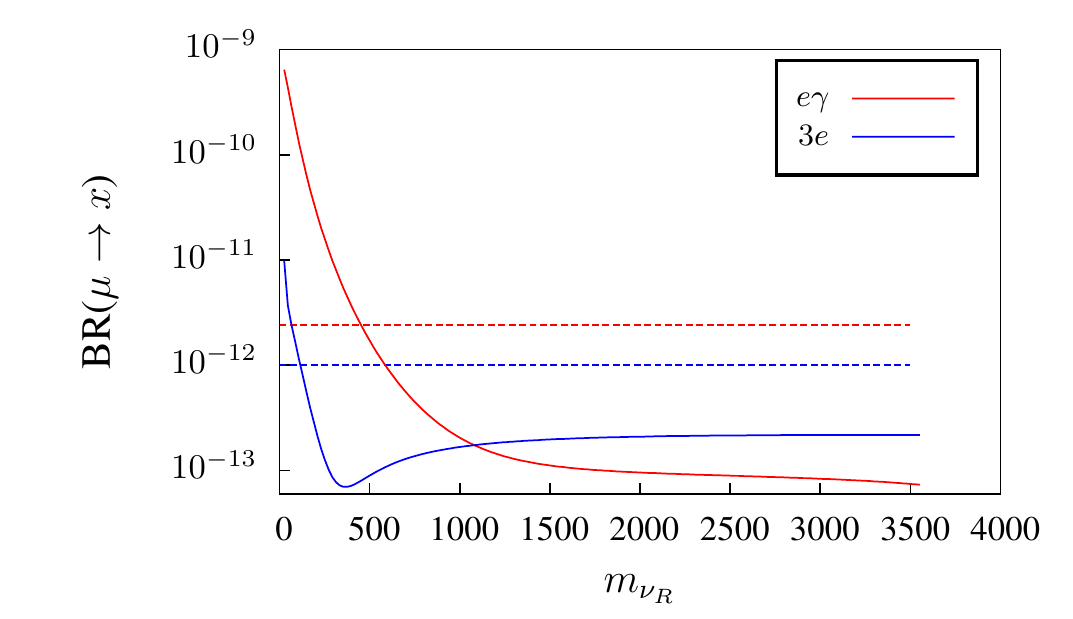}
	\caption{Supersymmetric $SU(3)_c \times SU(2)_L \times U(1)_{B-L} \times
  U(1)_R$ Model: Branching ratios of lepton flavour violating processes as a function of $m_{\nu R}$ for $m_0 =
\SI{800}{GeV}$, $M_{1/2} = \SI{1200}{GeV}$, $\tan \beta = 10$, $A_0 = 0$, $v_R = $\SI{10}{TeV}, $\tan \beta_R = 1.05$, $\mu_R = \SI{-500}{GeV}$,
$m_{A_R} = \SI{1000}{GeV}$. The dashed red line is the predicted branching ratio for \meg
 and the dashed blue line for \mte. Taken from \protect{\cite{Hirsch:2012kv}}.}
	\label{fig:hirsch}
\end{figure}
This model represents a supersymmetric version of the SM, minimally extended by 
additional $U(1)_{B-L}  \times U(1)_R$ symmetry groups~\cite{Malinsky:2005bi,DeRomeri:2011ie}. 
This model includes the generation of light neutrino masses by the seesaw
mechanism, can explain the
observed large neutrino mixing angles and can be easily
embedded into a $SO(10)$ grand unified theory. 
This model predicts an additional light Higgs
particle, which is expected to mix with the lightest MSSM Higgs particle,
and has been recently studied also in the context
of LFV processes \cite{Hirsch:2012kv}. 
Also in this study it is found that at high SUSY mass scales
the photon-mediated LFV penguin diagrams are
more suppressed than the $Z$-mediated LFV penguin diagrams 
and that this suppression scales with $m_Z^4/m_{SUSY}^4$ as naively expected
from a dimensional analysis. 
Branching ratio predictions for the processes \mte and \meg are shown in Figure~\ref{fig:hirsch}
as function of the right-handed neutrino mass $m_{{\nu}_R}$ for the SUSY
model parameters as given in the figure caption.
Also here it can be seen that for high masses $m_{{\nu}_R}>300$~GeV the
$Z$-mediated penguin diagram starts to contribute dominantly to the \mte process
and that for $m_{\nu_R}>1000$~GeV the \mte process is expected to have an even higher
branching fraction than \meg. 
For even higher masses the non-decoupling behaviour is visible in the \mte prediction.

\begin{figure*}[tb!]
	\centering
		\includegraphics[width=0.49\textwidth]{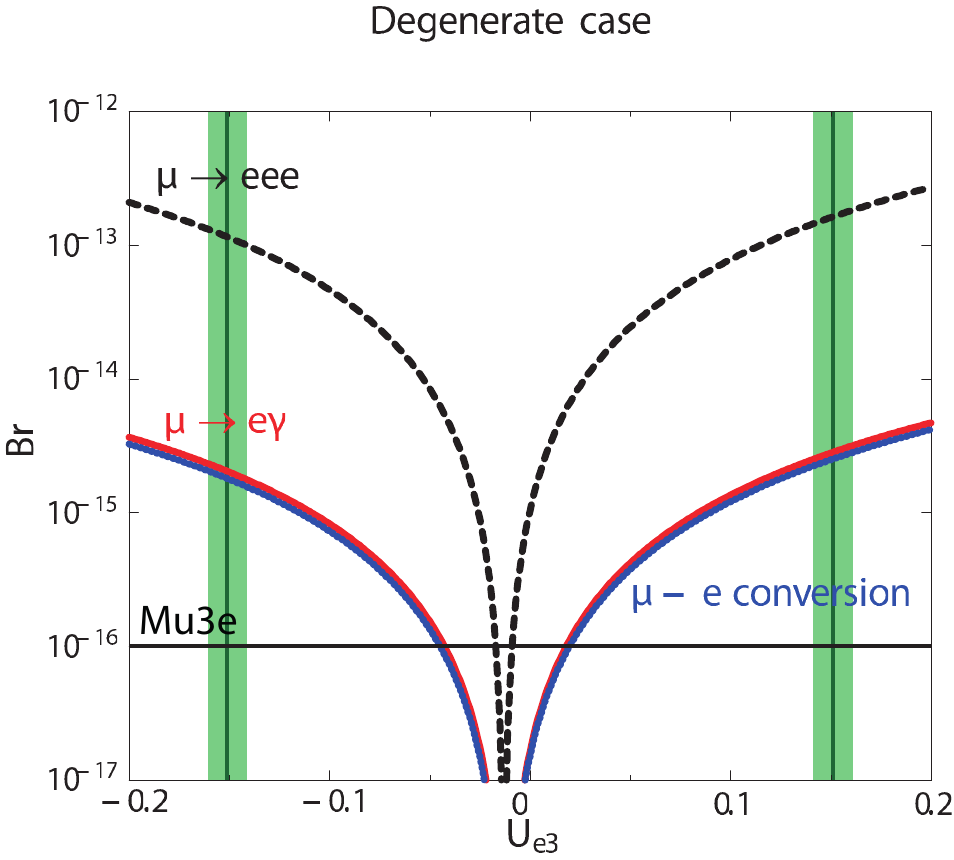}
		\includegraphics[width=0.49\textwidth]{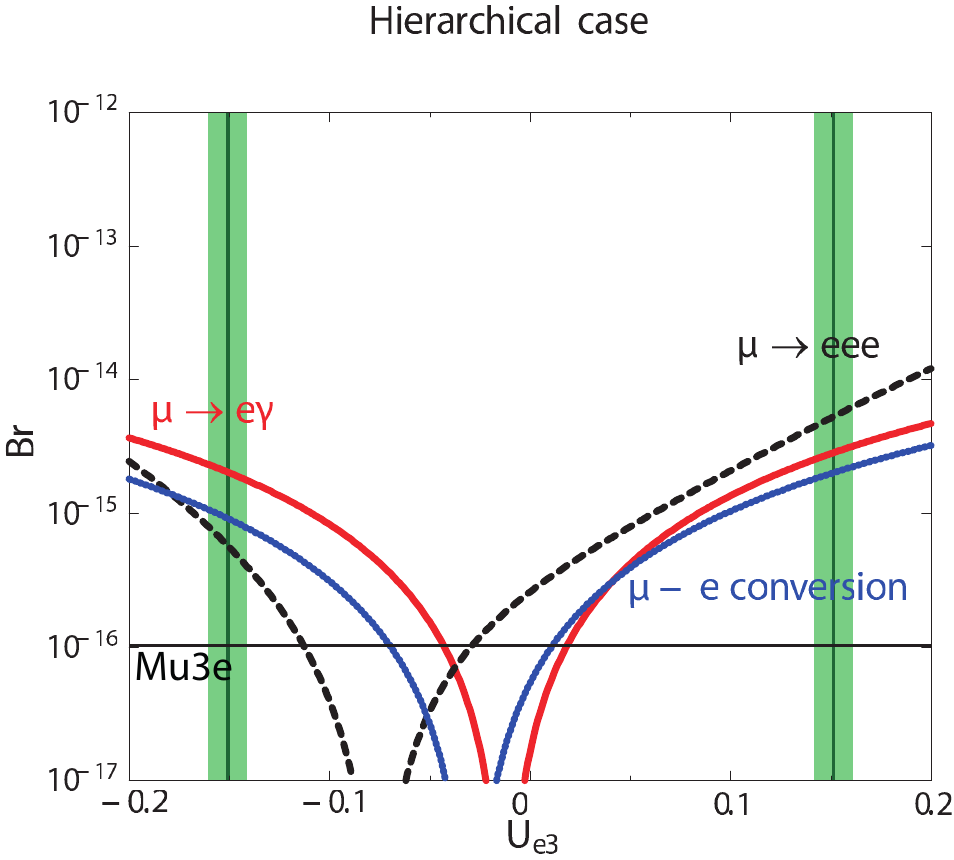}
		\includegraphics[width=0.49\textwidth]{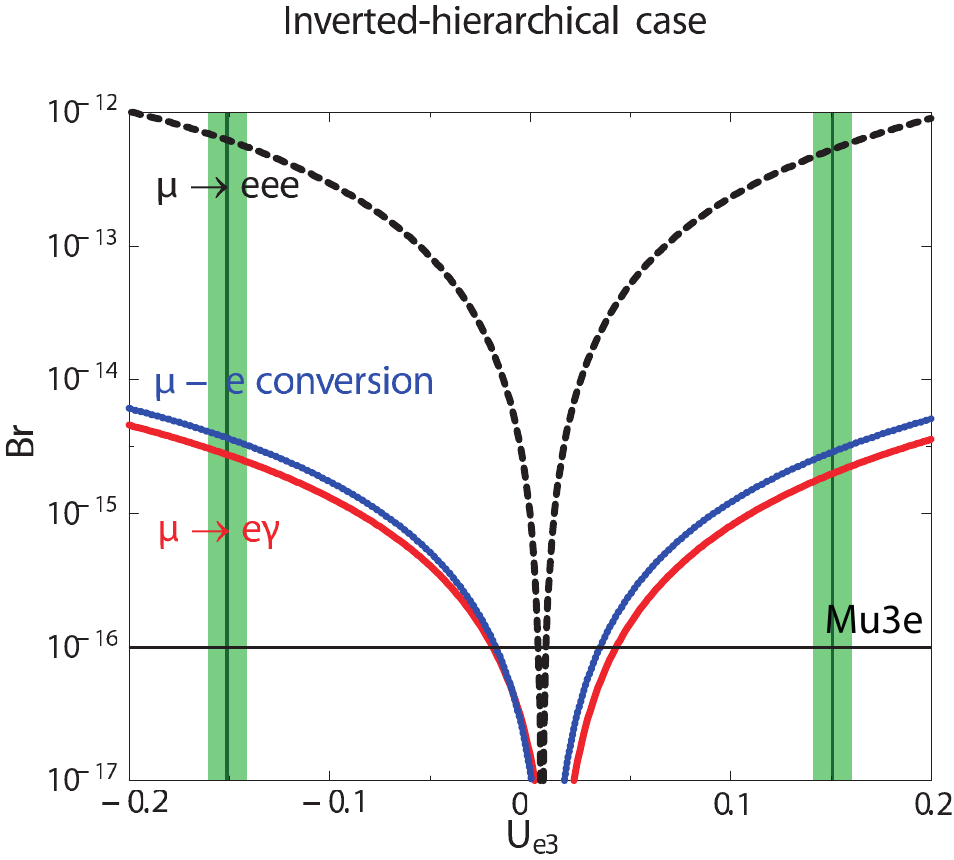}	
	\caption{Branching ratios of $\textrm{B}(\mu \rightarrow eee)$, 
$\textrm{B}(\mu \rightarrow e \gamma)$ and muon conversion $\textrm{B}(\mu \rightarrow e)$
in different Higgs triplet scenarios with a degenerated, hierarchical or
inverted mass hierarchy of the neutrinos as function of the neutrino
mixing matrix element $U_{e3}$ and for the model parameters: $M = 200$~GeV, $A =
25$~eV, and $m_\nu= 0.1$~eV for the degenerate case.
These plots were taken from \protect{\cite{Kakizaki:2003jk}} and the $U_{e3}$
constraints (green bands) as obtained from  \protect{\cite{DayaBay2}} were added posterior. 
}	\label{fig:kakizaki}
\end{figure*}

\subsection{Other Models}
The above discussed enhancement of the $Z$-mediated penguin diagram appears also in Little
Higgs Models with T-parity (LHT) where ratios 
$\textrm{B}(\mu \rightarrow eee)/\textrm{B}(\mu \rightarrow e \gamma) \approx
0.02-1$ have been predicted \cite{delAguila:2008zu,delAguila:2011wk,Blanke:2007db,Blanke:2009am},
or in Left-Right Symmetric models with additional Higgs triplets.
LFV interactions in Higgs-triplet models can be also generated directly in
tree diagrams, see Figure~\ref{fig:m3e_feyn3}.

In \cite{Kakizaki:2003jk}, these LFV violating effects are studied in a model
where the Higgs triplet is responsible for neutrino mass generation.
Figure~\ref{fig:kakizaki} shows the predicted branching ratios
for each of the three LFV muon processes and for different realisations
of the neutrino mass hierarchy. 
Note that the absolute value of the branching ratios depends on the mass scale
$M$ and can vary. 
For the hierarchical case, Figure~\ref{fig:kakizaki}~b), all branching ratios are expected to be of similar
size whereas for the degenerate, Figure~\ref{fig:kakizaki}~a), and the inverted case,
Figure~\ref{fig:kakizaki}~c), the $\mu \rightarrow eee$ branching ratio
dominates in the allowed region of $U_{e3}$.
As the LFV-mediating Higgs triplet boson does not couple to quarks, the $\mu
\rightarrow eee$ decay is enhanced compared to the $\mu \rightarrow e \gamma$
decay and the muon-to-electron conversion processes, which are both loop
suppressed.

This enhancement of the LFV tree diagram 
is also found to be large in extra dimension models  
\cite{Chang:2005ag,Randall:1999ee} or models
with new heavy $Z$ bosons.  
In Randall-Sundrum (RS) models \cite{Randall:1999ee}, flavor changing neutral currents (FCNCs) arise
already at the tree level. This is caused by the flavor-dependent couplings of
these gauge bosons, due to their non-trivial profiles in the extra dimension.
Moreover, FCNCs arise through the exchange of the Higgs boson, as due to the
contribution to the fermion masses from compactification,
there is a misalignment between the masses and the Yukawa couplings.

\balance

Electroweak precision observables suggest that for RS models 
featuring the Standard Model gauge group, the new-physics mass scale $M_{\rm KK}$ 
(the scale of the Kaluza Klein excitations) should not be lower than $\ord(\SI{10}{\tera\electronvolt})$ 
at $\SI{99}{\percent}$ CL \cite{Casagrande:2008hr,Carena:2003fx,Delgado:2007ne}.
Thus, without additional structure/symmetries, the experimental situation suggests 
that it could be challenging to find direct signals from RS models at the LHC.
In such a situation, precision experiments, like the measurement of the decay
$\mu\to eee$, will furnish the only possibility to see the impact 
of warped extra dimensions. 

\section{Theory Summary}

The search for the decay $\mu \rightarrow eee$ is in itself of
fundamental interest and might reveal surprises not foreseen
by any of the models discussed above. This search is largely
complementary to other LFV searches, in particular to the decay $\mu \rightarrow e\gamma$
and to the $\mu \rightarrow e$ conversion in muon capture experiments. 
In a wide range of models for physics beyond the standard model, 
highest sensitivity in terms of branching ratio is expected for the 
$\mu \rightarrow eee$ decay process.

\chapter{Experimental Situation}
\label{sec:ExperimentalSituation}
\nobalance

\section{\emph{SINDRUM} Experiment} \label{sec:Sindrum}

\begin{table*}
\begin{center}
\begin{tabular}{lrr}
% \hline
\toprule
\sc SINDRUM parameter & \sc Value &\\ % \hline \hline
\midrule
rel. momentum resolution $\sigma_p/p$  & $\SI{5.1}{\percent}$ & ($p=\SI{50}{\mega\electronvolt\per\c}$)\\
rel. momentum resolution $\sigma_p/p$  & $\SI{3.6}{\percent}$ & ($p=\SI{20}{\mega\electronvolt\per\c}$)\\
polar angle $\sigma_\theta$  & $\SI{28}{\milli\radian}$ & ($p=\SI{20}{\mega\electronvolt\per\c}$)\\
vertex resolution $\sigma_{dca}$ & $\approx \SI{1}{\milli\metre}$ & \\
MWPC layer radiation length in $X_0$   & \SI{0.08}{\percent} - \SI{0.17}{\percent} & \\
% \hline
\bottomrule
\end{tabular}
\end{center}
\caption{\emph{SINDRUM} tracking parameters taken from \protect{\cite{Bellgardt:1987du}}.
} \label{tab:sindrum}
\end{table*}

The \emph{SINDRUM} experiment was in operation at PSI from 1983-86 to search for
the process $\mu \rightarrow eee$. No signal was found and the limit
$\textrm{B}(\mu \rightarrow eee) \num{< e-12}$ was set at $\SI{90}{\percent}$ CL \cite{Bellgardt:1987du},
assuming a decay model with a constant matrix element.

The main components of the experiment were a hollow double-cone shaped target 
of dimension $\num{58 x 220}~\si{\milli\metre\squared}$ to stop surface muons of $\SI{28}{\mega\electronvolt\per\c}$ in a solenoidal magnetic field of $\SI{0.33}{\tesla}$,
five layers of multiwire proportional chambers and a trigger hodoscope.
The main tracking parameters which were most relevant for the
search sensitivity of the experiment are shown in Table~\ref{tab:sindrum}.

The time resolution obtained by the hodoscope of less than $\SI{1}{\nano\second}$ was, 
together with the achieved momentum resolution,
sufficient to suppress the accidental background completely.

After all selection cuts, no candidate event was seen by the \emph{SINDRUM} 
experiment. The sensitivity of the experiment was mainly determined
by the $\mu \rightarrow eee \nu \nu$ background process and
estimated as $\num{5e-14}$ \cite{bertl2008}; the obtained
limit was basically given by the limited number of muon stops.

\section{\emph{MEG} Experiment} \label{sec:MEG}

The \emph{MEG} experiment at PSI is in operation since 2008 
and is searching for the LFV decay
$\mu \rightarrow e \gamma$. 
The main components used for event reconstruction 
are drift chambers for positron detection and a liquid xenon 
calorimeter for photon detection.

In the
first running period in the year 2008 about $\num{e14}$ muons were stopped on target 
\cite{Adam:2009ci}. 
No signal was found and a limit on the decay of
$\textrm{B}(\mu \rightarrow e \gamma) \num{< 2.8e-11}$ ($\SI{90}{\percent}$ C.L.) was set.

After upgrading the detector the search sensitivity and the limit was
improved using data taken in the years 2009/2010 to 
$\textrm{B}(\mu \rightarrow e \gamma) \num{< 2.4e-12}$ ($\SI{90}{\percent}$ C.L.) \cite{Adam:2011ch}.
%\cite{Sawada:2010zz} compared to an expected
%sensitivity of $6.1 \times 10^{-12}$ (90\% C.L.). 
%Five events have been observed in the signal region whereas three events were
%expected.

The dominant background contribution for $\mu \rightarrow e \gamma$ comes from accidentals where 
a high energy photon from a radiative muon decay or from a
brems\-strahlung process 
is recorded, overlayed with a positron from the upper edge of the Michel spectrum. 
This accidental background mainly determines the final sensitivity of the
experiment. 

The amount of background is predominantly determined by the timing, 
tracking and energy resolution. Selected resolution parameters as
achieved in the 2009 run are summarized in Table~\ref{tab:meg}.
The \emph{MEG} experiment will continue operation until middle of 2013. The 
final sensitivity is expected to be a few times  $\num{e-13}$. The collaboration has started to discuss possible upgrades to further improve 
the sensitivity by about one order of magnitude.
These numbers are to be compared to the bound from the earlier \emph{MEGA} experiment
of
$\textrm{B}(\mu \rightarrow e \gamma) \num{< 1.2e-11}$ \cite{Brooks:1999pu}.

\begin{table}[b!]
\begin{center}
\begin{tabular}{lr}
% \hline
\toprule
\sc \emph{MEG} parameter 2011 publ.& \sc Value \\ % \hline \hline
\midrule
rel. momentum resolution $\sigma_p/p$  & $\SI{0.7}{\percent}$ (core) \\
polar angle $\sigma_\theta$  &$\SI{9}{\milli\radian}$ \\
azimuthal angle $\sigma_\phi$  & $\SI{7}{\milli\radian}$ \\
radial vertex resolution $\sigma_{R}$ & $\SI{1.1}{\milli\metre}$  \\
long. vertex resolution :  $\sigma_{Z}$ &  $\SI{1.5}{\milli\metre}$  \\
% \hline
\bottomrule
\end{tabular}
\end{center}
\caption{Best \emph{MEG} tracking parameter resolutions achieved in the year 2009/2010.
The resolutions are given for positrons of $\SI{53}{\mega\electronvolt\per\c}$ momentum.  Values taken from
\protect{\cite{Adam:2011ch}}.}
\label{tab:meg}
\end{table}

The study of the $\mu \rightarrow e \gamma$ decay sets stringent bounds
on models predicting new heavy particles mediating LFV dipole couplings. 
These dipole couplings can also be tested in the process 
$\mu \rightarrow eee$, where the sensitivity is reduced by
a factor of $\frac{\alpha}{3\pi} (\ln(m_\mu^2/m_e^2)-11/4)=0.006$ 
(note however that for \mte also box diagrams, $Z^0$-mediated penguin
diagrams and tree digrams contribute  as described in chapter \ref{sec:Theory}).
In the case that the LFV dipole couplings are dominant,
the projected sensitivity of $\num{e-13}$ of the \emph{MEG} experiment
 corresponds accordingly to a sensitivity of about $\num{e-15}$ 
in the search for the $\mu \rightarrow eee$ decay
and the envisaged sensitivity of  $\textrm{B}(\mu \rightarrow eee)=\num{e-16}$
corresponds to more than one order of magnitude higher sensitivity compared to
the \emph{MEG} experiment.

%The \emph{MEG} experiment, in contrast to the proposed $\mu \rightarrow eee$ experiment,
% has no sensitivity to lepton flavor violating four-fermion contact interaction couplings.

\begin{figure*}[t!]
	\centering
		\includegraphics[width=0.90\textwidth]{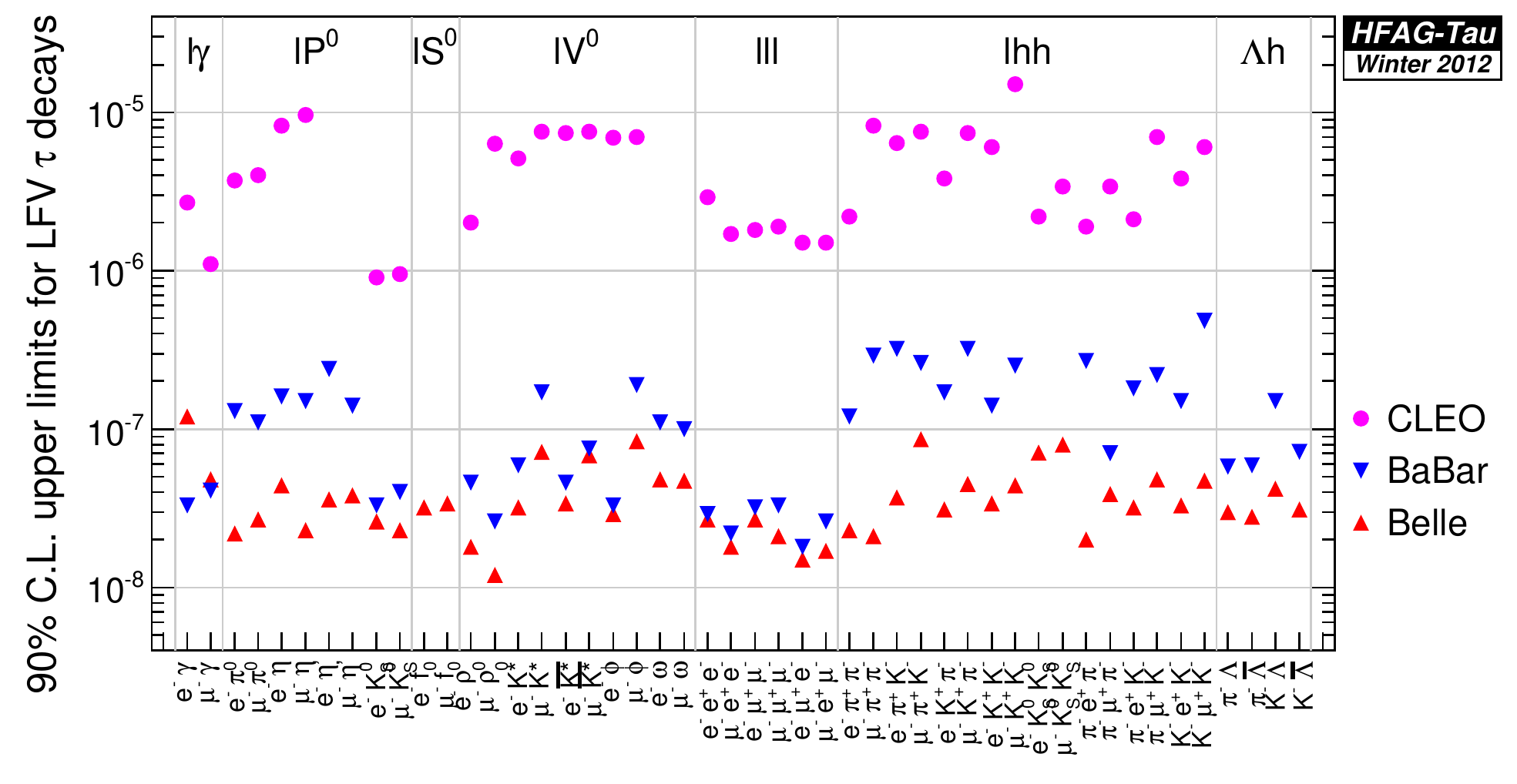}
	\caption{Limits on LFV $\tau$ decays. Taken from \cite{Amhis:2012bh}}
	\label{fig:TauLFV_UL_2012001}
\end{figure*}

\begin{table*}[b!]
	\centering
		\begin{tabular}{lccccc}
		\toprule
		\sc Decay  																	& \sc Belle limit & \sc Babar limit & \sc Belle II proj. & \sc Belle II proj. & \sc SuperB proj.$^1$\\
		\sc channel              										&             &             & ($\SI{5}{ab^{-1}}$) & ($\SI{50}{ab^{-1}}$)& ($\SI{75}{ab^{-1}}$)\\
		\midrule
		$\tau \rightarrow \mu \gamma$ 		& \num{4.5e-8} \cite{Hayasaka:2007vc}	
																			& \num{4.4e-8}\cite{Aubert:2009ag}
																			& \num{10e-9} \cite{Aushev:2010bq,Abe:2010sj}
																			& \num{3e-9} \cite{Aushev:2010bq,Abe:2010sj}
																			& \num{1.8e-9} \cite{Bona:2007qt}\\
		$\tau \rightarrow e \gamma$ 			& \num{12e-8}	\cite{Hayasaka:2007vc}
																			& \num{3.3e-8}\cite{Aubert:2009ag}
																			& 
																			& 
																			& \num{2.3e-9} \cite{Bona:2007qt}\\
		$\tau \rightarrow \mu \mu \mu$ 		&	\num{2.1e-8}\cite{Hayasaka:2010np}
																			& \num{3.3e-8}\cite{Lees:2010ez}
																			&\num{3e-9} \cite{Aushev:2010bq,Abe:2010sj}
																			&\num{1e-9} \cite{Aushev:2010bq,Abe:2010sj}
																			&\num{2e-10} \cite{Bona:2007qt}\\
		$\tau \rightarrow e e e$ 					&	\num{2.7e-8}\cite{Hayasaka:2010np}
																			& \num{2.9e-8}\cite{Lees:2010ez}
																			& 
																			& 
																			&\num{2e-10}  \cite{Bona:2007qt}\\
		$\tau \rightarrow \mu \eta$ 			&	\num{2.3e-8}\cite{Hayasaka2011}
																			& \num{15e-8}\cite{Aubert:2006cz}
																			&\num{5e-9} \cite{Aushev:2010bq,Abe:2010sj}
																			&\num{2e-9} \cite{Aushev:2010bq,Abe:2010sj}
																			&\num{4e-10}  \cite{Bona:2007qt}\\
		$\tau \rightarrow e \eta$ 	  		&	\num{4.4e-8}\cite{Hayasaka2011}
																			& \num{16e-8}\cite{Aubert:2006cz}
																			& 
																			& 
																			&\num{6e-10}  \cite{Bona:2007qt}\\
		$\tau \rightarrow \mu K_S^0$ 	  	&	\num{2.3e-8} \cite{Miyazaki:2010qb}
																			& \num{4.0e-8} \cite{Aubert:2009ys}
																			& 
																			& 
																			&\num{2e-10} \cite{Bona:2007qt}\\
		$\tau \rightarrow e K_S^0$ 	  		&	\num{2.6e-8} \cite{Miyazaki:2010qb}
																			& \num{3.3e-8} \cite{Aubert:2009ys}
																			& 
																			& 
																			&\num{2e-10} \cite{Bona:2007qt}\\																	
		\bottomrule	
		\end{tabular}
	\caption{Measured and projected limits on selected lepton flavour violating $\tau$ decays ($\SI{90}{\percent}\,\, C.L.$).\\
	{\footnotesize $^1$ The SuperB projections assumed a polarized
          electron beam; they also assumed that all backgrounds except initial
          state radiation can be suppressed to the desired level. The
          SuperB project was canceled in November 2012.
	}}
	\label{tab:LimitsOnLeptonFlavourViolatingTauDecays}
\end{table*}

\section{Muon Conversion Experiments}
%\textrm{B}($\mu^+e^âˆ’ \rightarrow  \mu^âˆ’e^+$) $< 8.3 \times 10^{âˆ’11}$ Willmann, et
%al. (1999)

Muon to electron conversion experiments $\mu \rightarrow e$ on nuclei exploit the clear
signature of monochromatic electrons. Differently to the search for 
LFV muon decays, which are performed using DC anti-muon beams in order to
reduce accidental backgrounds, muon conversion experiments are performed 
using pulsed muon beams to reduce the rapidly decaying pion background.
A limitation of this type of experiment is the background from 
ordinary decays of captured muons with large nuclear recoil and from pions.
 
The most stringent limits for muon-electron conversion on various nuclei
have been obtained by the \emph{SINDRUM~II} collaboration
\cite{Kaulard:1998rb,Dohmen:1993mp,Bertl:2006up}. 
The strongest limit has been set using a gold target
$\textrm{B}(\mu \; Au \rightarrow e \; Au ) \num{< 7e-13}$ \cite{Bertl:2006up}.

Similar to the $\mu \rightarrow eee$ process, the sensitivity to 
dipole couplings in muon conversion is reduced by about $\alpha_{em}$ compared to
the more direct $\mu \rightarrow e \gamma$ search. However,
new experiments planned at Fermilab (\emph{Mu2e} \cite{Bartoszek2012,Carey:2008zz, Tschirhart:2011zzb}) and
at J-PARC (\emph{COMET} \cite{Cui:2009zz, Kuno:2010zz,Akhmetshin2012} and \emph{PRISM} 
\cite{Kuno:2008zz,Pasternak:2010zz}) aim for branching ratios of $\num{e-16}$ or smaller
relative to the captured muon decay and have a higher sensitivity
to LFV dipole couplings than the running \emph{MEG} experiment.
Similar to the  $\mu \rightarrow eee$ process, also four-fermion couplings
are tested in $\mu \rightarrow e$ conversion experiments. 
These couplings involve light quarks and are thus complementary to
all other LFV search experiments.

The \emph{Mu2e}  and \emph{COMET}  experiments are ambitious 
projects and are expected to come into
operation at earliest by the end of this decade. In a few years time the
\emph{DeeMe} experiment
at J-PARC \cite{Aoki:2012zza} will start taking data, aiming for a sensitivity
for muon-to-electron conversions of $\num{e-14}$.

At Osaka university, the \emph{MuSIC} project \cite{Yamamoto:2011zb, Ogitsu:2011rg} aims for a very high intensity DC muon beam using 
a high-field capture solenoid around a thick conversion target. One of many possible users of that beam is
a \mte experiment.
However, an experimental concept has not yet been presented.

\section{LFV in $\tau$ Decays}

A wide variety of LFV decay channels are open in $\tau$ decays. These decay modes have been 
extensively explored at the $B$-factories, producing limits on branching ratios of a few $\num{e-8}$, 
see Table~\ref{tab:LimitsOnLeptonFlavourViolatingTauDecays} and Figure~\ref{fig:TauLFV_UL_2012001}. The next generation of $B$
experiments at $e^+e^-$ colliders could push these limits down by one to two orders of magnitude.
For certain channels such as $\tau \rightarrow \mu\mu\mu$, the LHCb experiment could also be
competitive given the luminosity expected in the coming years \cite{LHCbtau3mu,Seyfert:2012ug}.

\section{LFV at the Large Hadron Collider}

\balance

LFV signatures might be observed at the LHC if e.g.~supersymmetric
particles are discovered, which naturally
generate LFV couplings in slepton mass mixing. Consequently,
if sleptons are light enough to be produced in pairs, different lepton
flavors might show up in decay chains such as: 
$\tilde{\ell}^+ \tilde{\ell}^- \rightarrow 
\ell^+ \ell^{-\prime} \chi^0 \chi^0$.

Known and new scalar or vector particles could also have lepton violating tree
couplings and might be directly reconstructed from resonance peaks: 
$H \rightarrow \ell \ell^\prime$ or $Z^\prime \rightarrow \ell \ell^\prime$. Due to the existing
bounds on flavor changing processes, these LFV decays are
small and difficult to detect above the large background 
from $WW$-production with subsequent leptonic decays. It seems however, 
that with high enough luminosities, the LHC can e.g.~go beyond the LEP
bounds \cite{Akers:1995gz,Abreu:1996mj,Adriani:1993sy,Decamp:1991uy} on LFV $Z$ decays \cite{Davidson2012}.

%The typical mass scale of lepton flavor violating couplings
%tested by \emph{MEG} of the order of 25~TeV \cite{}WHERE FROM??? INKONSISTENT exceeds the mass reach
%of the LHC by one order of magnitude and demonstrates the complementarity 
%of low energy precision experiments and the LHC.

If new particles exist at the TeV mass scale, i.e.~in the discovery reach
of the LHC, it is very likely that precision experiments will discover lepton flavor 
violation via radiative loops. 
Dedicated LFV search experiments like the proposed \mte experiment
would then allow one to measure the LFV couplings of the new particles, 
complementary to the $\si{\tera\electronvolt}$ scale experiments at the LHC.

Conversely, in the case that no new physics (excluding the SM Higgs boson \cite{Atlas:2012gk,CMS:2012gu}) 
were discovered at the LHC, the discovery of LFV in precision experiments
is not excluded as e.g. rare muon decays are testing the mass scale $\SI{>1}{\peta\electronvolt}$, 
three orders of magnitude higher than at LHC.

\chapter{The Decay {$\mu \rightarrow eee$}}
\label{sec:DecayMu3e}

\nobalance

\section{Kinematics}
The decay $\mu \rightarrow eee$ proceeds promptly.
For discriminating signal and background, energy and momentum conservation
can be exploited. The vectorial sum of all decay particle momenta should
vanish:
\begin{equation}
\left|\vec{p}_{tot}\right| \; = \; \left|\sum \vec{p}_i\right| \; = \; 0  \quad
\end{equation}
and the total energy has to be equal to the muon mass.

The energies of the decay electrons (positrons) are in the range $\SIrange{0}{53}{\mega\electronvolt}$.
All decay particles must lie in a plane and the decay is described by two
independent variables in addition to three global rotation angles, which
describe the orientation in space.

\section{Detector Acceptance}

\begin{figure}[tb!]
	\centering
		\includegraphics[width=0.49\textwidth]{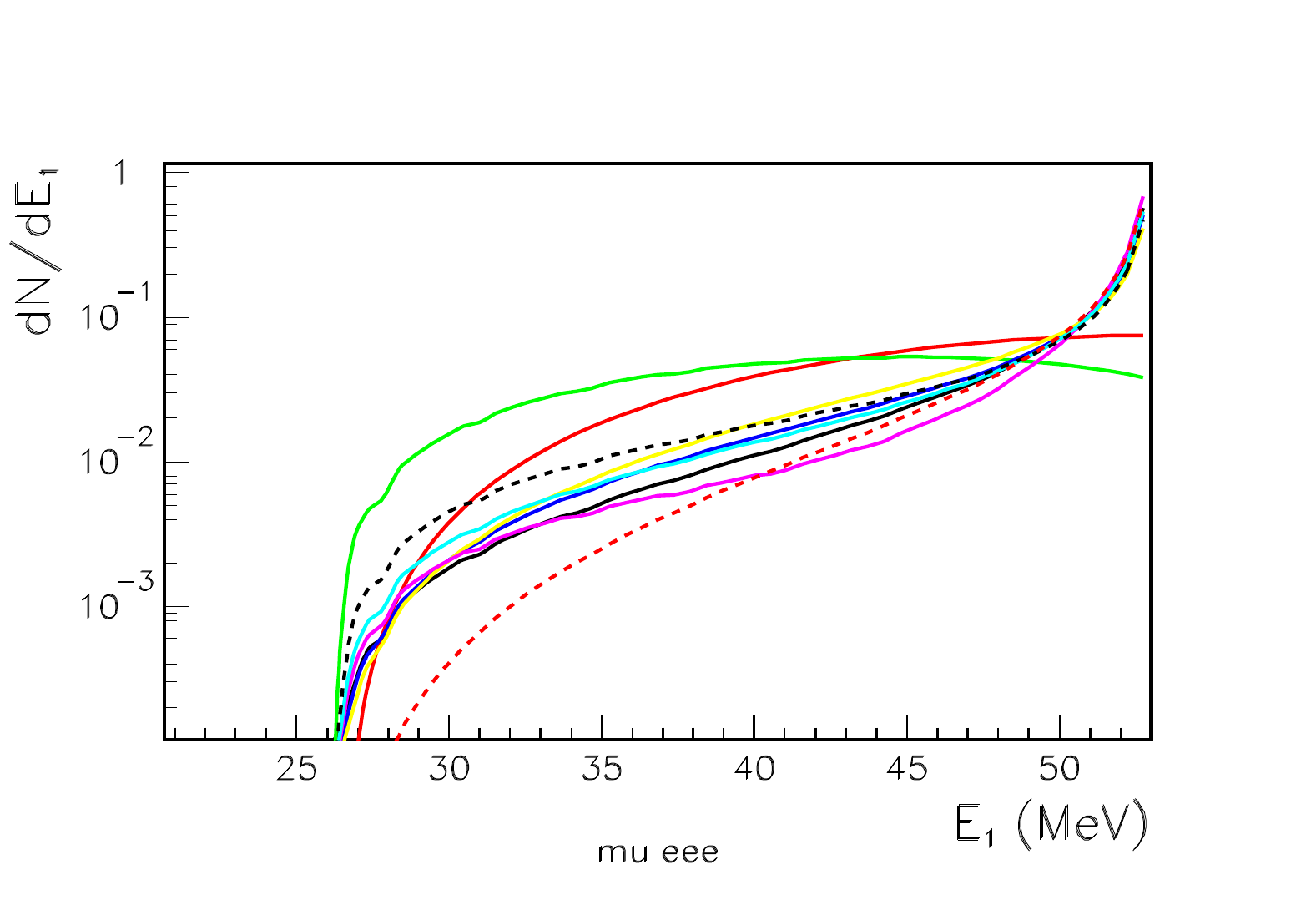}
	\caption{Energy distribution of the highest energy positron in the
          decay $\mu^{+} \rightarrow e^{+}e^{-}e^{+}$
for different
effective LFV models. The solid red and the green lines correspond to
pure four-fermion contact interaction models (no penguin) contribution.}
	\label{fig:brmee_x1_10.norm}
\end{figure}

\begin{figure}[tb!]
	\centering
		\includegraphics[width=0.49\textwidth]{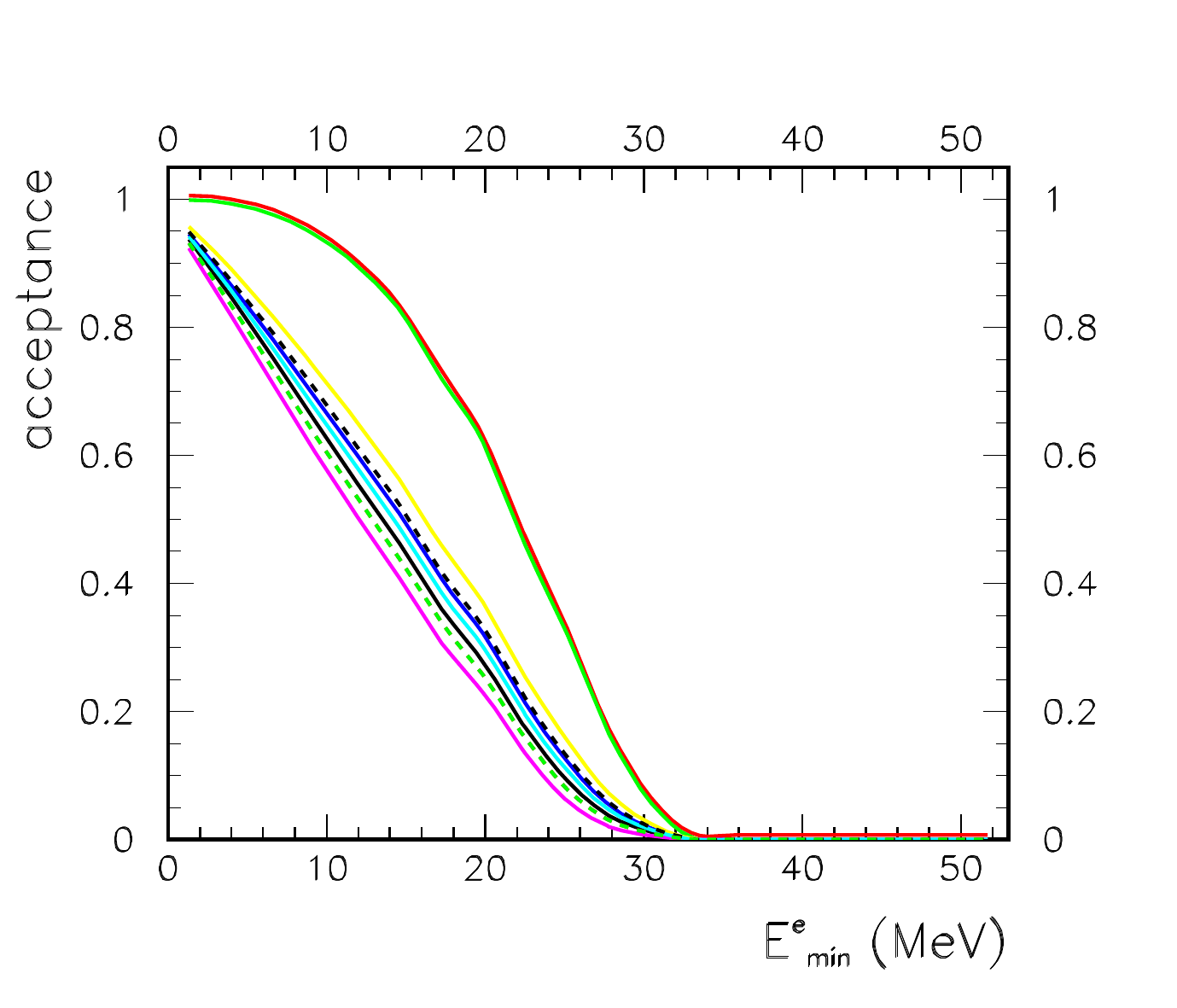}
	\caption{Acceptance of the lowest energy decay electron (positron) for different
effective LFV models as function of the minimum transverse momentum. 
The solid red and green lines correspond to
pure four-fermion contact interaction models (no penguin) contribution.}
	\label{fig:brmee_ptmin_10.summed}
\end{figure}

The acceptance of the proposed \mte experiment is determined by its
geometrical acceptance and energy coverage. 
For various coupling assumptions about the LFV amplitude, see also
equation~\ref{eq:lagrangian}, 
the energy spectrum of the highest energy, $E_1$, and lowest energy decay
particles, $E^e_{min}$,
are shown in Figures \ref{fig:brmee_x1_10.norm} and
\ref{fig:brmee_ptmin_10.summed}, respectively. In order to achieve a high
acceptance, the detector must be able to reconstruct tracks with momenta
ranging from half the muon mass down to a few $\si{\mega\electronvolt}$ with large solid angle
coverage. 
The proposed experiment should cover the energy range 
$>\SI{10}{\mega\electronvolt}$ to provide 
acceptances of $\SI{50}{\percent}$ or more for all models.

\section{Backgrounds}

\begin{figure}
	\centering
		\includegraphics[width=0.48\textwidth]{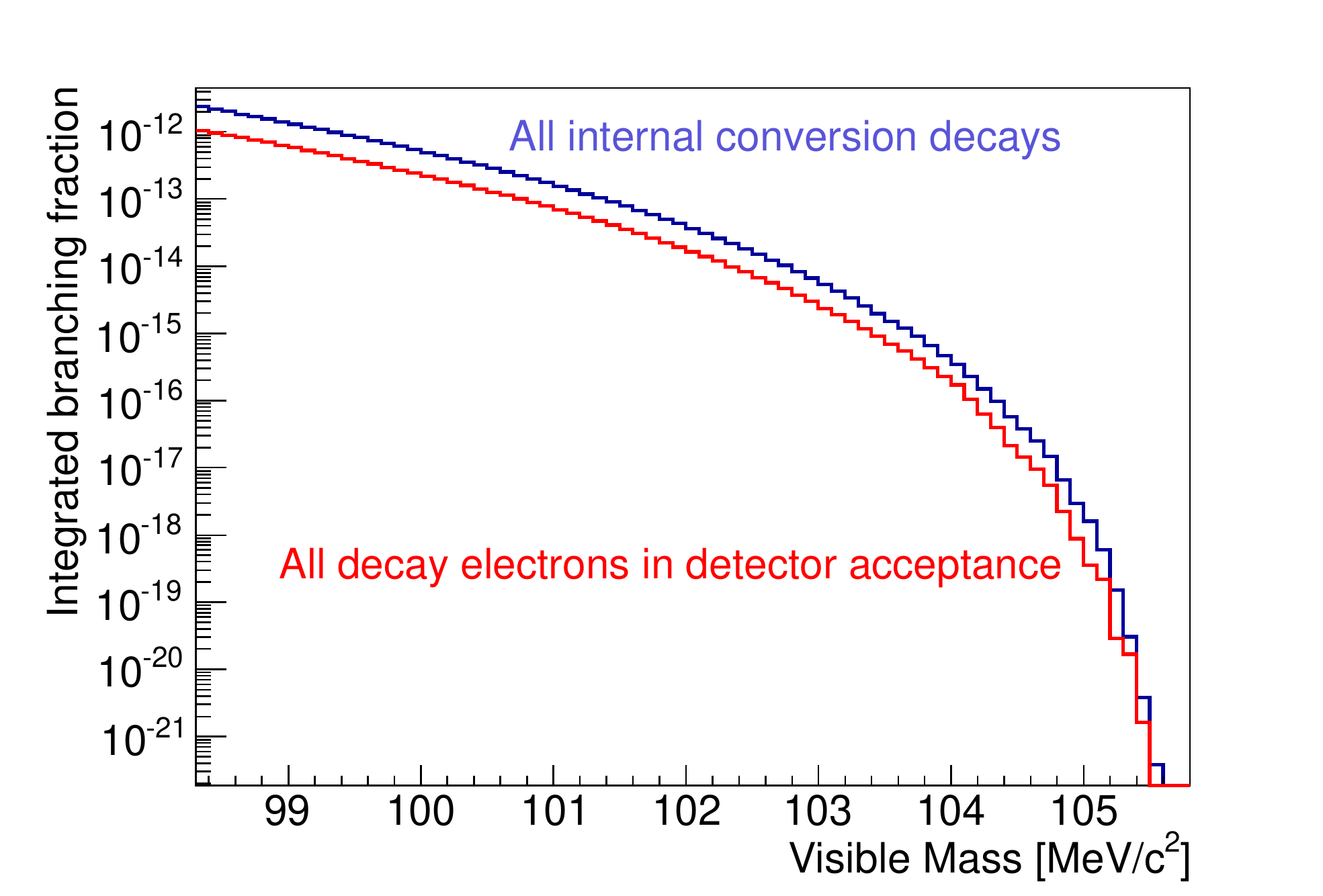}
	\caption{Integrated branching fraction for the decay $\mu \rightarrow eee \nu \nu$ in dependence of the visible mass for all internal conversion decays and those with all three decay particles in the detector acceptance. The matrix element was taken from \cite{Djilkibaev:2008jy}.}
	\label{fig:icfraction}
\end{figure}
	
%\begin{figure*}
%	\centering		%\includegraphics[width=0.9\textwidth]{DecayMu3e/figures/Both_Djilkibaev} 
%	\caption{Summed energy of the charged leptons normalised to the muons
%         mass, $y$, in the decay $\mu
%          \rightarrow eee \nu \nu$  (left) and branching ratio of the same
%          process as a function of the missing energy $m_\mu-E_{tot}$ cut %(right). Figures adapted from \cite{Djilkibaev:2008jy}.}
%	\label{fig:InternalConversionSpectra}
%\end{figure*}

\begin{figure}
	\centering
		\includegraphics[width=0.48\textwidth]{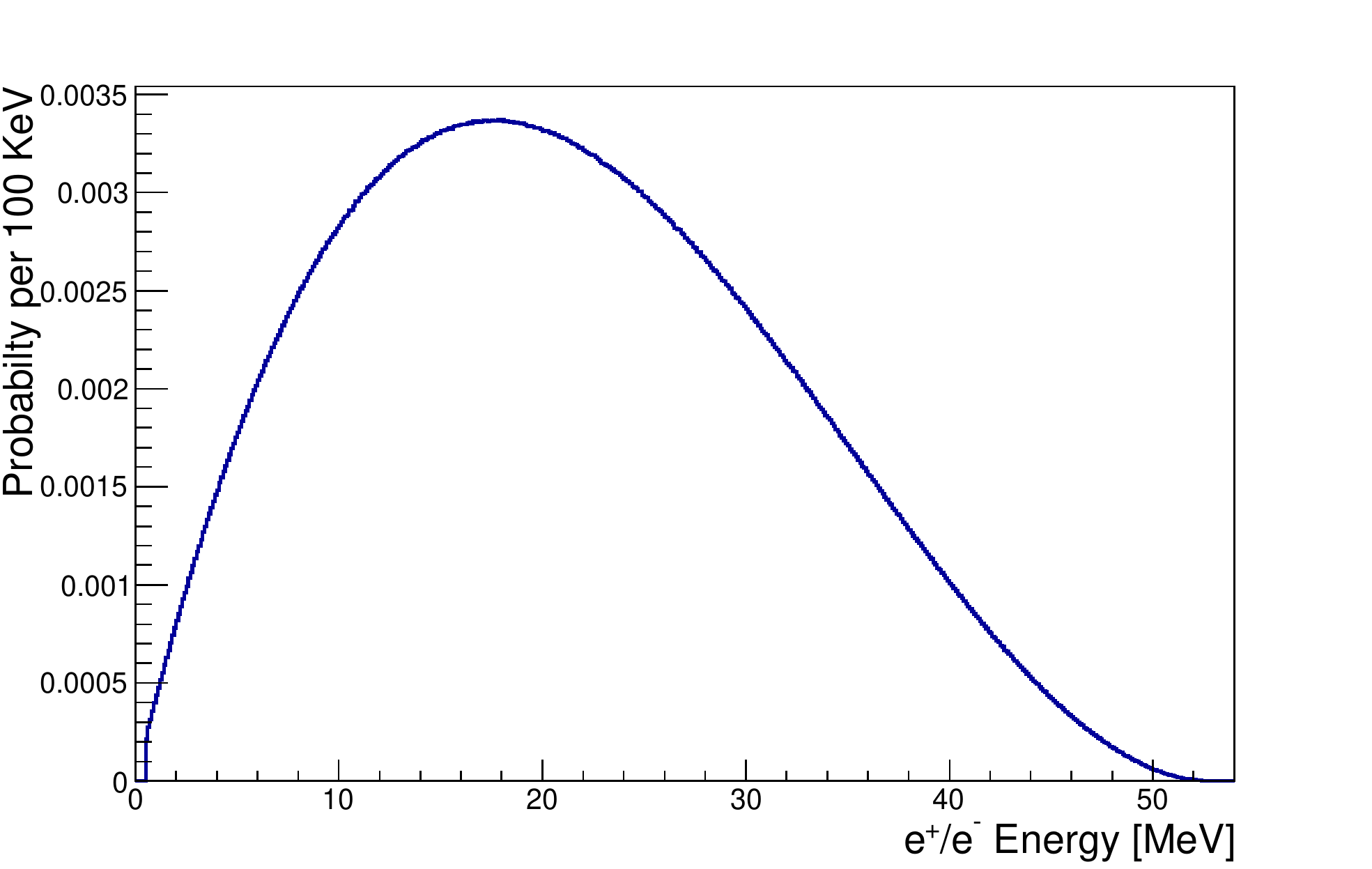}
	\caption{Spectrum of electrons from internal conversion decays.}
	\label{fig:ic_e_spectrum}
\end{figure}

\begin{figure}
	\centering
		\includegraphics[width=0.48\textwidth]{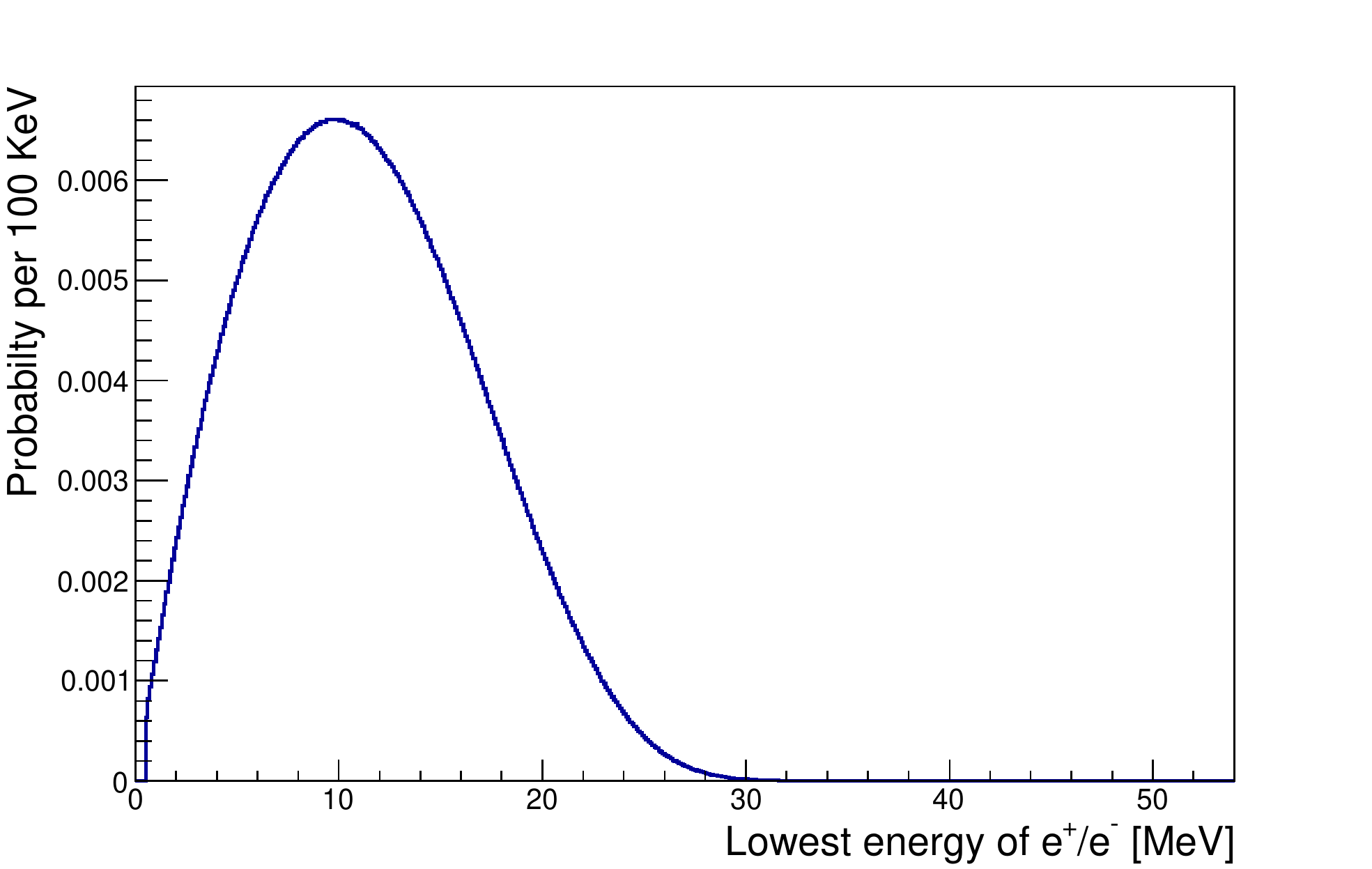}
	\caption{Spectrum of the electron with minimum energy from internal conversion decays.}
	\label{fig:ic_emin_spectrum}
\end{figure}

\begin{figure}[tb!]
	\centering
		\includegraphics[width=0.48\textwidth]{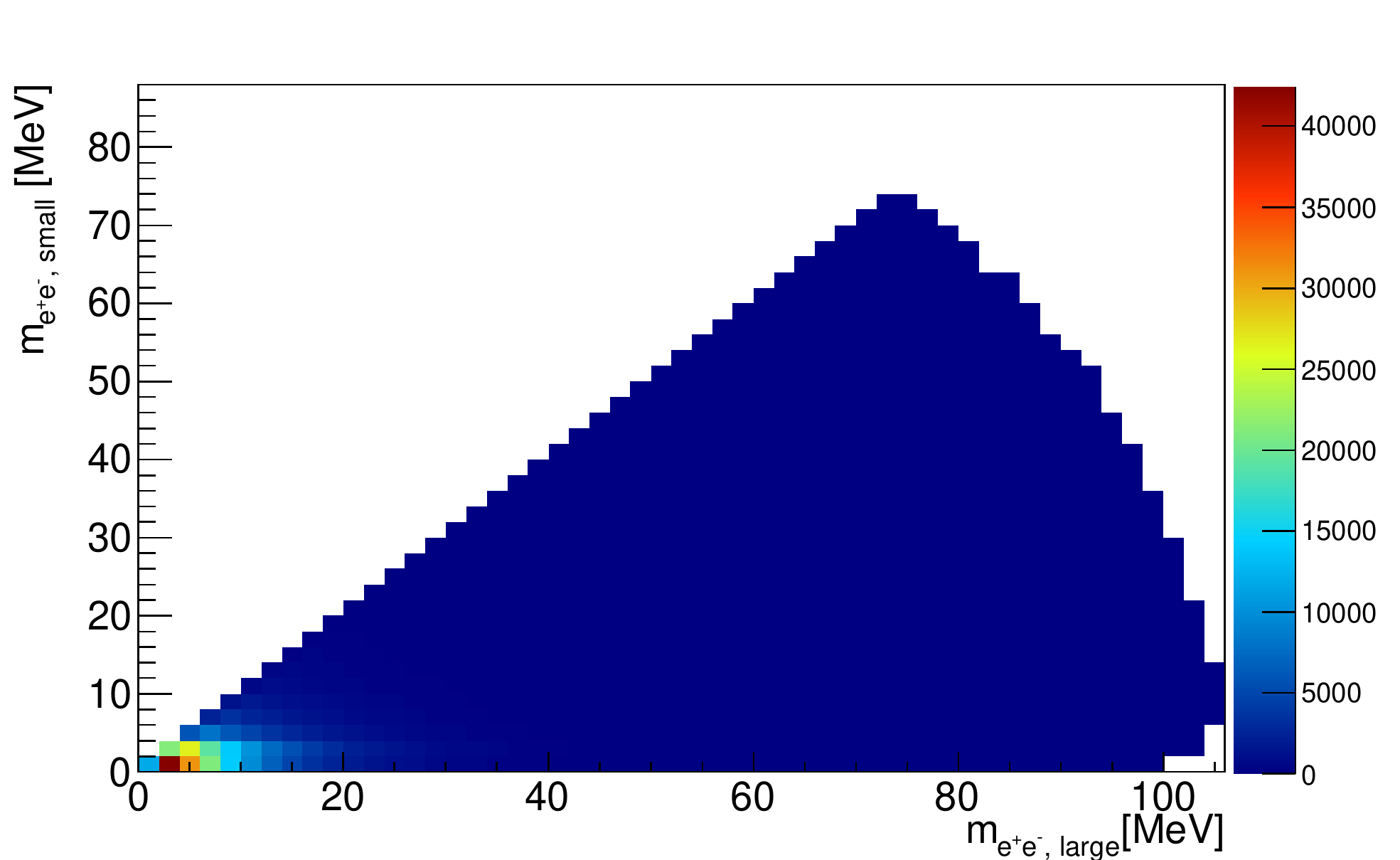}
	\caption{Invariant masses of the two possible $e^+e^-$ combinations for internal conversion decays.}
	\label{fig:ic_m1_m2_all}
\end{figure}

\begin{figure}[tb!]
	\centering
		\includegraphics[width=0.48\textwidth]{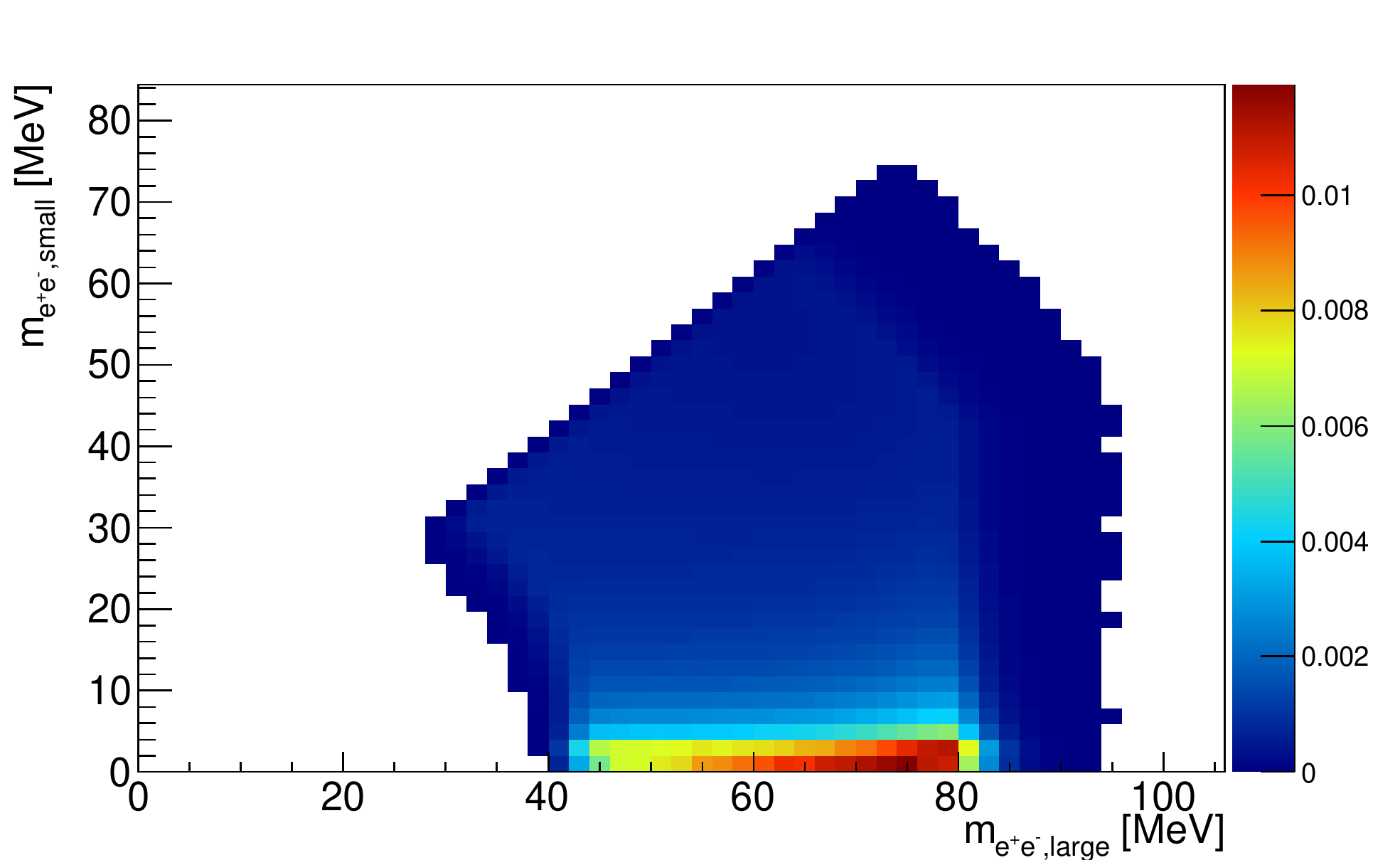}
	\caption{Invariant masses of the two possible $e^+e^-$ combinations for internal 
	conversion decays with a visible mass above $\SI{90}{MeV}$ and the electrons and 
	positrons in the detector acceptance ($E >\SI{10}{MeV}$, $|\cos\theta| < 0.8$).}
	\label{fig:ic_m1_m2_90}
\end{figure}

The final sensitivity of the proposed experiment depends on the ability
to reduce backgrounds from various sources. Two categories of backgrounds are considered; 
irreducible backgrounds, such as  $\mu^+
\rightarrow e^+ e^+ e^- \nu \bar{\nu}$, which 
strongly depend on the granularity and resolution of the detector, and 
accidental backgrounds that scale linearly or with the square of the beam intensity.

In the following sections, the main background sources considered are discussed.

\subsection{Internal Conversions}

The decay $\mu \rightarrow eee \nu \nu$ occurs with a branching fraction of
$\num{3.4e-5}$ \cite{PDG10}. It can be distinguished from the \mte process
by making use of energy and momentum conservation to reconstruct the
undetected neutrinos; in order to separate the
\mte events from $\mu \rightarrow eee \nu \nu$ events, the total momentum in
the event is required to be zero and the energy equal to the muon rest
energy. The branching fraction as a function of the
energy cut of the $\mu \rightarrow eee \nu \nu$ process
\cite{Djilkibaev:2008jy} is shown in Figure~\ref{fig:icfraction}. 
Figures.~\ref{fig:ic_e_spectrum} and \ref{fig:ic_emin_spectrum} show the energy spectrum of
all and the lowest energy electron from internal conversion decays, Figs.~\ref{fig:ic_m1_m2_all} and \ref{fig:ic_m1_m2_90} the invariant masses of $e^+e^-$ combinations calculated with the matrix element from \cite{Djilkibaev:2008jy}.
This process is the most serious background for the \mte search and can only
be resolved by a very good energy resolution.

\subsection{Michel Decays}

Using a beam of positive muons, one of the main 
processes contributing to accidental background is that of the ordinary Michel 
decay $\mu^+ \rightarrow e^+ \nu \bar{\nu}$. This process does not produce 
a negatively charged particle (electron), which is one of the main
characteristics of the $\mu^+ \rightarrow e^+e^+e^-$ decay, and can therefore only contribute 
as potential background if a track is wrongly reconstructed. Other processes
which ``naturally'' provide negatively charged tracks (electrons) are 
radiative decays with internal or external photon conversions or Bhabha
scattering.

\subsection{Radiative Muon Decays}

\balance

The process $\mu^+ \rightarrow e^+ \gamma \nu \nu$ (branching fraction $\num{1.4e-2}$ 
for photon energies above $\SI{10}{\mega\electronvolt}$ \cite{PDG10}) can deliver an oppositely charged 
electron if the photon converts either in the target region or in the 
detector. Contributions from conversions outside of the target are greatly
suppressed if a vertex constraint is applied and by minimizing the material 
in both the target and detector. Photon conversion in the target generates an event
topology similar to the radiative decay with internal conversion: $\mu \rightarrow eee \nu \nu$, 
which is discussed above.

Due to the missing energy from the neutrinos, this process mainly contributes 
to the accidental background in combination with an ordinary muon decay.

\subsection{Bhabha Scattering}

Positrons from the ordinary muon decay or beam-positrons can undergo Bhabha
scattering with electrons in the target material, leading to an
electron-positron pair from a common vertex. 
Due to the missing energy, this process mainly contributes 
to the accidental background in combination with an ordinary muon decay.

\subsection{Pion decays}
\label{sec:PionDecay}

Certain decays of pions, especially $\pi \rightarrow eee \nu$ (branching
fraction $\num{3.2e-9}$ \cite{PDG10}) and $\pi \rightarrow \mu \gamma
\nu$ (branching fraction $\num{2.0e-4}$ \cite{PDG10}) with subsequent
photon conversion are indistinguishable from signal events if the momenta of
the final state particles fit the muon mass hypothesis; a low pion contamination of
the primary beam (estimated to be in the order of $\num{e-12}$ for the high intensity beamline), the small branching fraction and the small slice of the
momentum is assumed to lead to negligible rates in the
kinematic region of interest.

\subsection{Summary of Background Sources}
\balance
First simulation studies have been performed to calculate the different
background contributions. Their results indicate that purely accidental backgrounds for $\sim \num{e9}$ muons stops per second are small for the proposed high resolution detector. 

The main concern are irreducible backgrounds, such as the process
 $\mu \rightarrow eee \nu \nu$, which can only be reduced by a very good
tracking resolution resulting in total energy resolution of $\sigma_E < \SI{1}{\mega\electronvolt}$
for the aimed sensitivities $\num{<e-15}$.

\newpage

\part{The Mu3e Experiment}

\chapter{Requirements for Mu3e}
\label{sec:Requirements}
	
\nobalance
	
\section{Goals of the Experiment}
\label{sec:GoalsOfTheExperiment}
	
	The goal of the Mu3e experiment is to observe the process \mte if its branching 
fraction is larger than $\num{e-16}$ or otherwise to exclude a branching fraction of 
$>\num{e-16}$ at the $\SI{90}{\percent}$ certainty level. In order to achieve 
these goals, $>\num{5.5e16}$ 
% 1e16/0.3*1.64 = 5.5e16
muon decays have to be observed\footnote{Assuming 
a total efficiency of $\SI{30}{\%}$.} and any background mimicking the signal 
process has to be suppressed to below the $\num{e-16}$ level. The additional 
requirement of achieving these goals within a reasonable measurement time of 
one year of data taking dictates a muon stopping rate of $\SI{2e9}{\Hz}$ and a 
high geometrical acceptance and efficiency of the experiment.
	
We plan to perform the experiment in two phases. The exploratory phase I will 
make use of existing muon beams at PSI and serve to commission the detectors, 
gain experience with the new technologies and validate the experimental concept, 
whilst at the same time producing a competitive measurement. The goal for this 
first phase is to reach a sensitivity of $\num{e-15}$, thus pushing the existing 
limit by three orders of magnitude. For this level of sensitivity, the demands on 
the detector are somewhat relaxed, thus allowing for cross-checks between detectors 
also on analysis level or running without the full instrumentation. The lower data 
rates also will not require the full read-out and filter farm system. The second 
phase of the experiment on the other hand will aim for the ultimate sensitivity 
and thus require that the detector works as specified and a new beamline delivers 
$> \SI{2e9}{\Hz}$ of muons.
	
The expected rate at an existing beamline is $1-\SI{1.5e8}{\Hz}$ of muons on target. 
In order to have a safety margin, we usually assume $\SI{2e8}{\Hz}$ for phase I 
background studies, except where the running time is concerned.
	
This proposal discusses the phase I experiment in detail and shows the path 
leading to full rate capability. We also discuss alternative approaches.
	
\section{Challenges for the Experiment}
\label{sec:ChallengesForTheExperiment}
	
\subsection{Backgrounds}
\label{sec:Backgrounds}
	
There are two kinds of backgrounds: Overlays of different processes producing 
three tracks resembling a \mte decay (\emph{accidental background}) and 
radiative decays with internal conversion (\emph{internal conversion background}) 
with a small energy fraction carried away by the neutrinos. Accidental background 
has to be suppressed via vertexing, timing and momentum measurement, whereas momentum 
measurement is the only handle on internal conversion.
	
\subsection{Geometric acceptance}
\label{sec:GeometricAcceptance}
	
For a three-body decay with a priori unknown kinematics such as \mte, 
% the acceptance is given by the single track acceptance to the third power.
% THIS IS NOT CORRECT A.S.
the acceptance has to be as high as possible in order to test new physics in
all regions of phase space.
There 
are two kinds of acceptance losses, losses of tracks downstream or upstream, 
where beam entry and exit prevent instrumentation, and losses of low
transverse momentum tracks, 
which do not transverse a sufficient number of detector planes, and are not reconstructed.  
	
\subsection{Rate capability}
\label{sec:RateCapability}
	
The Mu3e detector should be capable of running with $\SI{2e9}{\Hz}$ of muon 
decays. This poses challenges for the detectors, the data acquisition and the readout.
	
\balance
	
\subsection{Momentum resolution}
\label{sec:MomentumResolution}
	
\begin{figure}
	\centering
		\includegraphics[width=0.48\textwidth]{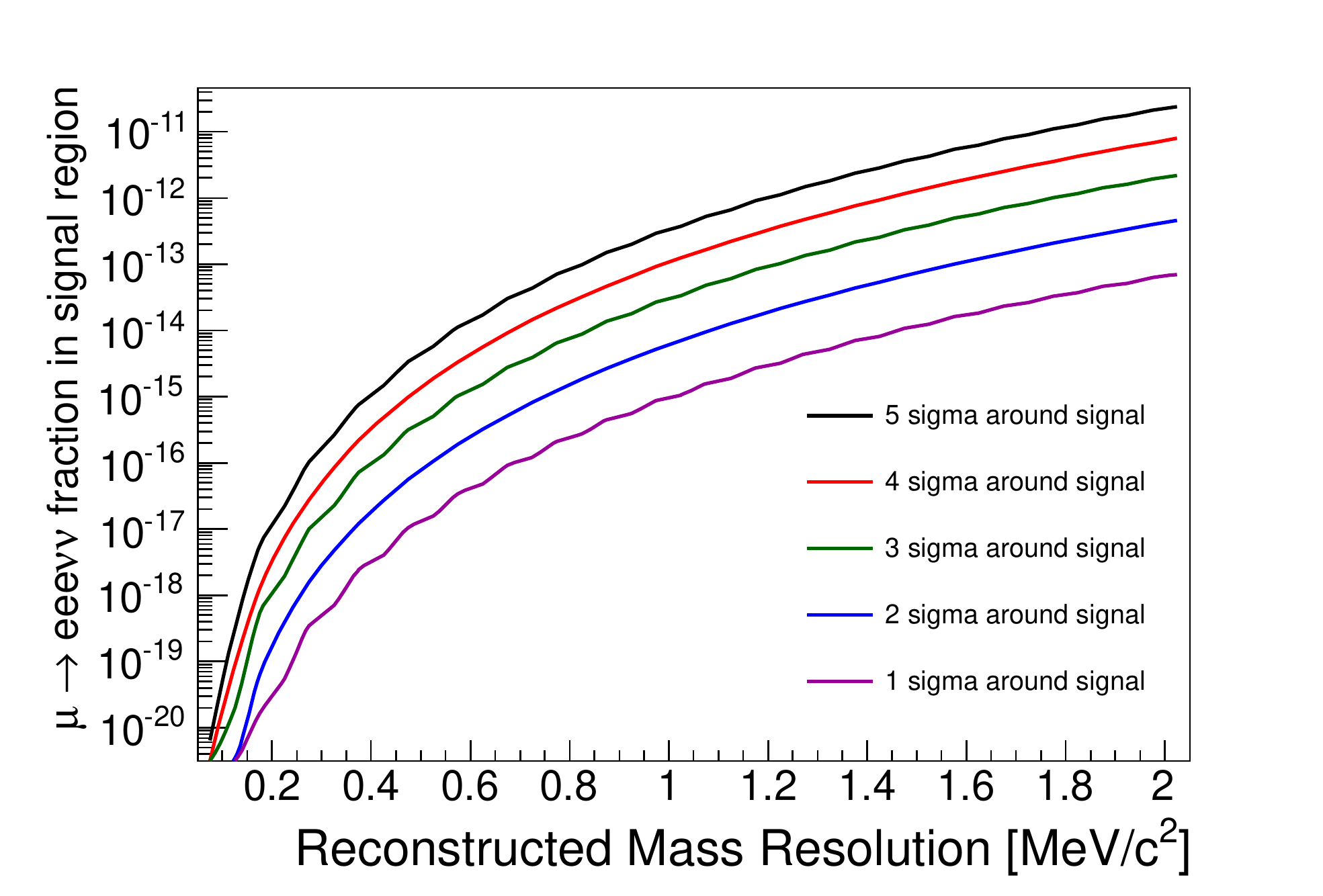}
	\caption{Contamination of the signal region (one sided cut) with internal conversion 
	events as a function of momentum sum resolution.}
	\label{fig:ic_contamination}
\end{figure}

The momentum resolution directly determines to what level internal conversion 
background can be suppressed and thus to which level the experiment can be ran
background free. In order to reach a sensitivity of $\num{e-16}$ with a 
$2 \sigma$ cut on the reconstructed muon mass, the average momentum 
resolution has to be better than $\SI{0.5}{MeV}$. For the phase I experiment 
aiming at $\num{e-15}$, this requirement is relaxed to $\SI{0.7}{MeV}$, see 
Figure~\ref{fig:ic_contamination}.
	
\subsection{Vertex resolution}
\label{sec:VertexResolution}
	
Keeping apart vertices from different muon decays is a key tool in suppressing 
accidental background. The vertex resolution is essentially determined by the 
amount of multiple scattering (and thus material) in the innermost detector 
layer. Ideally the vertex resolution is sufficient to eliminate almost all 
combinatorial backgrounds; for the phase I rates, this appears achievable, 
whereas in the phase II experiment, very good timing is needed in addition.
	
\subsection{Timing resolution}
\label{sec:TimingResolution}
	
Good timing is essential for reducing combinatorial background at rates which lead to more than about 10 muon decays per frame on average.

\chapter{Experimental Concept}
\label{sec:Concept}

\nobalance

\begin{figure}[b!]
	\centering	\includegraphics[width=0.40\textwidth]{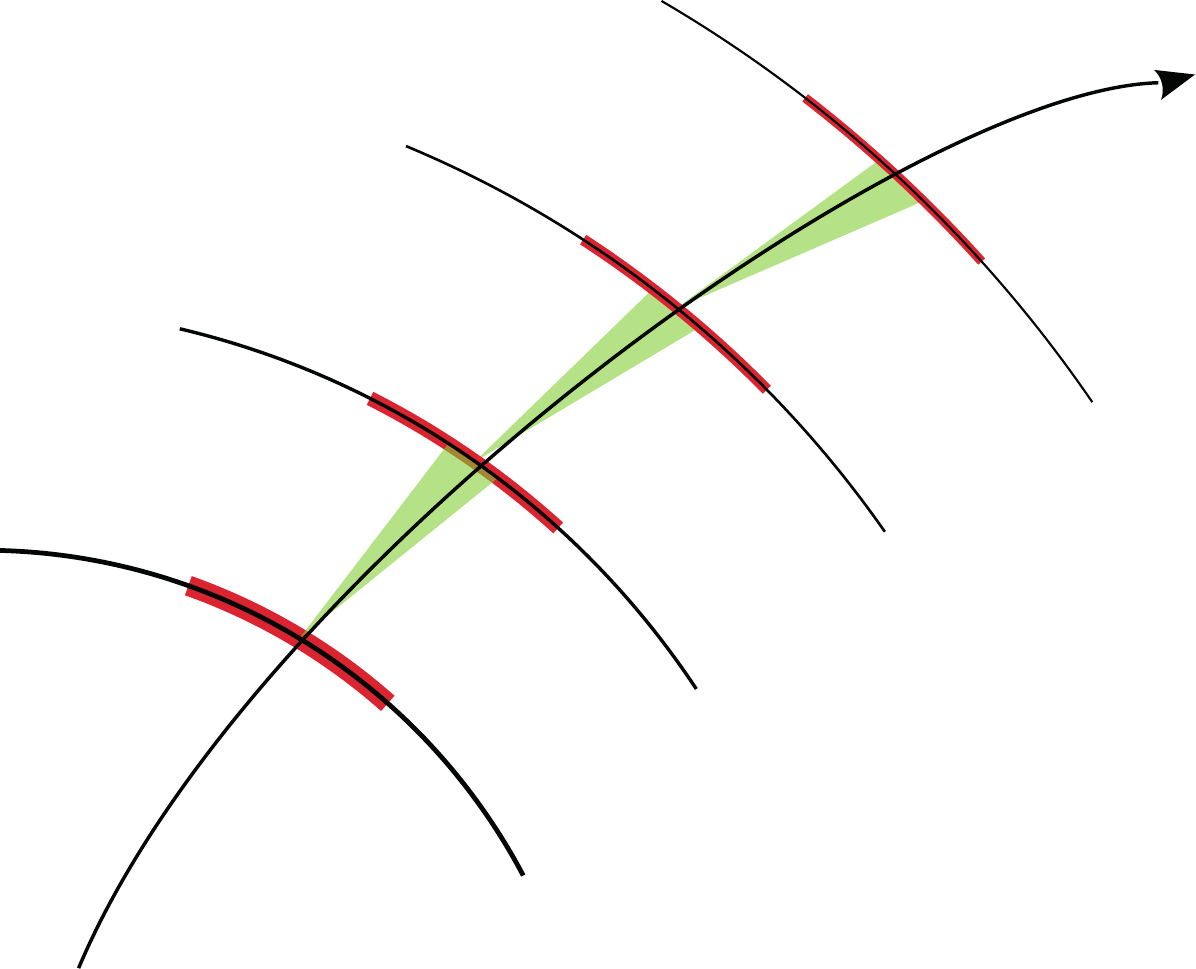}
	\caption{Tracking in the spatial resolution dominated regime}
	\label{fig:tracking_spatial_regime}
\end{figure}

The Mu3e detector is aimed at the background free measurement or exclusion of the decay \mte at the level of $\num{e-16}$. As discussed in more detail in the preceding chapter \ref{sec:Requirements}, these goals require to run at high muon decay rates, an excellent momentum resolution in order to suppress
background from the internal conversion decay \mtenunu and good vertex and timing resolution in order to efficiently suppress combinatorial background. 

We intend to measure the momenta of the muon decay electrons in a solenoidal magnetic field using a silicon pixel tracker. At the electron energies of interest, multiple Coulomb scattering in detector material is the dominating factor affecting momentum resolution. Minimizing this material in the active detector parts is thus of utmost importance.  

\begin{figure}[b!]
	\centering
		\includegraphics[width=0.40\textwidth]{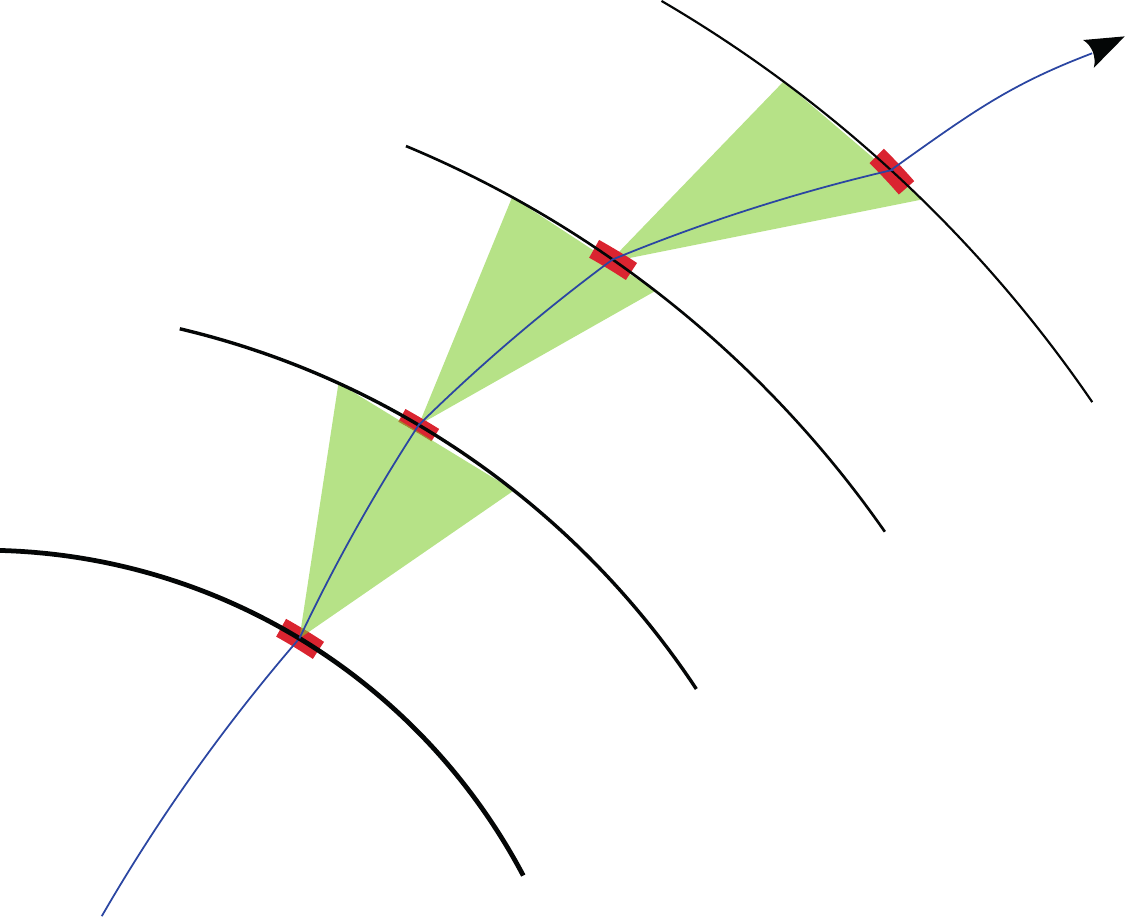}
	\caption{Tracking in the scattering dominated regime}
	\label{fig:tracking_scattering_regime}
\end{figure}

The proposed detector consists of an ultra thin silicon pixel tracker, made possible by the High-Voltage Monolithic Active Pixel (HV-MAPS) technology (see chapter \ref{sec:Pixel}). Just four radial layers around a fixed target in a solenoidal magnetic field allow for precise momentum and vertex determination. Two timing detector systems guarantee good combinatorial background suppression and high rate capabilities. 
 
The Mu3e experiment is designed to have a sensitivity four orders of magnitude better than the current limit on \mte ($\num{e-12}$), so it is reasonable to plan for a staged detector design, with each stage roughly corresponding to an order of magnitude improvement. 
    
\section{Momentum Measurement with Recurlers}
\label{sec:MomentumMeasurementWithRecurlers}

Due to the low momenta of the electrons from muon decay, multiple scattering is the dominating effect on momentum measurement. With our fine-grained pixel detector, we are thus in a regime where scattering effects dominate over sensor resolution effects, see Figs.~\ref{fig:tracking_spatial_regime} and 
\ref{fig:tracking_scattering_regime}. Thus adding additional measurement points does not necessarily improve the precision.

\begin{figure}
	\centering
		\includegraphics[width=0.4\textwidth]{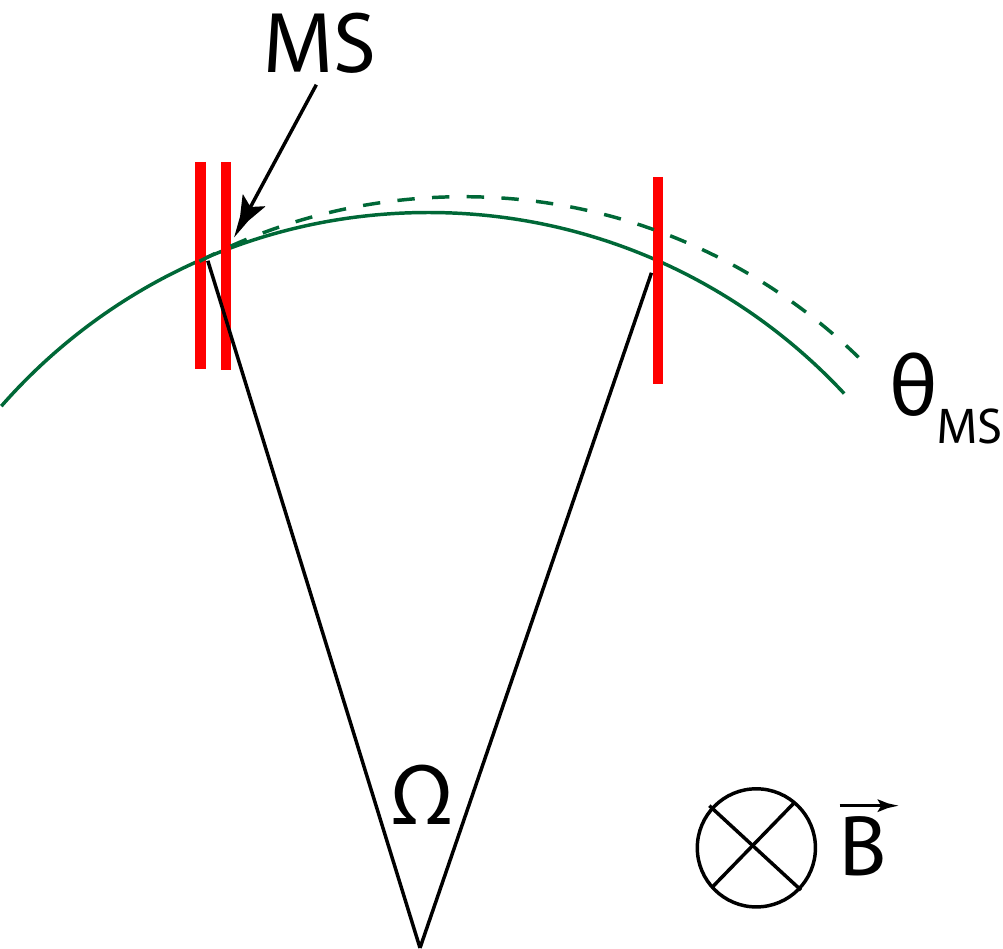}
	\caption{Multiple scattering as seen in the plane transverse to the magnetic field direction.}
	\label{fig:MS}
\end{figure}

\begin{figure}
	\centering
		\includegraphics[width=0.4\textwidth]{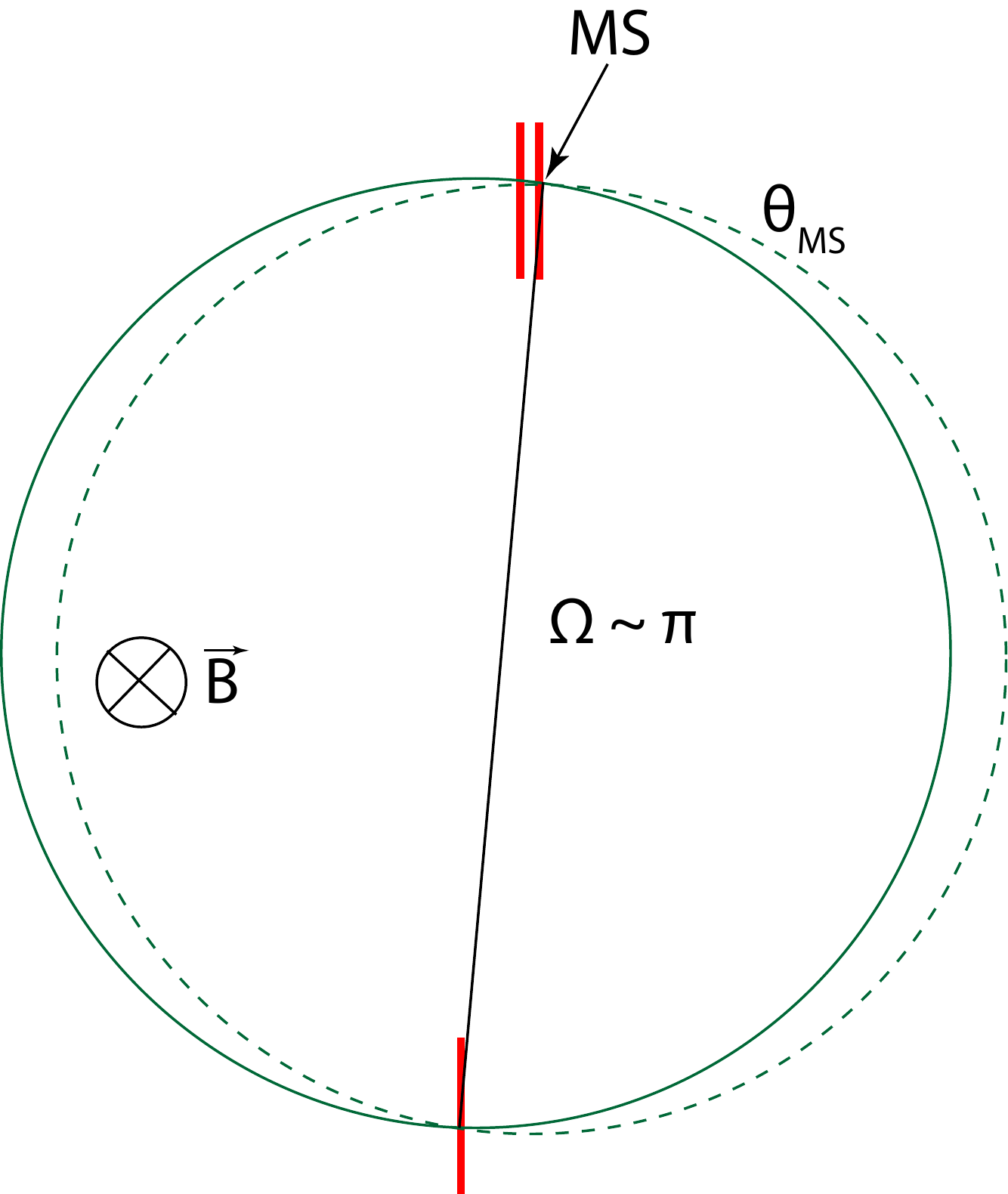}
	\caption{Multiple scattering for a semi-circular trajectory.}
	\label{fig:MScircle}
\end{figure}

The precision of a momentum measurement depends on the amount of track
deflection $\Omega$ in the magnetic field $B$ and the multiple scattering angle $\Theta_{MS}$, see Figure~\ref{fig:MS};
to first order:
\begin{equation}
	\frac{\sigma_p}{p} \propto \frac{\Theta_{MS}}{\Omega}.
\end{equation}
So in order to have a high momentum precision, a large lever arm is
needed. This can be achieved by moving tracking stations to large radii, which
however compromises the acceptance for low momentum particles. 
In the case of muon decays, all track momenta are below
$\SI{53}{\MeV}$ and all tracks will thus curl back towards the magnet axis if
the magnet bore is sufficiently large. 
After exactly half a turn, effects of multiple scattering on momentum measurement cancel in first order, see
Figure~\ref{fig:MScircle}. 
To exploit this feature we optimized the experimental design specifically for
the measurement of re-curling tracks, leading to a narrow, long tube layout.

Measuring the momentum from bending outside of the tracker also allows us to place timing detectors inside, without strongly affecting the resolution.

\section{Baseline Design}
\label{sec:BaselineDesign}

\begin{figure*}[tb!]
	\centering
		\includegraphics[width=1.00\textwidth]{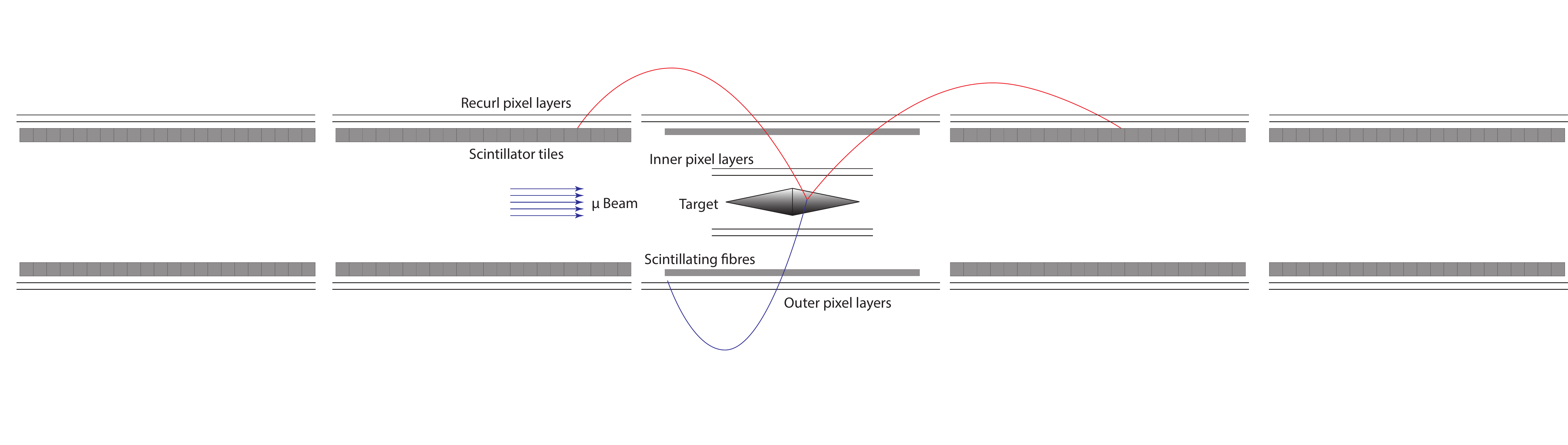}
	\caption{Schematic view of the experiment cut along the beam axis in
          the phase~II configuration.}
	\label{fig:schematic_longitudinal}
\end{figure*}

\begin{figure*}[tb!]
	\centering
		\includegraphics[width=0.70\textwidth]{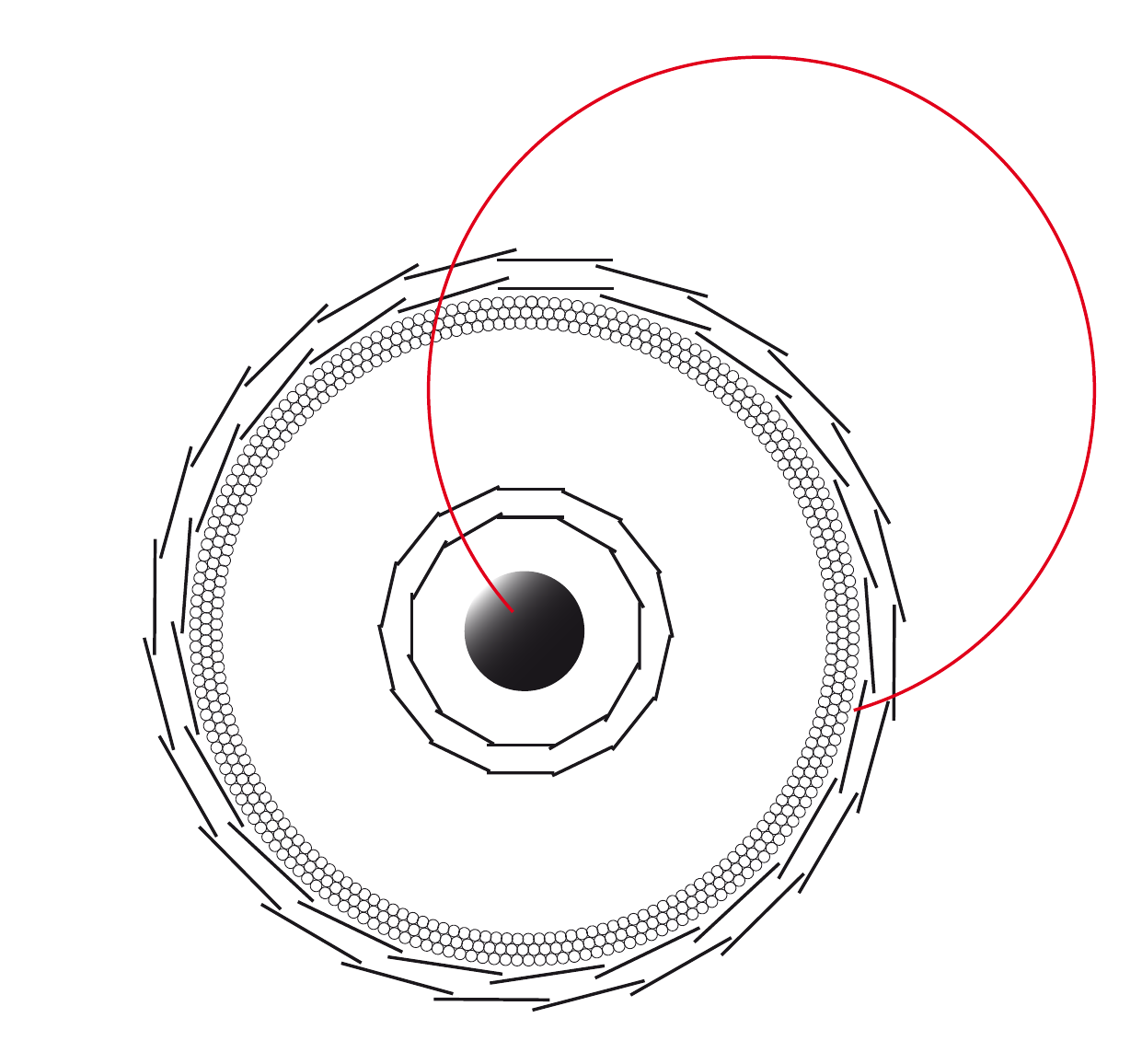}
	\caption{Schematic view of the experiment cut transverse to the beam axis. Note that the fibres are not drawn to scale.}
	\label{fig:schematic_transverse}
\end{figure*}

The proposed Mu3e detector is based on two double layers of HV-MAPS
around a hollow double cone target, see Figures \ref{fig:schematic_longitudinal} and \ref{fig:schematic_transverse}. 
The outer two pixel sensor layers are extended upstream and
downstream to provide precise momentum measurements in an extended region with the help of re-curling electrons.
The silicon detector layers (described in detail in chapter~\ref{sec:Pixel}) are supplemented by two timing systems, 
a scintillating fibre tracker in the central part (see chapter~\ref{sec:Fibre}) and scintillating tiles (chapter~\ref{sec:Tiles}) inside the recurl layers.
Precise timing of all tracks is necessary for event building and to suppress accidental combinatorial background.

The entire detector is built in a cylindrical shape around a beam pipe, with a total length of
approximately $\SI{2}{\m}$, inside a $\SI{1}{\tesla}$ solenoid magnet with $\SI{1}{\m}$ inside diameter and $\SI{2.5}{\m}$ total length (chapter~\ref{sec:Magnet}).
In the longitudinal direction the detector is sub-divided into five stations, the central detector with target, inner silicon double layer, fibre tracker and outer silicon double layer, and two forward and backward recurl stations with two silicon recurl layers surrounding a tile timing detector. In order to separate tracks coming from
different muon decays, the target has a large surface with $\SI{10}{\cm}$ length and $\SI{2}{\cm}$ diameter. The target shape is a hollow double cone, see chapter \ref{sec:Target}. 
Around the target the two inner silicon pixel layers, also referred to
as the vertex layers, cover a length of $\SI{12}{\cm}$. The innermost layer will have 12, the second one 18 sides
of $\SI{1}{\cm}$ each, corresponding to an average radius of $\SI{1.9}{\cm}$ and $\SI{2.9}{\cm}$, respectively. 
The inner silicon layers are supported by two half cylinders made from $\SI{25}{\micro\meter}$ thin Kapton foil mounted on 
plastic end pieces. All silicon sensors are thinned to
$\SI{50}{\micro\meter}$, resulting in a radiation length of
X/X$_0\leq\SI{0.1}{\%}$ per layer. The detector will be cooled with gaseous helium.

The hit information from the silicon sensors is read out at a rate of
$20$~MHz using timestamps providing a time resolution of $20$~ns.
%The two innermost silicon layers are used for the precise decay vertex %determination which is necessary for suppression of combinatorial background %and are part of the tracking detectors for
%precise momentum measurement.

The fibre tracker sits inside silicon pixel layer three at around $\SI{6}{\cm}$, providing timing information for 
decay positrons and electrons. It is composed from three to five layers of $\SI{250}{\micro\meter}$ thick $\SI{36}{\cm}$ long
scintillating fibres, see Figure~\ref{fig:schematic_transverse}.
The fibre tracker is read out by fast silicon photo multipliers and can provide timing information with $\leq\SI{1}{ns}$ accuracy.

The silicon pixel layers three and four are just outside the fibre tracker at a mean radius of $\SI{7.6}{\cm}$ and $\SI{8.9}{\cm}$.
The active area has a cylindrical shape of $\SI{36}{\cm}$ length. The layer three has 24 sides, layer four
28 sides of $\SI{1.9}{\cm}$ width each. Both outer layers are constructed as modules of 4 sides, six modules for layer three
and seven modules for layer four. Similar to the inner two layers the mechanical frames of these modules are build from
$\SI{25}{\micro\meter}$ Kapton foil with plastic end pieces. 

Copies of silicon pixel layer three and four are also used in the recurl stations.
Two recurl stations each are covering the upstream and downstream regions.
These recurl stations add further precision to the momentum measurement of the
electrons,
% leaving the central part and to form re-curling tracks in the forward and
% backward region
see section~\ref{sec:MomentumMeasurementWithRecurlers}.
While the silicon layer design is (almost) identical to the central part, 
the timing detector in the recurl region can be much thicker compared to the fibre tracker, as the particles can and should be stopped here. This is done by using scintillating tiles of about
$7.5\times 7.5\times \SI{5}{\mm^3}$ size. 
These tiles provide a much better time resolution than the thin fibre tracker in the center.
Following the dimensions of the recurl silicon layers, the tile station have a active length of $\SI{36}{\cm}$ and a cylindrical shape
with  a radius of $\approx \SI{6}{\cm}$. 
All central detector components are mounted on spokes providing a light stiff support.
The recurl silicon layers and tiles are mounted on the beam pipe support.
In addition to the silicon and scintillating tile sensors the beam pipe support also carries the services
and the PCBs equipped with the front-end electronics (chapter~\ref{sec:DAQ}). 
Signal and power connection to the silicon layers is provided by flex prints which are also part of the mechanical support of the silicon sensors.

\section{Building up the Experiment}
\label{sec:BuildingUp}

\begin{figure*}
	\centering
		\includegraphics[width=1.00\textwidth]{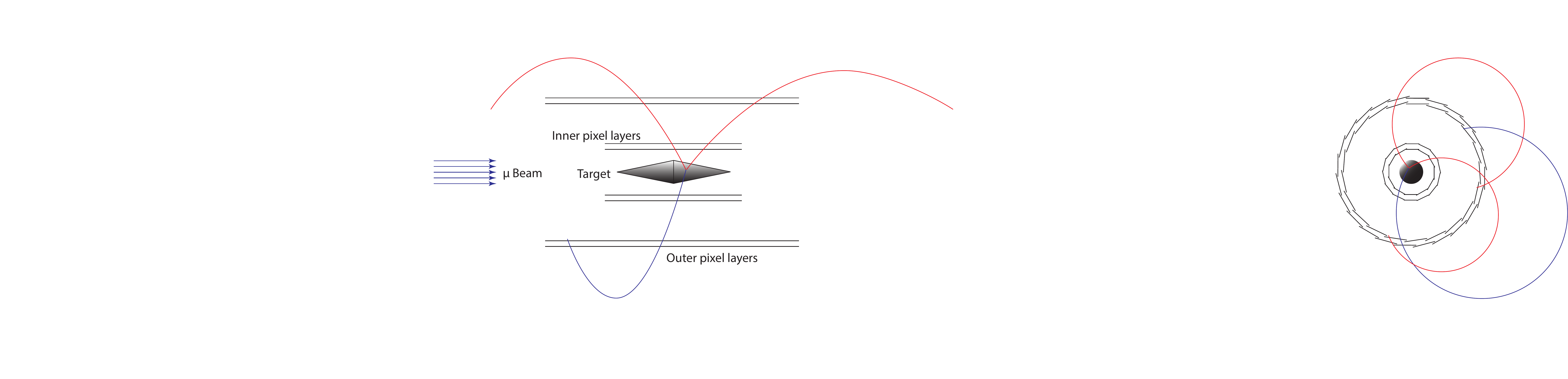}
	\caption{Minimum detector cofiguration for early commissioning with central silicon only (phase IA).}
	\label{fig:Schematic10_NoFibres}
\end{figure*}

%\begin{figure*}
	%\centering
		%\includegraphics[width=1.00\textwidth]{Concept/figures/Schematic10_NoRecurl}
	%\caption{Central detector only for early physics runs and fibre commissioning.}
	%\label{fig:Schematic10_NoRecurl}
%\end{figure*}

\begin{figure*}
	\centering
		\includegraphics[width=1.00\textwidth]{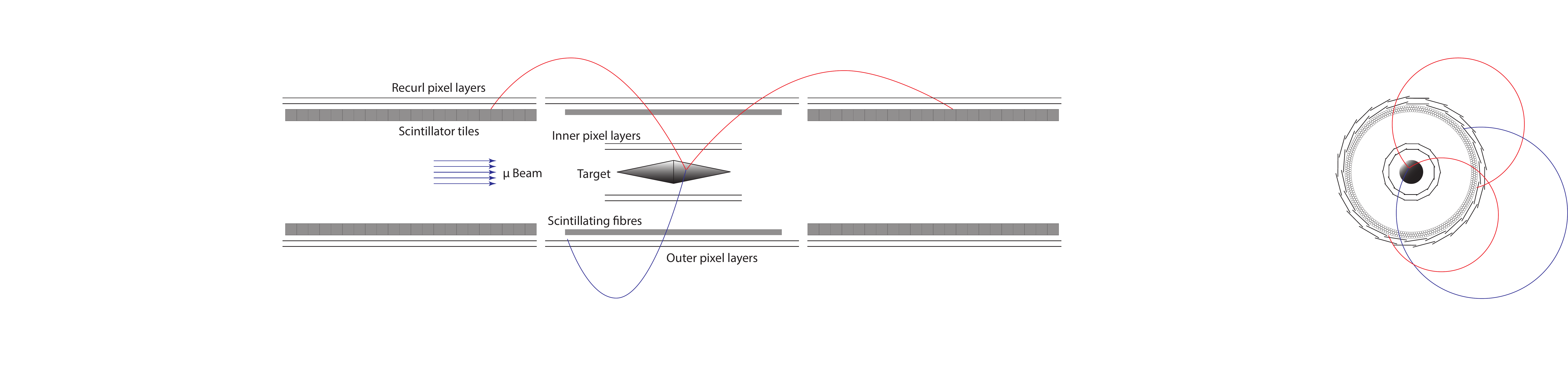}
	\caption{Detector with one set of recurl stations for physics runs and tile detector commissioning (phase IB).}
	\label{fig:Schematic10_ShortRecurl}
\end{figure*}

\begin{figure*}
	\centering
		\includegraphics[width=1.00\textwidth]{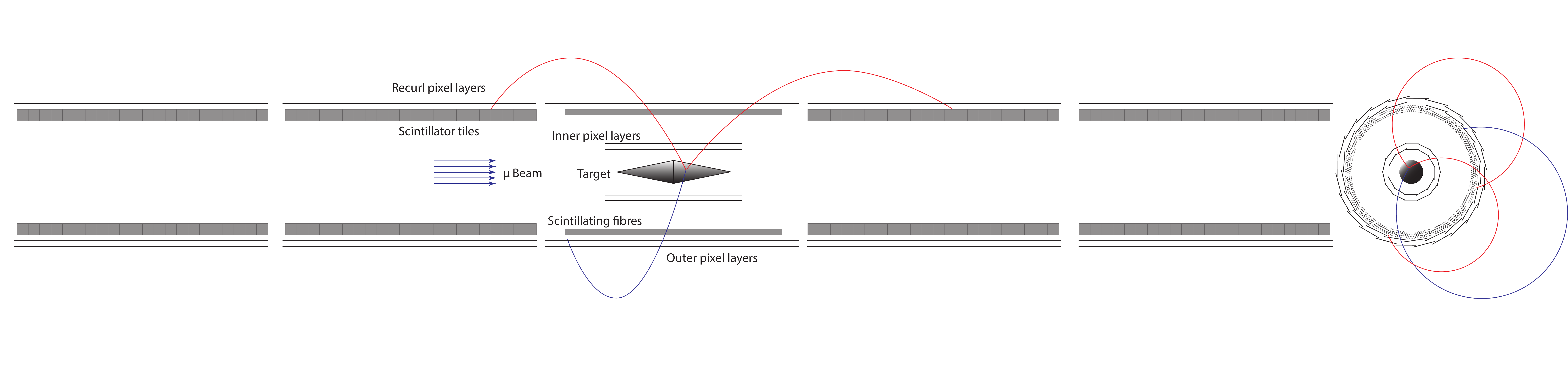}
	\caption{Final detector with two sets of recurl stations for high intensity physics runs (phase II).}
	\label{fig:Schematic10_Full}
\end{figure*}

One of the advantages of the design concept presented is its modularity. Even
with a partial detector, physics runs can be taken. The full instrumentation
is only required for achieving the final sensitivity of $\num{e-16}$ at muon rates
above $\SI{1e9}{\Hz}$. 
On the other hand, in an early commissioning phase at smaller muon stopping rates, 
the detector could run with the central silicon detector only (see Figure \ref{fig:Schematic10_NoFibres}). 
%A next step is adding the fibre tracker, in the beginning possibly with readout of only parts of the fibres (Figure \ref{fig:Schematic10_NoRecurl}).
The silicon detectors of the recurl stations are essentially copies of the
central outer silicon detector; 
after a successful commissioning of the latter, they can be produced and added to the experiment as they become available together with the connected tile detectors.
The fibre tracker can also be added later, since it is only needed to resolve
combinatorial background at higher event rates and track multiplicities.
The loss of momentum resolution due to multiple scattering at the
additional material of the fibre tracker will be fully
compensated by the improved momentum measurement with re-curlers.  
The configuration with two recurl stations (Figure
\ref{fig:Schematic10_ShortRecurl}) defines a medium-size setup, well
suited for phase~I running.  
The configuration with four recurl stations (Figure
\ref{fig:Schematic10_Full}) defines the full setup for phase~II running. 

%This will lead to a short configuration with one recurl station per side (Figure \ref{fig:Schematic10_ShortRecurl}), followed by the full detector for high rate running (Figure \ref{fig:Schematic10_Full}).

In the following sections, the experimental configurations for running at the existing
$\pi$E5  beam-line (the \emph{Phase I Experiment}) and the final detector for running at $> \SI{1e9}{\Hz}$ muon stopping rate (the \emph{Phase II Experiment}) are outlined.

\section{The Phase I Experiment}
	\label{sec:ThePhaseIExperiment}

The phase I of the Mu3e experiment will start with a minimum configuration
(phase IA detector) with the target regions surrounded by double layers
of inner and outer silicon pixel detectors, see Figure 
\ref{fig:Schematic10_NoFibres}. 
This configuration defines the minimal configuration as it allows to determine
the momentum, the vertex position and the time of the decay precise enough to produce very
competitive physics results with a sensitivity down to $\mathcal
O(\num{e-14}$). It is foreseen to run in the first year in this configuration
at a muon stopping rate on target at around $\SI{2e7}{\Hz}$. The number of decays in one readout frame of the pixel tracker of $\SI{50}{\ns}$ will be around one on average and combinatorial background can be suppressed with the help of the vertex reconstruction. The precision of the momentum resolution will be somewhat limited, as most tracks do not recurl within the instrumented volume, see chapter~\ref{sec:SensitivityStudies}. 

\balance

In the phase IB the detector will be complemented by the first pair of recurl
stations, the corresponding tile detectors and the fibre tracker, see Figure
\ref{fig:Schematic10_ShortRecurl}. Adding the recurl stations will
significantly enhance the momentum resolution and thus improve the suppression
of internal conversion background. The insertion of the fibre tracker and the
tile detector stations gives a much better time resolution in comparison to
the silicon pixel only. The fibre tracker will deliver a time resolution of
about $200$-\SI{300}{\ps}, while the tile detector will have $<\SI{100}{\ps}$
resolution for the tracks passing the recurl stations. The high time
resolution will allow running at the highest possible rate at the $\pi$E5 muon
beam line at PSI of $\approx\SI{1e8}{\Hz}$. 
The sensitivity reach in this phase of the experiment of $\mathcal
O(\num{e-15})$ will be limited by statistics only. limited by the available
muon decay rate.

\section{The Phase II Experiment}
	\label{sec:ThePhaseIIExperiment}

A new high intensity muon beam line \cite{Kettle:2010}
delivering  $\approx\SI{2e9}{\Hz}$ muon stops is crucial for the phase II of the proposed experiment.
To fully exploit the new beam facility the limited detector acceptance at
phase IB will be further enhanced by adding another a second pair of recurl and tile detector stations,
see Figure \ref{fig:Schematic10_Full}. These extra stations will allow to
measure precisely the momentum of all particles in the acceptance of the inner
tracking detector. 
At the same time the extra tile detector stations with their high time
resolution and small occupancy will help to fight the increased combinatorics
at very high decay rates. 
The combined performance of the final detector setup together with the high
stopping rate will allow to search for the \mte decay with a sensitivity of B(\mte)$\leq\num{e-16}$.

\chapter{Muon Beam}
\label{sec:MuonBeam}

\nobalance

\section{General Beam Requirements}
\label{sec:GeneralBeamRequirements}

\begin{figure}[b!]
\centering
\includegraphics[width=0.49\textwidth]{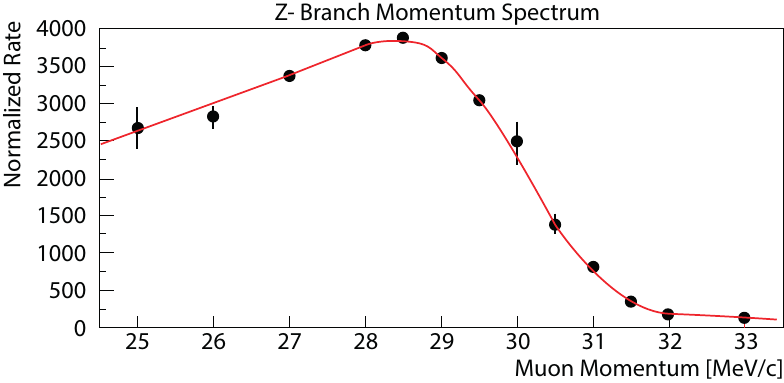}
\caption{$\pi$E5 measured muon momentum spectrum with fully open slits. Each point is obtained by
optimizing the whole beam line for the corresponding central momentum and measuring the full beam-spot intensity.
The red-line is a fit to the data, based on a theoretical $p^{3.5}$ behaviour, folded with a Gaussian resolution
function corresponding to the momentum-byte plus a constant cloud-muon background. }
\label{fig:pspect}
\end{figure}

      The general beam requirements for a high intensity, low-energy, stopped muon coincidence experiment such as Mu3e are six-fold: an abundant supply of low-energy surface muons (from stopped pion decay at rest, at the surface of the production target \cite{Pifer:1976ia}) capable of being stopped in a thin target; high transmission optics at \SI{28}{\MeV\per c}, close to the kinematic-edge of stopped pion decay and hence close to the maximum production rate of such muons, as shown in Figure~\ref{fig:pspect}; a small beam emittance to minimize the stopping-target diameter; a momentum-byte of less than \SI{10}{\%} with an achromatic final focus, allowing an almost monochromatic beam with a high stopping density, to be stopped in a minimally thick target; minimization and separation of beam-related backgrounds such as beam $e^{+}$ originating from $\pi^{0}$-decay in the production target, or decay particles produced along the beam line and finally minimization of material interactions in the beam, for example such as those in windows, thus requiring  vacuum or helium environments to keep the multiple scattering under control.

  \begin{figure*}
\centering
\includegraphics[width=0.9\textwidth]{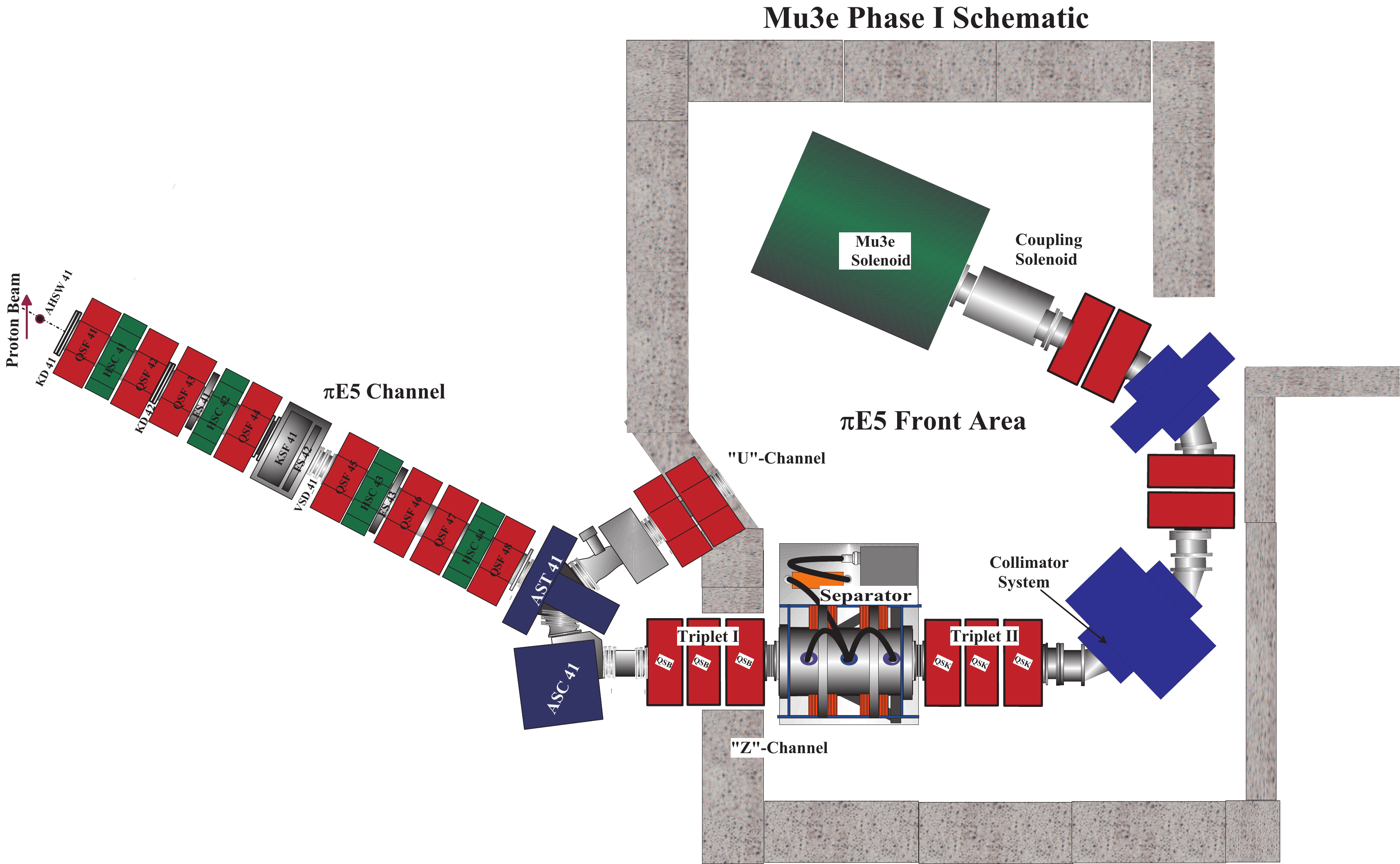}
\caption{Mu3e potential beam line layout in the front-part of the $\pi$E5  area.}
\label{fig:area}
\end{figure*}   

\begin{figure}
\centering
\includegraphics[width=0.49\textwidth]{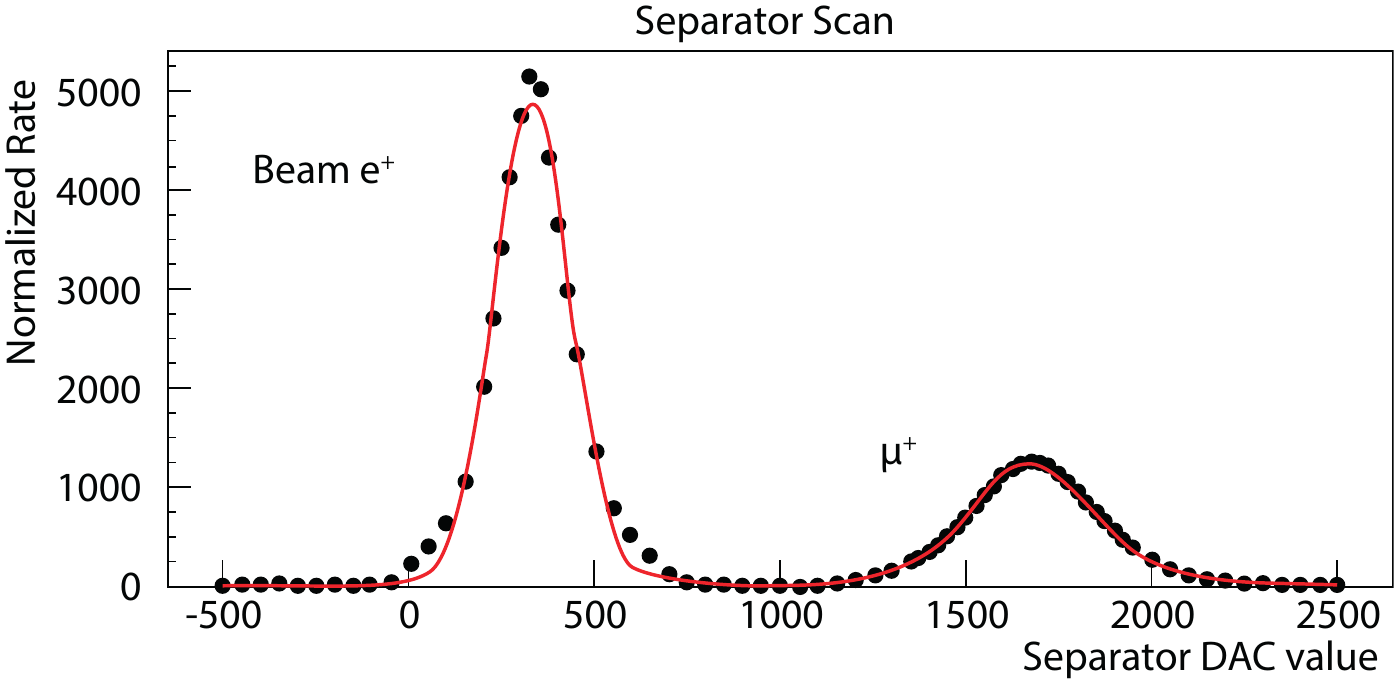}
\caption{Separator scan plot, measured post collimator system. The black dots represent 
on-axis, low-threshold intensity measurements at a given separator magnetic field value during 
the scan, for a fixed electric field value of \SI{-195}{\kilo\volt}. The red curve represents a double 
Gaussian fit to the data points, with a constant background. A separation of 8.1 muon beam 
$\sigma_\mu$ is found, corresponding to more than \SI{12}{\cm} separation of beam-spot centres at the collimator.}
\label{fig:sep}
\end{figure}  
     
\section{Beam for phase I running}
\label{sec:BeamForPhaseIRunning}

      As previously outlined, a multi-staged detector configuration will be sought for phase I running, this in turn requires muon beam intensities ranging between \SIrange{e7}{e8}{muons\per\s}  for the initial phase IA (central detectors), while for phase IB a maximal  intensity close to \num{1e8} muons/s will be sought. The quoted maximum beam intensity which includes a phase space reduction factor due to collimation in the central region of the Mu3 magnet is based on measured intensities at the centre of the MEG detector, without a degrader and normalized to a proton beam current of $\SI{2.3}{\mA}$. This demand for the highest intensities necessitates the selection of only one possible facility in the world, namely the $\pi$E5 channel at PSI. Based on the experience gained in the design of the MEG beam line, a similarly developed concept is also envisaged for phase I of Mu3e. This should allow the required muon intensity to be achieved. However, since the area is likely to be shared with other experiments, that of MEG (R-99-05), and the Lamb-shift experiment (R-98-03) a compact muon beam line designed specifically to fit into the front-part of the $\pi$E5 area is under design.  This would not only allow the beam line elements, such as the Wien-filter, triplet I and II, plus the MEG collimator system to be used by this experiment, but would also allow access to the MEG detector during running periods, by means of placing a shielding wall just upstream of the MEG detector, as previously adopted during the run periods of experiment R98-03.   
      
      Figure~\ref{fig:area} shows the potential area layout adopted. 
Surface muons of \SI{28}{MeV\per c} will be extracted from the $\pi$E5 ``Z-channel'' and the initial part of the current MEG beam line, including: Triplet I, the \SI{200}{\kilo\volt} crossed-field Wien-filter, Triplet II and the collimator system.
This combination of elements allows for an optimal beam correlated background suppression, as demonstrated
 in Figure~\ref{fig:sep}, which shows the separation quality post collimator,
 between muons and beam  positrons for the above mentioned section of beam
 line. 
Due to the severe restrictions imposed by space, a matching section, including two dipole magnets of \SI{90}{\degree} and \SI{66}{\degree} bending angles respectively, with an intermediate quadrupole doublet or triplet, is envisaged. A final doublet/triplet or intermediate transport solenoid would in turn couple the beam line to the Mu3e superconducting magnet. The beam line vacuum is currently planned to end close to the centre of the target and includes collimation to match the beam-spot to the target size and prevent beam interactions from occurring directly in the small radii inner silicon layers. The expected usable muon intensity at the Mu3e target is between \SIrange[fixed-exponent=8, scientific-notation=fixed, range-units=single]{0.7e8}{1.0e8}{muons\per\second}.

      Although it is understood that simultaneous running of prospective $\pi$E5 experiments is not possible, it is nevertheless clear that the Mu3e phase I beam line will in fact benefit from the availability of MEG beam elements upstream of the detector and that this option, together with provision of the available standard PSI magnets currently in storage or potentially sharable, would cover most of the beam line requirements, except that of  a short coupling solenoid, for which a potential solution is also currently under study and possible dipole vacuum chamber modifications.               
\begin{table*}[t!]
\centering
{\footnotesize 
\begin{center}
%\resizebox{0.8\textwidth}{!}{
\begin{tabular}{llll}
\toprule
\textbf{Laboratory}/  & \textbf{Energy}/ & \textbf{Present Surface} &  \textbf{Future estimated}  \\
\textbf{Beam line}    & \textbf{Power} &  \textbf{$\mu^+$ rate (Hz)} & \textbf{$\mu^+ /\mu^-$ rate (Hz)} \\
\midrule
\\
\textbf{PSI (CH)} &  (590 MeV, 1.3 MW, DC) &  & \\
LEMS & " & $4\cdot10^8 $ & \\
$\pi E5$ & " & $1.6\cdot10^8 $ & \\ 
HiMB & (590 MeV, 1 MW, DC)  &  & $4\cdot10^{10} (\mu^+)$\\
\midrule
\\
\textbf{J-PARC (JP)} &  (3 GeV, 1 MW, Pulsed) &  & \\
                                    &  currently 210 KW           &  & \\
MUSE D-line          &              "                             &$3\cdot10^7$ & \\                                   
MUSE U-line          &              "                             & & $2\cdot10^8 (\mu^+)$  (2012)\\                                                           
COMET                      & (8 GeV, 56 kW, Pulsed) & & $10^{11} (\mu^-)$   (2019/20)\\
PRIME/PRISM          & (8 GeV, 300 kW, Pulsed) & & $10^{11-12} (\mu^-)$   ($> 2020$)\\
\midrule
\\
\textbf{FNAL (USA)} & & & \\
Mu2e  &  (8 GeV, 25 kW, Pulsed) &  & $5\cdot10^{10} (\mu^-)$  (2019/20) \\
Project X Mu2e  &  (3 GeV, 750 kW, Pulsed) &  & $2\cdot10^{12} (\mu^-)$  ($>2022$) \\
\midrule
\\
\textbf{TRIUMF (CA)} & (500 MeV, 75 kW, DC)  & & \\
M20 & " & $2\cdot10^{6}$ &  \\
\midrule
\\
\textbf{KEK (JP)} & (500 MeV, 2.5 kW, Pulsed)  & & \\
Dai Omega & " & $4\cdot10^5$ & \\
\midrule
\\
\textbf{RAL -ISIS (UK)} & (800 MeV, 160 kW, Pulsed)  & & \\
RIKEN-RAL & & $1.5\cdot10^6$ & \\
\midrule
\\
\textbf{RCNP Osaka Univ. (JP)} & (400 MeV, 400 W, Pulsed)  & & \\
MUSIC & currently max 4W & & $10^8 (\mu^+)$  (2012)\\
              & & & means $> 10^{11}$ per MW\\
\midrule
\\
\textbf{DUBNA (RU)} & (660 MeV, 1.65 kW, Pulsed)  & & \\
Phasatron Ch:I-III & & $3\cdot10^4$ & \\              
\bottomrule
\end{tabular}

\caption{\label{tab:facilities} Currently running muon beam facilities around the world used for particle physics experiments and materials science $\mu$SR investigations. Also shown are the planned  next-generation facilities designed for cLFV experiments, together with an estimate of the starting date. The PSI HiMB solution is currently only under study and is included purely for completeness.}

\end{center}
}
\end{table*}
%\end{center}

\section{High intensity muon beamline for phase II running}
\label{sec:HighIntensityMuonBeamlineForPhaseIIRunning}
	
      In order to reach the ultimate sensitivity goal of
      $\mathcal{O}(\num{e-16})$ for the phase II experiment, an unpulsed muon
      stopping rate in the \si{\giga\hertz} region is required. As demonstrated in
      Table~\ref{tab:facilities}, there are no such (pulsed or unpulsed) high-intensity sources of muons currently available world-wide. Future intensity frontier facilities are however in the planning in the US and Japan and are also associated with LFV-experiments, more specifically Mu2e and Project X in the U.S. \cite{Bartoszek2012,Carey:2008zz} and the COMET and PRIME/PRISM experiments in Japan \cite{Cui:2009zz, Akhmetshin2012}. To meet the needs of such experiments a totally new concept is therefore necessary. One such concept, which is still in its infancy, though is proving to be a promising candidate, is the HiMB project at PSI \cite{Kettle:2010}, a next-generation high-intensity muon beam, currently under study. A detailed feasibility study is due to start at PSI in 2013. This concept would provide the basis for a new Mu3e beam line for the phase II measurements, based on the production of surface muons from the Swiss Spallation Neutron Source's (SINQ) spallation target window.
    
\begin{figure*}
\centering
\includegraphics[width=0.55\textwidth]{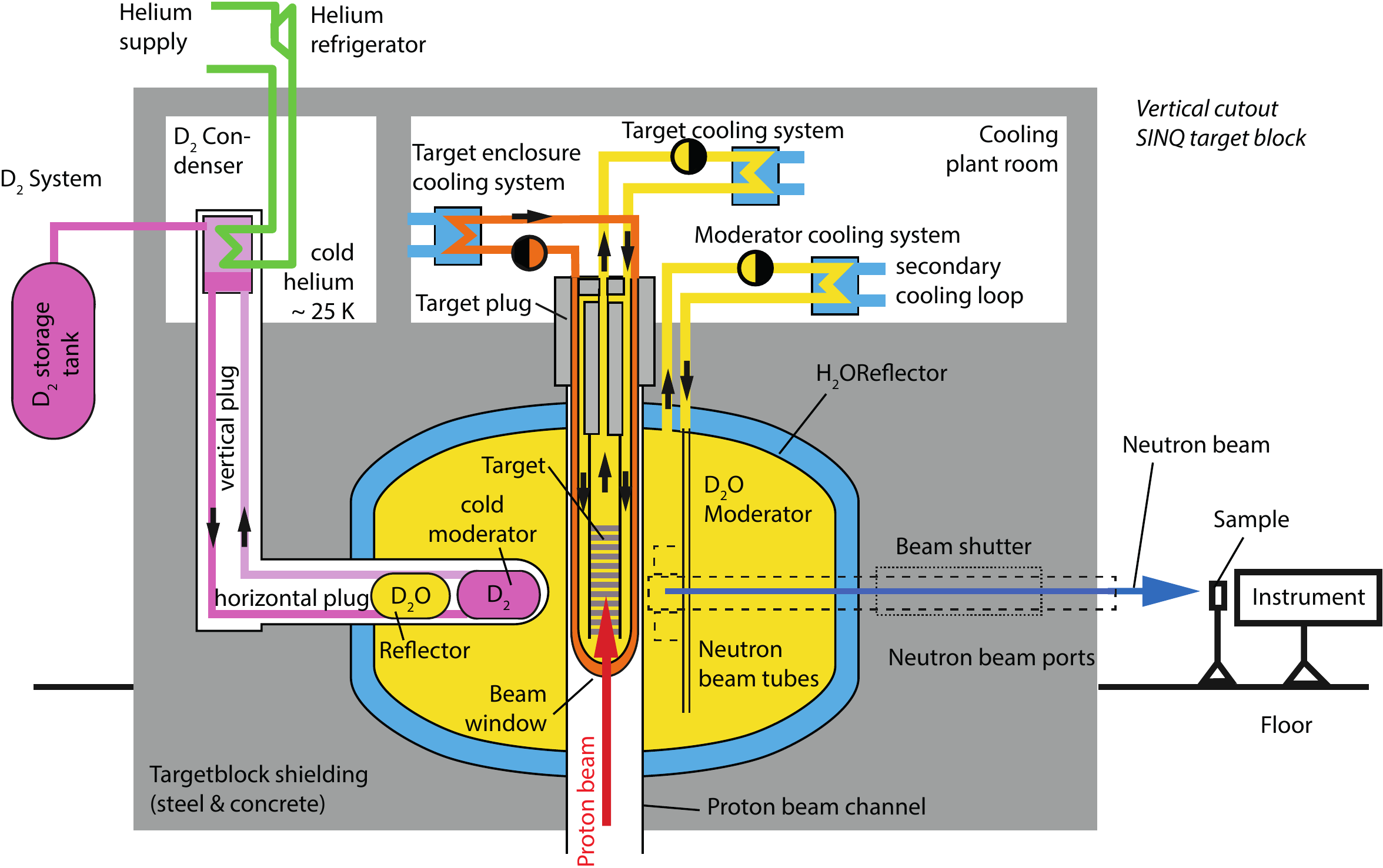}
\includegraphics[width=0.44\textwidth]{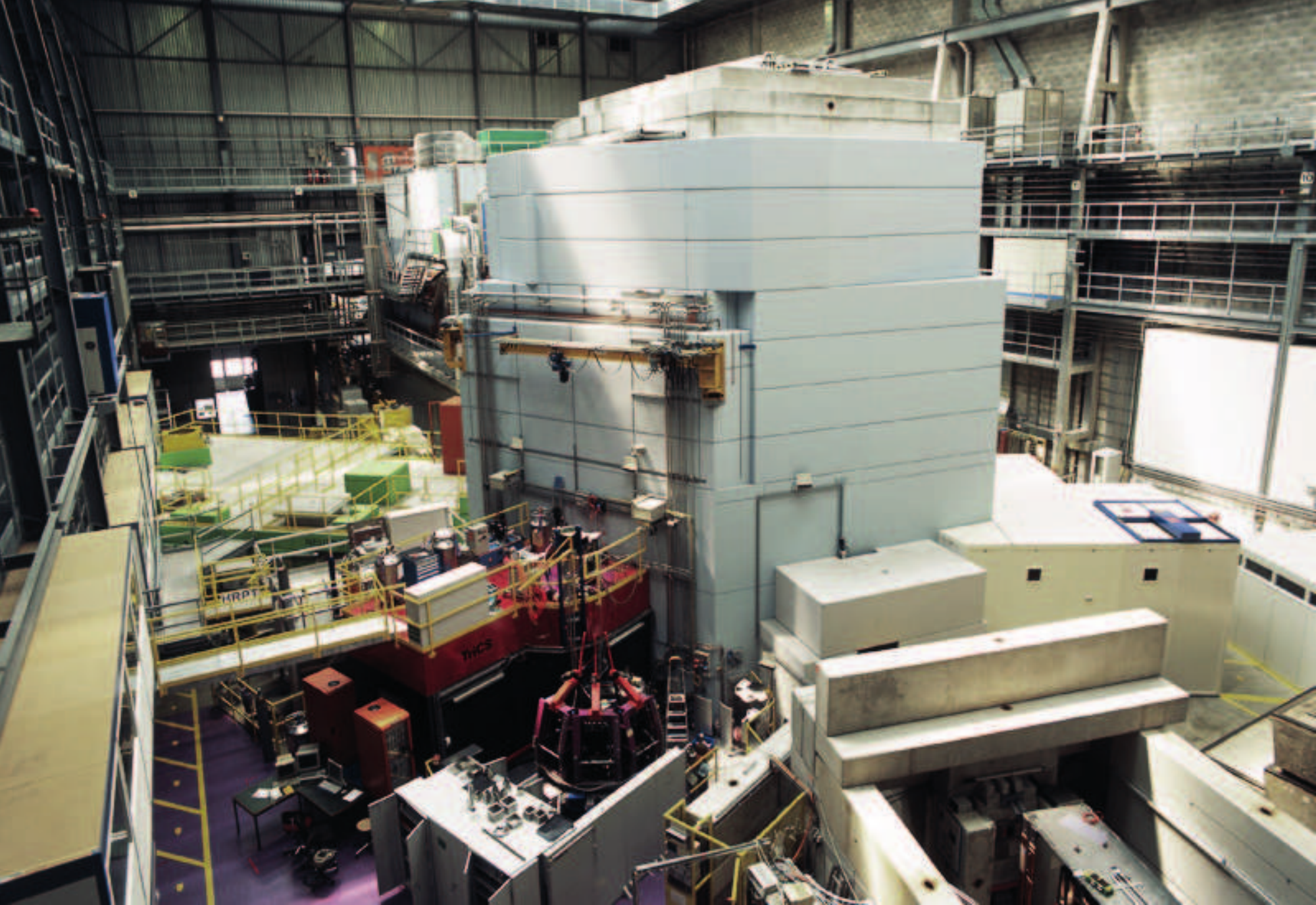}
\caption{(Right) The SINQ spallation neutron source as seen in the experimental hall. The tower in the  foreground, shown in section with the proton beam incident from below. (left) contains the main neutronics components, the spallation source, a Pb/Zr/Al structure cooled using D$_{2}$O, the D$_{2}$O moderator tank and reflector shield and a cold-source of solid deuterium. The upper-part of the tower deals with the cryogenic connections of the source.} \label{fig:SINQ}
\end{figure*}     
      
 The layout of the source in the SINQ hall \cite{Bauer2002}, together with a schematic diagram of the source with the proton beam injected from below is shown in Figure~\ref{fig:SINQ}. The characteristics of the source, which resembles a medium-flux reactor, are that the protons are injected from below and defocussed onto a double-layered aluminium window separated by a D$_{2}$O cooling layer, before being stopped in the target, a ``cannelloni'' construct of lead-filled zircaloy tubes.         

\begin{figure*}
\centering
\includegraphics[width=0.95\textwidth]{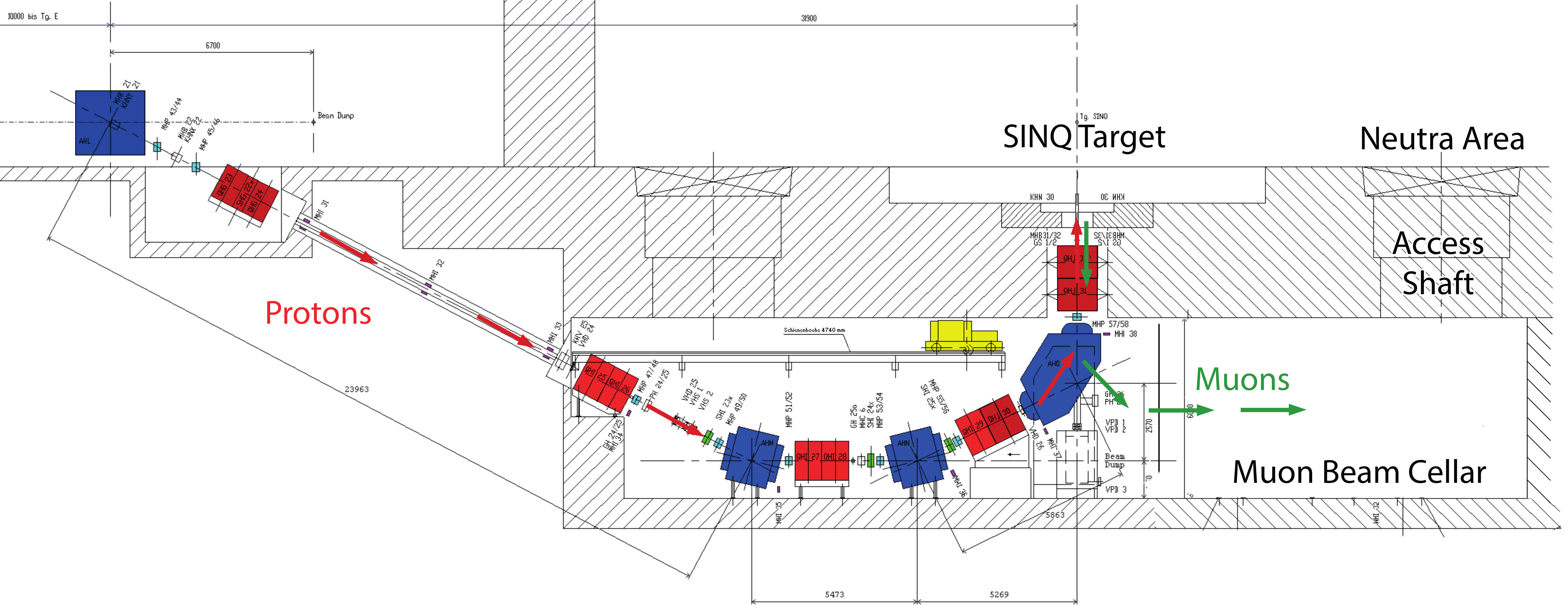}
\caption{ Shows the layout of the proton beam line at SINQ as well as showing the possibility for extracting the muons into the empty cellar region below the SINQ target. The blue elements are dipole bending magnets and the red elements focussing quadrupole magnets. Since the proton and muons are  like-sign charged particles but travelling in opposite directions they will be bent in opposite directions, allowing the muons to access the empty cellar region shown. This is where a muon beam line could be  placed.}\label{fig:cellar}
\end{figure*}

      The HiMB project plans to extract the downward travelling muons produced in the aluminium window via a two-stage channel, the first stage, a solenoidal one, which uses the same beam-tube as the upward travelling protons and extracts the muons, in the opposite direction, to a large collection solenoid connecting to the second stage, a conventional dipole and quadrupole channel planned for the empty service cellar under the SINQ target. The general layout of both the proton channel and the service area are shown in Figure~\ref{fig:cellar}
      
      There are several advantages of this concept, which would lead to a substantial enhancement compared to target E, namely: the increased number of primary proton interactions since \SI{70}{\%} of the beam stops in the target; a much larger pion energy range of up to \SI{150}{\MeV} can be exploited in the case of SINQ, above which the high energy tail of the pion production cross-sections becomes negligible, in the case of Target E this limit is around \SI{45}{\MeV} \cite{crawford1,crawford2}; a substantially larger pion-production volume contribution compared to Target E and finally a significantly larger surface muon production volume.

\begin{figure}
\centering
\includegraphics[width=0.4\textwidth]{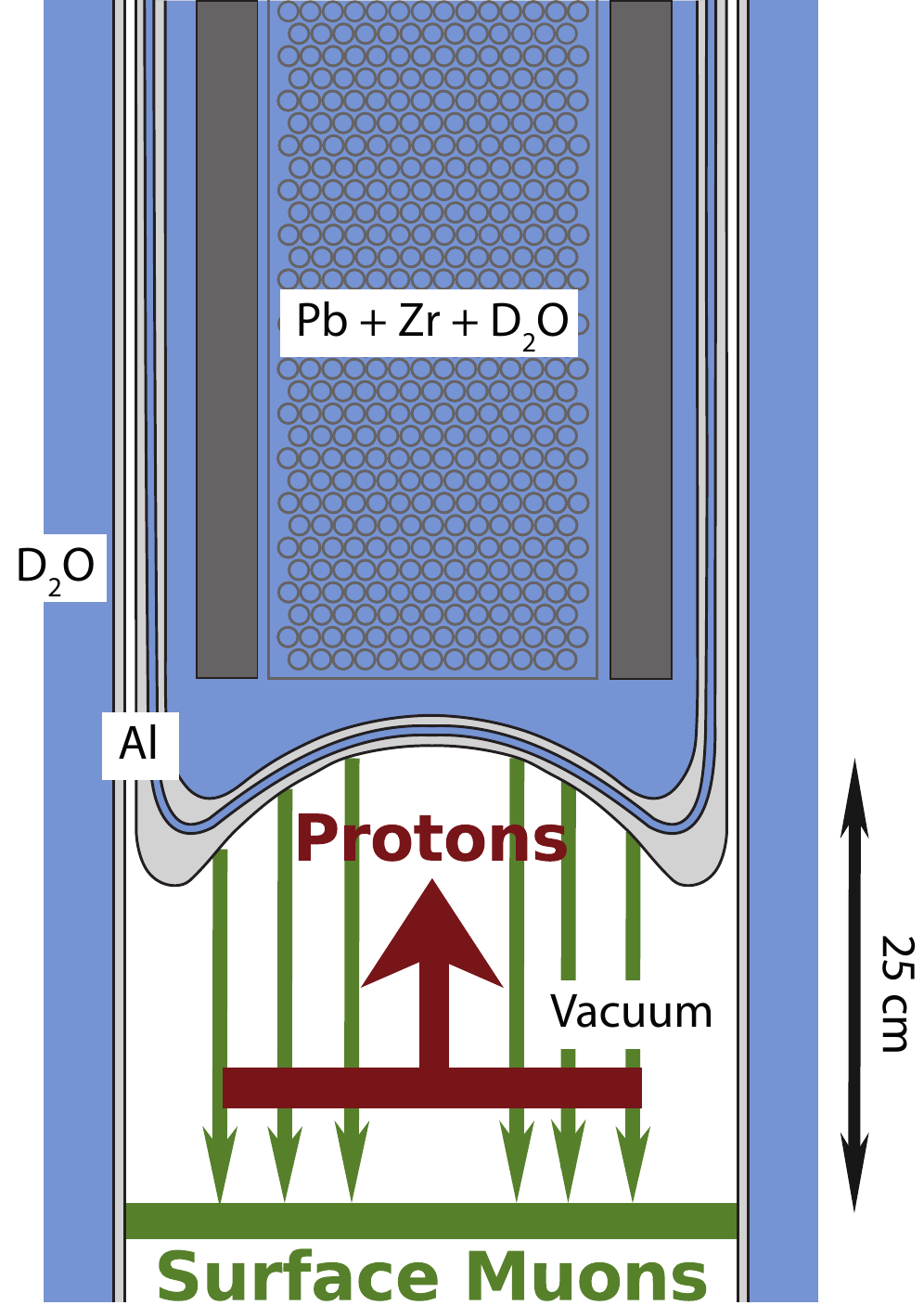}
\caption{The Monte-Carlo model for the SINQ target with all components taken into account. Light grey is the aluminium window, with the D$_2$O cooling channel in the middle. Protons are generated according to a measured 2-D distribution at the dark red plane. Surface muons leaving the target at the correct energy and crossing the light-green plane, \SI{25}{\cm} below the window and moving downwards within the beam-pipe, are counted. }
\label{fig:target}
\end{figure}

      Realistic Monte-Carlo studies were undertaken together with M.~Wohlmuther (Head of the Target Development Group at PSI) using the Los Alamos Laboratory MCNPX code (Monte Carlo n-particle extended code), used also to design the SINQ target. A simulation of the surface muon production rate was made using the complete model of the SINQ target environment. Based on a total of \num{4e8} generated upward moving protons, corresponding to the measured 2D beam profile at SINQ, a complete particle tracking was done using three different event generators. For surface muons, the simulated fluences were determined for the conditions of a particle leaving the target with the correct energy, travelling downwards within the beam-pipe and crossing a horizontal plane \SI{25}{\cm} below the window. This is shown schematically in Figure~\ref{fig:target}. The calculated fluences from the three event generators agreed to within \SI{35}{\%} of each other. Based on the standard event generator, which also has the smallest statistical uncertainty, a summed fluence (E$\leqslant \SI{4.12}{\MeV}$) of surface and sub-surface muons of \SI{1e11}{muons\per\s} is calculated at a proton current of \SI{3}{\mA} on target E, which corresponds to \SI{2.1}{\mA} on SINQ. However, on the assumption that the proton current on Target E will only rise to a maximum of \SI{2.4}{\mA} in the future, a value that has already been achieved during routine test periods since 2010 and taking into account the variation of event generators in the simulation, a conservative estimated fluence of \num[separate-uncertainty=true]{3(1)e10} good surface muons within a \SI{10}{\%} FWHM momentum-byte could clearly be extracted from the SINQ target (c.f. Table~\ref{tab:beamline}).
       
\begin{table} 
\begin{center}
\begin{tabular}{lcc}
\toprule
$P_{0}$ & $\Delta P/P$ & Rate  \\
MeV/c & $\%$ FWHM & Hz \\
\midrule
28 & Full & $(7\pm1)\cdot10^{10}$ \\
28  & 10& $(3\pm1)\cdot10^{10}$ \\
26  & 10 & $(3\pm1)\cdot10^{10}$ \\

\bottomrule
\end{tabular}
\caption{\label{tab:beamline}Estimated surface and sub-surface muon rates based on a proton current of \SI{2.4}{\mA} on Target E and full transmission efficiency.}
\end{center}
\end{table}

\begin{figure}
\centering
\includegraphics[width=0.47\textwidth]{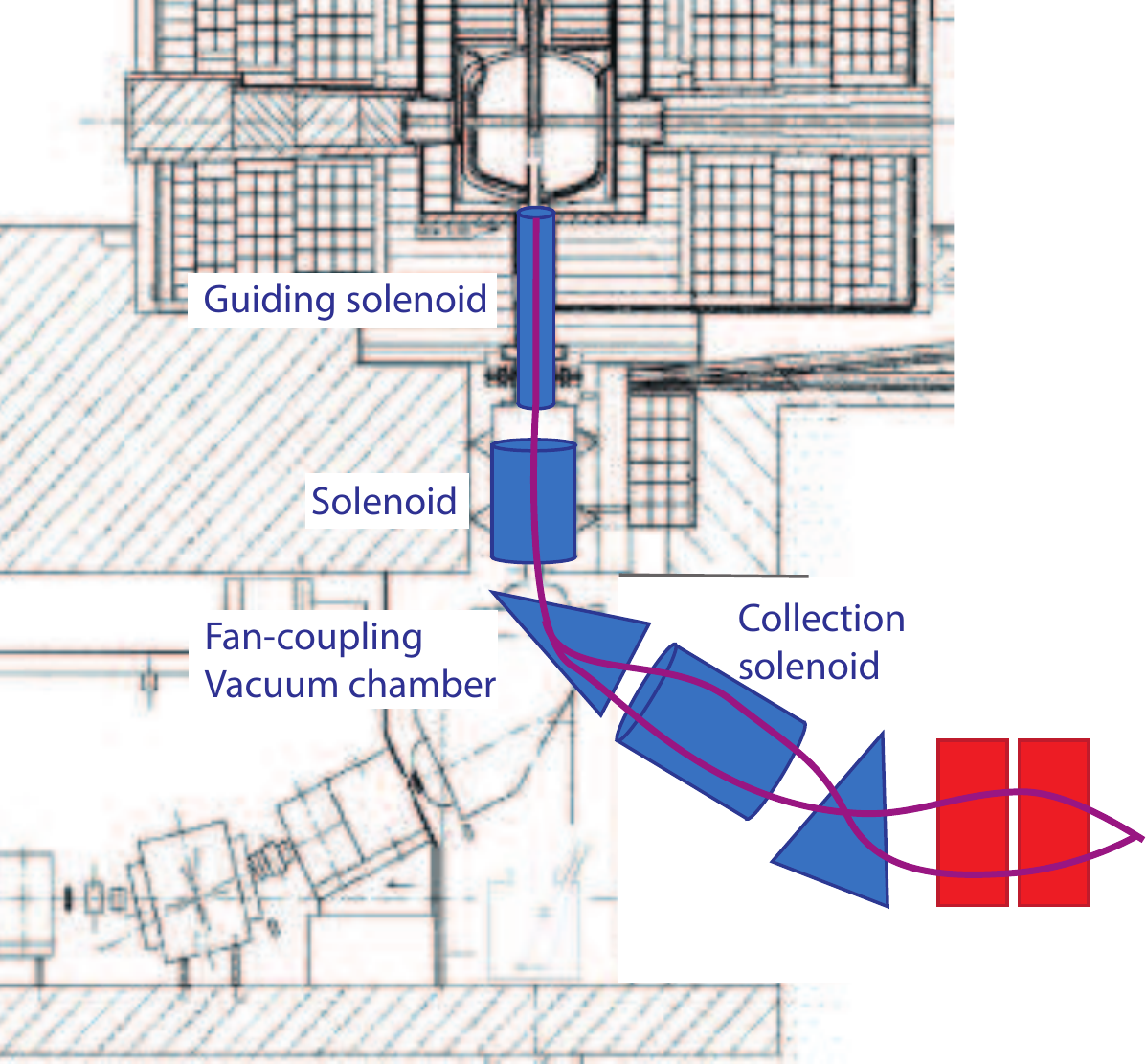}
\caption{Schematic of HiMB muon extraction principle, with a guiding solenoid, followed by a focussing solenoid to satisfy proton and muon transmission. The extraction is done in the fringe-field of the AHO magnet, with a strong-focussing collection solenoid. This is followed by a conventional beam line consisting of a dipole magnet and quadrupole channel.}
\label{fig:channel}
\end{figure}

\begin{figure*}
\centering
\includegraphics[width=0.6\textwidth]{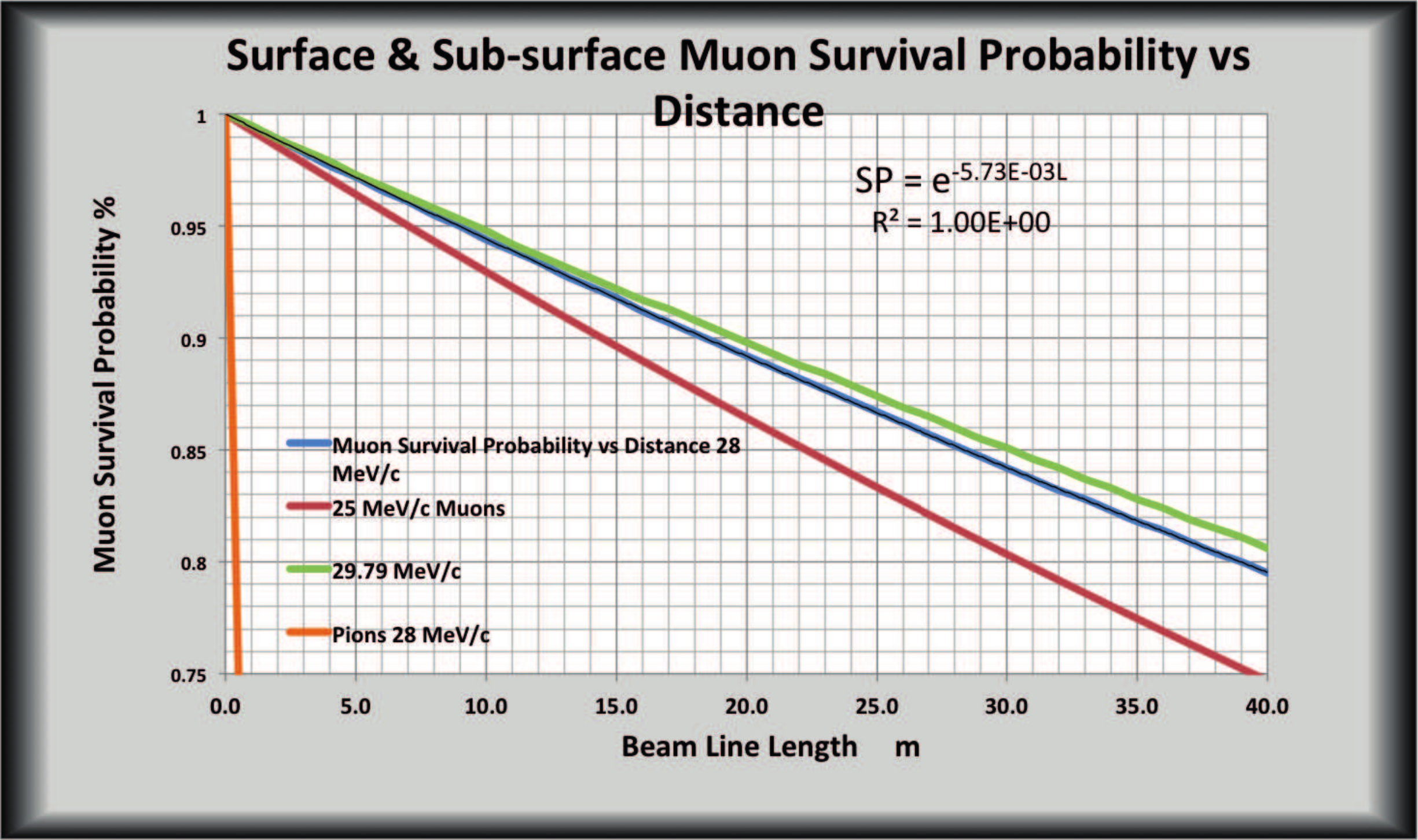}
\caption{Survival probability for muons at the kinematic edge of \SI{29.79}{\MeV\per c} (green curve ), surface muons of \SI{28}{\MeV\per c} (blue curve) and sub-surface muons of \SI{25}{\MeV\per c} (red curve). Also shown (orange curve) is the survival  probability for pion contamination at \SI{28}{\MeV\per c}, which is at the \num{e-12} level.}
\label{fig:survival}
\end{figure*}    
       
      Conclusions seen from the source point of view look very promising and providing the beam can indeed be transmitted without significant losses, the phase II rates of \SI{2}{\giga\hertz} muons could be realized. 
      Finally, the current extraction principle, to be studied extensively in the feasibility study, is demonstrated in Figure~\ref{fig:channel}. Muons originating from the target window are to be guided in a downward direction
using a low-field guiding solenoid. Since the muon momentum is about a factor of 42 times smaller than that of the protons, the low-field for the muons should not dramatically affect the protons. The present radiation-hardened, defocusing quadrupoles QTH 31 and 32, just below the SINQ target, must be replaced by a rotationally symmetric element such as a solenoid, since the muons will not traverse a set of quadrupoles which only focus alternately in the horizontal and vertical planes but have field strengths that are 42 times too high for the muons. The extraction will be done in the fringe-field of the last dipole magnet AHO, which means that the following strong-focussing collection solenoid must be placed close to the ``fan-coupling'' in order to fully collect the beam. Following this solenoid is a dipole magnet whose  function is to bend the beam onto the horizontal plane, where a conventional secondary beam line quadrupole channel could be constructed. As the current cellar ends within just a few meters of the above SINQ hall wall, one could  envisage bending the beam upwards again at the end of the cellar and  extracting to a hall, exterior to SINQ, on the East-side. This would imply a relatively long beam line of order \SIrange{30}{35}{\m}, which is not too problematic from the muon loss (decay-in-flight) point-of-view where \SI{80}{\%} transmission for \SI{28}{\MeV\per c} muons is expected at \SI{40}{\m}, as shown in Figure~\ref{fig:survival}. Also shown are similar plots for the maximum momentum at the kinematic edge of pion-decay, as well as for a lower momentum sub-surface muon beam. Finally, the beam would naturally be free of pion contamination to a level of about \num{e-12}, which is also important for backgrounds relevant to the \mteee decay such as $\pi\rightarrow eee\nu$ and $\pi\rightarrow \mu\nu\gamma$.

 The HiMB project is in its infancy at present and there are many aspects that \balance will be studied in detail within the scope of the planned feasibility study, to show the feasibility of this next generation high-intensity beam line. The initial step taken concerning the muon source intensity, the basis for the HiMB feasibility study, has been shown to be very promising. The next major steps to be studied are the optical extraction of the muons from the proton beam and the solenoidal replacement of the final proton defocussing quadrupole doublet in front of the SINQ target, while still maintaining the safety restraints on the SINQ target. The implementation of such a concept into the SINQ environment could only coincide with a major SINQ shutdown, which is currently planned for the period of 2016-2017.

\chapter{Magnet}
\label{sec:Magnet}

The magnet for the Mu3e experiment has to provide a homogeneous solenoidal
magnetic field for the precise momentum determination of the muon decay
products. In addition it will
also serve as beam optical element guiding the muon beam to the target. 
The basic parameters of the superconducting solenoid magnet, which are
currently being specified for the preparation of order placements, are given in Table~\ref{tab:MagnetProperties}.
The outer dimensions include also an iron field shield.
% are $\SI{1}{\tesla}$ field-strength, $\SI{1}{\m}$ inner diameter, $\SI{2.5}{\m}$ overall length and $\SI{2.5}{\m}$ width and height including an iron field shield, see Table~\ref{tab:MagnetProperties}.

\begin{figure}[tb!]
	\centering
		\includegraphics[width=0.50\textwidth]{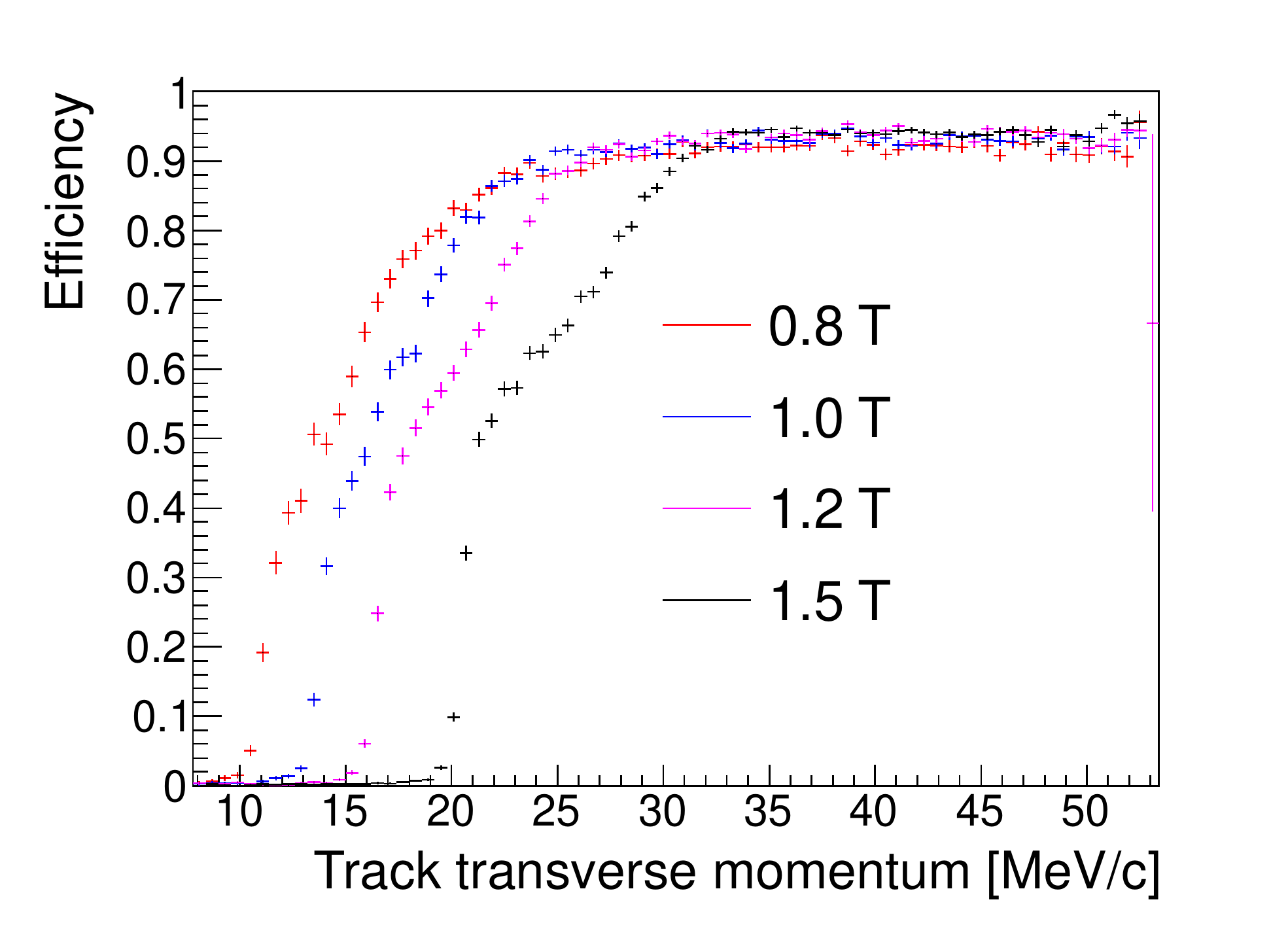}
	\caption{Reconstruction efficiency as a function of track transverse  momentum for different magnetic fields, with recurl stations and without fibre detector.}
	\label{fig:Effcomp}
\end{figure}

\begin{figure}[tb!]
	\centering
		\includegraphics[width=0.50\textwidth]{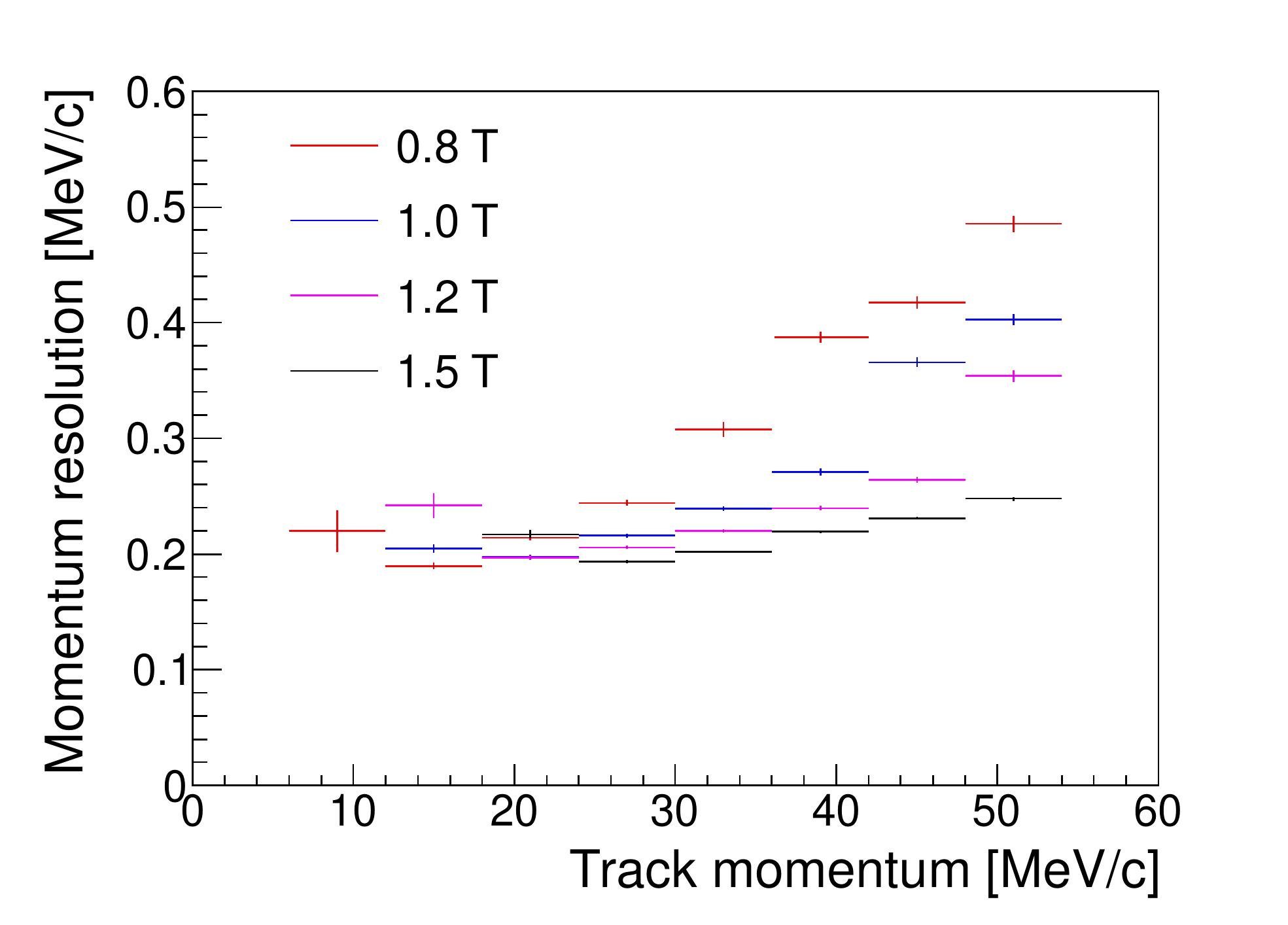}
	\caption{Momentum resolution as a function of track momentum for different magnetic fields, with recurl stations and without fibre detector.}
	\label{fig:Resocomp}
\end{figure}

The nominal magnetic field strength is $\SI{1}{\tesla}$ in the central part,
providing the optimum bending radius in terms of resolution for the proposed
experimental design. 
A higher magnetic field would lead to a loss of acceptance as the low momentum
particles would not reach the central outer pixel layers (see
Figure~\ref{fig:Effcomp}). A lower magnetic field would lead to less magnetic
deflection at constant multiple scattering, leading to worse momentum
resolution (see Figure~\ref{fig:Resocomp}). For systematic studies and to
allow for possible reuses of the magnet for other experimental measurements, 
the field can be varied between 0.8 and $\SI{2}{\tesla}$. 

The dimensions of the cylindrical warm bore of the magnet are $\SI{1}{\m}$ in
diameter and $> \SI{2}{\m}$ in length. The minimum diameter is given by four
times the bending radius of the highest momentum ($\SI{53}{\MeV/\c}$) decay products at the
lowest possible field of $B=\;$\SI{0.8}{\tesla} plus the target diameter. 
In addition the detector support and extraction rail system has to be taken into account when choosing the warm bore diameter.

The total length is a compromise between geometric acceptance for recurling particles and the very tight space constraints for the phase I experimental area at $\pi$E5. In principle a longer solenoidal magnet would provide an intrinsically more homogenous field. At both ends of the magnet it is foreseen to have full access by means of removable flanges. 

\begin{table}[b!]
	\centering
		\begin{tabular}{lr}
			% \hline
			\toprule
			\sc Magnet parameter 								& \sc Value \\ % \hline \hline
			\midrule
			field for experiment								& $\SI{1}{\tesla}$ \\
			field range 												& $0.8 < 1 < \SI{2}{\tesla}$ \\
			warm bore diameter 									& $\SI{1}{\m}$ \\
			warm bore length      							& $> \SI{2}{\m}$ \\
			field description	$\Delta B/B$						& $\leq\num{e-4}$ \\
			field stability $\Delta B/B$ (100 days)  	& $\leq\num{e-4}$ \\
			outer dimensions: length				    & $< \SI{2.5}{\m}$\\
						\hspace{2.8cm}	width					  & $< \SI{2.5}{\m}$\\
						\hspace{2.8cm}	height          & $< \SI{3.5}{\m}$\\
			% \hline
			\bottomrule
			
		\end{tabular}
	\caption{Properties of the Mu3e magnet}
	\label{tab:MagnetProperties}
\end{table}

While the ideal magnet would have a constant field throughout the inner
volume, real solenoid magnets show a drop in field to $\SI{50}{\%}$ at the end
of the coil. The simplest solution would be a longer magnet, which however
does not fit inside the phase I area. Another possibility is to introduce
correction coils at both ends of the magnet, such that the high field region
can be extended. The insertion of several compensating coils would make the
magnet system more complex both in construction and operation due to the need
of additional current settings and power supplies. 
%At the same time a simple analytical description of the magnetic field would
%not be possible anymore.  Verstehe ich nicht A.S.?!
At present the baseline magnet concept foresees three equal coils with a
single power supply. The field change along the z-axis has to be taken into
account for the reconstruction of tracks in the recurl stations by using a
look up table for the field map plus interpolation between these
points. Choosing the right granularity for the look-up table a linear
interpolation of the field will be enough to reach an approximation of
$\Delta B/B\leq\num{e-4}$. For the fast online selection of events the assumption of \balance
a constant field in the active part of the experiment will be sufficient. 
Though the assumption of a constant (maximum) field leads to an systematic bias
towards larger momenta and an increase of online selected background events
from internal radiative muon decays with internal conversions, no signal
events would be lost.
%Those events can be separated from true signal events calculating the missing %energy with the help of the field map plus interpolation method described %above. 

The superconducting magnet is made from three coils of equal size, which has advantages over one long coil in terms of mechanical stability. The small dips in the magnetic field can be treated numerically in the same way as the roll-off of the field to the ends of the magnet. The choice for the superconducting wires or conductors will be driven by commercial availability, since standard components allow for the desired $\SI{2}{\tesla}$ maximum field strength. A warm normal conducting magnet is no option because of size, cost (copper price) and operational stability. Superconducting magnets have an intrinsic immunity against absolute field changes, as they have to run at a constant (low) temperature. If feasible in terms of number of cooling compressors, a dry cooled system will be chosen. 

There will be a magnetic shielding around the magnet. The shielding is
required since the experimental hall is densely populated with other
experiments and infrastructure. Also for the read-out of the proposed experiment it will be much easier to work in a low field environment. A beneficial side effect of the shielding is a gain of field homogeneity inside the magnet and less field dependence on variation of outside parameters. 

The long term stability of the magnetic field should be $\Delta B/B\leq\num{e-4}$ over each 100 day data taking period. This can be achieved by using state of the art magnet power supplies and by permanent measurement of the absolute field with a hall probe inside the experiment. 

The cool-down time for a system of the projected size will be one week and the ramp time will be in the order of one hour. The number and power of the dry compressors will be chosen to fulfill these requirements, in the case of a dry cooled magnet. 
 
The D0 Magnet \cite{Brzezniak:1994rd} fulfills most requirements of the future Mu3e magnet and serves as a prototype for the magnet design process.

\chapter{Stopping Target}
\label{sec:Target}

\nobalance

The main challenge for the design of the stopping target is to optimize the
stopping power on one hand and to minimize the impact on the track measurement on the other hand.
Therefore the stopping target should contain just enough material in the beam
direction to stop most of the $\SI{29}{MeV}$ surface muons 
but should be as thin as possible in the flight direction of decay electrons
measrued in the detector acceptance. 
Usage of a low $Z$ material is advantageous as tails from large angle Coulomb
scatterering are suppressed. 
In addition, the decay vertices should be spread out as widely as possible in
order to reduce accidental coincidences of track vertices and to produce a
more or less even occupancy in the innermost detector layer.

\section{Baseline Aluminium Design}
\label{sec:BaseleineAluminiumDesign}

\begin{figure}[hb!]
	\centering
		\includegraphics[width=0.48\textwidth]{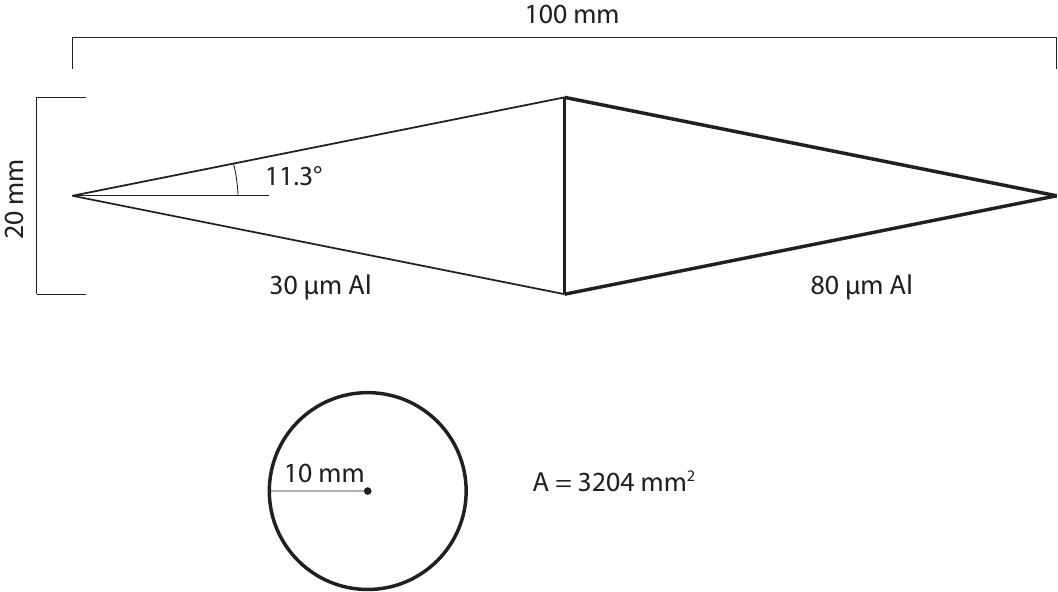}
	\caption{Dimensions of the baseline design target. Note that the material thickness is not to scale.}
	\label{fig:Target}
\end{figure}

These requirements can be met by a hollow double cone target \`a la SINDRUM
\cite{Bellgardt:1987du, bertl2008}. In our baseline design (see
Figure~\ref{fig:Target}), the target is made from $\SI{30}{\micro \meter}$ of
aluminium in the front part and $\SI{80}{\micro \meter}$ aluminium in the back
part, with a total length of $\SI{100}{mm}$ and a radius of $\SI{10}{mm}$. 
This results in an total area of $\SI{3204}{mm^2}$ and an effective target thickness
in beam direction of $\SI{560}{\micro \meter}$ corresponding to 0.063
radiation lengths $X_0$ of Aluminium. 
The target can be suspended from the innermost tracking layer by e.g.~nylon
fishing wire (which we assume in the simulation) and which does not
significantly add material in the beam line.

In the Geant4~\cite{Agostinelli2003250} simulation (see \ref{sec:Simulation}),
about 83.3\% of the muons\footnote{Muons are generated with an energy spectrum
  modeling the one observed in MEG.} that reach the target are
stopped. Obviously, this fraction can be increased by adding material, which
will however lead to additional multiple scattering and thus a reduced
momentum resolution. 
For the phase I experiment, where muon rates rather than momentum resolution is limiting the sensitivity, a thicker target could be envisaged.

Stopping $\SI{2e9}{Hz}$ muons in the target corresponds to about $\SI{1}{\milli\watt}$ of power. Compared with the power dissipation of the sensor chips, this is negligible and easily taken care of by the helium cooling.

\section{Vertex distribution}
\label{sec:VertexDistribution}

The distances between tracks on the target can be reconstructed already online
and used by the event filter farm to reject frames containing only background,
see section~\ref{sec:Farm}.
The only physics process exhibiting three tracks from the same vertex is the
radiative muon decay with an internal conversion at a rate of \num{3.5e-5}.

\begin{figure}
	\centering
		\includegraphics[width=0.4\textwidth]{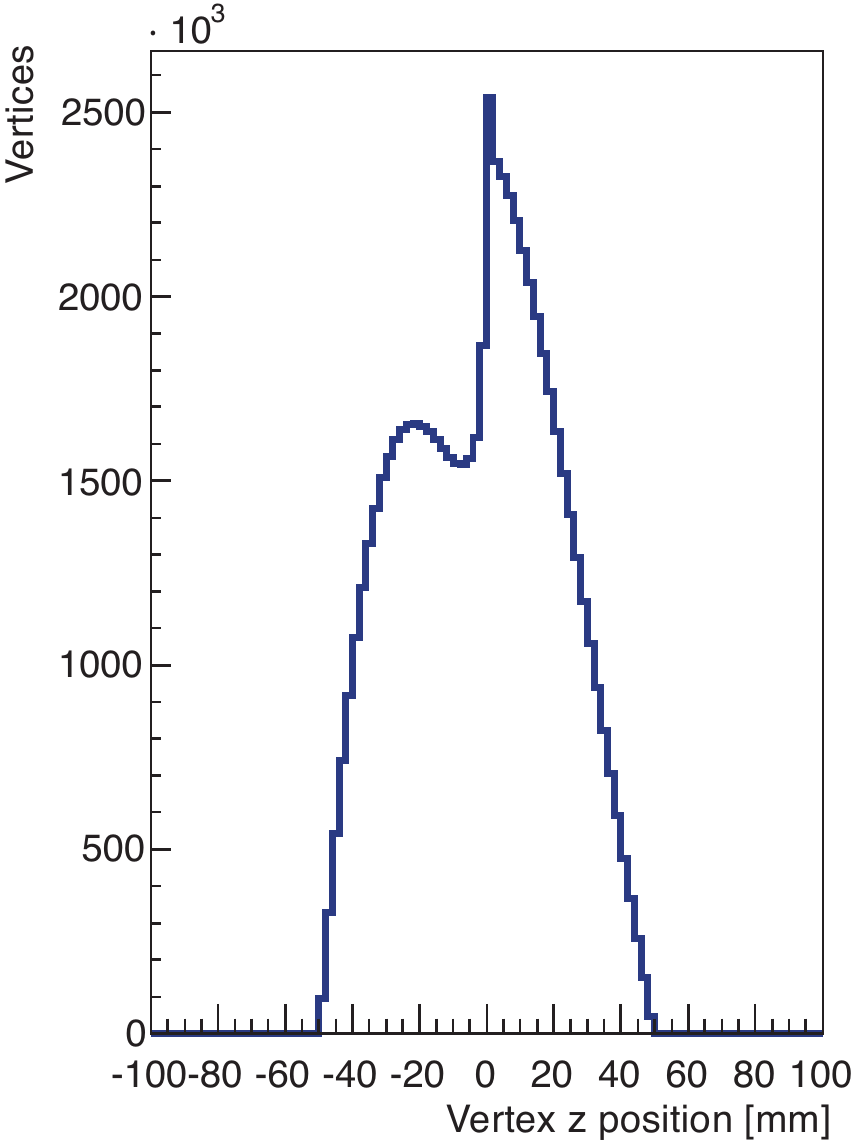}
	\caption{Vertex distribution along the beam direction.}
	\label{fig:Vertex_z}
\end{figure}

\begin{figure}
	\centering
		\includegraphics[width=0.48\textwidth]{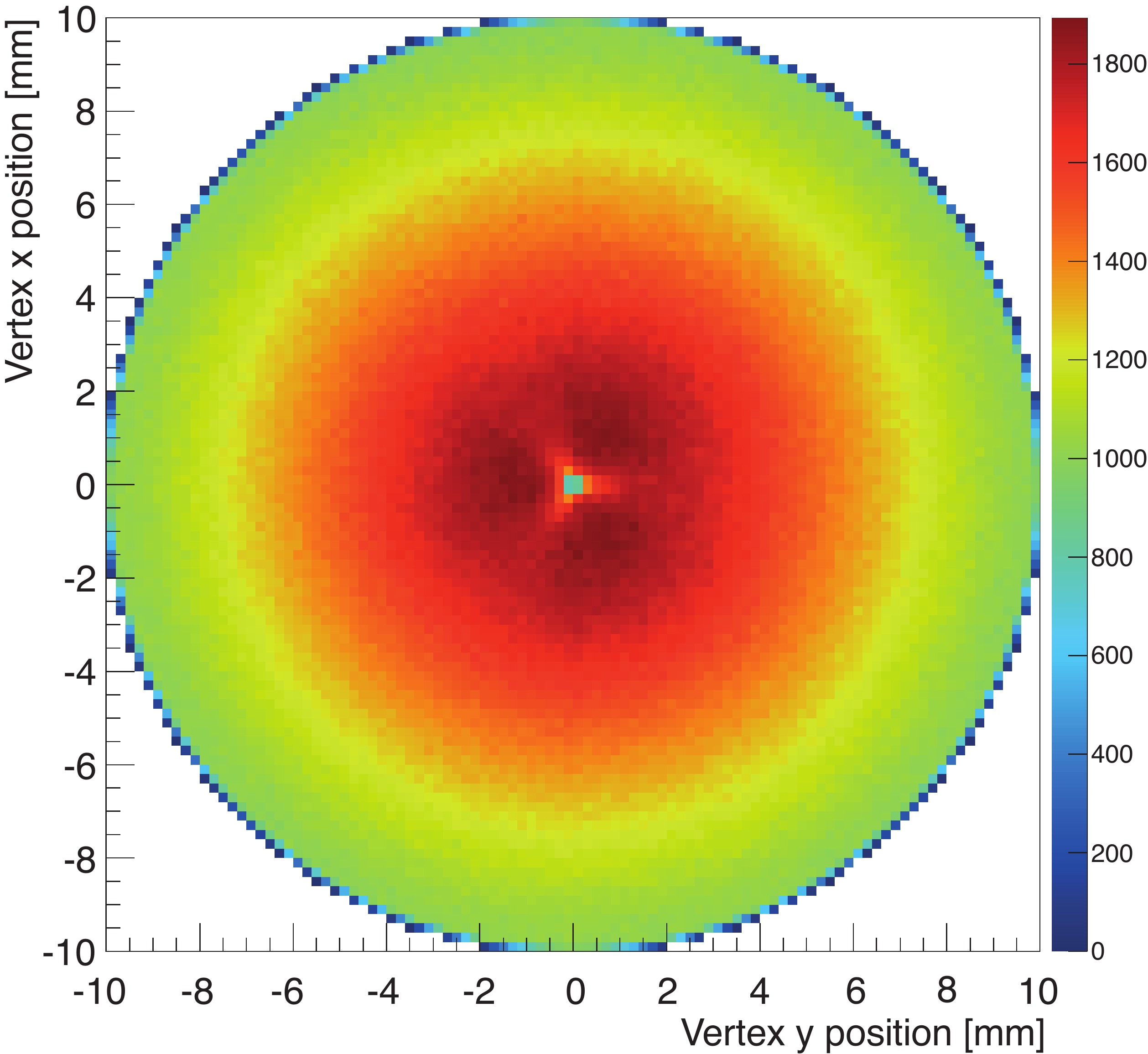}
	\caption{Vertex distribution transverse to the beam direction.}
	\label{fig:Vertex_xy}
\end{figure}

The simulated distribution of vertices (more precisely: intersections of
simulated particles with the target) 
is shown in Figures~\ref{fig:Vertex_z} and \ref{fig:Vertex_xy}. In the longitudinal direction, the effect of the thicker material in the back part can be clearly discerned, whereas for the transverse view, the ``shadows'' of the target suspension are visible in the projected beam profile.

\begin{figure}[p!]
	\centering
		\includegraphics[width=0.48\textwidth]{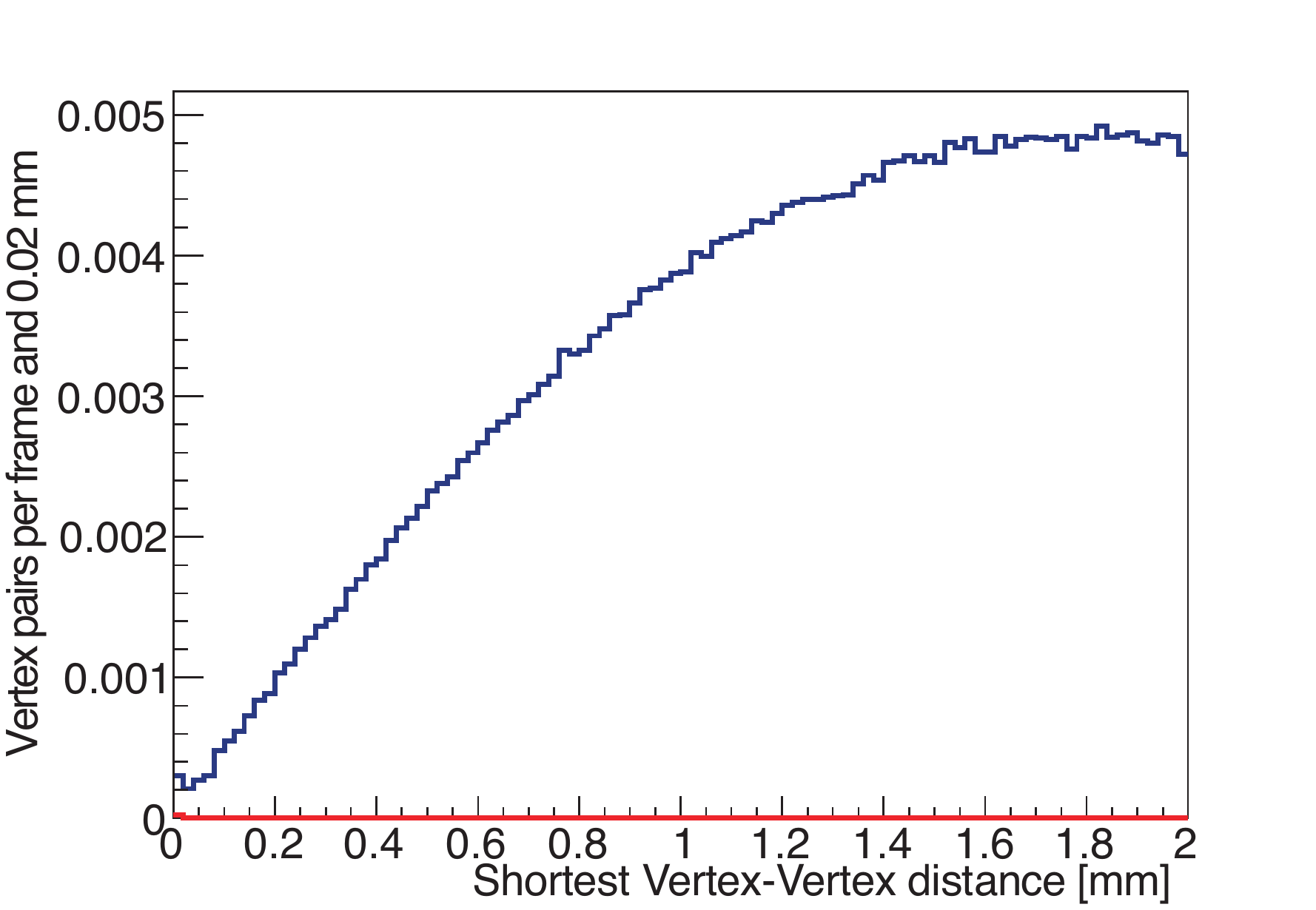}
	\caption{Shortest vertex-vertex distance inside a readout frame with 10 tracks on average (phase IB). Note that every crossing of a simulated electron/positron track is counted as a vertex. The red histogram shows the contribution from internal conversion decays.}
	\label{fig:shortdist_phase1}
\end{figure}

\begin{figure}[p!]
	\centering	\includegraphics[width=0.48\textwidth]{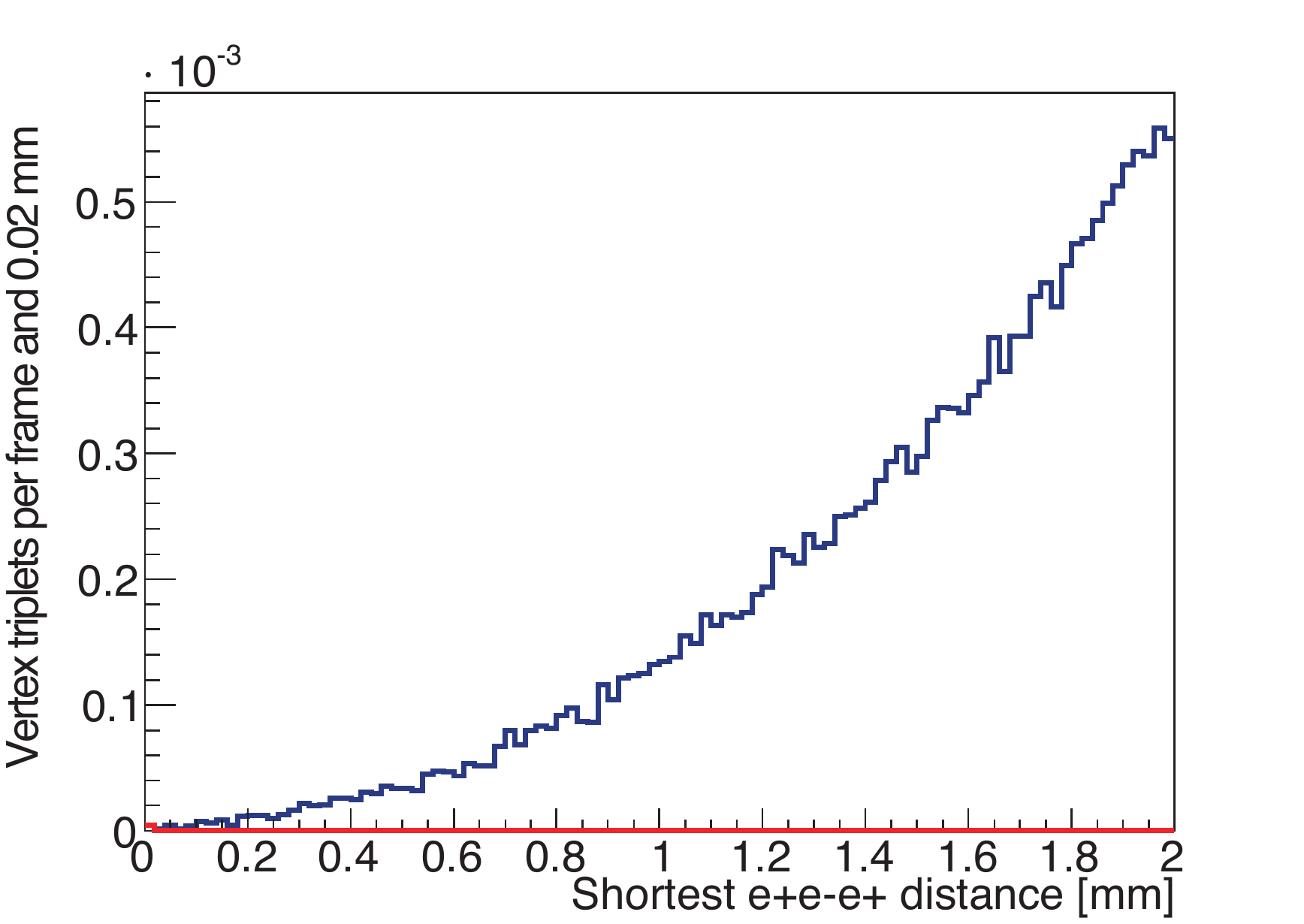}
	\caption{Shortest distance containing three vertices consistent with $e^+e^-e^+$ inside a readout frame with 10 tracks on average (phase IB). Note that every crossing of a simulated electron/positron track is counted as a vertex; charge assignments are made purely on the apparent curvature, i.e.~recurling positrons are counted as electrons. The red histogram shows the contribution from internal conversion decays.}
	\label{fig:shortdist3_phase1}
\end{figure}

\begin{figure}[p!]
	\centering
		\includegraphics[width=0.48\textwidth]{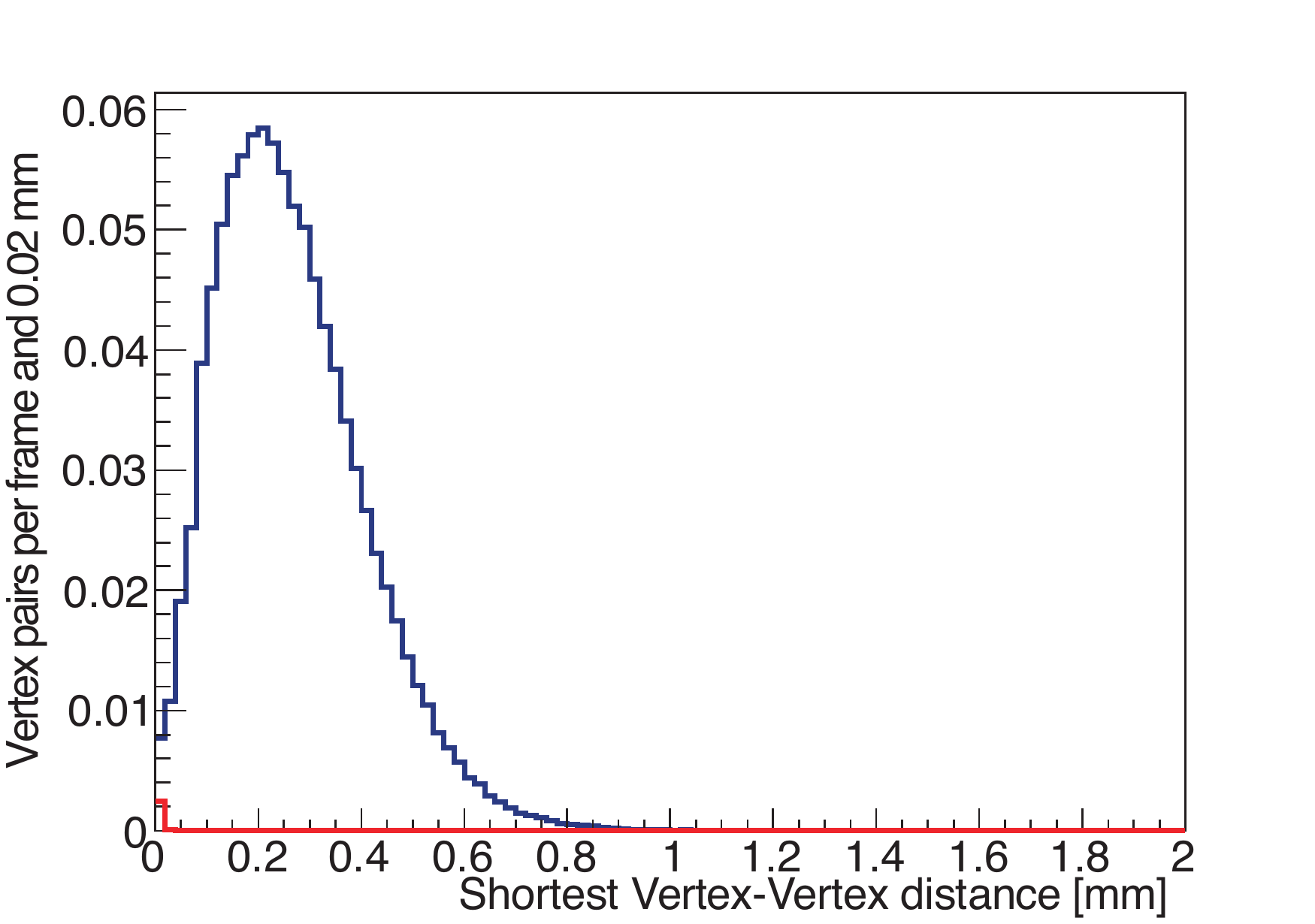}
	\caption{Shortest vertex-vertex distance inside a readout frame with 100 tracks on average (phase II). Note that every crossing of a simulated electron/positron track is counted as a vertex. The red histogram shows the contribution from internal conversion decays.}
	\label{fig:shortdist_phase2}
\end{figure}

Figure \ref{fig:shortdist_phase1} shows the shortest distance between two vertices in a frame for $\SI{2e8}{Hz}$ muon stop rate. Less than 10\% of the frames have tracks that come within $\SI{1}{mm}$ of each other on the target. Figure \ref{fig:shortdist3_phase1} shows the shortest distance within which three tracks approach on the target surface; one of the tracks has to be assigned negative charge, either because it is a true electron or a recurling positron track. Figures \ref{fig:shortdist_phase2} show the same distributions for $\SI{2e9}{Hz}$ muon stop rate. Here all frames have two tracks approaching to closer than a millimeter, but a three track coincidence requirement still has a considerable suppression power, which can surpass a factor of $\num{e-3}$, if recurlers can be identified with high efficiency (see also chapter \ref{sec:Farm}).

\section{Alternative Designs}
\label{sec:AlternativeDesigns}

\subsection{Material Alternatives}
\label{sec:MaterialAlternatives}

If the aluminium foil design proves unworkable or not mechanically stable enough, it could be replaced by an equivalent design in carbon fibre reinforced plastic (CFRP), where the material thickness would be approximately double. $\SI{200}{\micro \meter}$ thin CFRP structures were built e.g.~for the CMS pixel detector upgrade \cite{Erdmann:2010zz}. 

Another material option would be the use of a low density foam-like material such as Rohacell, as used in the SINDRUM experiment \cite{Grab1985}.

\subsection{Active target}
\label{sec:ActiveTarget}

\begin{figure}[p!]
	\centering
		\includegraphics[width=0.48\textwidth]{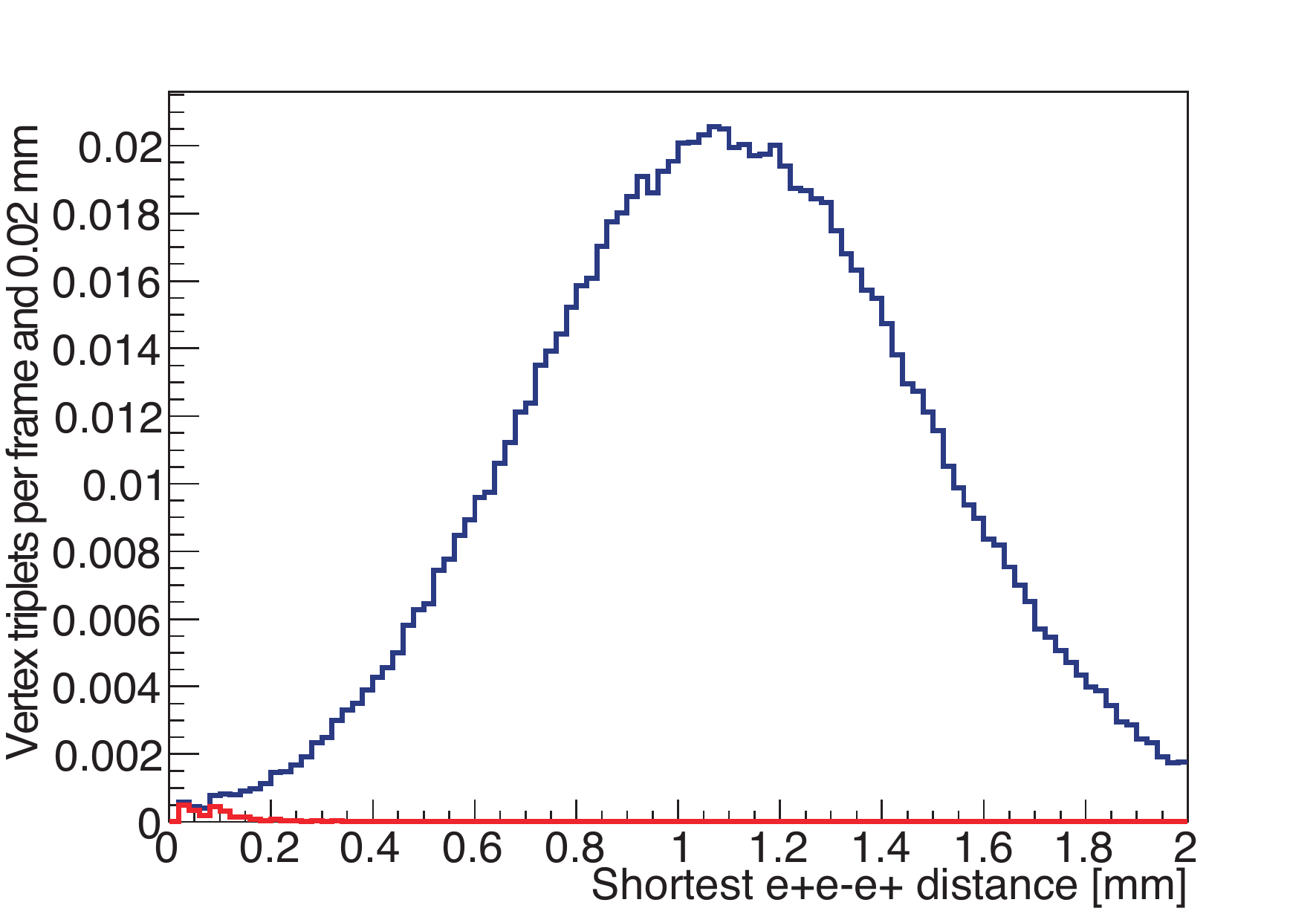}
	\caption{Shortest distance containing three vertices consistent with $e^+e^-e^+$ inside a readout frame with 100 tracks on average (phase II). Note that every crossing of a simulated electron/positron track is counted as a vertex; charge assignments are made purely on the apparent curvature, i.e.~recurling positrons are counted as electrons. The red histogram shows the contribution from internal conversion decays.}
	\label{fig:shortdist3_phase2}
\end{figure}

We have also considered the use of our Kapton-sensor assemblies (see chapter \ref{sec:Pixel}) as an active target. This would lead to a vertex separation ability in the order of the pixel size ($\SI{80}{\micro \meter}$), as opposed  to the $\approx \SI{200}{\micro \meter}$ expected from track extrapolation. At an appropriate inclination angle, a wedge of two chips with $6 \times \SI{2}{cm^2}$ active area and $\SI{50}{\micro \meter}$ thickness presents a comparable amount of material to the beam (however, the cabling and cooling required would add significant additional material at least in the downstream direction). Such an arrangement would sacrifice $\phi$-symmetry.

The hit rate expected in an active target exceeds $\SI{3}{GHz}$ for high intensity running, corresponding to about $\SI{8}{KHz}$ per pixel, or $\SI{500}{MHz}$ per reticle, far beyond the $\SI{80}{MHz}$ expected in the innermost sensor layer. This in turn would necessitate the development of a completely new read-out block for the sensor chips. The gain in selectivity when going from $\approx \SI{200}{\micro \meter}$ vertex resolution to \balance $\SI{80}{\micro \meter}$ does not justify the cost and technical risks associated with the active target.

\chapter{The Mu3e Pixel Detector}
\label{sec:Pixel}

\nobalance

The Mu3e pixel tracker is to be built from High-Voltage Monolithic Active Pixel Sensors (HV-MAPS) thinned to $\SI{50}{\micro\meter}$. Signal and power lines are aluminum traces on a Kapton flex-print, which, together with a Kapton prism, also serves as a support structure. The detector should be cooled with gaseous helium.

\section{HV-Maps Sensor}
\label{sec:HVMapsSensor}

We propose to use Monolithic Active Pixel Sensors (MAPS) as tracking detectors as they integrate sensor and readout functionalities in the same device and thus greatly reduce the material budget. Classical concepts like hybrid designs have usually a higher material budget due to additional interconnects (bonds) and extra readout chips, which compromise the track reconstruction performance especially at low track momentum. 

\begin{figure}[hb]
	\centering	\includegraphics[width=0.48\textwidth]{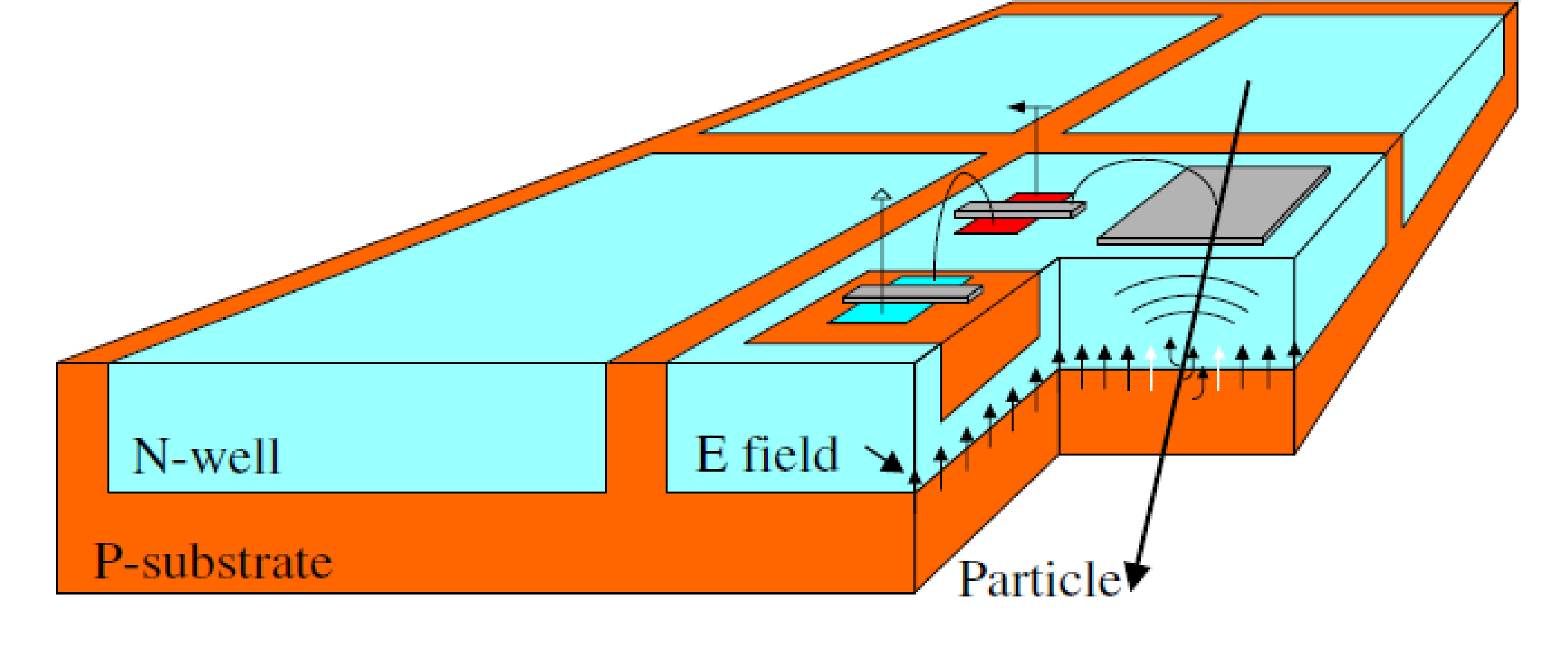}
	\caption{Sketch of the MAPS detector design from \cite{Peric:2007zz}.}
	\label{fig:cmos_sketch}
\end{figure}

In the first MAPS designs ionization charges were collected mainly by 
diffusion with a timing constant of several hundreds of nanoseconds.
MAPS designs with high bias voltages exceeding $\SI{50}{V}$, however 
overcome this problem by collecting charges via drift and provide timing resolutions of better than $\SI{10}{ns}$. 

We propose to use the High Voltage MAPS design with the amplifier electronics 
completely implemented inside the deep pixel N-well, which was first proposed
in~\cite{Peric:2007zz} and since successfully tested~\cite{Peric2010504,Peric2010}, see also section \ref{sec:CharacterisationOfThePrototypes}. 

Figure~\ref{fig:cmos_sketch} shows a sketch of a Monolithic Pixel Detector. The readout circuitry allows an efficient zero suppression of pixel information and the implementation of timestamps to facilitate the assignments of hits between different pixel layers. 

For readout designs providing $50-\SI{100}{ns}$ timing resolutions
power consumptions of about $\SI{150}{mW\per cm\squared}$ are expected~\cite{PericPriv}.

Because of the small size of the active depletion zone,  
the detectors can be thinned down to $\SI{50}{\micro\meter}$ or less. 
By thinning, the material budget can be significantly reduced and becomes,
averaged over the tracking volume
comparable to ordinary gaseous detectors.

A further advantage is that HV-MAPS can be implemented in a ``cheap'' commercial process. We use the AMS/IBM $\SI{180}{nm}$ HV-CMOS process \cite{IBM7HV}, which was developed mainly for the automotive industry, and thus offers long-term availability as well as being specified for a very wide range of operating conditions. The process offers a maximum reticle size of $2\times\SI{2}{\cm^2}$.
One of the few disadvantages of the process is the fact that the first metalization layer is in copper, thus introducing a small amount of medium $Z$ material. There are however plans for replacing also that layer with an aluminium metalization in a future version of the process.

Radiation-tolerance studies of the HV-MAPS sensors in $\SI{180}{nm}$
technology are ongoing also on other projects (e.g.~ATLAS pixel R\&D). Several
test chips with similar pixel electronics as MUPIX have been irradiated at PS
(CERN) up to doses between 80 and $\SI{430}{MRad}$ (the latest corresponds to
a fluence of nearly $\SI{1e16}{neq\per cm\squared}$). The results are
promising. Despite of the use of standard NMOS layouts, the chip irradiated to
80 MRad still detects the particles radiated by a $^{90}$Sr source. The setup
irradiated to $\SI{430}{MRad}$ is strongly activated, and no accurate
detection of $^{90}$Sr signals is possible. The chip was able to detect
particles from the beam up to $\SI{410}{MRad}$ dose. The main radiation effect
is that the electronics suffers from ionizing effects such that it is 
difficult to find a proper operation point for the pixel amplifier.
% Too much of details for proposal
%It is difficult to find a proper bias point for the pixel amplifier. This can
%be expected since several important NMOS transistors operate in the weak
%inversion region with bias current of only $\SI{100}{pA}$. Radiation induced
%leakage currents can influence these devices strongly. We expect that the use
%of radiation tolerant layout techniques, such as enclosed NMOS gates will
%improve radiation tolerance of the electronic components.
As the Mu3e experiment is performed at a muon beam-line the requirements on
radiation hardness are not comparable to those at hadron colliders like the
LHC.
The radiation tests of the HV-MAPS sensors done so far all indicated that there
will be no radiation damage even at highest muon rates at phase~II.

\section{Sensor specification}
\label{sec:SensorSpecification}

\begin{table}
	\centering
	\small
		\begin{tabular}{lrr}
			\toprule
			 & Small Sensor & Large Sensor\\
			\midrule
			 Pixel Size [$\SI{}{\micro\meter\squared}$]& $80\times80$ & $80\times80$\\
			 Sensor Size [$\SI{}{\cm\squared}$]& $1.1\times2$ & $2\times2$\\
			 Assembly & $1\times3$ & $1\times3$\\
			 Assembly size [$\SI{}{\cm\squared}$] & $1.1\times6$ & $2\times6$\\
			 Max.~LVDS links & 4 & 2\\
			 Bandwidth [$\SI{}{Gbit/s}$] & 3.2 & 1.6\\
			\bottomrule
		\end{tabular}
	\caption{Sensor specifications}
	\label{tab:SensorSpecifications}
\end{table}

We plan to use two types of sensors in the Mu3e experiment, a smaller one for the inner layers and a larger one for the outer layers, see Table~\ref{tab:SensorSpecifications}. The pixel size is $80\times\SI{80}{\micro\meter\squared}$, much smaller than the multiple scattering contribution. 

The wafers are to be thinned to $\SI{50}{\micro\meter}$. If the yields permit it, we will cut strips of three subsequent sensors from the wafers and mount them in one piece.

The sensor output is zero suppressed and consist of time-stamps and addresses of hit pixels, serialized on a $\SI{800}{Mbit/s}$ low voltage digital signaling (LVDS) link. The sensor is configured via a JTAG interface \cite{JTAG}. Including supply voltages, we expect about 30 pads (and thus bond wires) to connect the chip to the Kapton flex-print. 

\section{Path towards the Full Sensor}
\label{sec:PathTowardsTheFullSensor}

\subsection{The MUPIX Prototypes}
\label{sec:TheMUPIXPrototypes}

First purpose-built sensor prototypes (the MUPIX series of chips) 
became available in 2011. 

\subsubsection{MUPIX1 and 2}
\label{sec:MUPIX1And2}

\begin{figure}
	\centering
		\includegraphics[width=0.30\textwidth]{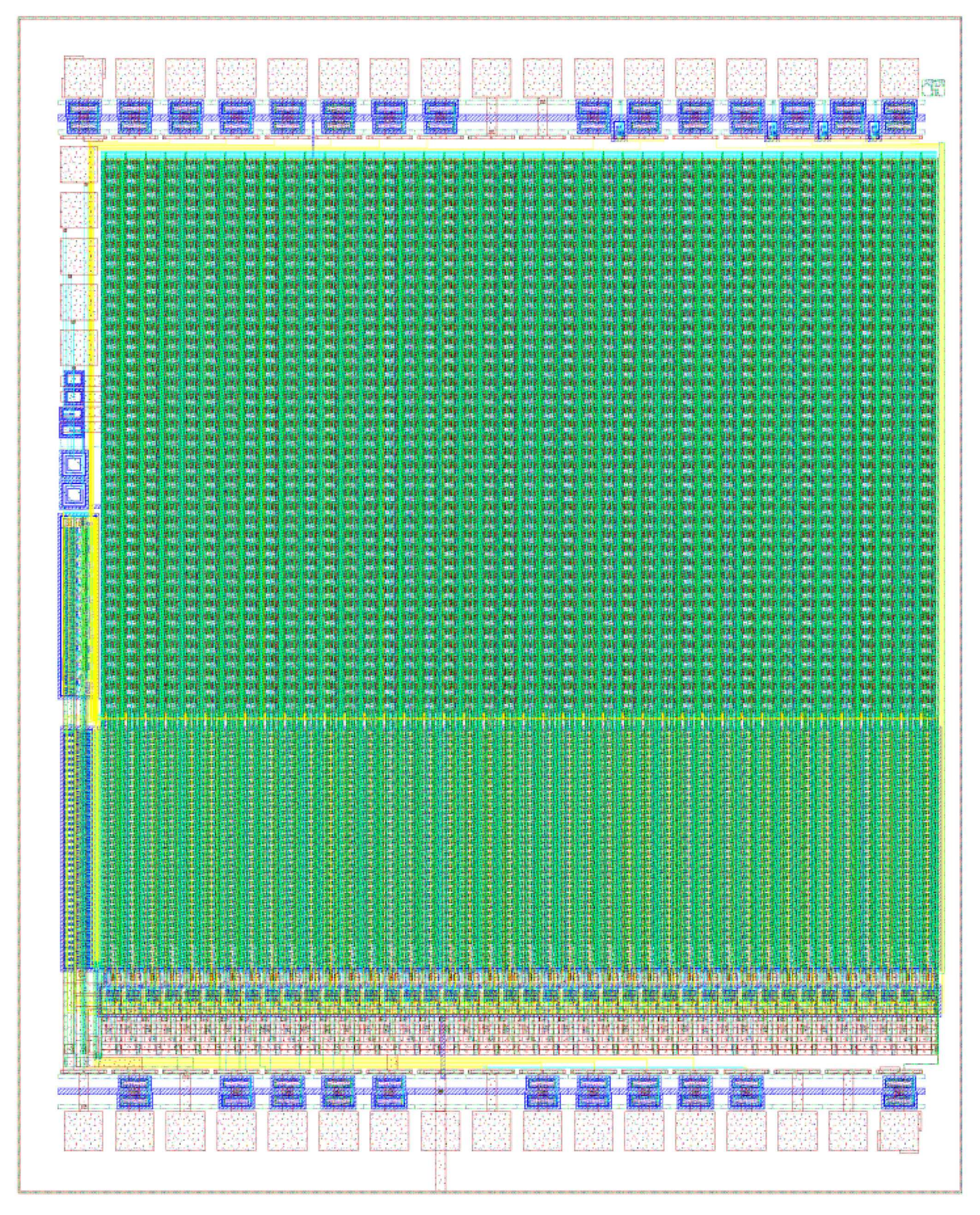}
	\caption{Design view of the MUPIX2 chip (actual size about 1.8 by $\SI{2.5}{mm}$).}
	\label{fig:Mupix2}
\end{figure}

\begin{figure}
	\centering
		\includegraphics[width=0.50\textwidth]{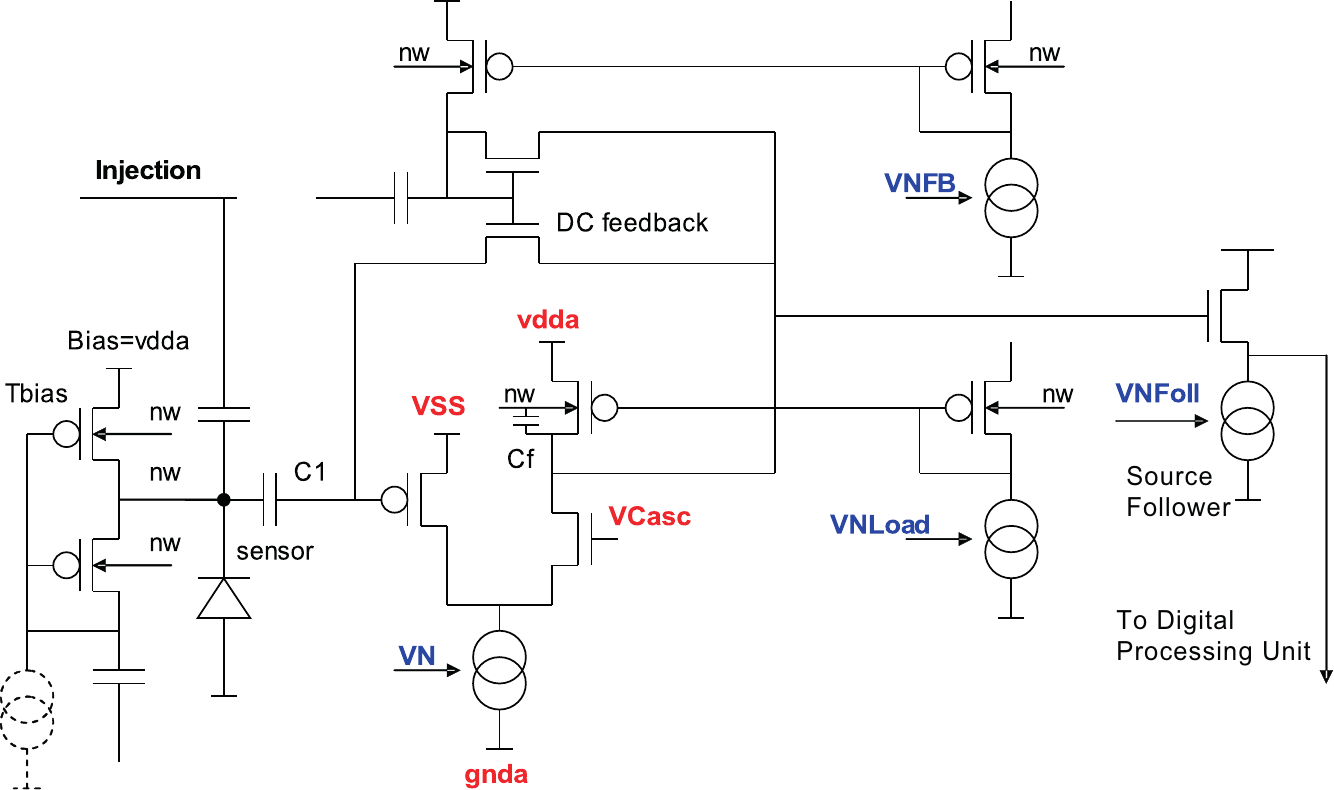}
	\caption{Schematic of the pixel cell analog electronics in the MUPIX chips. \emph{nw} stands for n-well.}
	\label{fig:PixelElectronics}
\end{figure}

\begin{figure}
	\centering
		\includegraphics[width=0.50\textwidth]{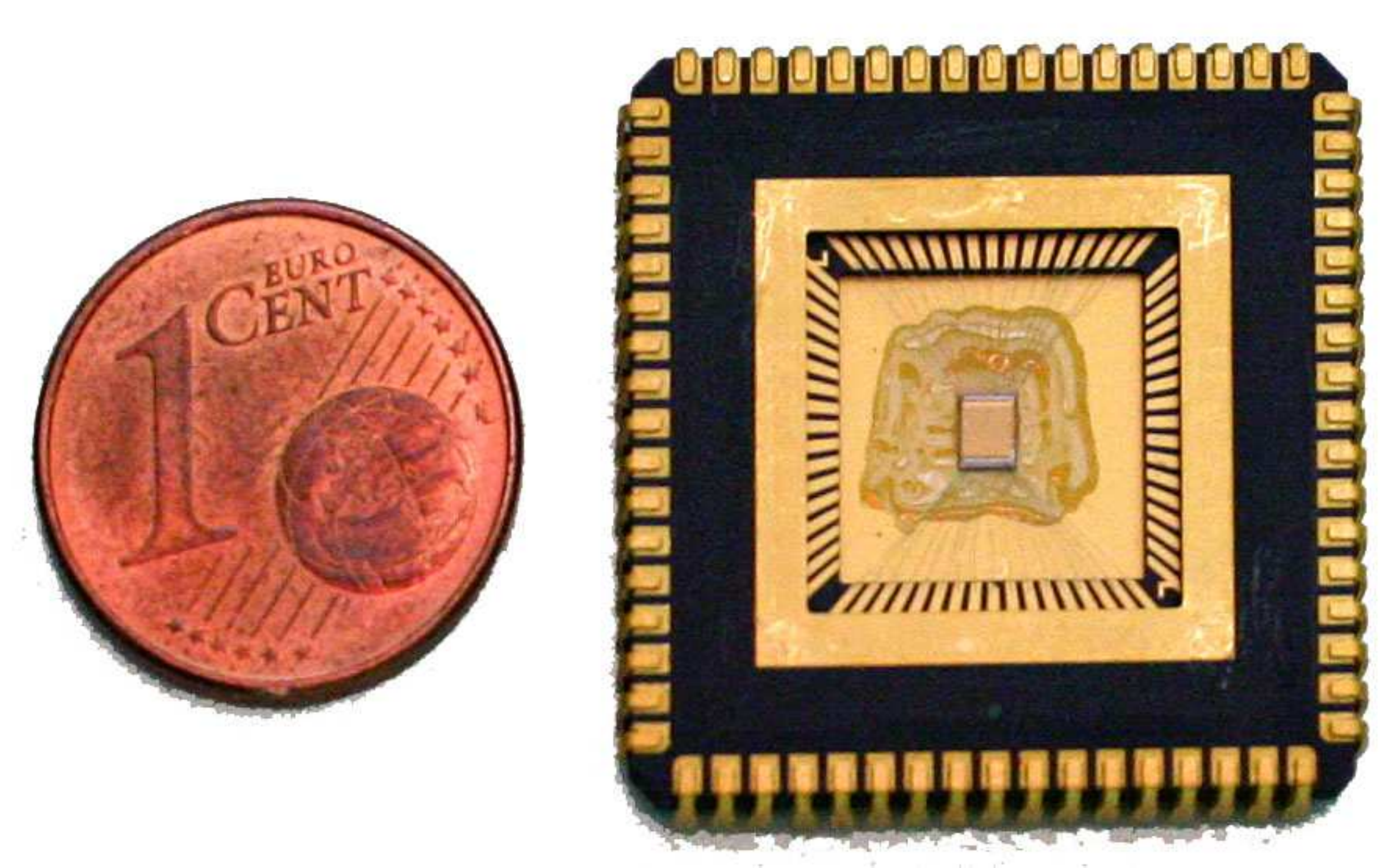}
	\caption{MUPIX2 sensor on ceramic carrier with one Euro-cent for scale comparison.}
	\label{fig:SensorandCent}
\end{figure}

The MUPIX 1 and 2 are small demonstration prototypes with a matrix of $42 \times 36$ pixels of $30\times\SI{39}{\micro\meter\squared}$ size for an active area of approximately $\SI{1.8}{\milli\meter\squared}$, see Figures~\ref{fig:Mupix2} and \ref{fig:SensorandCent}. Each pixel consists of the sensor diode, a charge-sensitive amplifier and a source follower to drive the signal to the chip periphery. In addition there is a capacity allowing to inject test charges. On the periphery, a comparator turns the analog signal into a digital time-over-threshold (ToT) signal. The threshold of the comparators is set globally for the chip and adjusted pixel-per-pixel with a 4-bit tune digital to analog converter (DAC). See Figure~\ref{fig:PixelElectronics} for an overview of the pixel electronics.
 In the test chips, the comparator output of an individual pixel can be observed via a dedicated output line. Alternatively, the whole chip can be read out via a shift register, where all the available information is whether a particular pixel saw a signal during an active gate.

The MUPIX 1 chip had an issue with feedback in the comparator that occasionally led to double pulses. This issue has been fixed in MUPIX 2, which in addition contains temperature sensors.

\subsubsection{MUPIX3}
\label{sec:MUPIX3}

\begin{figure}
	\centering
		\includegraphics[width=0.50\textwidth]{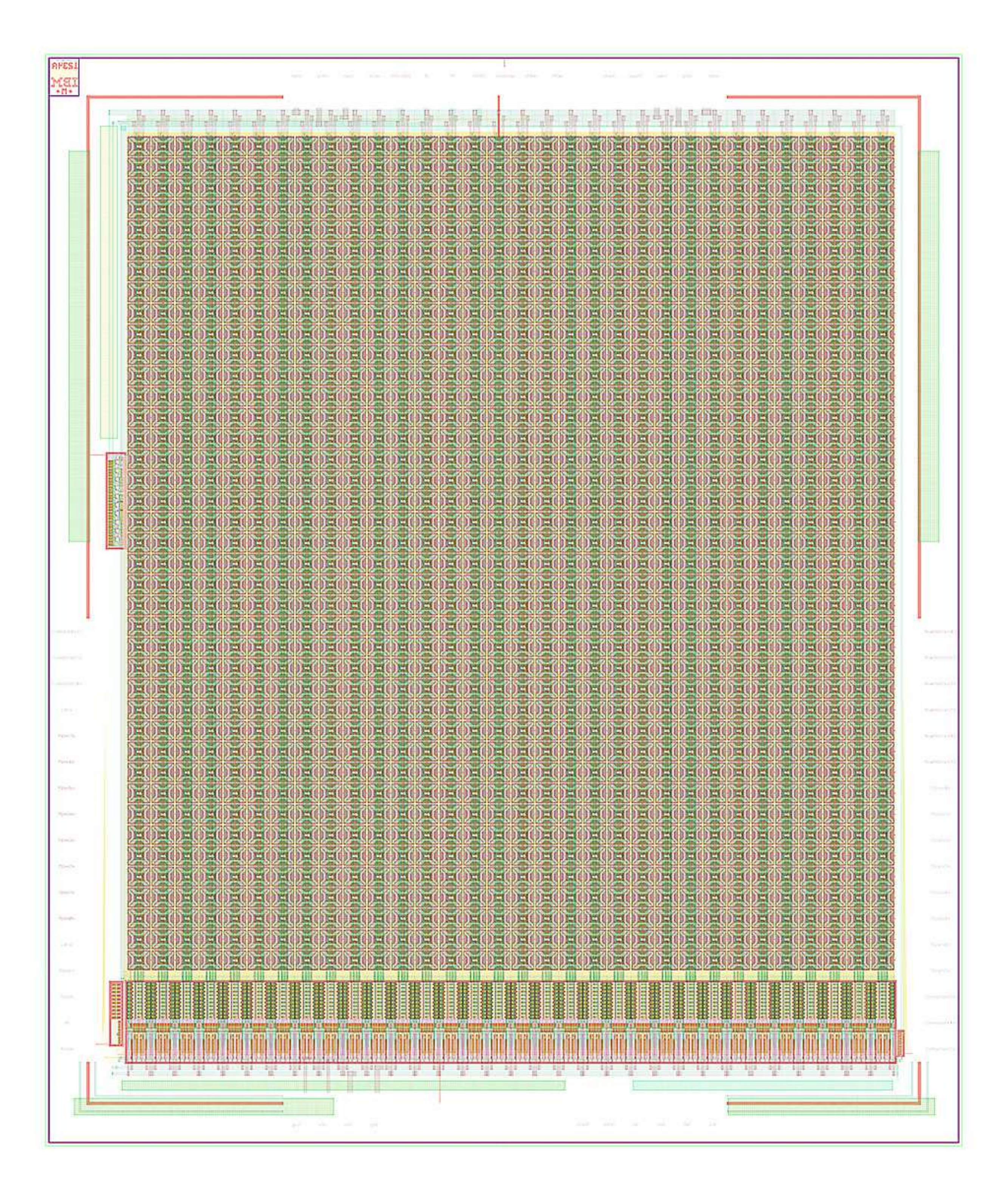}
	\caption{Design view of the MUPIX3 chip (actual size about 4 by $\SI{5}{mm}$).}
	\label{fig:Mupix3}
\end{figure}

In August 2012, we submitted the MUPIX 3 chip, a major step towards the final sensor. The new chip has $40 \times 32$ pixels of $92\times\SI{80}{\micro\meter\squared}$ size for an active area of approximately $\SI{9.4}{\milli\meter\squared}$, see Figure~\ref{fig:Mupix3}. It implements the full digital column logic, allowing for address generation and serial readout of zero-suppressed data. In addition, MUPIX3 has faster signal shaping.

The main differences with the final sensor are the lack of a high-speed LVDS output, buffers in the columns and the chip-wide hit collection logic. For this prototype, the corresponding logic will be emulated off-chip in an FPGA. Also several control voltages which should be generated on-chip in the final version are currently produced on a test printed circuit board (PCB) in order to allow for easier debugging. 

We have just received the first MUPIX 3 samples. First test results will become available early 2013. 

\subsection{Plans for 2013}
\label{sec:PixelPlans2013}

As soon as the first results from the MUPIX~3 sensor are available, we will prepare another multi-project-wafer (MPW) run, implementing the remaining digital logic and addressing potential issues discovered with MUPIX 3. This should clear the path for an engineering run in the second half of 2013, opening up the possibility to build a full scale tracker prototype.

\section{Characterization of the Prototypes}
\label{sec:CharacterisationOfThePrototypes}

We have studied the properties of the MUPIX 1 and 2 prototypes using injection pulses, LEDs, laser diodes, X-rays, radioactive sources and test beam measurements. In the following, we will outline the core results; details of the findings can be found in a master \cite{Perrevoort2012} and a bachelor thesis \cite{Augustin2012}.

The test setup was based on a test board housing the chip itself and a logic interface card. The chip test board has voltage regulators for the supply voltages, digital to analog converters for the threshold and injection pulse height and a flat ribbon connector to the logic interface card. This interface card uses a FPGA both to program and readout the MUPIX chip and to communicate with a PC via USB. These two boards together with a C++ control software have been used for all tests described here.

\subsection{Signal to Noise Ratio}
\label{sec:SignalToNoiseRatio}

\begin{figure}
	\centering
		\includegraphics[width=0.48\textwidth]{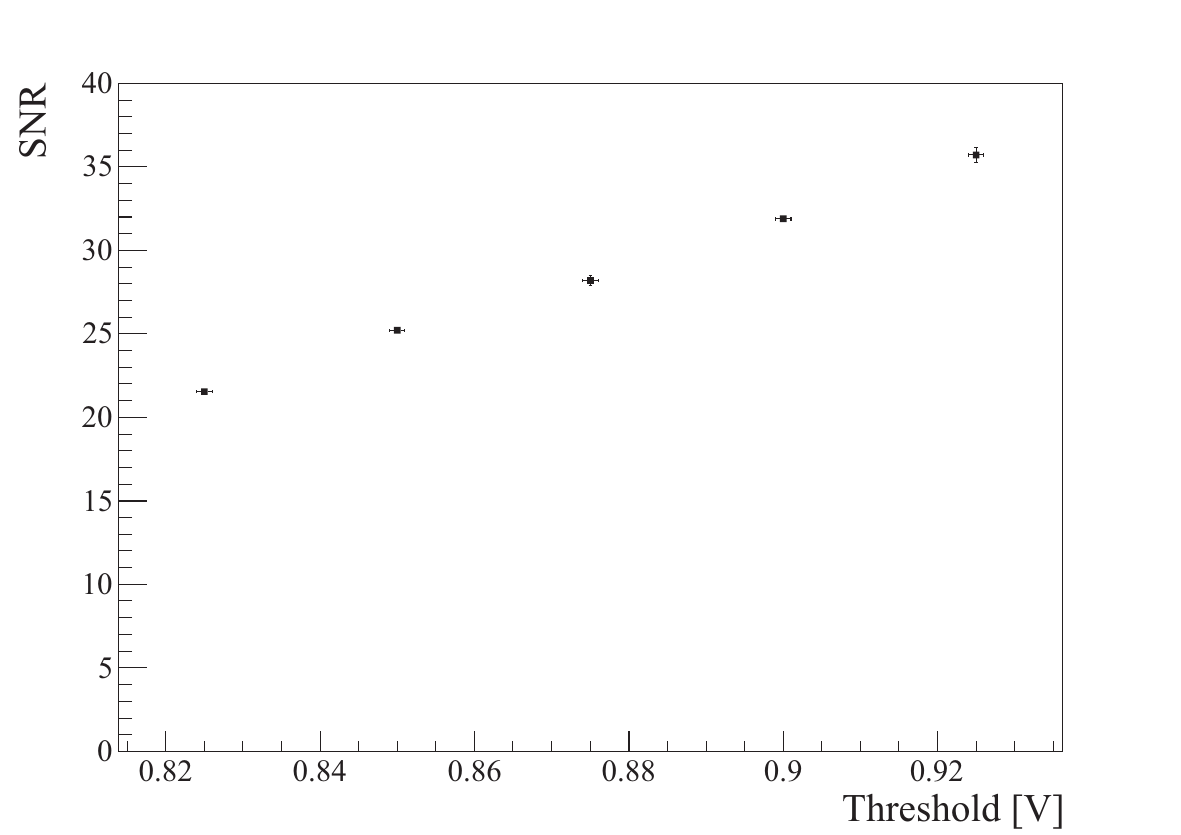}
	\caption{Signal to noise rate for the MUPIX 2 chip. The signal size is taken from measurements with a $^{55}$Fe source, whilst the noise is measured in a injection pulse scan, see \cite{Perrevoort2012} for details.}
	\label{fig:SignalToNoise}
\end{figure}

Injection pulses can simulate the charge deposition of ionizing particles. A scan of the injection pulse height at constant threshold would deliver a step function if no noise were present. Due to the additional noise on top of the injection pulse the step function is transformed to an error function with a finite slope. In turn the width of the step function can be used to determine the noise as a function of the applied threshold. A $^{55}$Fe source has been used as reference signal for the determination of the signal to noise ratio (SNR). The injection pulse height corresponding to the  $^{55}$Fe signal is found by matching the time over threshold of both signals at a given threshold. For the MUPIX~2 chip the signal to noise ratio as a function of threshold ranges between 21.5 and 35.7, see Figure~\ref{fig:SignalToNoise}. Any signal to noise ratio >9 is considered good.       

\subsection{Pixel to Pixel Uniformity}
\label{sec:Uniformity}

\begin{figure*}
	\centering
		\includegraphics[width=1.00\textwidth]{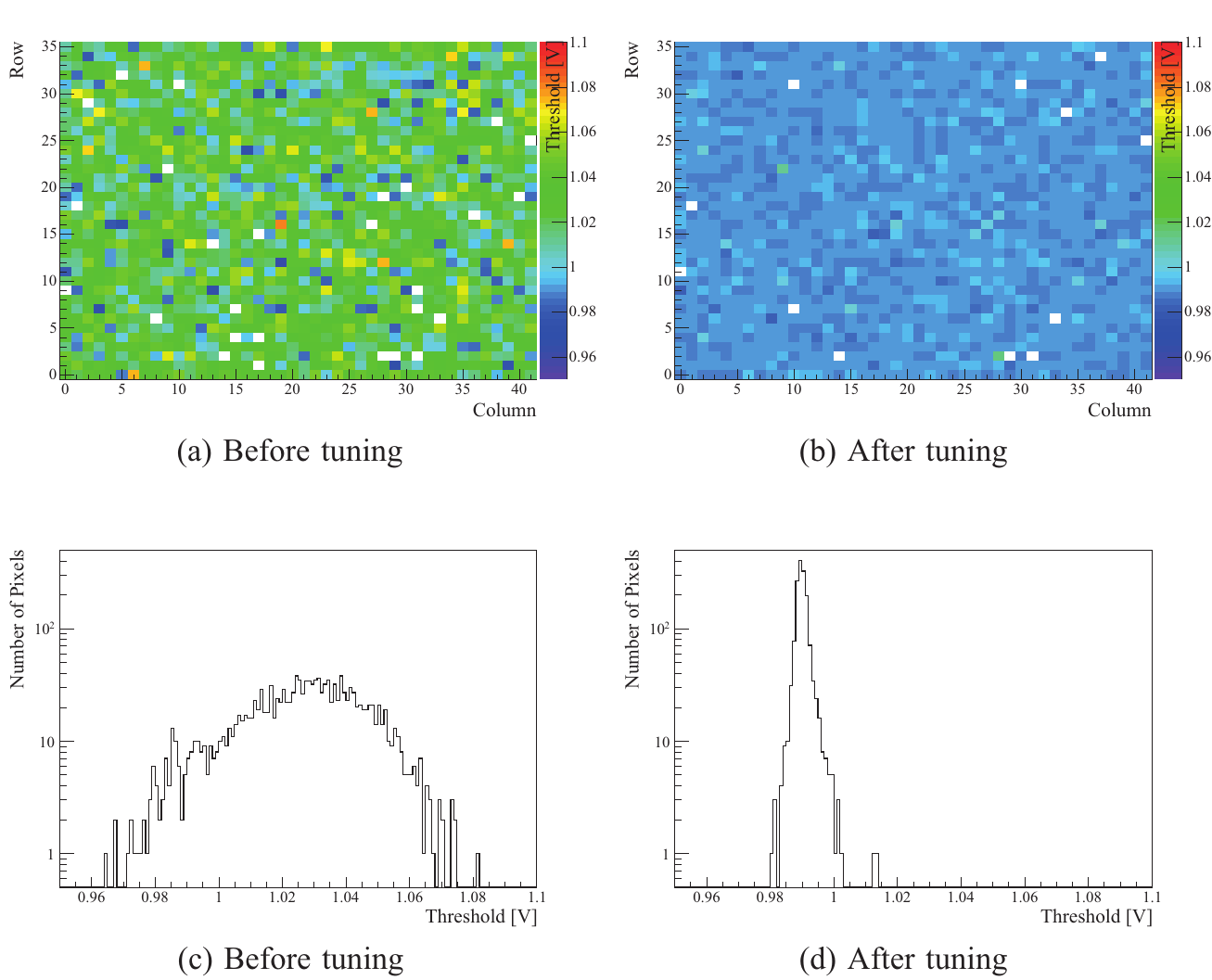}
	\caption{Uniformity of the pixel response before and after tuning of the pixel DAC values, from \cite{Perrevoort2012}.}
	\label{fig:Uniformity}
\end{figure*}

Using a similar method as for the measurement of the signal to noise ratio, the pixel to pixel uniformity can be determined and optimized. At a constant injection pulse height a scan of the threshold setting for the entire pixel matrix is carried out. For each pixel the error function fit is performed and the $\SI{50}{\%}$ value of the error function rising edge plotted, see Figure~\ref{fig:Uniformity}a). As can be seen in Figure~\ref{fig:Uniformity}c) the distribution of the threshold values for the pixels of the same chip is quite broad. In order to compensate for these inequalities, a threshold offset, the so called tune DAC (TDAC) value can be set for each pixel. The best TDAC values for the pixels are found by running an automated procedure, see \cite{Perrevoort2012}. Figure \ref{fig:Uniformity}b) and d) show that the threshold spread (reflecting the response uniformity) after tuning the threshold offsets has decreased by almost an order of magnitude from $\SI{19.6}{\mV}$ to $\SI{2.3}{\mV}$ RMS.         

\subsection{Pixel Response Time}
\label{sec:PixelResponseTime}

The timing characteristics of the MUPIX chip is very important because of the requirement to run at high muon decay rates. A fast response of the MUPIX chip leads to precise timing of hits and helps suppressing combinatorial background. In the inner detector layers the double pulse resolution should be good as the hit rate is up to $\SI{2}{\kHz}$ for each pixel. 

\begin{figure}
	\centering
		\includegraphics[width=0.48\textwidth]{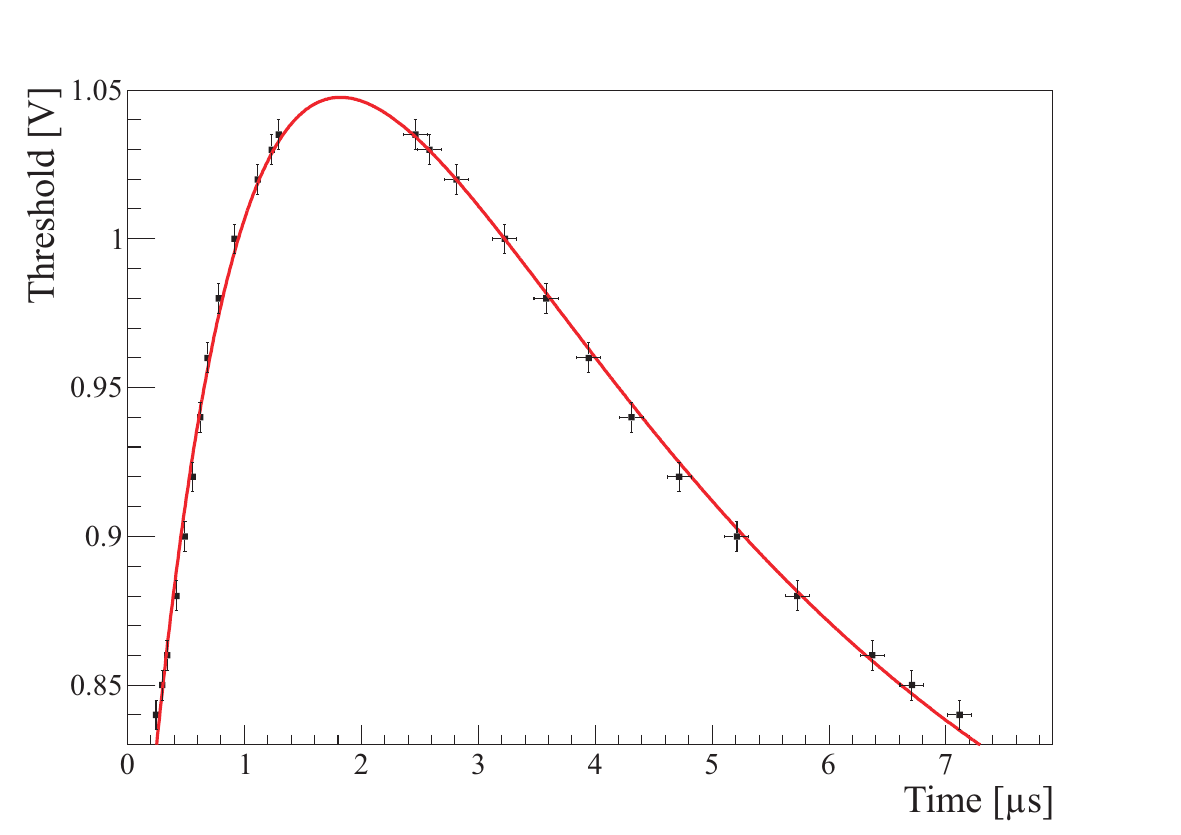}
	\caption{Measured pulse shape of a MUPIX pixel in response to an LED pulse, fitted with the expectation from $RC-CR$ shaping, from \cite{Perrevoort2012}.}
	\label{fig:PulseShape}
\end{figure}

Timing studies have been carried out using a LED driven by a pulse generator to stimulate the sensor. The discriminator output of a single pixel is then compared to the second output of the pulse generator on an oscilloscope. In this configuration it is possible to measure the latency between generator and pixel output and the time over threshold (ToT) of the pixel. By plotting latency and ToT for many threshold values one obtains the pulse shape, see Figure~\ref{fig:PulseShape}. The measured pulse shape reflects the RC-CR shaping characteristic of the pixel electronics. The charge collection process in the depletion zone is much faster and has little influence on the measured timing.     

\begin{figure}
	\centering
		\includegraphics[width=0.48\textwidth]{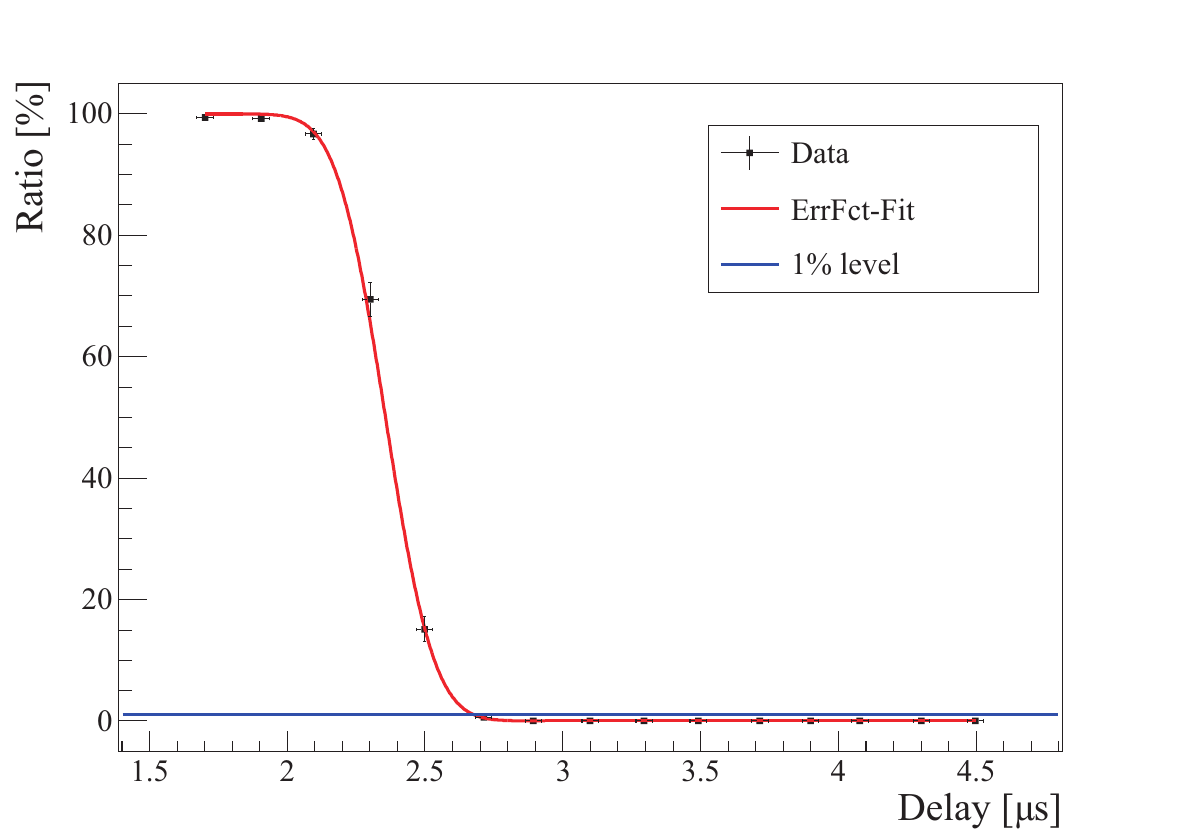}
	\caption{Double pulse resolution of the MUPIX2 chip, from \cite{Perrevoort2012}.}
	\label{fig:DoublePulse}
\end{figure}

By generating double pulses and dividing the number of unresolved pulses at
the pixel output by the number of all pulses, the double pulse resolution can
be determined. Figure \ref{fig:DoublePulse} shows this ratio as a function of
pulse to pulse delay. An error function has been fitted and the point of
$\SI{1}{\%}$ un-resolved pulses gives a double pulse resolution of
$\SI{2.7}{\us}$
consistent with the expectation from the chosen shaper characteristics.

\subsection{Measurements with Radioactive Sources}
	\label{sec:MeasurementsWithRadioactiveSources}
	
\begin{figure}
	\centering
		\includegraphics[width=0.48\textwidth]{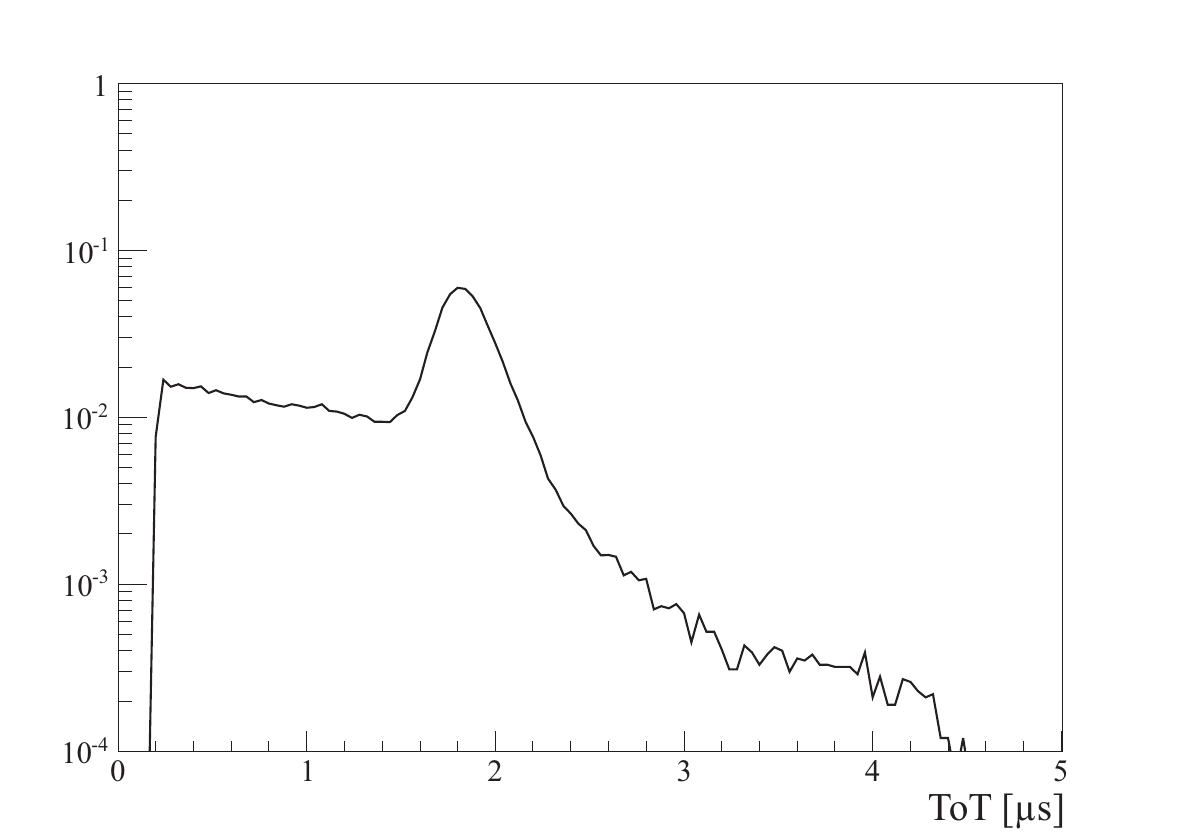}
	\caption{Time-over-threshold (corresponding to energy) spectrum of a $^{55}$Fe radioactive source, from \cite{Perrevoort2012}.}
	\label{fig:IronPeak}
\end{figure}

Radioactive sources allow to test the MUPIX chips with real particles. 
For a $^{55}$Fe source the time over threshold (ToT) distribution, which is an
measure for the energy distribution, is shown in
Figure~\ref{fig:IronPeak}.
X-ray fluorescence in the range of 4 to $\SI{18}{\keV}$ has been used to derive a relative energy resolution of 10 to $\SI{20}{\%}$.    

\begin{figure}[t!]
	\centering
		\includegraphics[width=0.48\textwidth]{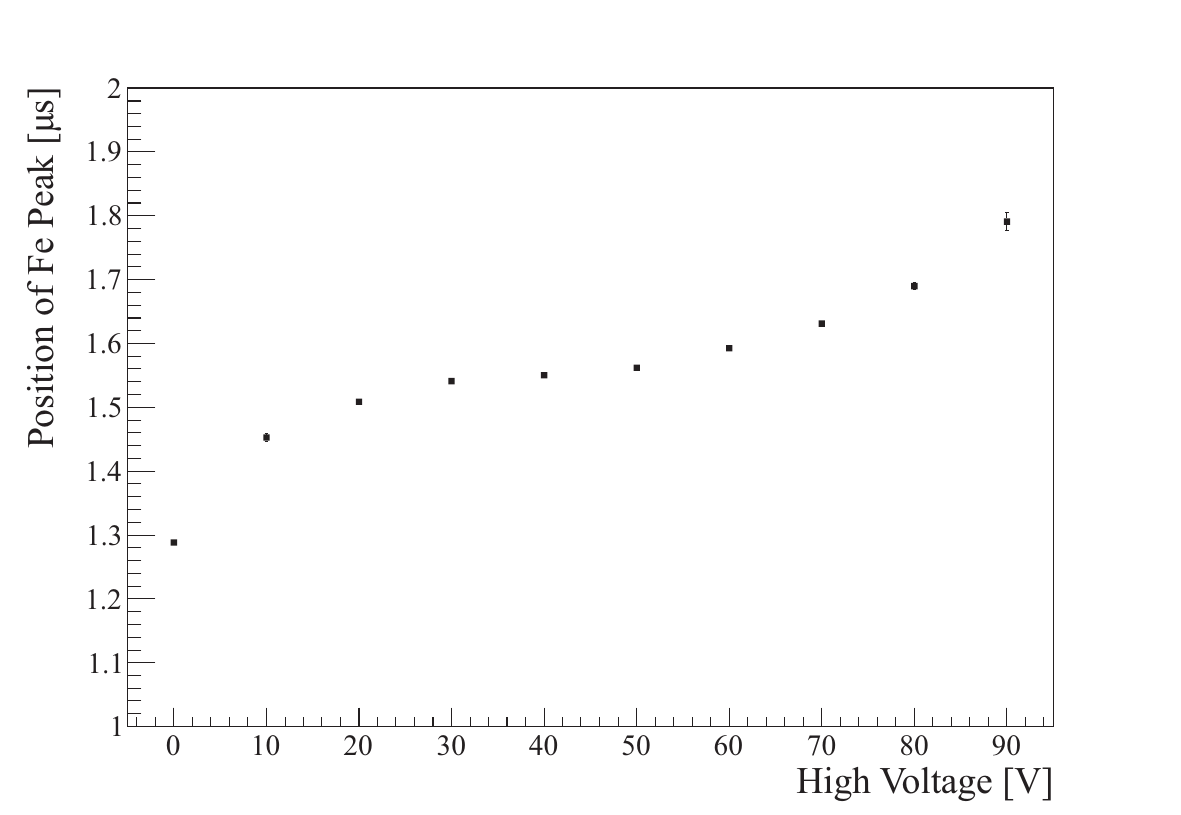}
	\caption{Position of the $^{55}$Fe peak in dependence of the applied high voltage, from \cite{Perrevoort2012}. }
	\label{fig:HighVoltage}
\end{figure}

\begin{figure}[t!]
	\centering
		\includegraphics[width=0.48\textwidth]{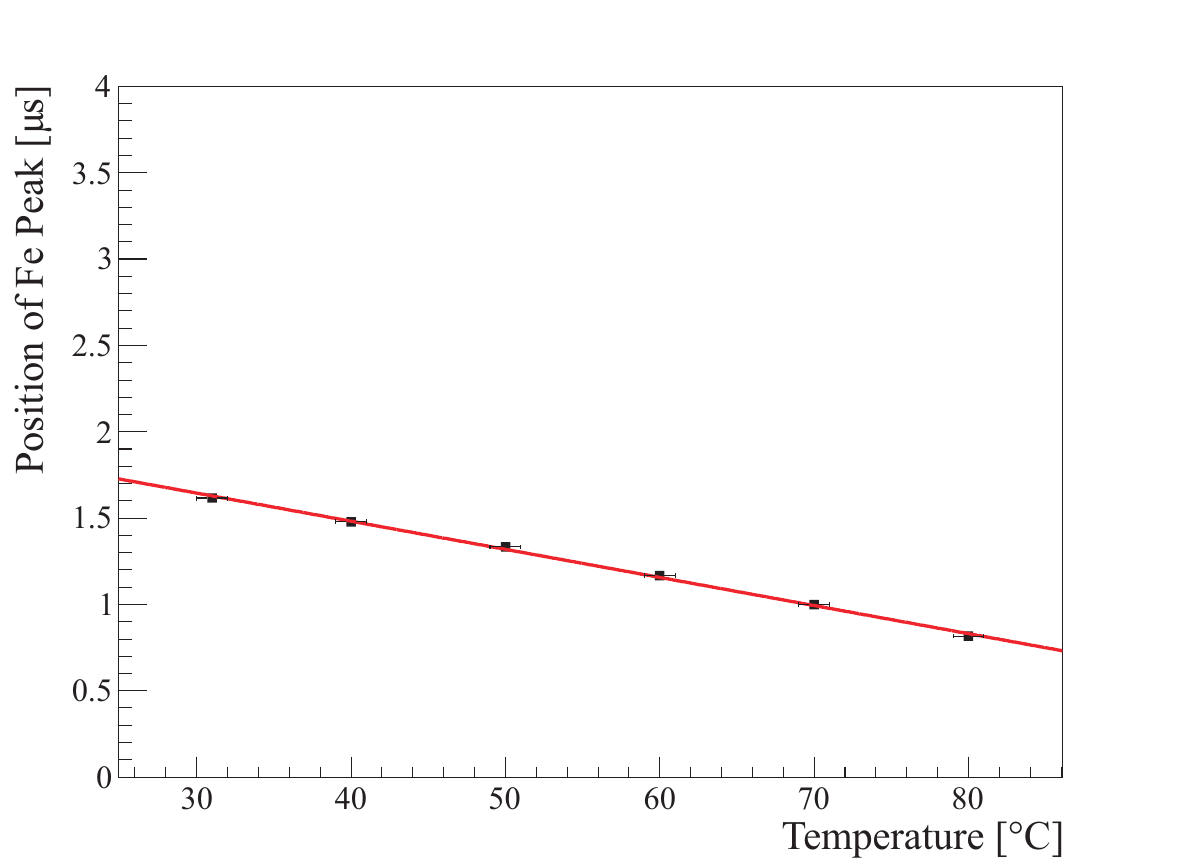}
	\caption{Temperature dependence of the position of the $^{55}$Fe peak, from \cite{Perrevoort2012}. }
	\label{fig:Temperature}
\end{figure}

The $^{55}$Fe peak in the ToT spectrum has been used to study the influence high voltage (HV) and temperature on the pixel signal. 
Figure \ref{fig:HighVoltage} shows the $^{55}$Fe peak as a function of the HV, revealing first a steep rise as the depletion zone grows. Between 20 and $\SI{60}{\V}$, the ToT stays almost constant. At even higher voltages the electric field in the depletion zone becomes strong enough for the creation of secondary electron-hole pairs, which results in a signal amplification. 
The temperature dependence of the pixel sensor is important, because in the
later experiment the different parts of the detector will operate at different
temperatures varying by $30$-$\SI{40}{\degreeCelsius}$. In Figure~\ref{fig:Temperature} the $^{55}$Fe peak of the ToT spectrum is plotted as a function of temperature between 20 to $\SI{60}{\degreeCelsius}$. As can be seen the ToT goes down from 1.6 to below $\SI{1}{\us}$, with a slope of $\SI{16.3}{\ns}$/$\SI{}{\degreeCelsius}$. Measurements with test pulse injection and simulation studies confirm that the observed temperature dependence is a feature of the charge sensitive amplifier of the pixel analog stage.  

\begin{figure}
	\centering	\includegraphics[width=0.45\textwidth]{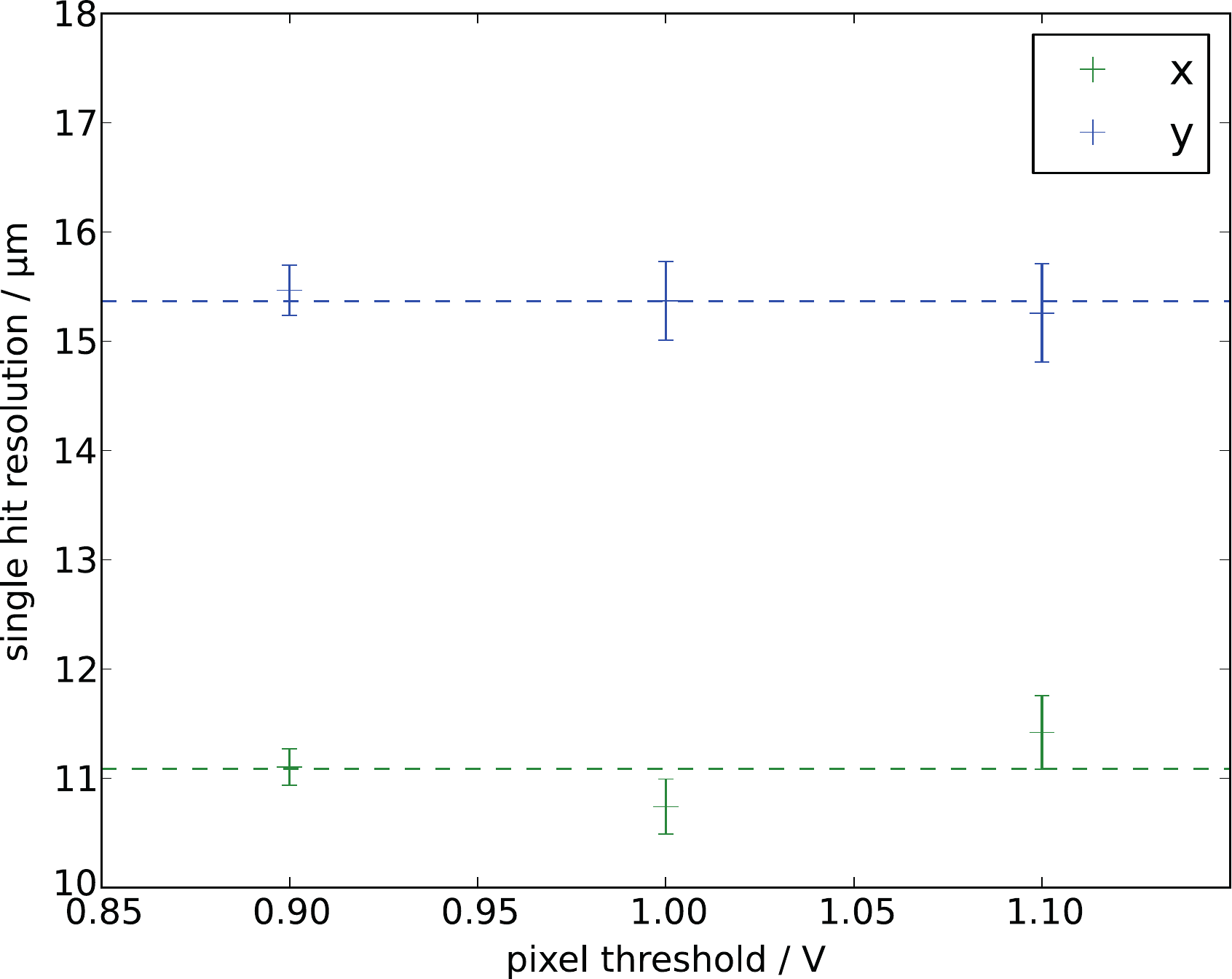}
	\caption{Resolution of the MUPIX 2 chip measured in a testbeam with $\SI{170}{GeV}$ pions and the TIMEPIX telescope. The measured resolutions correspond very well to the expectation from the $\SI{5}{\micro\meter}$ resolution of the telescope and the pixel pitch in $x$ and $y$ direction, respectively.}
	\label{fig:resolutionTestbeam}
\end{figure}

\subsection{Testbeam results}
	\label{sec:TestbeamResults}

In August 2012 the MUPIX2 chip was tested at the CERN SPS. The beam from the SPS was a $\SI{170}{\GeV}$ pion beam, chosen for little multiple scattering effects due to the device under test (DUT). In two separate data taking periods the MUPIX2 was mounted inside the TimePix telescope \cite{Akiba:2011vn}. This silicon pixel telescope has four layers before and four layers after the DUT, providing very precise pointing resolution of $\SI{5}{\um}$. The MUPIX2 chip was tested facing the beam and under an angle of $\SI{45}{\degree}$. For both beam periods a threshold scan was performed in order to derive the efficiency as a function of threshold, but no threshold calibration to equalize the gain of the pixels was performed. Figure \ref{fig:resolutionTestbeam} shows the resolution in $x$ and $y$ direction of the MUPIX2 chip facing the beam and tested at three different threshold values. The measured resolution of $\SI{11.2}{\um}$ in $x$ and $\SI{15.4}{\um}$ in $y$ corresponds to the expected resolution given the telescope resolution and the pixel size. Further results are still subject to studies and will follow soon.

%We have studied the properties of the MUPIX 1 and 2 prototypes using injection pulses, LEDs, laser diodes, X-rays, radioactive sources and test beam measurements. In the following, we will outline the core results; details of the findings can be found in a master \cite{Perrevoort2012} and a bachelor thesis \cite{Augustin2012}.

\section{Mechanics}
\label{sec:Mechanics}

\begin{figure}
	\centering
		\includegraphics[width=0.48\textwidth]{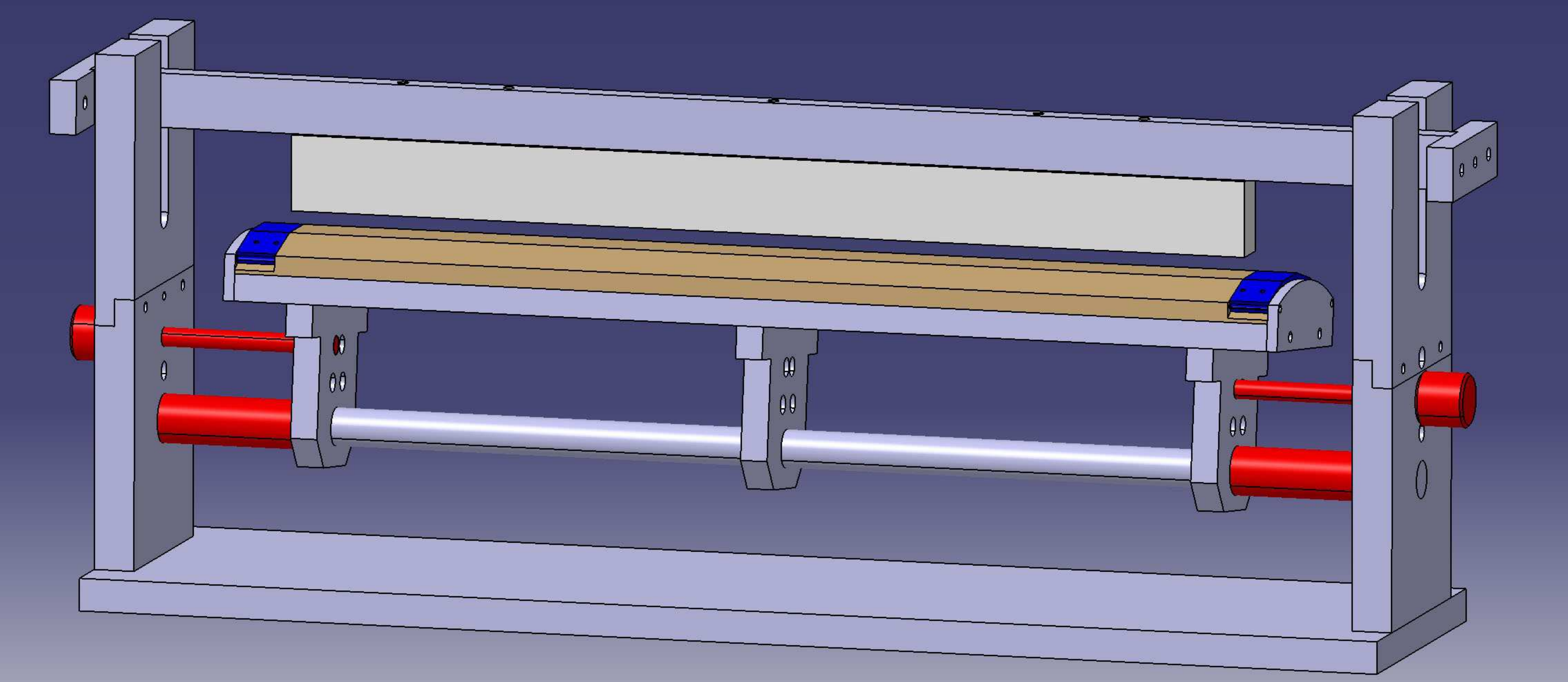}
	\caption{Tool for layer 3 segment assembly.}
	\label{fig:Layer3GlueingTool_Drawing}
\end{figure}

\begin{figure}
	\centering
		\includegraphics[width=0.48\textwidth]{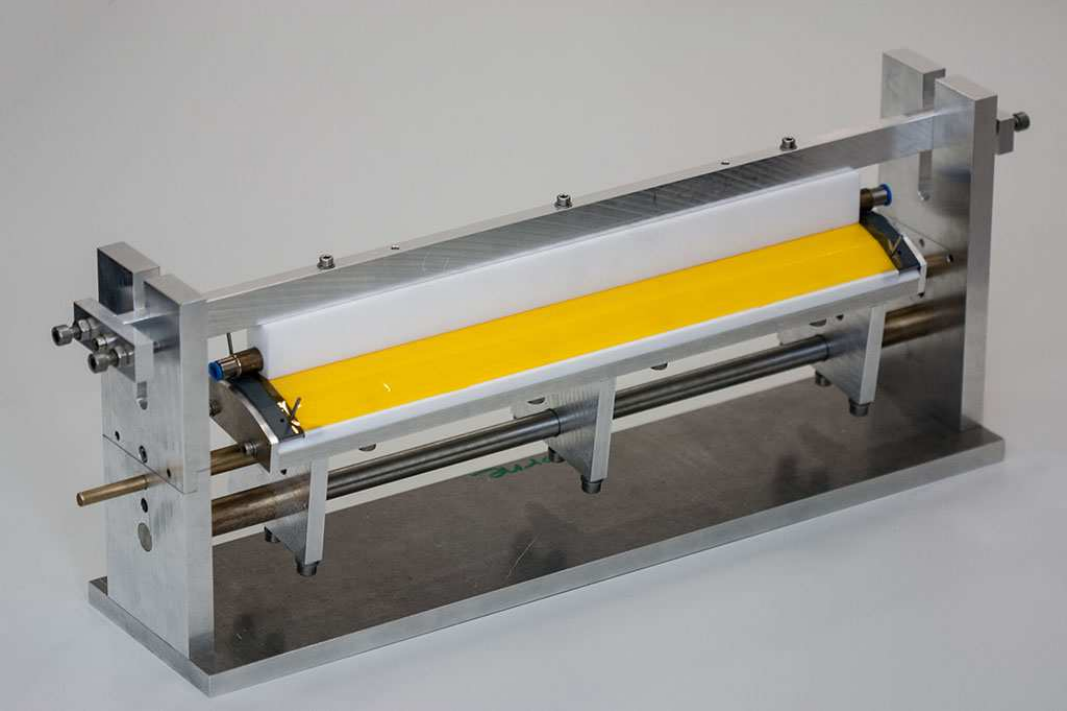}
	\caption{Tool for layer 3 segment assembly.}
	\label{fig:SegmentAssemblyTool}
\end{figure}

\begin{figure}
	\centering
		\includegraphics[width=0.48\textwidth]{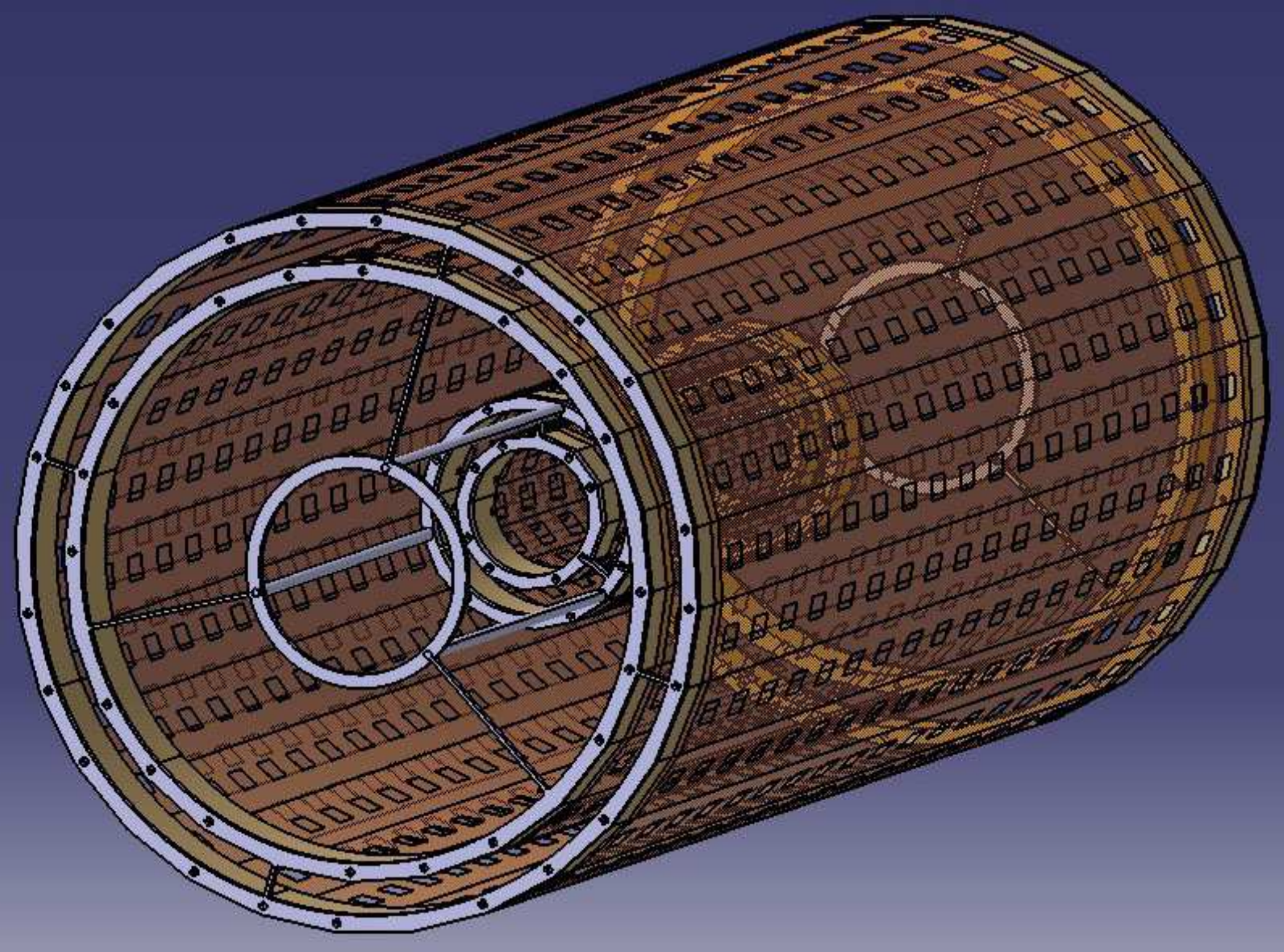}
	\caption{Mechanics of the central pixel detector}
	\label{fig:CentralMechanics}
\end{figure}

\begin{figure}
	\centering
		\includegraphics[width=0.48\textwidth]{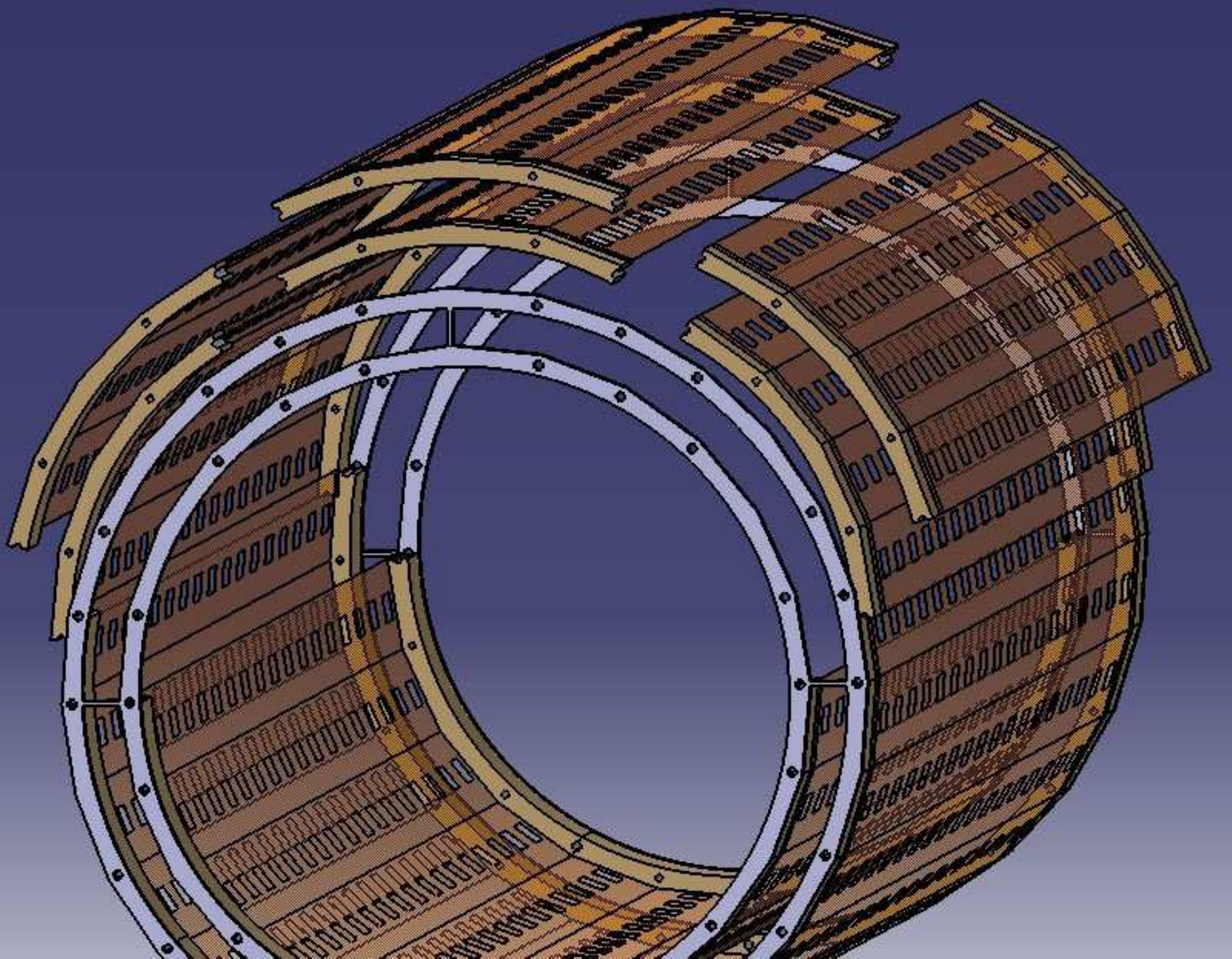}
	\caption{Segmentation of the central outer layers.}
	\label{fig:Segments}
\end{figure}

\begin{figure}
	\centering
		\includegraphics[width=0.48\textwidth]{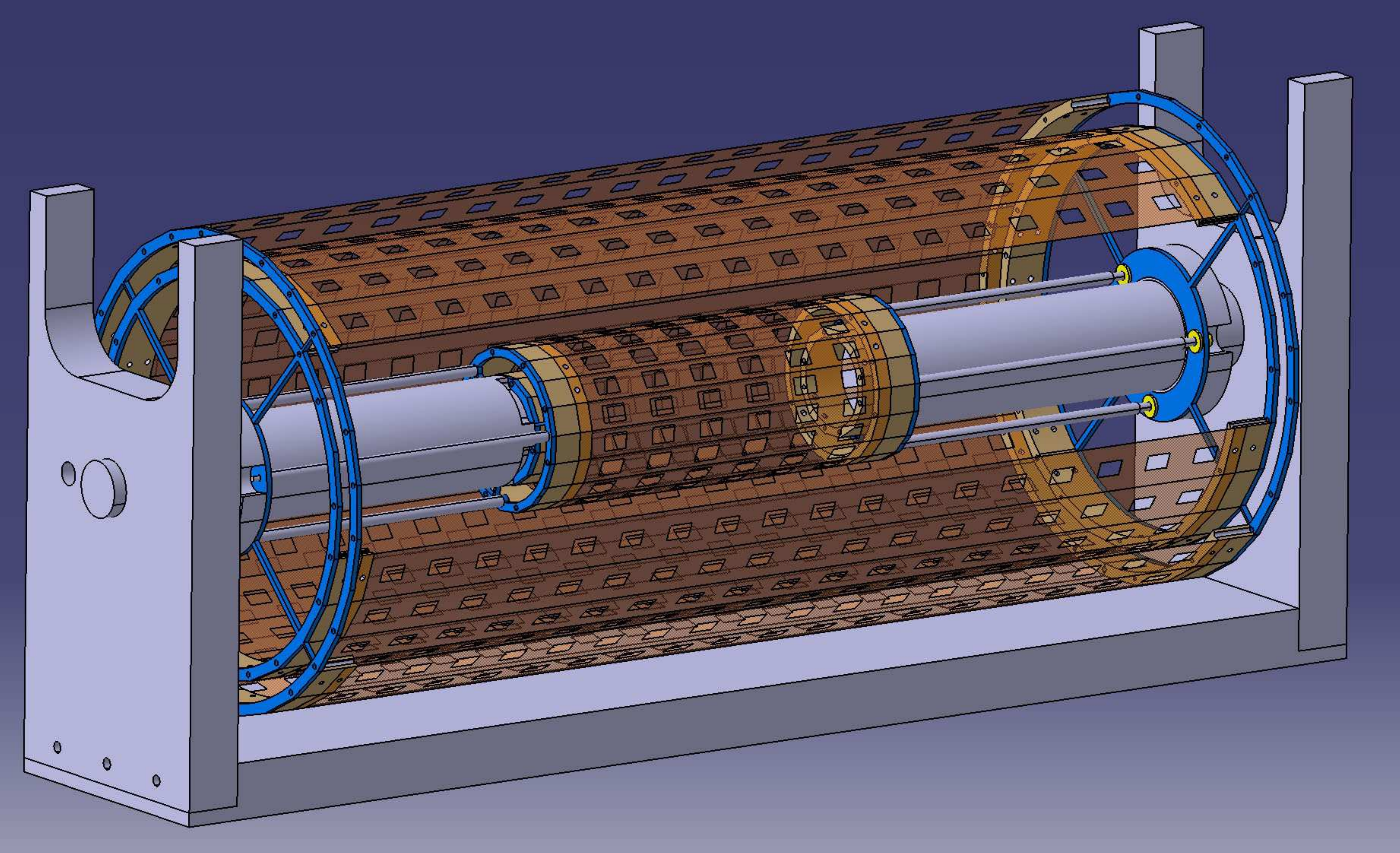}
	\caption{Mounting tool for the central pixel detector.}
	\label{fig:Mu3eCentralPixelMountingTool}
\end{figure}

The pixel detector mechanics has been optimized for very low material budget in the active detector region. Additional requirements are mechanical stability, resistance to temperatures over a wide range and a modular design for ease of assembly and repair. It is proposed to build the frame for the silicon pixel detector from thin Kapton foil. The sensors are glued and bonded on a flex-print and then mounted onto the Kapton frame.

The Kapton foil used for the mechanical frame construction is $\SI{25}{\um}$ thin. It gains in mechanical stability as it is folded around 
a prism-shaped template and glued to plastic end-pieces. For this process tools have been built, which support the detector modules until they are ready for full station assembly, see Figures~\ref{fig:Layer3GlueingTool_Drawing} and \ref{fig:SegmentAssemblyTool}. In a separate tool the $\SI{50}{\um}$ thin pixel chips are glued on the single layer Kapton flex-print. Alignement groves secure the correct position of the chips. In the next step the chips are wire bonded to the flex print. With a vacuum lift tool the flex-prints are then positioned and glued to the mechanical frames. The radiation length of a pixel detector layer is summarized in Table~\ref{tab:LayerRadLength}.  
  
\begin{figure*}
	\centering
		\includegraphics[width=0.80\textwidth]{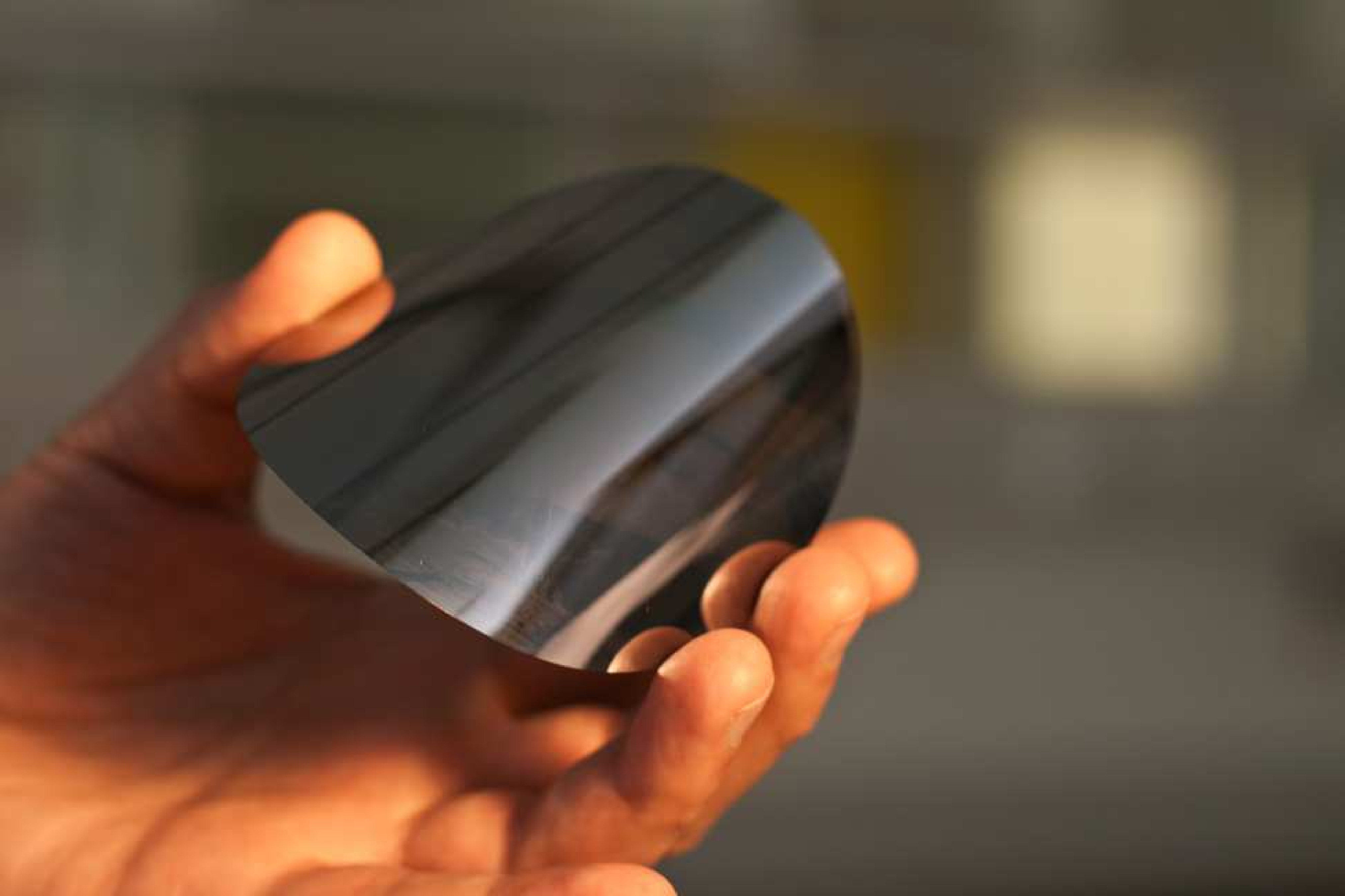}
	\caption{A $\SI{50}{\micro\meter}$ thin silicon wafer.}
	\label{fig:ThinWafer}
\end{figure*}

\begin{figure*}
	\centering
		\includegraphics[width=0.80\textwidth]{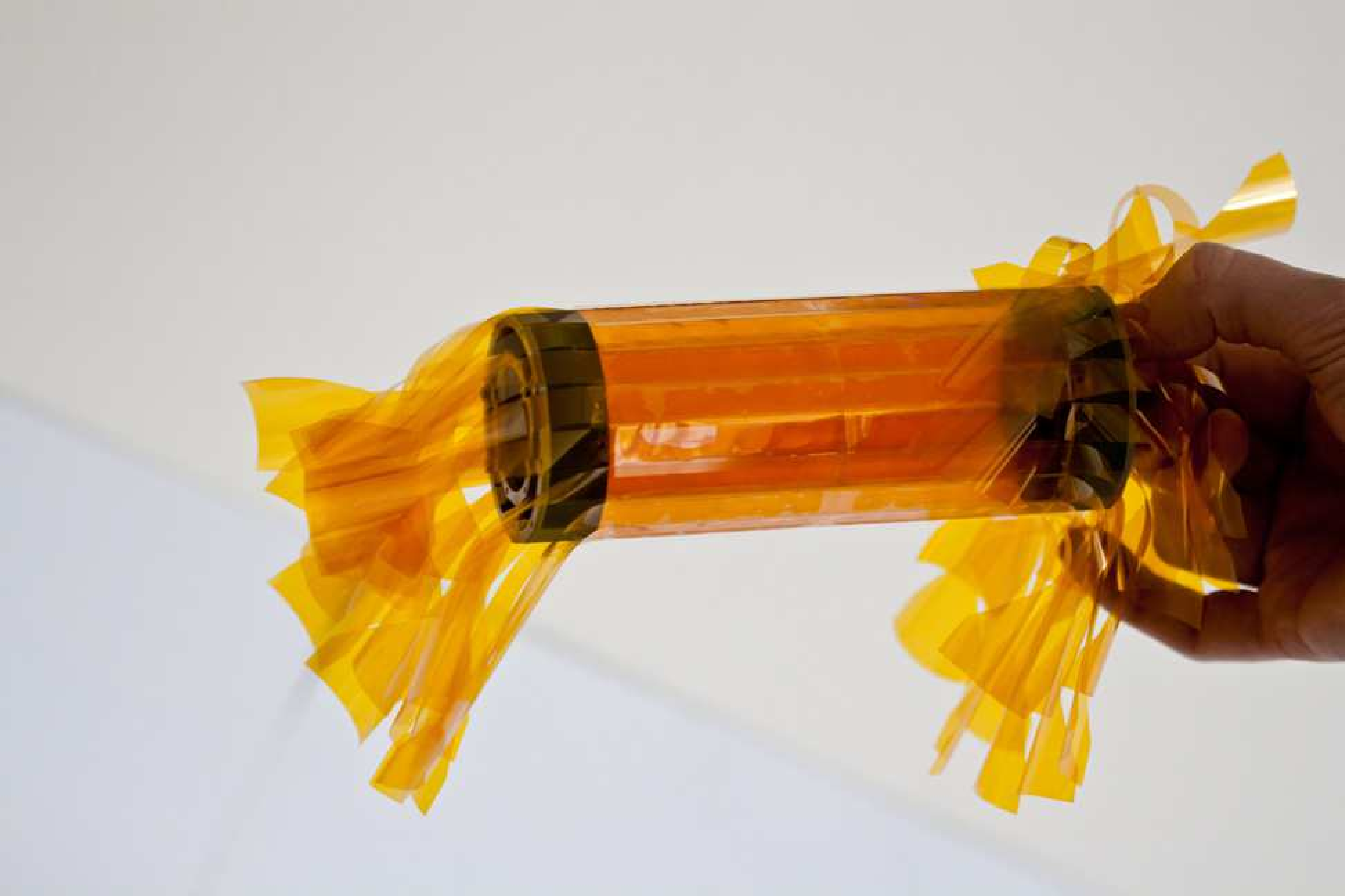}
	\caption{Mechanical prototype of the inner pixel layers. Thin glass plates replace the silicon chips.}
	\label{fig:InnerLayerPrototypes}
\end{figure*}

The pixel detector is build from layers of four different sizes and prism shapes, see Figure~\ref{fig:CentralMechanics}. As the digital readout circuits of the pixel chips create an approximately $\SI{0.5}{mm}$ wide dead area, there is a $\SI{1}{mm}$ overlap to the adjacent sensor. The inner double layers have $\SI{12}{\cm}$ active length. Layer 1 has 12  and layer 2 18 sides of $\SI{1}{\cm}$ width. Each inner layer is assembled from two half-modules. As a consequence the plastic end-pieces are half moon shaped. When all four half modules of the inner detector layers are produced and tested they are mounted to two thin rim wheels. A mechanical prototype for the inner double layer has been constructed from $\SI{25}{\um}$ Kapton foil both for the frame and the flex-print layer, while the pixel chips have been simulated with $\SI{100}{\um}$ thick glass plates. Glass of $\SI{100}{\um}$ thickness is of comparable flexibility as thinned silicon, which is shown in Figure~\ref{fig:ThinWafer}. The resulting mechanical unit is surprisingly sturdy and fully self supporting, as can be seen in Figure \ref{fig:InnerLayerPrototypes}.
Layer 3 and layer 4 have a three times larger active length of $\SI{36}{\cm}$. The sides of these outer layers are $\SI{19}{\mm}$ wide. The layers 3 and 4 have 24 and 28 sides. An outer double layer module combines four sides, so layer 3 consists of 6 and layer 4 consists of 7 modules, see Figure~\ref{fig:Segments}. The station assembly is done in a special mounting frame, combining the inner two layers with the modules of the outer two layers on large rim wheels (Figure~\ref{fig:Mu3eCentralPixelMountingTool}). Re-curl stations will be assembled accordingly.

\begin{table}
	\begin{center}
	\begin{tabular}{lcr}
% \hline
	\toprule
	\sc Component								& \sc thickness 			& \sc X/X$_0$				\\ % \hline \hline
															& \sc [$\SI{}{\um}$] 	& \sc [$\SI{}{\%}$]	\\ % \hline \hline
	\midrule
	Kapton frame								&		25								& 	0.018 	\\
	Kapton flex-print    				&		25								&		0.018 	\\
	Aluminum traces 						&		15/2							&		0.008		\\
	($\SI{50}{\%}$ coverage)	 	&	 										&  			 		\\
	HV-MAPS 										&  	50								&  	0.053 	\\
	Adhesive  									&  	10								&	 	0.003 	\\
	\midrule
	Full detector layer 				& 	125								&		0.100		\\ 
	% \hline
	\bottomrule
	\end{tabular}
	\end{center}
	\caption{The pixel detector layer radiation length is dominated by the HV-MAPS chips.}
	\label{tab:LayerRadLength}
\end{table}

\section{Cooling}
\label{sec:Cooling}

The cooling system for the pixel detector must be capable of keeping the temperatures at a reasonable level ($\leq \SI{70}{\degreeCelsius}$) and should at the same time add very little extra material to the active volume of the detector. Cooling with gaseous helium has been chosen as it appears to offer a reasonable compromise between cooling potential and radiation length. 

Given its size, the heat load introduced by the pixel detector is
considerable. Table \ref{tab:heatload} details the contributions of the
different parts of the pixel detector assuming a realistic heat load of
$\SI{150}{\mW\per\cm\squared}$. It is worth mentioning that the heat load is
strongly dependent on the clock speed at which the pixel chips are
running. Pixel chips with very low occupancy, for example those installed in
the recurl stations, could optionally be operated at half the clock speed which would largely reduce the total thermal power of the pixel detector. 

In order to keep the temperature of the coolant equally low in all parts of the detector, it is foreseen to place nozzles between the support structures pointing at different parts of the sensor layers. 

In order to study the potential of gaseous helium cooling a bachelor thesis
has been carried out \cite{Zimmermann2012}. In this thesis a thin sandwich of
aluminum foil, Kapton and silicon was heated inductively and cooled with a
helium flow. For the $\SI{8.5}{\cm}$ long sample heated with
$\SI{100}{\mW\per\cm\squared}$ a flow of $\SI{0.4}{\m\per\s}$ was necessary to
maintain a temperature difference between helium and sensor of
$\SI{32}{\K}$. Further studies using a full scale central detector prototype
in a wind tunnel are on the way. This full scale prototype will be assembled
from aluminized Kapton foil such that Ohmic heating can be applied. It is planned to run the upcoming tests with the helium heat exchanger acquired for phase I of the experiment. 

\begin{table}
	\begin{center}
	\begin{tabular}{lcr}
% \hline
	\toprule
	\sc Detector Part& \sc size [$\SI{}{\cm\squared}$] & power [$\SI{}{\W}$]\\ % \hline \hline
	\midrule
	layer 1 													&		158.4		& 	23 	\\
	layer 2     											&		237.6		&		35 	\\
	vertex layers (1+2)								&		396			&		59		\\
	layer 3	 													&	 1728			&  259 		\\
	layer 4 													&  2016			&  302 		\\
	outer layers (3+4)  							&  3744			&	 561 		\\
	central detector (1-4)						&  4140			&  621			\\
	detector with                     &           &         	\\         
	2 recurl stations
        &	11628			&	1744 		\\ 
	full detector 										& 19116			&	2867		\\ 
	% \hline
	\bottomrule
	\end{tabular}
	\end{center}
	\caption{Pixel detector heat load for $\SI{150}{\mW\per\cm\squared}$}
	\label{tab:heatload}
\end{table}

%\section{Performance Studies}
%\label{sec:PerformanceStudies}

\section{Alternative Technologies}
\label{sec:AlternativeTechnologies}

Since the Mu3e pixel detector plays the key role in the experiment, it is
important to carefully compare the chosen HV-MAPS technology with
alternatives. To do so one has to have the detector requirements in mind, see
chapter \ref{sec:Requirements}. The goal of a branching ratio sensitivity of
$\num{e-16}$ leads to a high muon decay rate on target of $\SI{2e9}{\Hz}$. In
order to suppress combinatorial background a very good vertex resolution of
$\mathcal{O}(\SI{200}{\micro\meter}$) is needed. The suppression of the
background from the conversion decay \mtenunu requires very precise momentum
reconstruction of better than $\SI{0.5}{\MeV}$, for the $\num{e-16}$ branching
ratio sensitivity. The momentum resolution for the low energetic decay
products of $\SI{10}{\MeV}$ to $\SI{53}{\MeV}$ will be governed by the
multiple scattering in the detector. Only concepts with extremely low
radiation length of <$\SI{1}{\%}$ in total can achieve this.

In the following different technology options for the \emph{Mu3e} tracking
detector are compared. The result of this comparison is summarized in table~\ref{tab:AlternatevTechnologies}.

\subsection{Time-Projection Chamber}
\label{sec:TimeProjectionChamber}

Time projection chambers are a very attractive concept for tracking detectors
with good spacial resolution and very low radiation length. The example of the
ALICE TPC \cite{Dellacasa:2000bm} has demonstrated that very high track
densities can be handled with such a detector. Over the last two decades many
studies for TPCs with very low radiation length, i.e.~$\SI{0.12}{\%}$ for a
proposed chamber at SLAC \cite{Boyarski:1992ch}, have been carried out. The
drift time of an electron in a possible TPC for the Mu3e experiment with
$\SI{2}{\m}$ total length would be in the order of $\SI{50}{\micro\second}$,
using the proposed gas mixture of Helium:CO$_2$:Isobutane (83:10:7$\%$). This
time of $\SI{50}{\micro\second}$ at $\SI{2e9}{\Hz}$ decay rate means that the
DAQ and event filter farm would have to handle frames containing >$\num{e5}$
muon decays, 
which is not realistic. Another problem would be strong space charge effects due to continuous very high track density, see \cite{2012arXiv1209.0482B}. These two aspects rule out the usage of a TPC in the phase IB and phase II of Mu3e, while it would be possible to use a TPC in phase IA.

\subsection{Other Pixel Technologies}
\label{sec:OtherPixelTechnologies}

\begin{table*}[bt!]
	\begin{center}
	\begin{tabular}{lcccccc}
	% \hline
	\toprule
	\sc  & &\sc Mu3e & \sc  	& \sc STAR & \sc BELLE & \sc LHCb\\
	\sc Parameter & required	&\sc HV-MAPS & \sc He TPC 	& \sc MIMOSA28 & \sc DEPFET & \sc VeloPix\\ % \hline \hline
	\midrule
	sensor thickness  X/X$_0$ [\%]	& $\leq$0.1	&	$0.075$ &	-        &	$0.079$	& $0.18$	&	$0.73$\\
	readout cycle [$\SI{}{\micro\second}$]	& $\leq 0.1$	&	$0.05$        &	$50$	&	$185.6$	&	$20$ &	$0.025$\\
	power [$\SI{}{\mW\per\cm\squared}$]	&	$\leq 200$ & $150$	&	-		&	$170$	&	$~100$	& $1500$ \\
% \hline
	\bottomrule
	\end{tabular}
\end{center}
\caption{Alternative technology comparison, only the crucial parameters are
  listed. }
\label{tab:AlternatevTechnologies}
\end{table*}

The usage of pixel detectors in particle and heavy-ion physics has recently lead to a variety of thin detectors capable of dealing with high particle fluxes. It is worth to discuss the solutions developed for the upgrades of the STAR, BELLE and LHCb vertex detectors. 

The STAR upgrade vertex detector \cite{Margetis:2011zz} is based on Monolithic Active Pixel Sensors (MAPS), more precisely the MIMOSA 28 sensor \cite{Valin:2012zz}. The vertex detector under construction is located around the interaction point at r=$\SI{2.5}{\cm}$, $\SI{8}{\cm}$ and has an intrinsic hit resolution of <$\SI{6}{\micro\meter}$. An ultralight carbon fibre support structure has been developed with a thickness of $\SI{200}{\micro\meter}$. The radiation length is X/X$_0$=$\SI{0.079}{\%}$ per layer in the active region. The power dissipation is about $\SI{170}{\mW\per\cm\squared}$ and forced air cooling at $\SI{10}{\m\per\s}$ is used \cite{Greiner2012}. The charge collection for the MIMOSA chips is based on electron diffusion, resulting in charge collection times of $\SI{50}{\nano\second}$. This has to be compared to the charge collection times of drift based silicon detectors like silicon strip and hybrid silicon pixel detectors of a few $\SI{}{\nano\second}$. The signal integration of the MIMOSA 28 chip is $\SI{185.6}{\micro\second}$. This long integration time corresponding to  >$\num{e5}$ decay events rules out the usage of the MAPS developed for STAR vertex upgrade in Mu3e,  

Belle2 is building a lightweight silicon vertex tracker based on DEPFET pixel sensors \cite{Abe:2010sj}. The Belle2 silicon vertex detector has two layers at r=$\SI{1.4}{\cm}$, $\SI{2.2}{\cm}$. The monolithic sensors are only $\SI{75}{\micro\meter}$ thick and self supporting, resulting in a radiation length of X/X$_0$=$\SI{0.18}{\%}$ per layer \cite{Kreidl2012}. The cooling system uses liquid CO$_2$ for the readout electronics and cold dry air with forced convection for the sensors. The sensor chip is equipped with an integrated read out amplification, the charge is then accumulated internally. The readout is enabled externally and done in a row wise rolling shutter mode. The readout time for the entire chip is $\SI{20}{\micro\second}$. As for the MIMOSA 28 sensor the relatively long readout time makes the Belle2 DEPFET sensor unsuitable for the Mu3e detector. On top of this the Belle2 DEPFET sensor produces analog output signals and requires further electronics for analog to digital conversion, baseline subtraction, discrimination and zero-suppression.

The proposed pixel detector for the LHCb vertex upgrade belongs to the TIMEPIX family of chips. The main goal of the LHCb upgrade is to move from a $\SI{1}{\MHz}$ readout to a $\SI{40}{\MHz}$ readout in order to make a track based L0-trigger possible. The power consumption per chip is $\SI{1.5}{\W\per\cm\squared}$ \cite{Beuzekom2012}, which is quite high. The sensor thickness is $\SI{150}{\micro\meter}$ to $\SI{200}{\micro\meter}$. The LHCb Velo upgrade pixel sensor is a hybrid design, so in addition to the sensor there is also the readout chip thickness, the solder bump bonds, aluminum traces and cooling elements summing up to X/X$_0$=$\SI{0.73}{\%}$ per layer. In LHCb the vertex detector is inside a vacuum tank around the interaction point, so the cooling scheme relies on a high thermally conductive spine (diamond) in combination with a liquid CO$_2$ cooling outside the active area. Alternative cooling studies look at micro-channel cooling which would extend the CO$_2$ cooling to the pixel chips in the active area. The pixel sensor is designed to have less than $\SI{25}{\nano\second}$ time-walk at 1000 electrons. The output bandwidth is > $\SI{12}{Gbit/s}$ with on-chip zero suppression. The readout is hit driven and asynchronous. While the readout speed would be ideal for the Mu3e detector, the radiation length per layer of X/X$_0$=$\SI{0.73}{\%}$ is too high.

\balance

\subsection{Conclusion}
\label{sec:AlternativeTechnologiesConclusion}   

In the field of tracking detectors with good resolution, low radiation length
and high rate capabilities there has been a very strong progress in the recent
years, see Table~\ref{tab:AlternatevTechnologies}. Time projection chambers
with helium gas mixtures would provide excellent momentum resolution. 
Because of the very long drift times in a large TPC and the constantly high
muon decay rates, it is not possible to do online track reconstruction in the
event filter farm for decay rates exceeding $\SI{1e7}{\Hz}$, which corresponds
to phase IA of the Mu3e running. 
As a change of technology between phases IA and~IB is not desirable, the TPC option can only be considered as a back-up
solution.

The above discussed pixel alternatives can be grouped in two categories. The systems based on the MIMOSA chip for the STAR upgrade and the DEPFET chip for Belle 2 have little radiation length of X/X$_0$=$\SI{0.079}{\%}$ and $\SI{0.18}{\%}$ per layer, but have long readout cycles. On the other hand the VeloPix based tracker for the LHCb upgrade can be read out at $\SI{40}{\MHz}$, but has a relatively large radiation length of X/X$_0$=$\SI{0.73}{\%}$ per layer. In all three cases further R$\&$D would be required to transform the existing designs into one fulfilling the requirements of all phases of the Mu3e experiment.

%!TEX root = ../RP.tex
\chapter{The Mu3e Fibre Detector}
\label{sec:Fibre}

\nobalance

\section{The time of flight detector}

%%%%%
\begin{figure}[h]
\centering
        \includegraphics[width=1.0\linewidth]{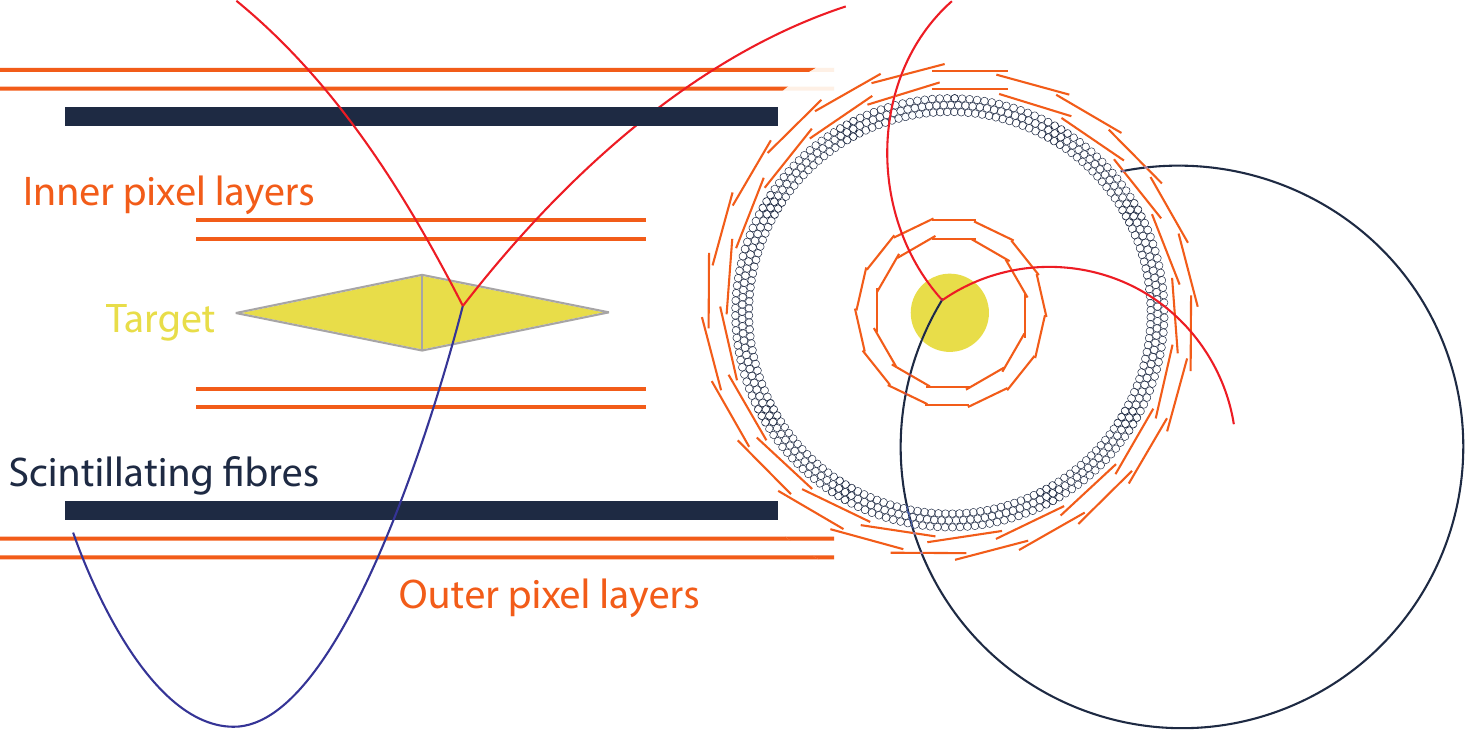}
\caption{Schematic side view \emph{(left)} and cross section \emph{(right)} of the detector components. The Sci-Fi system is highlighted in blue.}
\label{fig:persp}
\end{figure}
%%%%%

A cylindrical time of flight (ToF) detector complements the central silicon tracking system. It consists
of a scintillating fibre (Sci-Fi) hodoscope with a radius of \SI{6}{\centi\metre} and a length of \SI{36}{\centi\metre}.
The expected time resolution is of several \SI{100}{\pico\second}
and a detection efficiency close to \SI{100}{\percent}. 
The main purpose of the ToF system is to measure very precisely the arrival time of 
particles in order to allow for the matching with hits detected in the silicon detectors.
This will help to reject pile-up events (accidental backgrounds) and allow for a charge 
(direction of propagation) measurement for recurling tracks.
The ToF system will operate at very high particle rates up to several \si{\mega\hertz} per channel.

A detailed R\&D program is ongoing to prove the feasibility of the Sci-Fi detector
and to help optimizing the design of the Sci-Fi detector.
The R\&D activity covers all the aspects of the ToF detector development: scintillating 
fibres, SiPMs, amplifiers, and readout electronics. The challenging aspects of the ToF 
system are
(i) the high rates per Sci-Fi readout channel and
(ii) the large data flow generated by the readout digitizing electronics, which has to be 
handled in real time. Time resolutions of about \SIrange[range-units=single, range-phrase = { to }]{200}{300}{\pico\second} 
have been already achieved with Sci-Fi hodoscopes with single-ended readout using multi-anode PMTs~\cite{Abb07}.
We aim at a similar time resolution.

%%%%%
\begin{figure*}[t!]
\centering
	\includegraphics[width=0.32\textwidth]{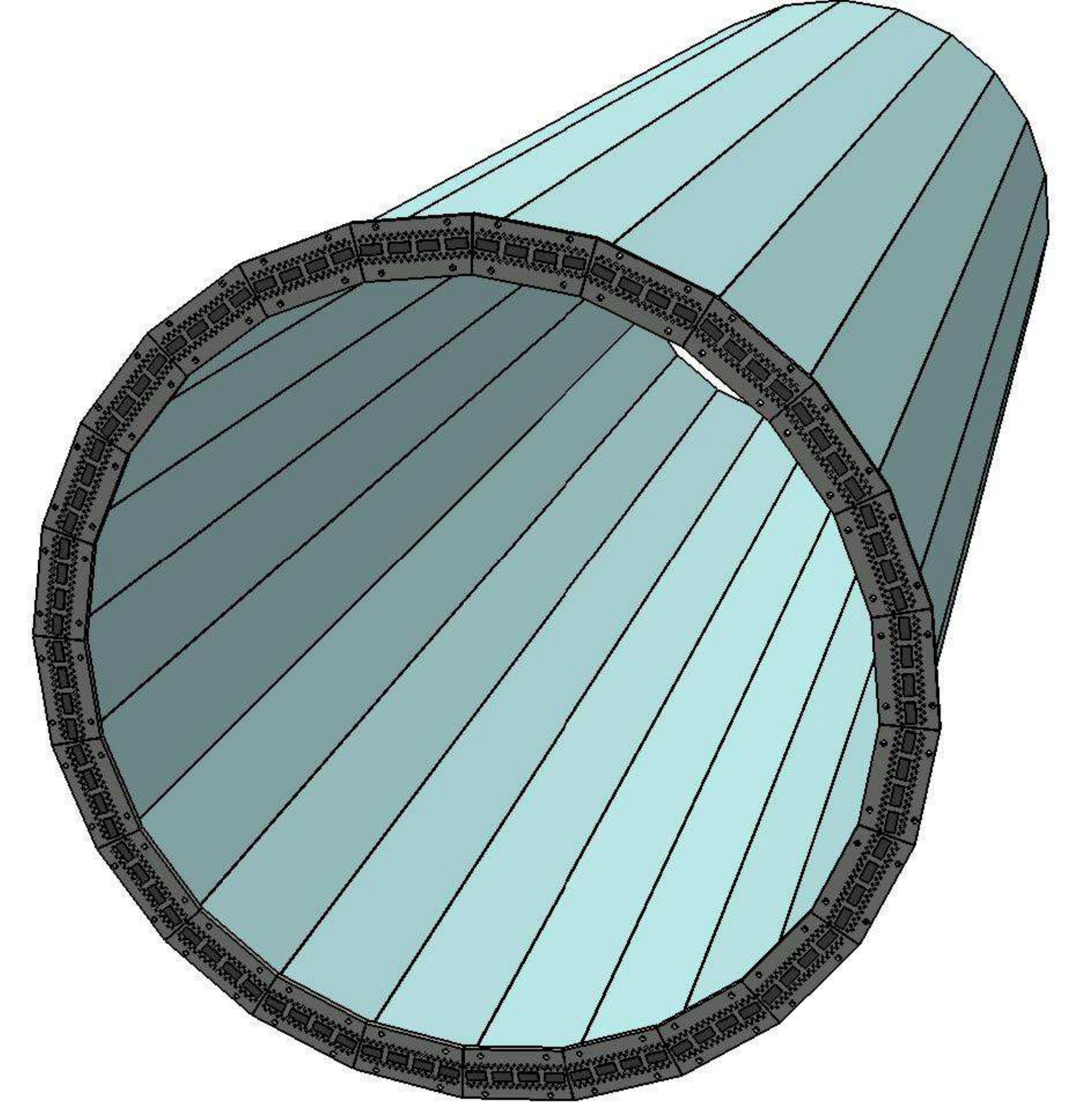}
         \includegraphics[width=0.32\textwidth]{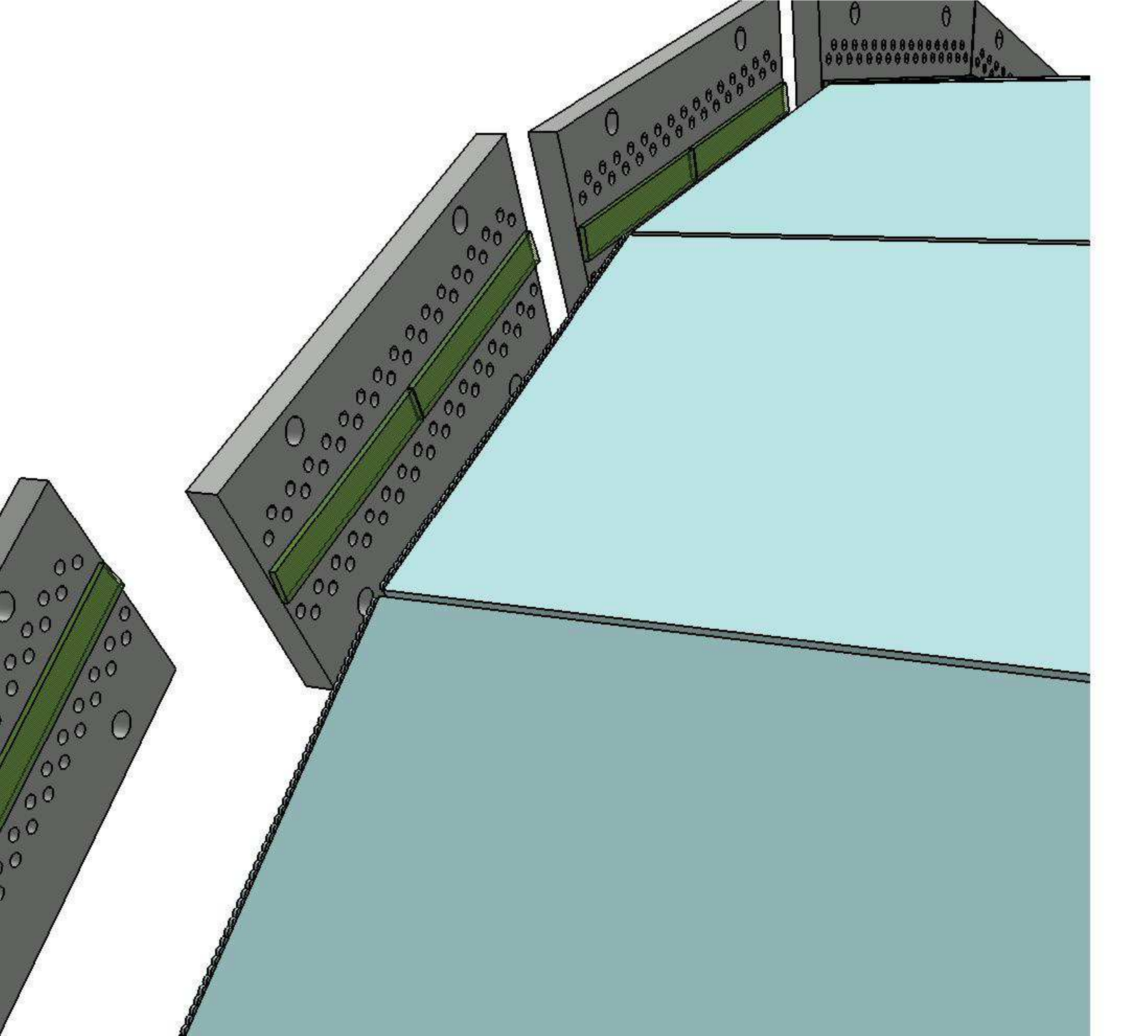}
	\includegraphics[width=0.32\textwidth]{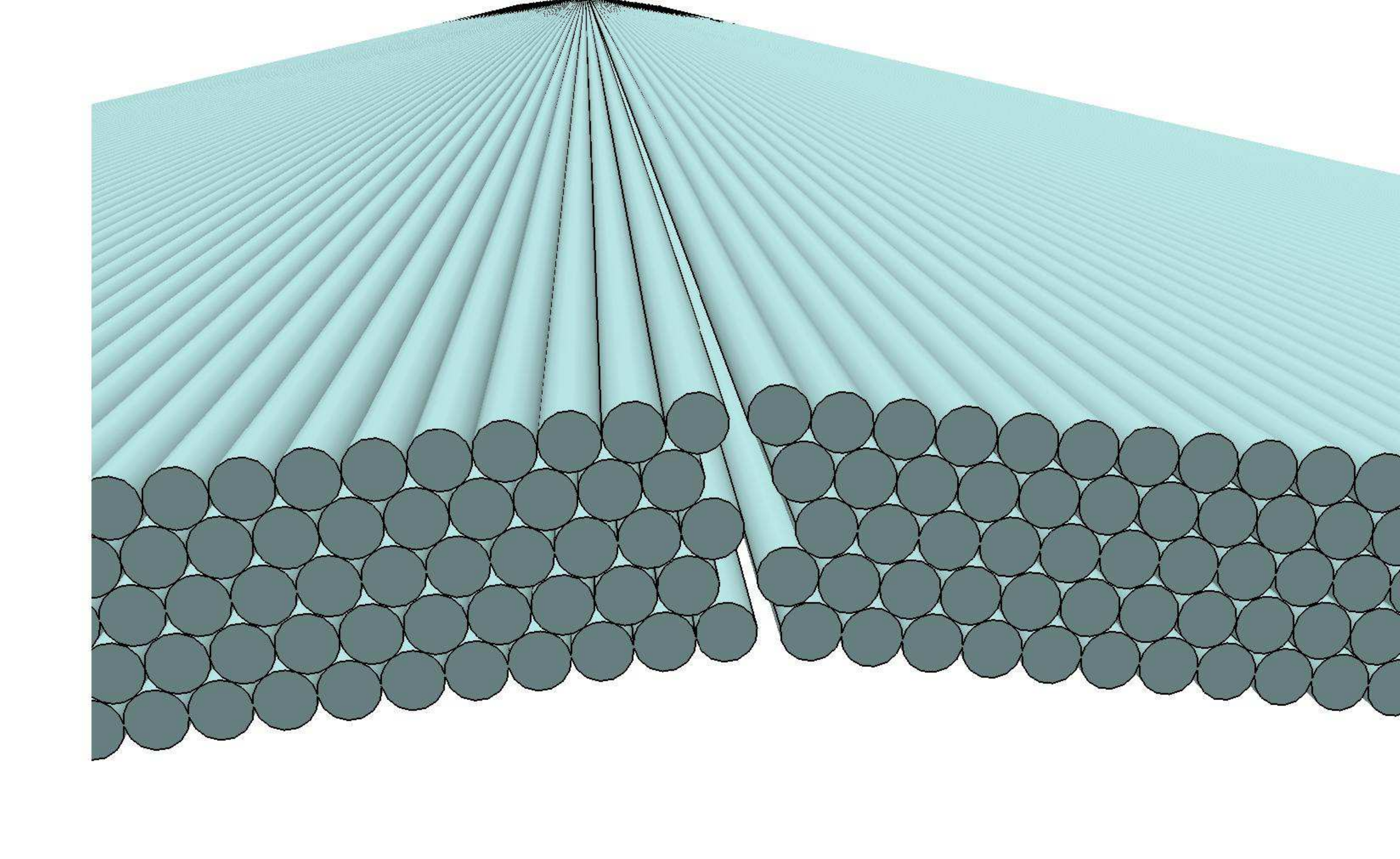}
\caption{{\it Left:} Overall view of the Sci-Fi barrel ToF detector.
The diameter of the detector is about \SI{12}{\centi\metre}, the length about \SI{36}{\centi\metre}.
{\it Middle:} Detail showing the SiPM ceramic supports and the Sci-Fi arrays.
{\it Right:} Details of the Sci-Fi ribbons. In this figure, the ribbons consist of five staggered Sci-Fi layers.
}
\label{fig:barrel}
\end{figure*}
%%%%%

\begin{figure*}[t!]
\centering
	\includegraphics[width=0.49\textwidth]{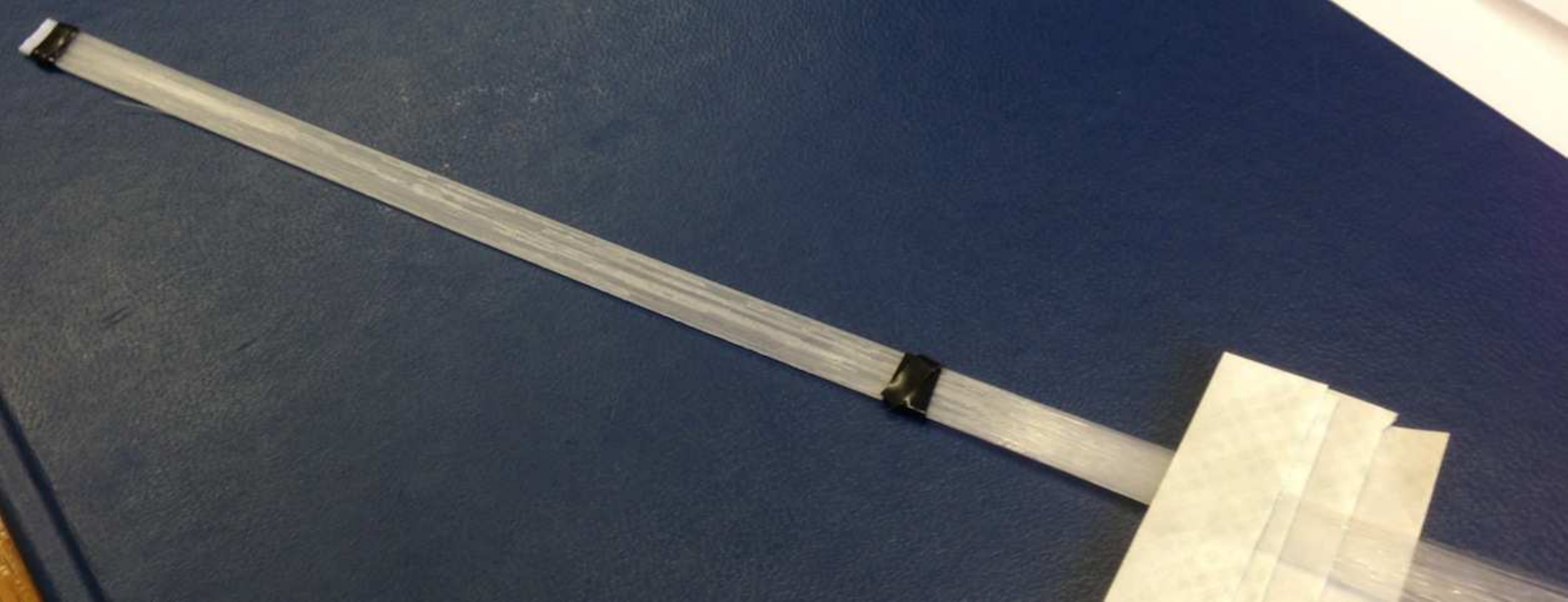}
	\includegraphics[width=0.49\textwidth]{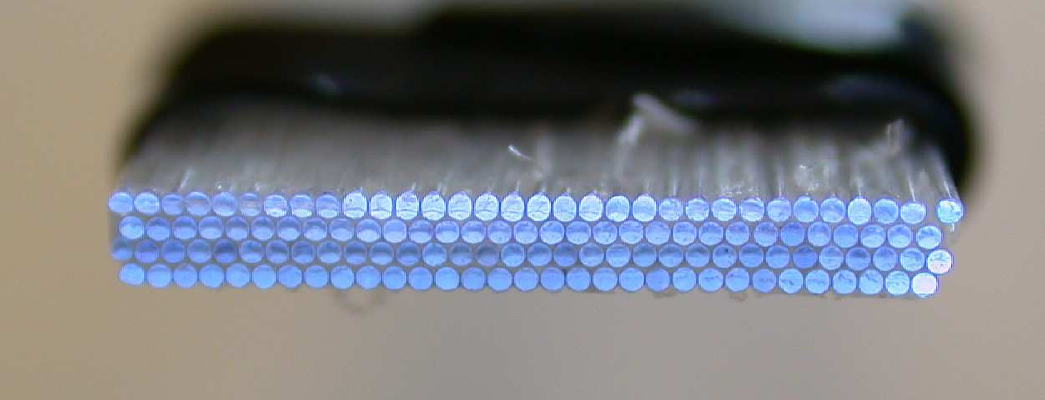}
\caption{Photos of a Sci-Fi ribbon ({\it left}) and of its cross section ({\it right}).
The uniform staggering of 4 Sci-Fi layers is clearly visible.}
\label{fig:scifi}
\end{figure*}

In the current design, the scintillating fibre detector is composed of 24 Sci-Fi
ribbons, each \SI{36}{\centi\metre} long and \SI{16}{\milli\metre} wide, as illustrated
in Figure~\ref{fig:barrel}. The ribbons will be supported by a carbon fibre mechanical
structure. The use of \SI{250}{\micro\metre} diameter multi-cladding scintillating fibres
produced by Kuraray is envisaged. Three to five Sci-Fi layers (Figure~\ref{fig:barrel}
right) are staggered such as to minimize empty spaces between fibres.
Figure~\ref{fig:scifi} shows a Sci-Fi ribbon and its cross section.
In the figure the uniform staggering of the 
Sci-Fi layers can be seen. The overall thickness of the Sci-Fi arrays varies depending on
the number of layers between \SIrange[range-phrase = { to },
range-units=single]{0.6}{1}{\milli\metre}. This thickness has to be kept as low as
possible, to reduce the multiple scattering to a minimum compatible with the required
performance (i.e. time resolution and efficiency). The empty space between adjacent
ribbons will be minimized in order to guarantee continuous coverage.

%%%%%
%%%%%

The scintillating light produced in the fibres will be detected with SiPM arrays at both
fibre ends. The choice of SiPMs as a photo-detection device is based on the fact that
they are very compact objects that can be operated in high magnetic fields with high gain
($\sim 10^6$) and at high counting rates. 

Two different readout schemes for the fibres are being considered and investigated:
1) fibres are grouped in vertical columns (this is implemented by the structure of the readout photo-device),
2) each fibre is read out individually. 
In the first design we envisage to use the 32 
channel SiPM arrays with \SI[product-units = power]{50 x 50}{\micro\metre} pixels
available from Hamamatsu~\cite{Beischer:2010zz,Beischer:2011zz}. The active size of the
sensor is \SI[product-units = power]{8 x 1.1}{\milli\metre} with a total surface of
\SI{9}{\milli\metre\squared}. The pixels are arranged in columns, corresponding to an
effective pitch of \SI{250}{\micro\metre}. 
Two such SiPM sensors will be assembled side by side with almost no dead region,
giving a \SI{16}{\milli\metre} wide photo-sensitive region with 64 readout channels, and
matching precisely the width of the Sci-Fi ribbons. The photo-detectors will be directly
coupled to the Sci-Fi arrays to maximize the light collection efficiency, and consequently
the timing performance. In total 96 SiPM arrays with 32 channels each will be required to
readout the Sci-Fi tracker at each end, for a total of \num{2 x 24 x 64} ($\sim
\num{3000}$) readout channels.

\begin{figure}[t!]
\centering
\includegraphics[width=1\linewidth]{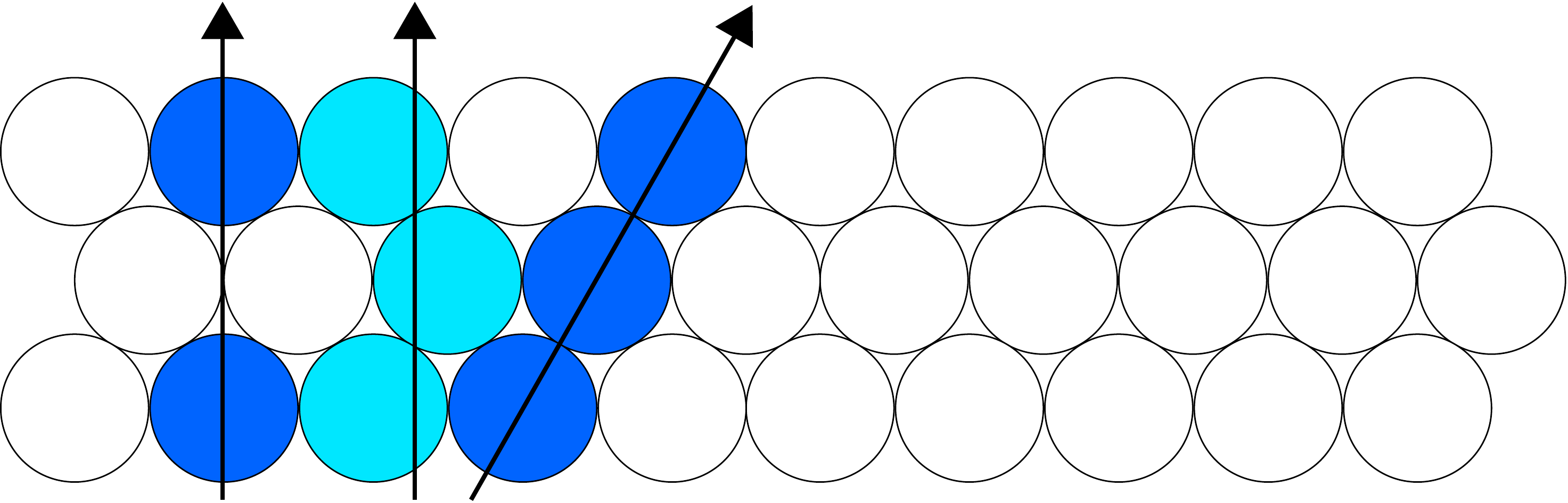}
%\vspace*{-20mm}
\caption{Sci-Fi occupancy vs. particle crossing angle and position. Dark fibres indicate higher energy deposit.}
\label{fig:occ}
\end{figure}

A detailed understanding of the expected occupancies in the the Sci-Fi
detector is required. The rates depend also on the background in the detector generated by
secondary interactions in the materials of the different components. A particle crossing
the Sci-Fi ribbons at \ang{90}, will excite one or two adjacent Sci-Fi columns, as
sketched in Figure~\ref{fig:occ}. A particle crossing the same Sci-Fi ribbon, for example
at \ang{45}, will excite 4 to 5 adjacent Sci-Fi columns for a 5 layer ribbon and 2 to 3 
for a 3 layer ribbon, thus increasing significantly the overall occupancy of the detector. 
At an expected rate of \num{1.5e9} muon decays per second using \num{1500} Sci-Fi
channels, the estimated rate in the Sci-Fi system is \SI{5}{\mega\hertz} on average.
At the same time the signal will be spread out over several SiPM channels. A well designed
clustering algorithm will be required to reconstruct the timing and position of the crossing
particle.
%%The granularity of the system will be such that the single
%%channel event rate will not exceed a few \si{\mega\hertz}.

The lowest possible detector readout occupancy can be achieved  by reading out each fibre individually.
In absence of dark current noise and background, the occupancy is given by the rate of detected 
electrons from muon decays and is expected to be \SI{1}{\mega\hertz} per fibre. 
When reading out each multi-cladding fibre individually, one has to take into account that the scintillating
light travels preferentially in the cladding of the fibre and exits the fibre in a cone with an aperture
of about \ang{45}.
In order to collect all the scintillating light exiting the fibre, and thus ensure high detection efficiency,
the photo-detector has to be wider then the fibre diameter, at least \SI{100}{\micro\metre} around the fibre.
This compensates also the misalignment, of the order of few tens of microns, between the fibres and the SiPMs. 
In order to avoid the light from one fibre to spill over to a neighboring photo-detector channel
(optical cross-talk) the fibres and photo-detectors have to be separated by \SI{500}{\micro\metre}
center-to-center.
To summarize, the estimated, optimal size of the photo-detector for reading out each \SI{250}{\micro\metre} fibre individually
is \SI[product-units = power]{400 x 400}{\micro\metre} (for a square device), and a center-to-center separation of \SI{500}{\micro\metre}.

A transition region between the highly compact Sci-Fi ribbons and the photo-detectors of about \SI{5}{\centi\metre}
will be required to fan-out the Sci-Fis to the desired center-to-center separation of \SI{500}{\micro\metre}
before coupling the fibres to the SiPMs.
To assure the precise alignment, the fibres will be glued in sockets with a precision of \SI{10}{\micro\metre}.
The SiPMs will be coupled to the Sci-Fis using the same sockets.
Reading out each fibre individually will require a total of about \num{9000} readout channels. 

In case the SiPMs could not be coupled directly to the Sci-Fis (because of
mechanical constraints inside the limited space), optical fibres coupled to the Sci-Fis
will be used to transmit the scintillating light outside of the solenoid, where
they will be coupled to the same type of SiPM arrays, however with some loss of light. 
In this case the use of multi-anode PMTs or multi-channel plate PMTs could be also envisaged
as alternative photo-detector device to the SiPMs for the readout of the Sci-Fi ribbons.

Different manufacturers are also being investigated. The reported
photo-detection efficiency of the Hamamatsu SiPMs is not that high (PDE $\sim$
\SI{25}{\percent}). Since the time resolution depends on the number of detected
scintillating photons ($\sigma_\tau \sim 1 / \sqrt{n_{\textnormal{ph}}}$) in order to obtain the best
timing performances we need to maximize the PDE of the SiPM arrays. Discussions with
Hamamatsu to explore various possibilities to increase the PDE are ongoing. We will
also discuss with other companies capable of producing such devices (we already contacted
FBK in Italy and Ketek in Germany).
Very likely it will  be required to design SiPM sensors matching precisely the requirements of the
Sci-Fi detector rather than using commercially available SiPMs.

Performance studies of Sci-Fi ribbons coupled to SiPM arrays will be carried out in test
beams at PSI. The test beam activities will include the study of the time resolution, rate
capabilities, detection efficiency, tracking resolution, and uniformity of the detector.
Different readout electronics will be also tested.

%%%%%
%%%%%

\section{Readout of photon detectors}

For the readout of the SiPMs different options are being investigated.
Details are given in chapter \ref{sec:DAQ}.

One possibility is to use the well-established waveform digitizing technology 
based on the DRS switched capacitor array chip developed at PSI.
The advantage of this technology compared to traditional constant fraction discriminators
and TDCs is that pile-up can be effectively recognized and corrected for.
Due to the bulkiness of this readout electronics, the digitizing electronics cannot
be located in the proximity of the Sci-Fi detector.
To transmit the electrical pulses from the SiPMs to the digitizing electronics
two options are being considered. In a first version the signals are transmitted without
amplification on shielded low attenuation coaxial cables.
Given the low amplitude of the pulses (\SIrange[range-units=single]{1.5}{2}{\milli\volt} per photo-electron)
and potentially high noise from the digital readout electronics,
the amplification of SiPM signals might be required.
In this second version the pulses are first amplified about a factor of ten and then transmitted.
Hybrid amplifier circuits will be installed on the other side of the SiPM mountings.
We already developed fast amplifiers using transistors
(rise time \SI{\sim 1}{\nano\second}, decay time \SI{\sim 10}{\nano\second}) matching the input
characteristics of SiPMs for optimal time performance and high rates.

A second possibility is to use the STiC chip, which includes a fast discriminator and a TDC
digitizing the time information.
The advantage of this solution is the compactness of the chip that could be installed very close to the SiPMs,
with no need to amplify and to transmit analog signals outside of the solenoid.
A third option could be offered by time to digital converters implemented in FPGAs.

\section{GEANT simulations}
\label{sec:FibreSimulation}

To optimize the overall design of the detector and of the Sci-Fi sub-system in particular,
extensive simulation studies are being carried out. We are exploring the propagation of
scintillating light in the fibres in a dedicated simulation that allows to compare the
optical properties for any possible fibre and ribbon geometry. Meanwhile a complete set of
simulation tools is available which allows a profound analysis of the different possible
layouts of the Sci-Fi system.

Ongoing simulation studies will help us to optimize and work out the details of the final
Sci-Fi ToF design. Given that a crucial requirement of the experiment is to minimize all
involved materials also a layer of reflecting paint or glue will add additional non active
material to the detector and thus increase the overall material budget seen by the
particles crossing the Sci-Fi ribbons, resulting in a deterioration of the overall
momentum resolution of the tracking system. This and similar effects can now be studied
easily with the available tools.

\balance

\begin{figure}
	\begin{center}
		\includegraphics[width=\linewidth]{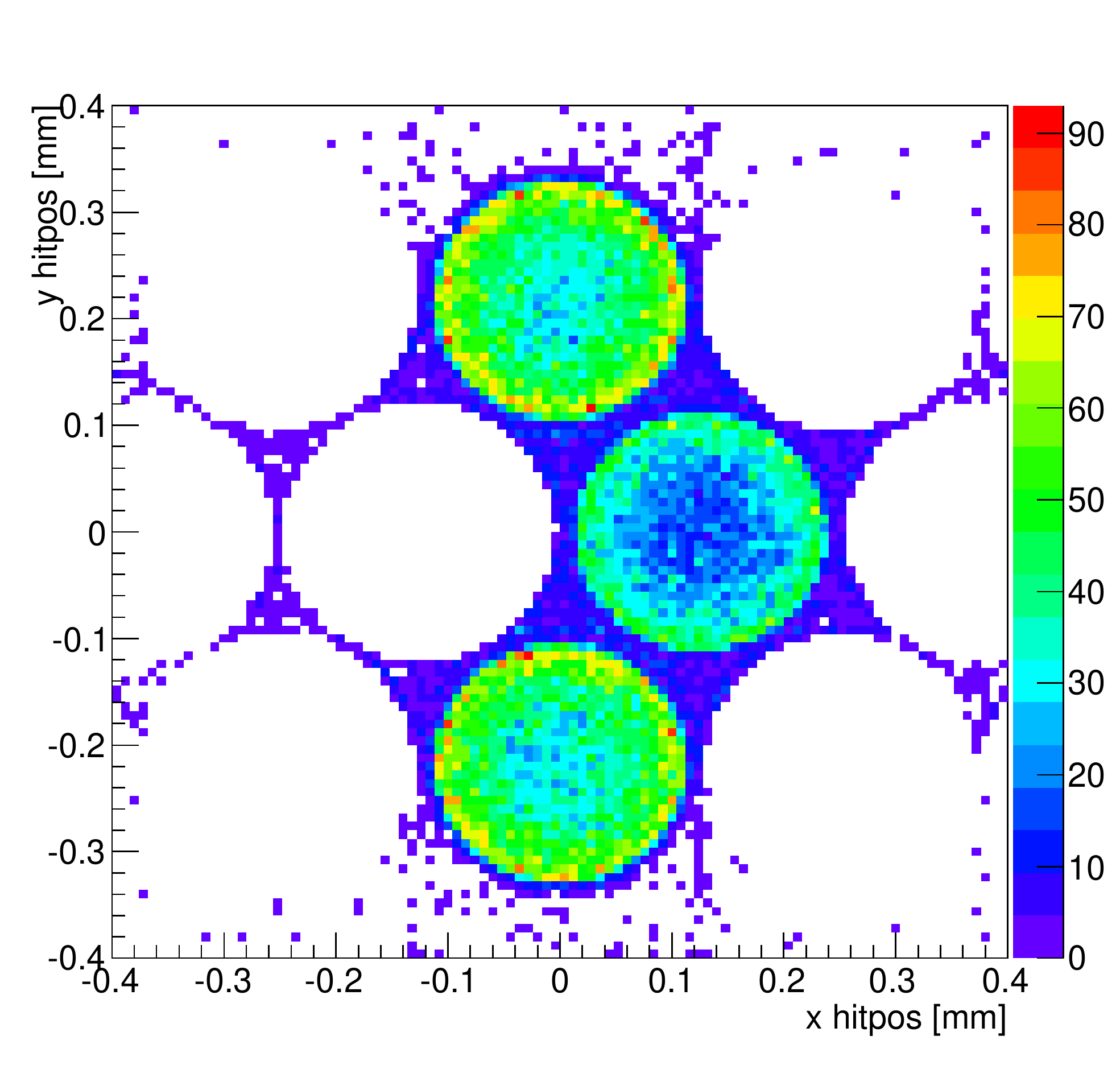}\\
		\includegraphics[width=\linewidth]{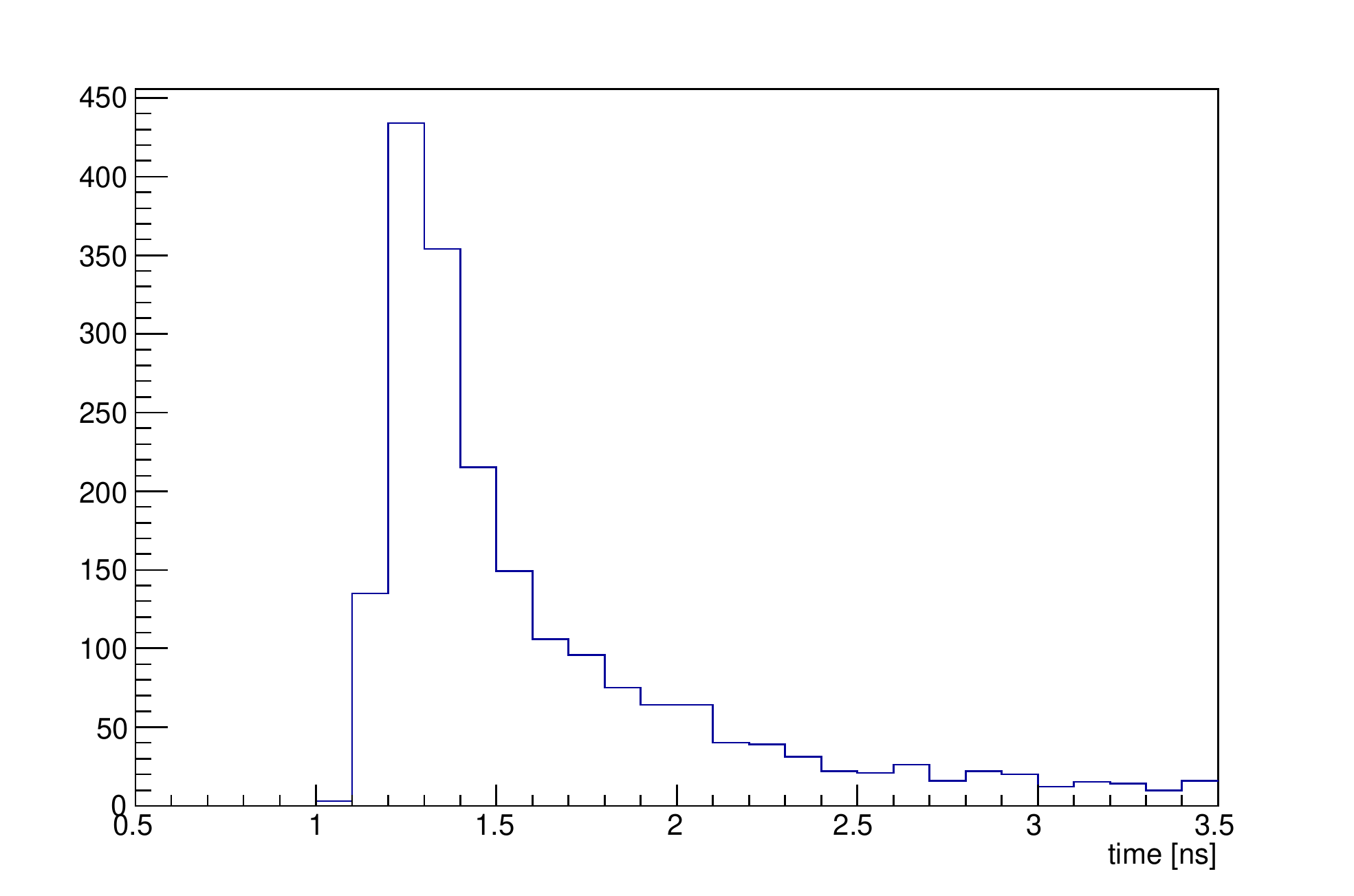}
		\caption{Simulated photon output integrated over \num{1000} positron crossing 
		events for a specific fibre geometry \emph{(top)} with the corresponding time
		distribution of the SiPM response using a constant fraction discriminator algorithm
		with a FWHM of about \SI{300}{\pico\second} \emph{(bottom)}.}
		\label{fig:simulatedCrossingEvent}
	\end{center}
\end{figure}

In combination with a simulation of the SiPM response \cite{Eckert:2012yr},
we are now able to estimate whether the aimed time resolution of a few hundred picoseconds is
feasible with the fibre geometry and the SiPMs under investigation.
First studies show that the resolution can be achieved with the baseline design
(Figure~\ref{fig:simulatedCrossingEvent}).
Further fibre geometries and readout combinations are
currently being evaluated. In addition to the simulation test setups are being developed
consisting of a complete ribbon with the SiPM detectors mounted on both ends to verify
the simulation results and to get the prove whether the desired geometry can be
constructed with reasonable effort. Furthermore the test stand will be used to evaluate
readout electronics.

An additional source of concern is the cross talk and after pulsing observed in most APDs
operated in Geiger mode. Both effects will lead to an increase of the occupancy of our
electronics channels and thus have to be minimized. Since the Sci-Fi ribbons are read-out
on both ends, this uncorrelated source of noise, in principle, can be rejected.

%%%%%
%\begin{table}[!t]
%\centering
%\begin{tabular}{lr}
%\hline
%item & estimated cost (CHF) \\
%\hline \hline
%Sci-Fi ribbons &       \num{7500} \\
%\hline
%support structure &  \num{5000} \\
%\hline
%SiPM arrays &  \num{30000} \\
%\hline
%amplifiers (?) & \num{10000} \\
%\hline
%DRS digitizers &  \num{75000} \\
%\hline
%power supplies (?) & \num{7500} \\
%\hline
%cabling (?) & \num{10000} \\
%\hline\hline
%TOTAL & \num{145000} \\
%\hline
%\end{tabular}
%\caption{Estimated cost of the Sci-Fi ToF-tracker for {\bf Phase I}.
%750 redout channels are assumed.}
%\label{tab:cost}
%\end{table}

\chapter{The Mu3e Tile Detector}
\label{sec:Tiles}

The second component of the Time-of-Flight hodoscope is a detector consisting  of scintillating tiles which are, as in the case of the Sci-Fi tracker, read out by SiPMs.
The tile detector is located in the recurl station on the inside of the recurl pixel layer (see Figure~\ref{fig:schematic_longitudinal}).
The detector aims at a time resolution of below $\SI{100}{ps}$ and an efficiency close to 
$100\%$ in order to effectively identify a coincident signal from three electrons and suppress combinatoric background. The main challenge of the detector design is to achieve these requirements under the high hit rate which is expected, especially in phase II.

\section{Detector Design}

The detector is divided into four modules (one for each recurl station) with a length of $\SI{36}{cm}$ and diameter of $\approx \SI{12}{cm}$. Each module is segmented into small tiles with a size of $\mathcal{O}(\SI{1}{cm\cubed})$, which are individually wrapped in reflective foil in order to increase the light yield.

\begin{figure}[t!]
	\centering
		\includegraphics[width=0.48\textwidth]{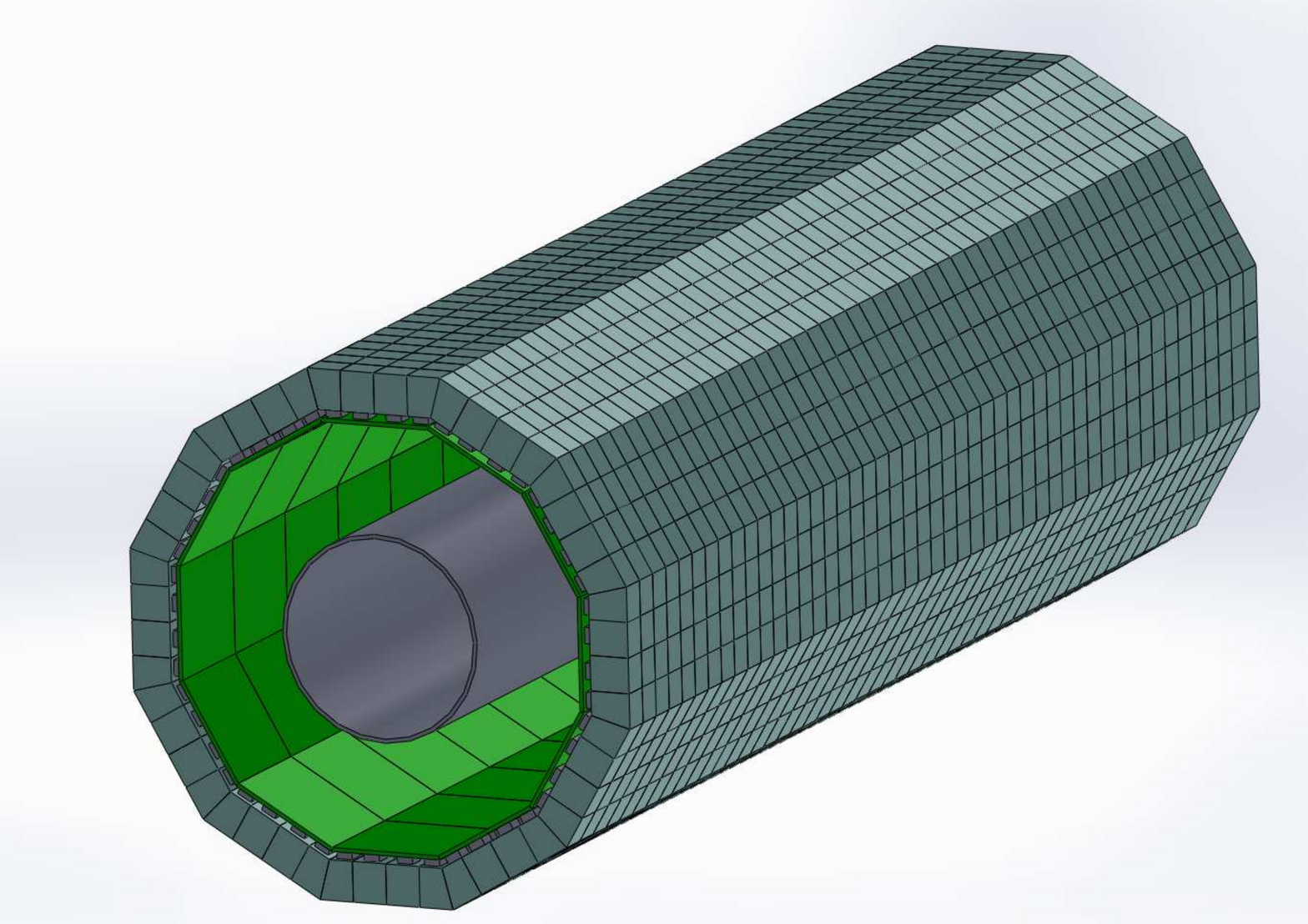}
	\caption{Drawing of a tile detector module with $48\times48$ tiles.}
	\label{fig:proto}
\end{figure}

In the current design, each module consists of 48 rings with a thickness of $\SI{5}{mm}$, and each ring consists of 48 tiles (see Figure~\ref{fig:proto}). This results in a total number of 2304 tiles per station, and a tile geometry of roughly $7.5\times 7.5\times \SI{5}{mm}$. For the outer two recurl stations, the number of tiles might be decreased to $36\times36$ due to the lower hit rate (see Figure~\ref{fig:rate}), in order to reduce the costs. This detector geometry is the result of extensive simulation studies which will be continued in order to further optimize the detector performance.

The tiles are made of plastic scintillator material which provides a fast light response. The main requirements for the scintillator are a high light yield, fast rise and decay time and a scintillation spectrum which approximately matches the spectral sensitivity of the SiPM. Different scintillators have been compared using a Geant4 simulation in combination with a simulation of the SiPM response. The best suited scintillator was found to be BC420 from Bicron. However, the differences between the individual scintillating materials were found to be small, and thus also more inexpensive alternatives are feasible.

The scintillation light of each tile is read out by a SiPM which is directly attached on the inside of the tile. SiPMs are well suited for this application due to the compact size, insensitivity to magnetic fields, high photon detection efficiency (PDE) and excellent timing properties.
The most important SiPM characteristics for this application are a high PDE and gain in order to obtain a large signal and therefore a good time resolution. Furthermore, a small signal decay time is desired in order to reduce signal pile up and thus increasing the signal efficiency.
The S10362-33-050 MPPC from Hamamatsu with an active area of $3\times\SI{3}{mm\squared}$ and 3600 pixels is a suitable sensor for this application.
%This sensor has a PDE of $\approx 35\%$, a gain of $\approx7.5\cdot10^5$ and a signal decay time of $\approx15\,\mathrm{ns}$.
A possible alternative sensor is the PM3350 from Ketek which offers a larger PDE and gain compared to the Hamamatsu device, but has a longer decay time.
A further option is the SensL MicroFB series, which will be available in the first quarter of 2013. These sensors are a very promising alternative, as the devices will have an additional fast output providing signals with a width of $\mathcal{O}(\mathrm{ns})$ and provide a high PDE and gain.
Measurements and simulation studies comparing the different sensor types will be continued, in order to find the devices which is best suited for this application.

For the readout of the SiPMs, two different options are currently discussed. One solution is to digitize the SiPM waveforms using the DRS chip. The waveforms then have to be further processed externally, in order to determine the time-stamps. The second option is the STiC chip which includes a fast discriminator and a TDC digitizing the time information. Details on both chips can be found in chapter~\ref{sec:DAQ}.

\section{Simulation}

Simulation studies using the full detector simulation described in section~\ref{sec:Simulation} have been carried out in order to determine the detector performance. The optical properties of a tile are parameterized using a separate Geant4 simulation of a single tile. The SiPM simulation GosSiP \cite{Eckert:2012yr} was used to generate the signal waveforms from the hit information of the full detector simulation.
The simulation has been done for a detector design with $48\times48$ $\SI{5}{mm}$ thick tiles per recurl station with BC420 scintillator and S10362-33-050 MPPCs.

\begin{figure}[htb!]
	\centering
		\includegraphics[width=0.48\textwidth]{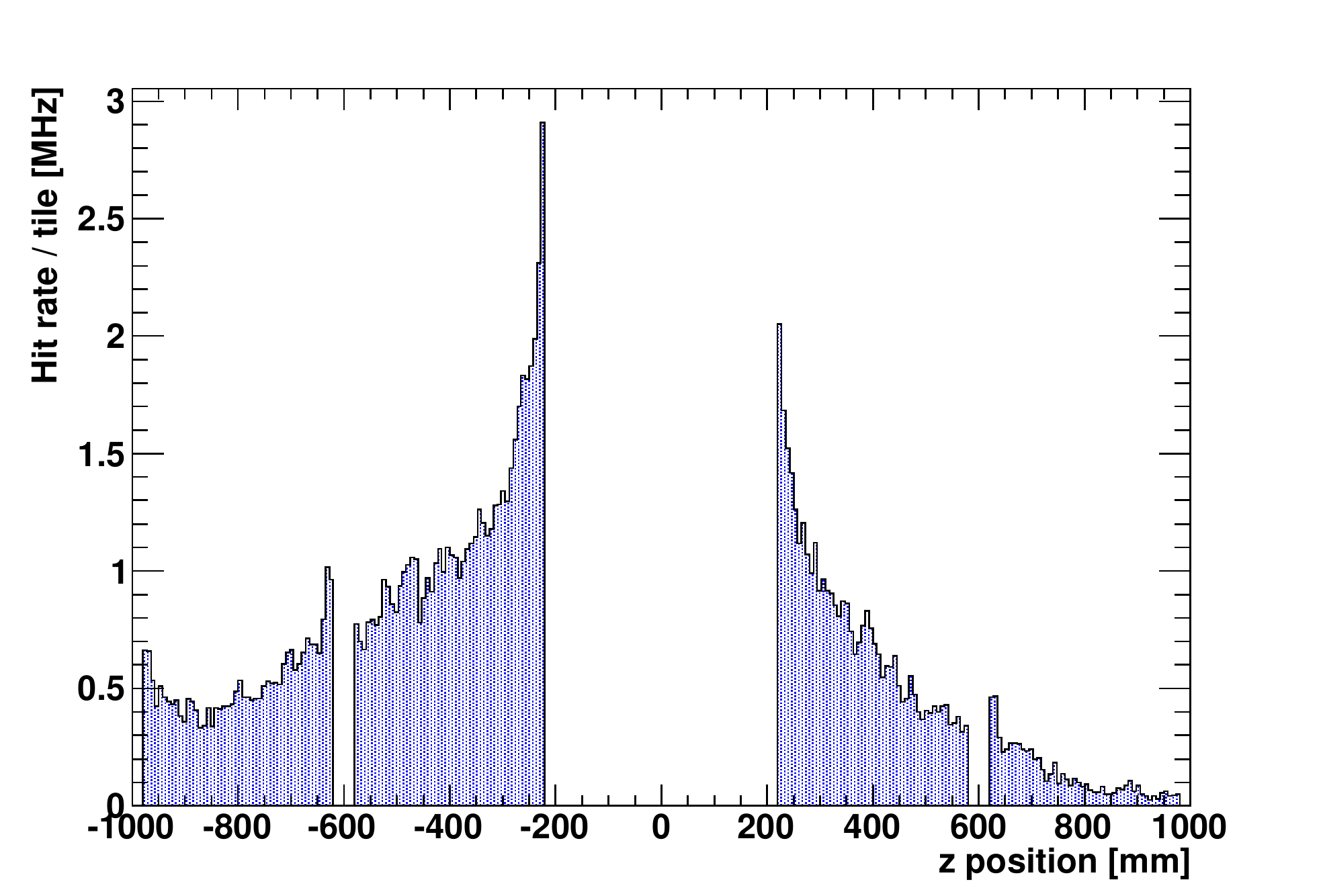}
	\caption{Hit rate per tile for phase II. The rate for phase I is a factor 10 to 20 smaller.}
	\label{fig:rate}
\end{figure}

Figure~\ref{fig:rate} shows the hit rate per tile as a function of the position along the beam direction for phase II, including also tracks which pass several tiles, as well as background events e.g.\ from the collimators. These rates can be handled by the proposed readout electronics (see section~\ref{sec:DAQ}). In order to further reduce the hit rate, and therefore the signal pileup, optimizations like adjusting the tile angle to match the mean incident angle of the tracks are currently studied.

%\begin{figure}[htb!]
%	\centering
%		\includegraphics[width=0.48\textwidth]{Tiles/figures/waveform}
%	\caption{Typical waveform for a single tile for phase II.}
%	\label{fig:waveform}
%\end{figure}

%Figure~\ref{fig:waveform} shows a simulated waveform for a single tile in the innermost ring ($z=-\SI{220}{mm}$). In most cases, the individual signal pulses are well separated and can clearly be distinguished.

The time-stamps for the individual hits are assigned using the signal waveforms generated by the SiPM simulation. A simple fixed threshold method is used which approximates the behaviour of the STiC chip. The same method can also be applied to the output of the DRS chip.

\begin{figure}[htb!]
	\centering
		\includegraphics[width=0.48\textwidth]{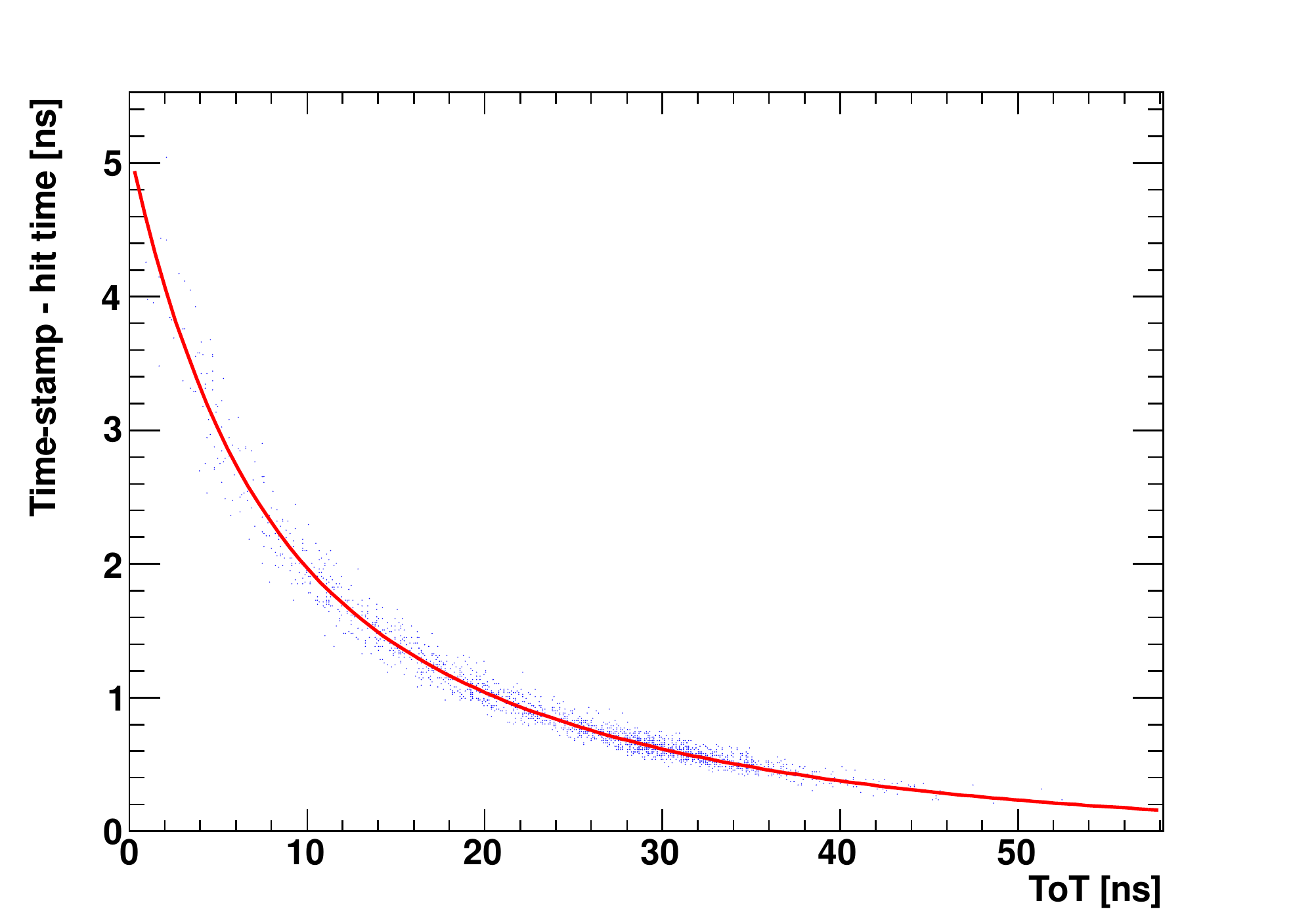}
	\caption{Timewalk correction.}
	\label{fig:twCorr}
\end{figure}

The optimal threshold was determined to be $\approx 10\%$ of the mean signal amplitude ($\approx 350$ pixels). Due to the varying signal amplitudes and the fixed threshold, the individual time-stamps have to be corrected for the timewalk effect. Figure~\ref{fig:twCorr} shows the timewalk as a function of the Time-over-Threshold.

\begin{figure}[htb!]
	\centering
		\includegraphics[width=0.48\textwidth]{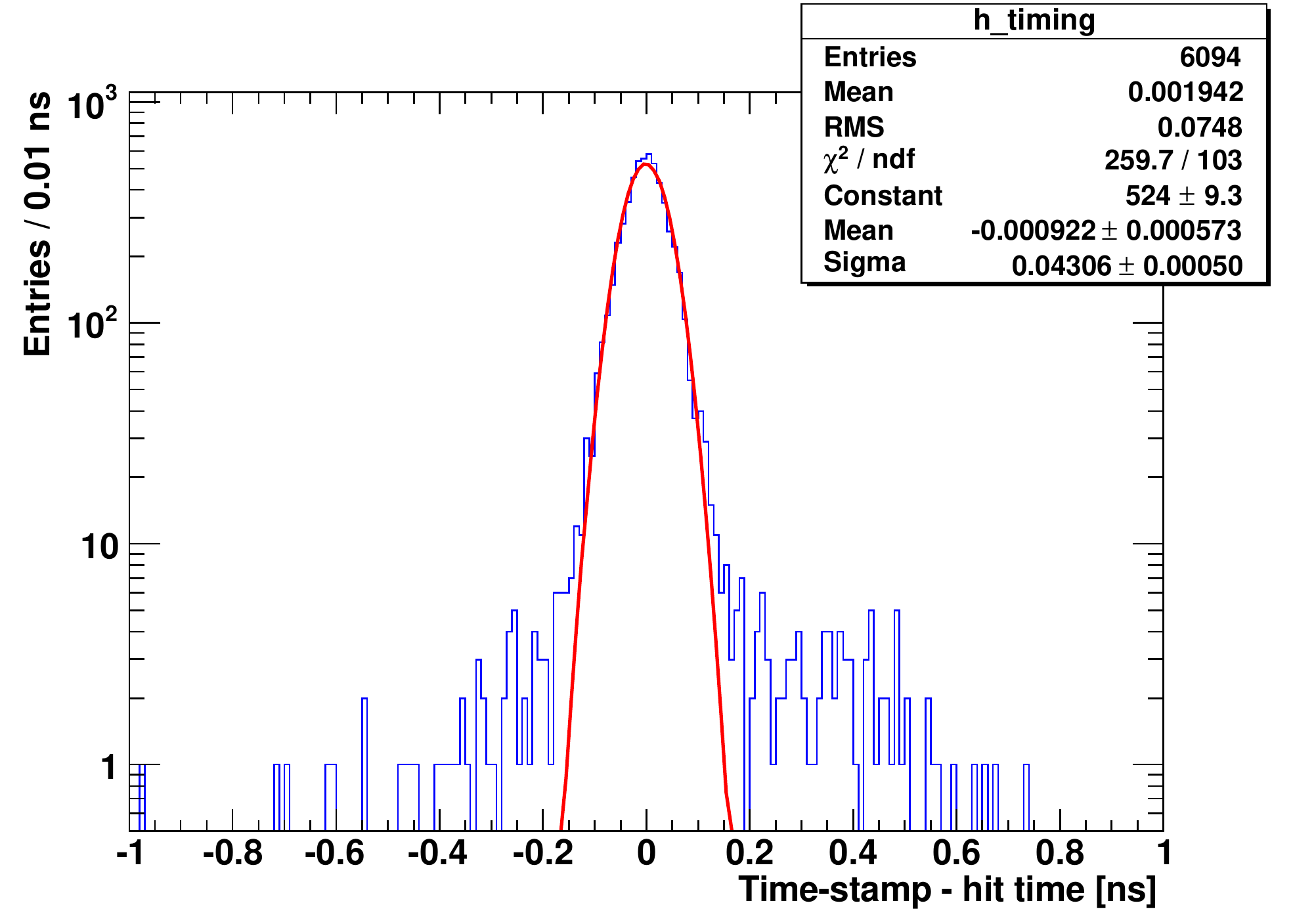}
	\caption{Timing for phase II. The timestamps have been corrected for timewalk.}
	\label{fig:timing}
\end{figure}

Figure~\ref{fig:timing} shows the distribution of the time-stamps relative to the true hit time given by the simulation for phase II. A fit of a Gaussian function yields a time resolution of $\sigma_t=\SI{45}{ps}$. The non-Gaussian tails of the distribution come from signal pileup.
%Assuming a coincidence window of $\pm 3\sigma_t$, the combinatorial background in a readout frame of $\SI{50}{ns}$ is hence suppressed by approximately $6\sigma_t / \SI{50}{ns} \approx 5\cdot 10^{-5}$.

The efficiency for assigning a time-stamp to an isolated track within $\pm 3\sigma_t$ of the true hit time is $\approx 100\%$. Inefficiencies at the edge of a recurl station, where a track deposits too little energy in a tile to pass the threshold can be neglected.

However, the efficiency is also influenced by signal pileup. For the simple fixed threshold model used in this analysis, an overall efficiency of $\approx 99.5\%$ for phase I and $\approx 98.0\%$ for phase II is achieved. With a more sophisticated peak-finding algorithm, it should be possible to resolve pileup signals which occur within a time interval of a few nanoseconds and consequently achieve an efficiency of $\approx 100\%$. Also, using a SiPM with a fast output, like the SensL MicroFB series, will significantly reduce the pileup and thus increase the efficiency.

\section{Time Resolution Measurements}

First measurements of time resolution of a single $1\times1\times\SI{1}{cm\cubed}$ tile and a Hamamatsu S10362-33-050 MPPC have been done using an oscilloscope with a bandwidth of $\SI{1}{GHz}$. The scintillator material used in this measurements is NE110. Compared to BC420, this scintillator has a lower light yield and a slightly slower rise and decay time; using a more suited scintillator will thus improve the results.
The scintillation light was triggered using a pulsed UV laser\footnote{The scintillator response to the UV light is a good approximation of the response to electrons.}.

\begin{figure}[htb]
	\centering
		\includegraphics[width=0.48\textwidth]{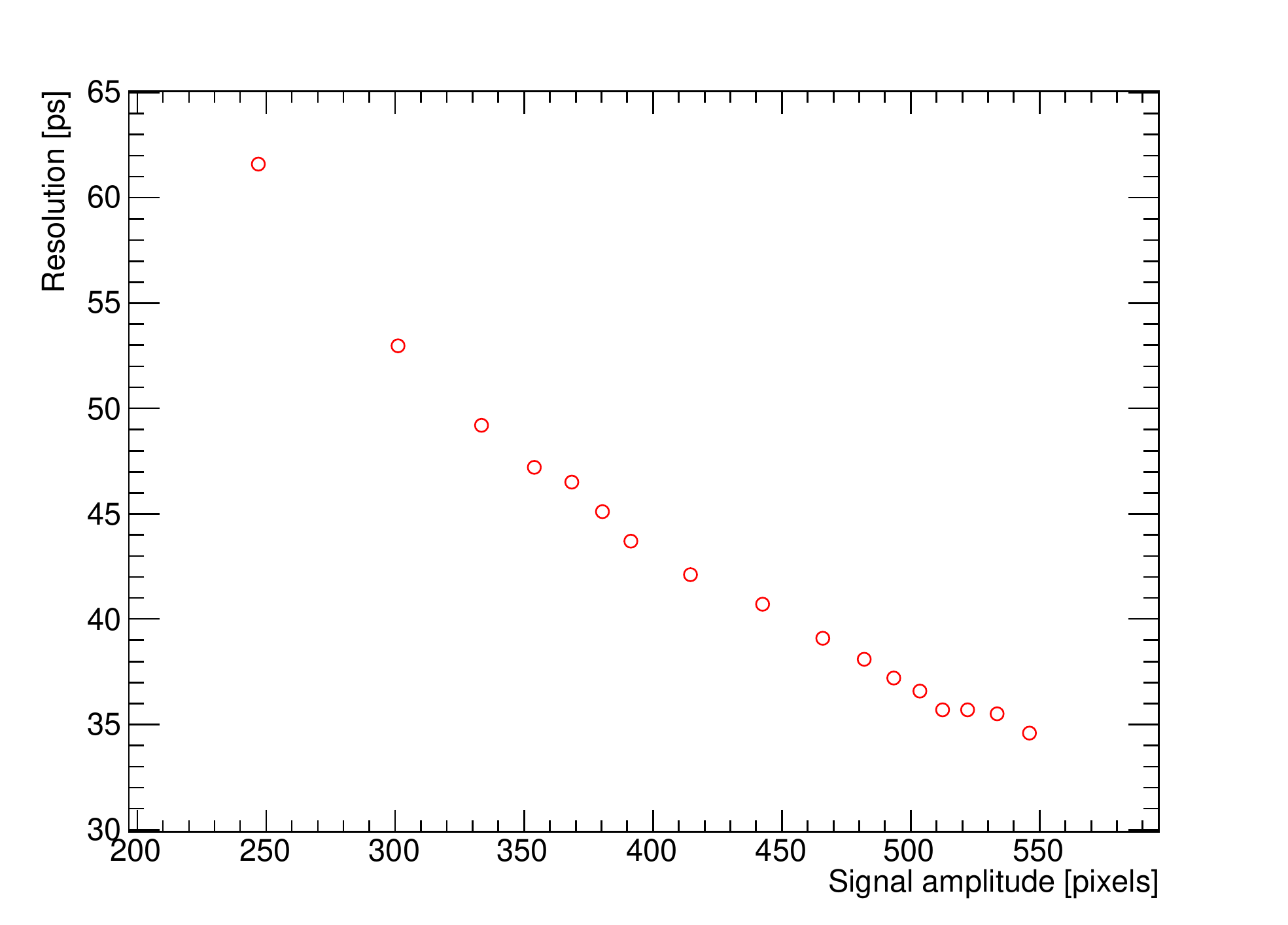}
	\caption{Time resolution as a function of the signal amplitude for a threshold of $10\%$. The amplitude expected in the tile detector is $\approx \SI{500}{pixels}$.}
	\label{fig:res_vs_int}
\end{figure}

\begin{figure}[htb]
	\centering
		\includegraphics[width=0.48\textwidth]{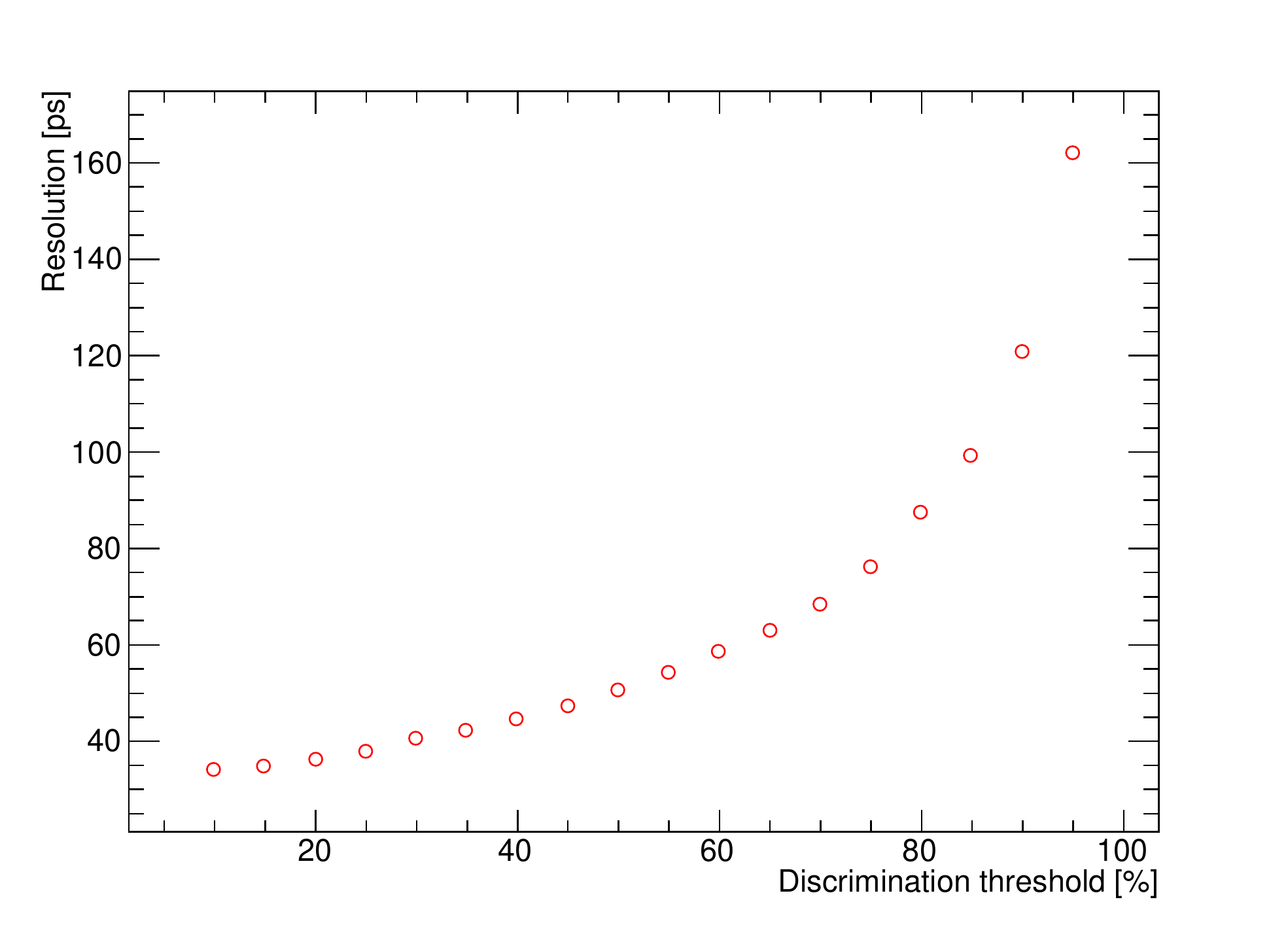}
	\caption{Time resolution as a function of the discrimination threshold for 500 pixel signal amplitude.}
	\label{fig:res_vs_thresh}
\end{figure}

\begin{figure}[htb]
	\centering
		\includegraphics[width=0.48\textwidth]{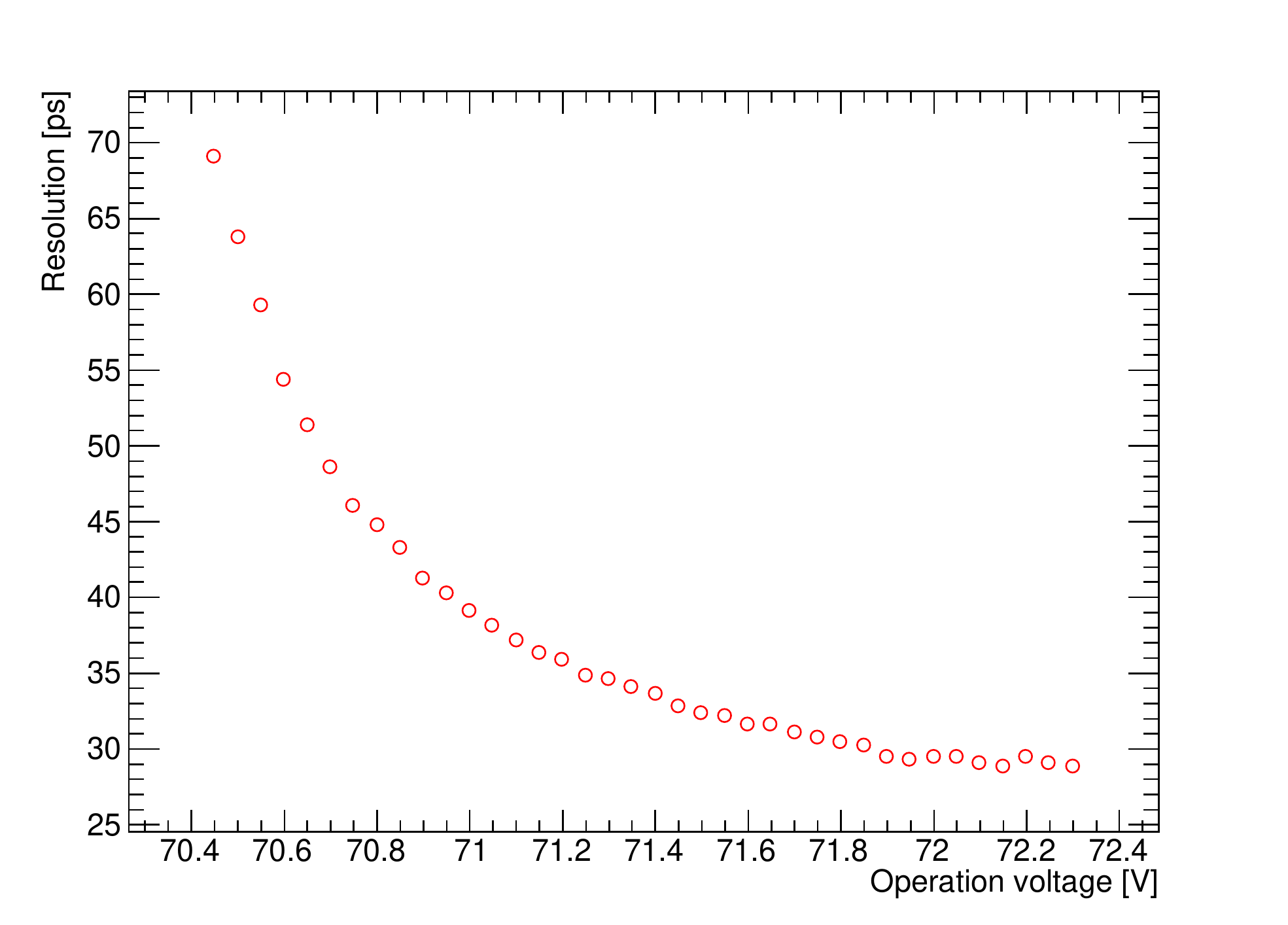}
	\caption{Time resolution as a function of the SiPM bias voltage for 500 pixel signal amplitude and a threshold of $10\%$.}
	\label{fig:res_vs_vbias}
\end{figure}

Figure~\ref{fig:res_vs_int} shows the measured time resolution as a function of the signal amplitude for a threshold of $10\%$. For a typical signal amplitude of $\approx 350$ pixels, a time resolution of $\approx \SI{45}{ps}$ is achieved, which is consistent with the simulation results. Figures~\ref{fig:res_vs_thresh} and~\ref{fig:res_vs_vbias} show the dependence on the applied threshold and SiPM bias voltage.

These results show, that a time resolution well below $\SI{100}{ps}$ can be achieved. It is expected, the resolution will not be degraded using the DRS or STiC readout chip, which will be verified in future measurements.

\section{Detector Prototype}

The design of a prototype of a tile detector module has been started (see Figure~\ref{fig:proto}) and the construction is expected to be completed within 2013. In this prototype the mechanical design of the support structures and readout infrastructure will be tested and measurements of the detector performance will be carried out.

%\balance

%\input{Services/Services}

\chapter{Data Acquisition}
\label{sec:DAQ}

\nobalance

\section{Overview}

\begin{figure}
	\centering
		\includegraphics[width=0.45\textwidth]{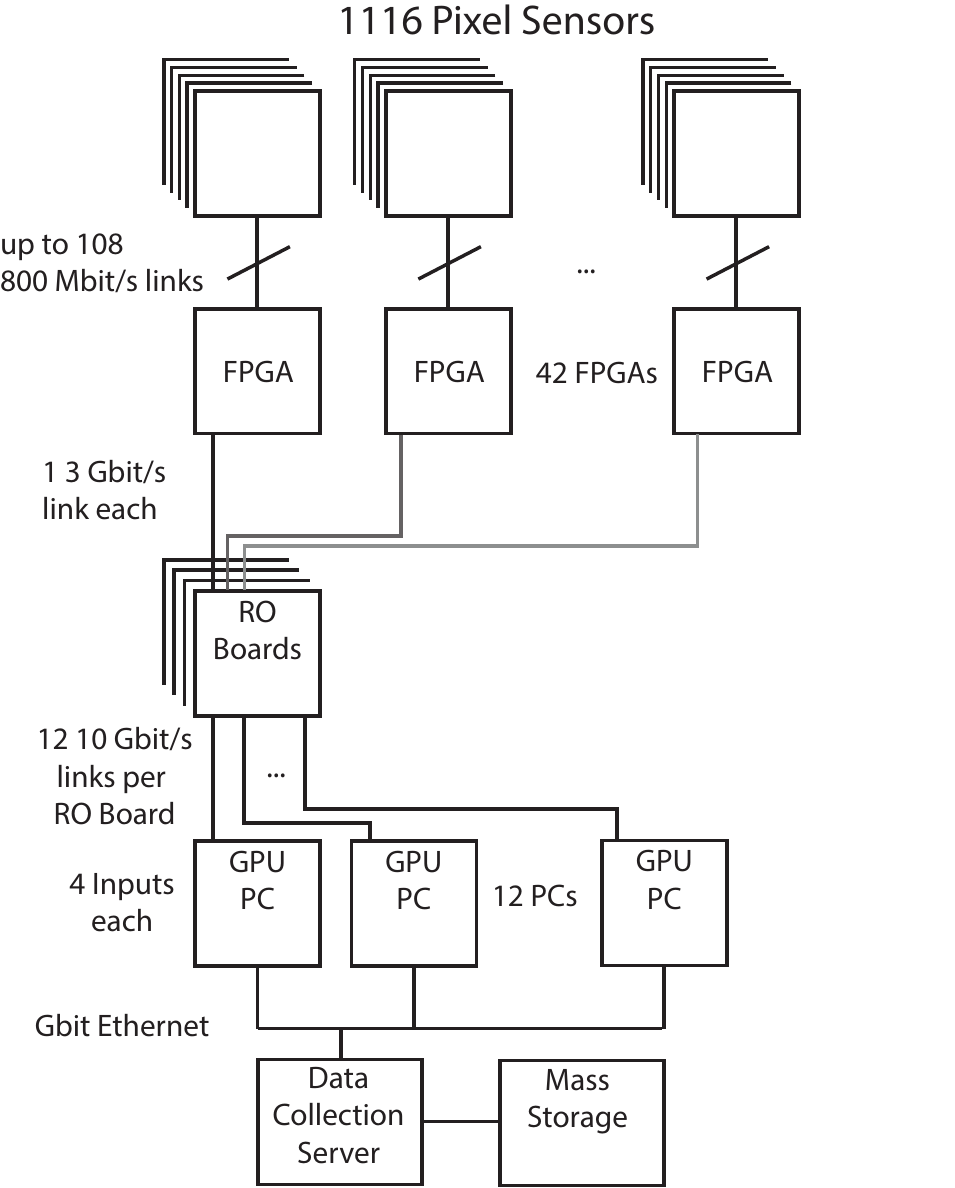}
	\caption{Mu3e readout scheme for the start-up detector.}
	\label{fig:RO_Scheme_Startup}
\end{figure}

\begin{figure*}
	\centering
		\includegraphics[width=1.00\textwidth]{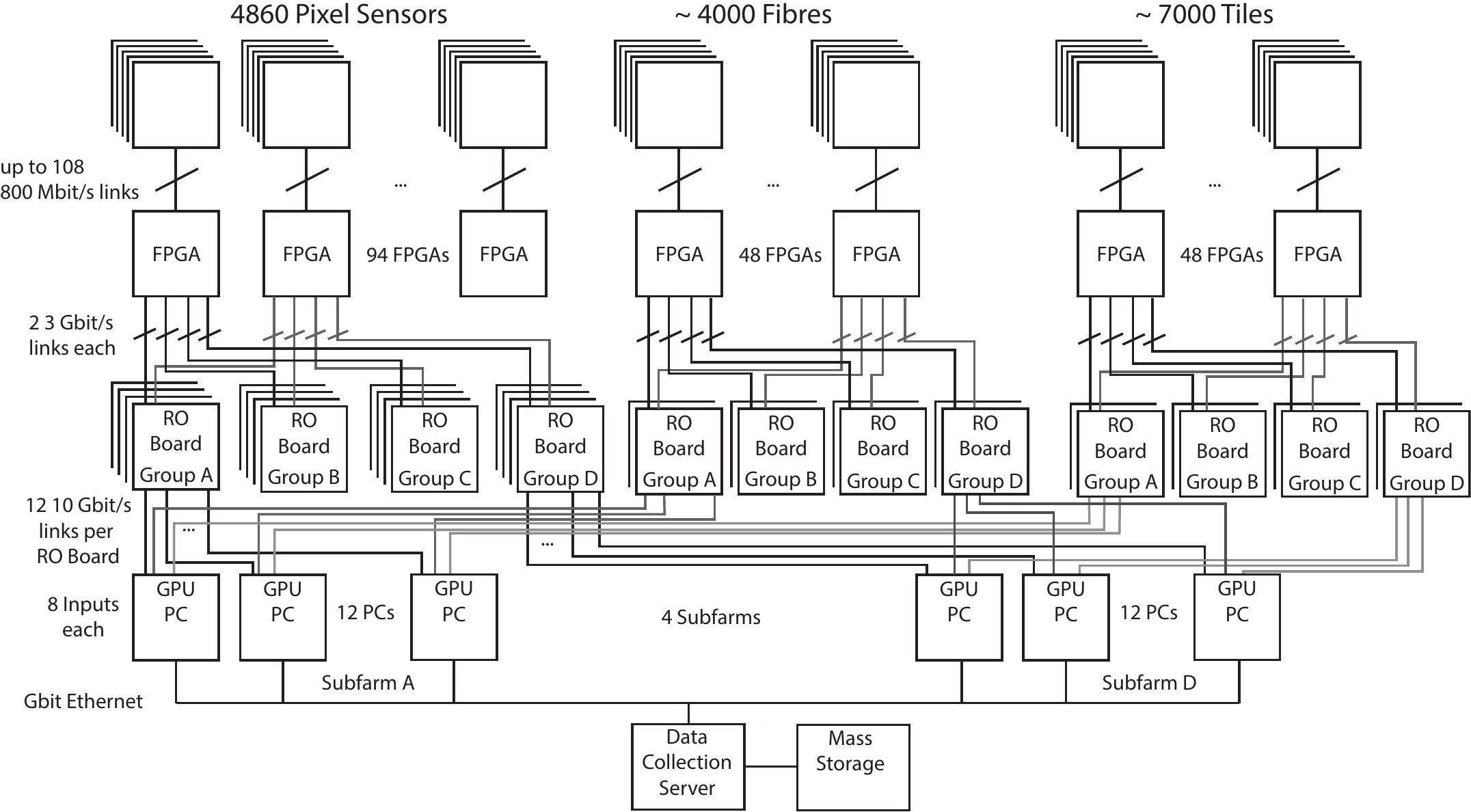}
	\caption{Overall Mu3e readout scheme}
	\label{fig:RO_Scheme}
\end{figure*}

The Mu3e data acquisition system works without a hardware trigger on a push
basis, i.e.~the detector elements continuously send hit information to the data
acquisition (DAQ) system. The DAQ consists of three layers, namely front-end
FPGAs, read-out boards and the filter farm. The topology of interconnects is
built such that every farm PC gets to see the complete detector information for
a select time slice. See Figure~\ref{fig:RO_Scheme} for an overview of the
readout scheme and Figure~\ref{fig:RO_Scheme_Startup} for the scheme at detector
start-up.

\section{Occupancy}
\label{sec:Occupancy}

\begin{figure}
	\centering
		\includegraphics[width=0.48\textwidth]{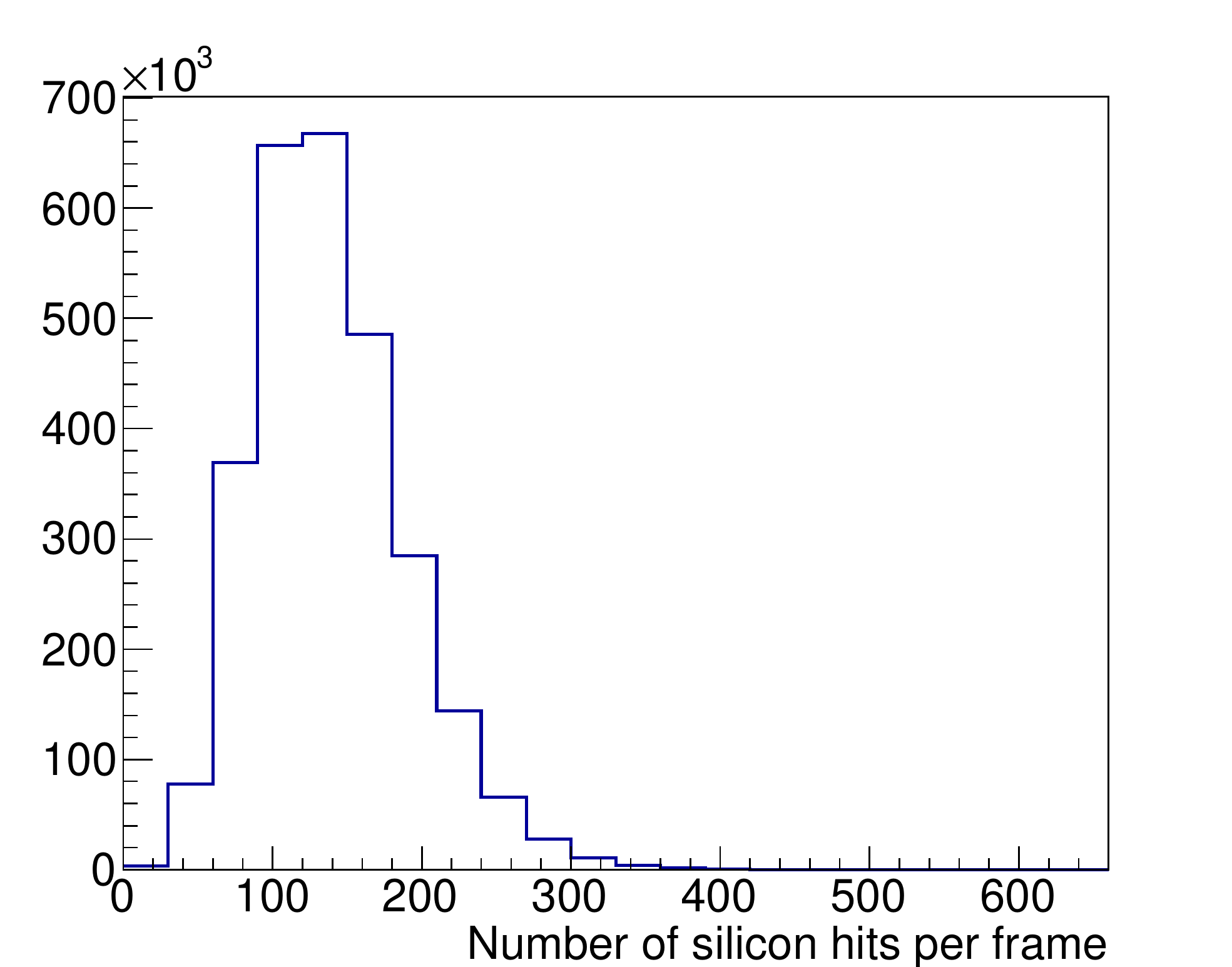}
	\caption{Number of pixel hits in the central detector per $\SI{50}{ns}$ frame in phase I running.}
	\label{fig:pixelhits_phase1}
\end{figure}

\begin{figure}
	\centering
		\includegraphics[width=0.48\textwidth]{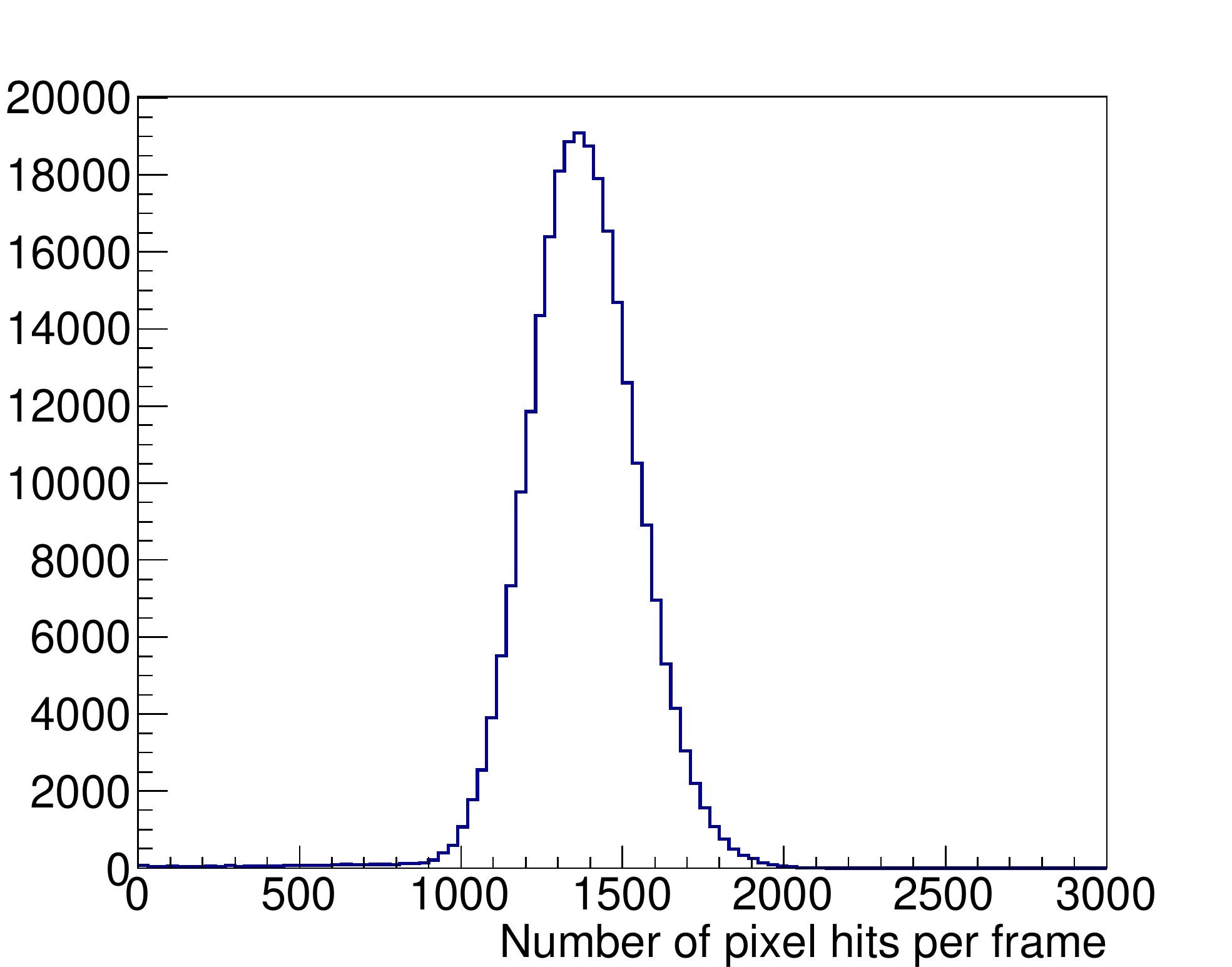}
	\caption{Number of pixel hits in the complete detector per $\SI{50}{ns}$ frame in phase II running.}
	\label{fig:pixelhits_phase2}
\end{figure}

\begin{figure}
	\centering
		\includegraphics[width=0.48\textwidth]{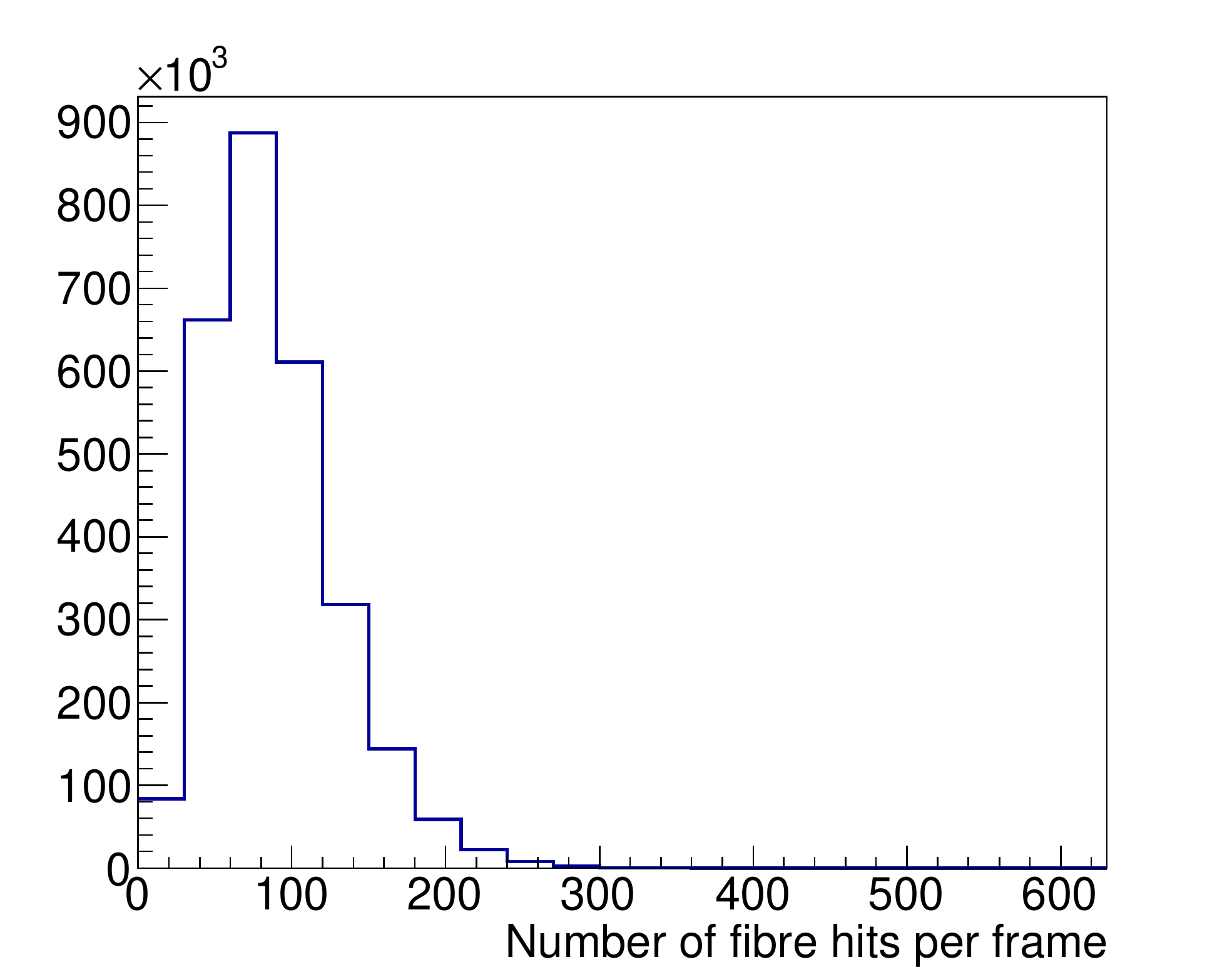}
	\caption{Number of fibre hits per $\SI{50}{ns}$ frame in phase I running.}
	\label{fig:fibrehits_phase1}
\end{figure}

\begin{figure}
	\centering
		\includegraphics[width=0.48\textwidth]{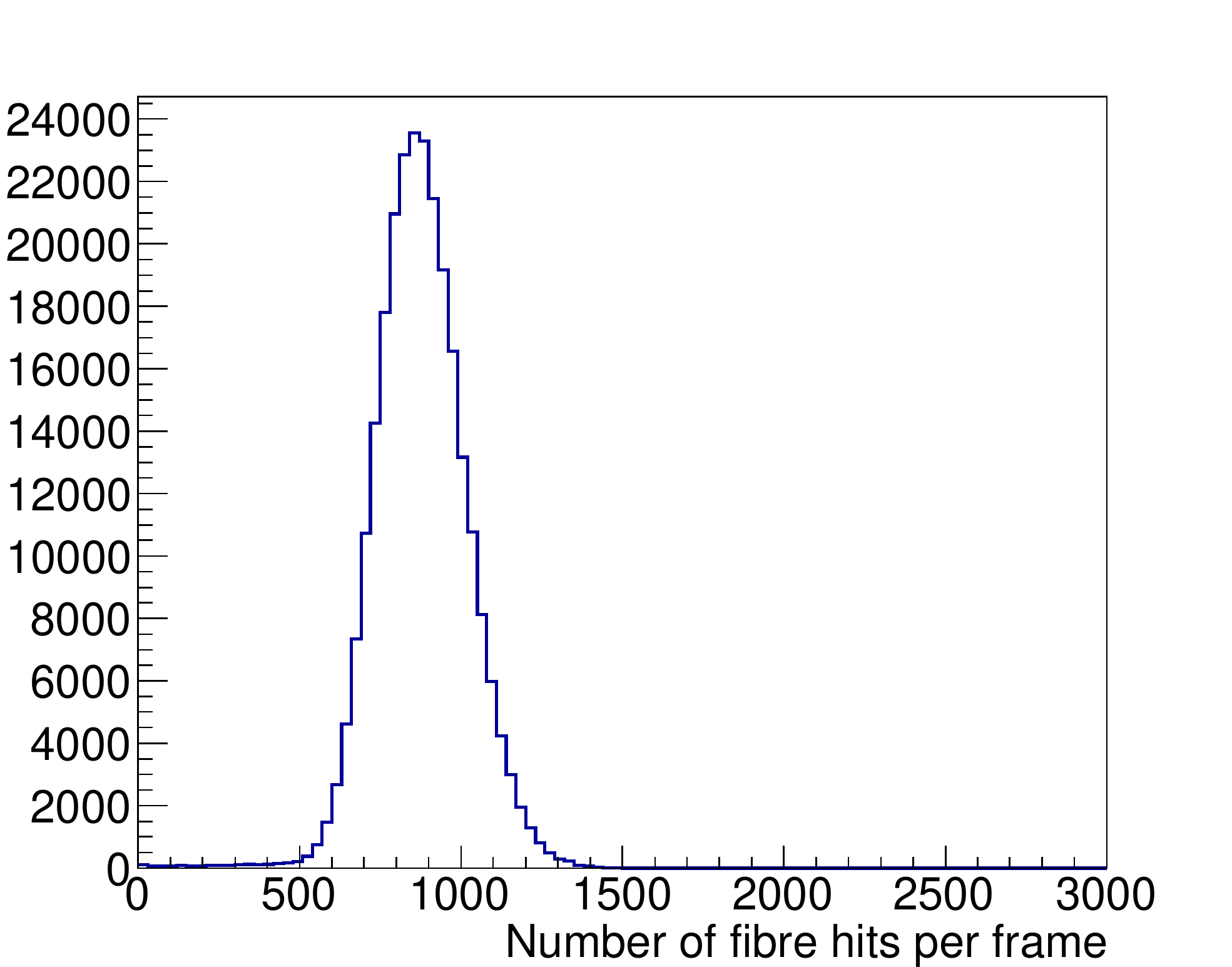}
	\caption{Number of fibre hits per $\SI{50}{ns}$ frame in phase II running.}
	\label{fig:fibrehits_phase2}
\end{figure}

\begin{figure}[t!]
	\centering	\includegraphics[width=0.48\textwidth]{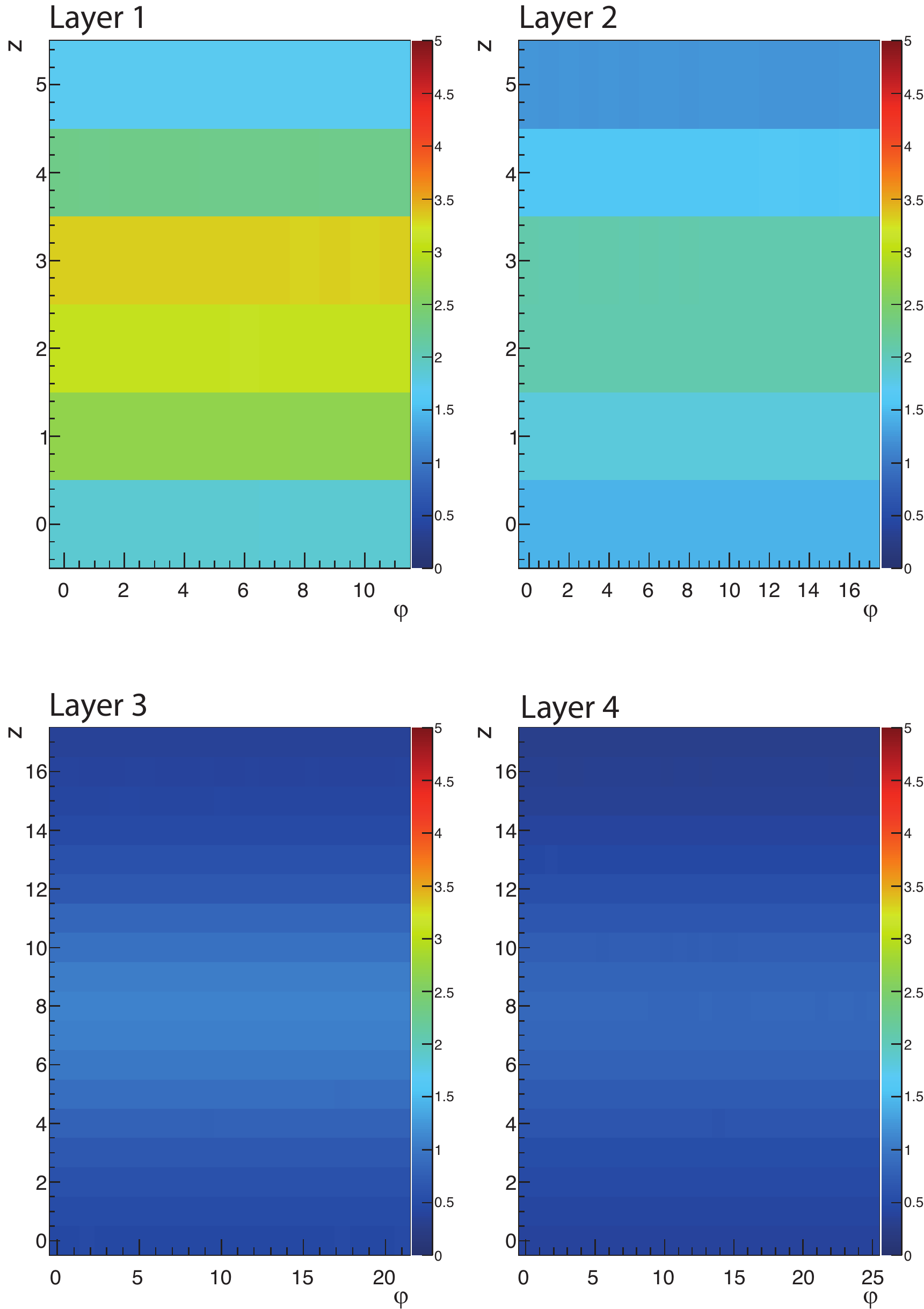}
	\caption{Occupancy in $\SI{50}{ns}$ frames of the central pixel sensors for phase II running. The axes enumerate sensor numbers. For phase I, the occupancy numbers have to be scaled down by a factor 10-20.}
	\label{fig:Occupancy_map_central_pixel}
\end{figure}

\begin{figure}[t!]
	\centering
\includegraphics[width=0.48\textwidth]{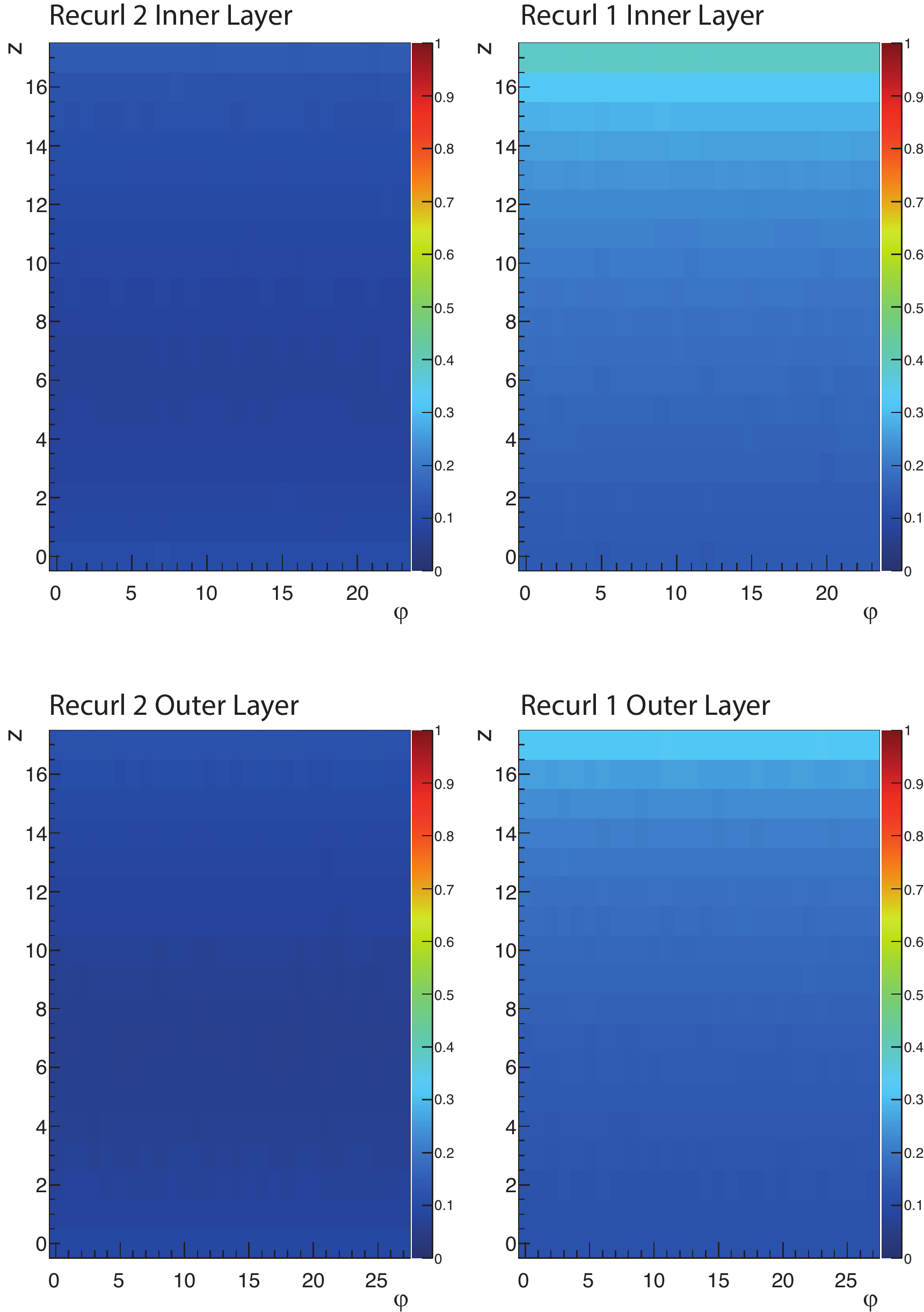}
	\caption{Occupancy in $\SI{50}{ns}$ frames of the recurl pixel sensors for phase II running. The axes enumerate sensor numbers.}
	\label{fig:Occupancy_map_recurl_pixel}
\end{figure}

The bandwidth requirements of the data acquisition are largely determined by the expected detector occupancy, as all the Mu3e sub-detectors produce zero-suppressed output. 

The occupancies shown are obtained with the full simulation running at a muon stop rate of $\SI{2e9}{Hz}$ ($\SI{2e8}{Hz}$ for phase I) and pessimistically estimating the beam related background by loosing another $\SI{4e9}{Hz}$ ($\SI{4e8}{Hz}$) of muons along the beam line. Figures~\ref{fig:pixelhits_phase1} and \ref{fig:pixelhits_phase2} show the expected number of hits per $\SI{50}{ns}$ frame in the pixel detector. Figures~\ref{fig:fibrehits_phase1} and \ref{fig:fibrehits_phase2} show the same for the fibre detector. The distribution of the occupancy over the pixel sensors is shown in Figures~\ref{fig:Occupancy_map_central_pixel} and \ref{fig:Occupancy_map_recurl_pixel}.

\section{Front-end}
\label{sec:FrontEnd}

\subsection{Pixel detector}

The pixel sensors contain electronics for hit detection and time as well as
address encoding. All hits assigned to the same (\SI{20}{MHz}) time-stamp
constitute a \emph{frame}. The sensors collect the data of 16 frames into a
\emph{superframe} and send it off chip via an \SI{800}{Mbit/s} low-voltage
differential signaling (LVDS) link. The signals travel over a maximum of
\SI{18}{cm} on a Kapton flex-print to the edge of the sensitive area, where they
are amplified by a driver chip. The Kapton prints then connects to a PCB located
between the recurl layers and the beam-pipe. On this PCB, up to 72 LVDS links are
fed into a FPGA. The FPGA provides buffering and collects a long stream of
frames (at least 1024) into a \emph{frametrain}. The assembled data are then
output to 8 \SI{3}{Gbit/s} links, such that the data of one frame-train are sent
on two links. On the PCB, the signals are converted to optical and sent
off-detector via fibres. An additional pair of optical links per FPGA is
required for slow control and monitoring.

\begin{table*}[tb!]
\begin{center}
\small
\begin{tabular}{lrrlrrrl}

\toprule
 & Sensor & Max & Average & \multicolumn{1}{l}{Chip$\rightarrow$FPGA} & \multicolumn{1}{l}{Chip$\rightarrow$FPGA} & \multicolumn{1}{l}{Front end} & FPGA$\rightarrow$RO \\ 
 & Chips & Hits & Hits & link capacity & total in Layer & FPGAs & capacity \\
 & & \multicolumn{1}{l}{/Chip} & \multicolumn{1}{l}{/Layer}   & \multicolumn{1}{r}{Mbit/s} & \multicolumn{1}{r}{Gbit/s} & \multicolumn{1}{r}{} & Gbit/s \\ 
 \midrule
Layer 1 & 72 & 0.35 & \multicolumn{1}{r}{18.0} & 220 & 16 & 8 & \multicolumn{1}{r}{17} \\ 
Layer 2 & 108 & 0.25 & \multicolumn{1}{r}{18.4} & 157 & 17 & 8 & \multicolumn{1}{r}{17} \\ 
Layer 3 & 432 & 0.15 & \multicolumn{1}{r}{31.0} & 94 & 40 & 12 & \multicolumn{1}{r}{29} \\ 
Layer 4 & 504 & 0.15 & \multicolumn{1}{r}{28.6} & 94 & 47 & 14 & \multicolumn{1}{r}{27} \\ 
\midrule
Total & 1116 & \multicolumn{1}{l}{} & \multicolumn{1}{r}{96} & \multicolumn{1}{l}{} & 120 & 42 & \multicolumn{1}{r}{90} \\ 
\bottomrule
\end{tabular}
\end{center}
\caption{Pixel readout requirements (Phase IB without recurl stations).}
\label{Tab:PixelRONBumbersPhaseI}
\end{table*}

\begin{table*}[tb!]
\small
\begin{tabular}{lrrlrrrl}

\toprule
 & Sensor & Max & Average & \multicolumn{1}{l}{Chip$\rightarrow$FPGA} & \multicolumn{1}{l}{Chip$\rightarrow$FPGA} & \multicolumn{1}{l}{Front end} & FPGA$\rightarrow$RO \\ 
 & Chips & Hits & Hits & link capacity & total in Layer & FPGAs & capacity \\
 & & \multicolumn{1}{l}{/Chip} & \multicolumn{1}{l}{/Layer}   & \multicolumn{1}{r}{Mbit/s} & \multicolumn{1}{r}{Gbit/s} & \multicolumn{1}{r}{} & Gbit/s \\ 
 \midrule
Layer 1 & 72 & 3.5 & \multicolumn{1}{r}{180} & 2203 & 155 & 8 & \multicolumn{1}{r}{166} \\ 
Layer 2 & 108 & 2.5 & \multicolumn{1}{r}{184} & 1574 & 166 & 8 & \multicolumn{1}{r}{170} \\ 
Layer 3 & 432 & 1.5 & \multicolumn{1}{r}{310} & 944 & 398 & 12 & \multicolumn{1}{r}{286} \\ 
Layer 4 & 504 & 1.5 & \multicolumn{1}{r}{286} & 944 & 465 & 14 & \multicolumn{1}{r}{264} \\ 
Recurl backward 1 inner & 432 & 0.5 & \multicolumn{ 1}{r}{|} & 315 & 133 & 6 & \multicolumn{ 1}{r}{|} \\ 
Recurl backward 1 outer & 504 & 0.5 & \multicolumn{ 1}{r}{|} & 315 & 155 & 7 & \multicolumn{ 1}{r}{|} \\ 
Recurl backward 2 inner & 432 & 0.25 & \multicolumn{ 1}{r}{|} & 157 & 66 & 6 & \multicolumn{ 1}{r}{|} \\ 
Recurl backward 2 outer & 504 & 0.25 & \multicolumn{ 1}{r}{|} & 157 & 77 & 7 & \multicolumn{ 1}{r}{|} \\ 
Recurl forward 1 inner & 432 & 0.3 & \multicolumn{ 1}{r}{|} & 189 & 80 & 6 & \multicolumn{ 1}{r}{|} \\ 
Recurl forward 1 outer & 504 & 0.3 & \multicolumn{ 1}{r}{|} & 189 & 93 & 7 & \multicolumn{ 1}{r}{|} \\ 
Recurl forward 2 inner & 432 & 0.2 & \multicolumn{ 1}{r}{|} & 126 & 53 & 6 & \multicolumn{ 1}{r}{|} \\ 
Recurl forward 2 outer & 504 & 0.2 & \multicolumn{ 1}{r}{$\Sigma$=490} & 126 & 62 & 7 & \multicolumn{ 1}{r}{$\Sigma$=452} \\ 
\midrule
Total & 4860 & \multicolumn{1}{l}{} & \multicolumn{1}{r}{1450} & \multicolumn{1}{l}{} & 1903 & 86 & \multicolumn{1}{r}{1515} \\ 
\bottomrule
\end{tabular}
\caption{Pixel readout requirements (Phase II), for the recurl stations only the sum of average hits per layer and FPGA$\rightarrow$RO capacity is given.}
\label{Tab:PixelRONBumbers}
\end{table*}

\subsubsection{Hardware}

The requirements for the on-detector FPGAs can be met by mid- or even low-price
devices (such as the ALTERA Cyclone IV family or the XILINX Artix VII family).
The FPGAs are to be mounted on PCBs that are placed between the recurl layers and the beam-pipe.

\subsubsection{Firmware}

The main task of the on-detector FPGAs is collecting the relatively short time
slices of 16 clock cycles assembled on the pixel chips to the long intervals 
treated by the individual filter farm PCs. During this buffering, the hits can
be time ordered inside a slice and the protocol overhead can be reduced. In
addition, hits can be clustered. 

A further task for the first line of FPGAs is the configuration and monitoring
of the pixel chips. A $\SI{32}{\bit}$ histogram of the hit counts in a single
sensor however requires $\SI{256}{\kilo\byte}$ of memory, thus exceeding the
capacity of the devices; an external memory interface would significantly
increase the pin count and the PCB complexity; the histograming task is thus
deferred to the readout boards.
%This requires the storage of hit frequency histograms and tune values. A
%$\SI{32}{\bit}$ histogram for one of the large pixel chips requires
%$\SI{256}{\kilo\byte}$ of memory, the tune values a fraction of that. Storage
%is probably workable without the use of external memory modules.

These tasks are all fairly standard and FPGAs that fulfill the bandwidth
requirements for the in- and output channels do provide enough logic for
implementing them. 

\subsection{Timing detector}
\label{sec:TimingRO}

For the timing detectors, three readout schemes are currently under investigation:
One based on a further development of the DRS switched capacitor array developed
at PSI, one based on the STiC chip developed at KIP, Heidelberg University and one
with FPGA-based TDCs (mainly for the fibres).

\subsubsection{DRS sampling readout}

The readout of the tile and fibre detectors requires high rate capability and extremely
good timing resolution. To achieve an overall detector timing accuracy below 100
ps, the associated electronics needs to be at least a factor of two better, i.e.
50 ps. The high rate environment causes significant pile-up, which limits the
usage of conventional techniques such as discriminators and TDCs. Therefore one option to read out the tiles is with the well-established waveform digitizing
technology developed at PSI, which is in use since many years in the MEG
experiment. It is based on the DRS4 switched capacitor array, which is capable
of sampling the SiPM signals with up to 5 Giga samples per second (GSPS) with a
resolution close to 12 bits. It has been shown in the MEG experiment that this
technology allows a timing accuracy in the order of 40 ps across many thousand
channels. The knowledge of the exact waveform of an event is very well suited to detect and suppress pile-up.

The tile readout electronics could be placed outside the detector in special
crates connected with a few meters of cable. This simplifies the design and
maintainability, while not compromising the signal quality dramatically. 

%It is
%planned to equip only one sector for the tile detector for the Phase I of the
%experiment, together with its associated electronics, a so-called "vertical
%slice". This is because the tile detector is not absolutely required for the
%first phase of the experiment running at lower rate, but the vertical slice %will
%allow the collaboration to gain experience with this and make a more solid
%design for Phase II.

A principal limitation arises from the DRS4 chip, which is capable of only a
limited event rate of about $\SI{100}{kHz}$. While this will be sufficient for Phase IB,
it has to be improved for Phase II. Therefore a new development has been started
to design a new version of this chip. The DRS5 chip will use an internal analog
memory (FIFO) to work in a dead-time less fashion up to an event rate of about $\SI{5}{MHz}$. A critical part of the DRS5 chip which is the inverter chain operating the
sampling circuitry has already been designed in the new $\SI{110}{nm}$ CMOS technology and
submitted. First test results are expected beginning of 2013. The dead-time less
operation of this chip will be combined with higher sampling speed (10 GSPS) and
a better timing accuracy, allowing for a time measurement well below $\SI{10}{ps}$.

In order to limit the amount of data to be read out, the FPGA connected to the
DRS chip will already analyze the waveform and extract its major parameters like
time and amplitude. Only a prescaled subset of events will contain the full
detector waveform in order to cross-check the analysis algorithms in the FPGA.
Methods have recently been published which obtain the timing information by
cross correlation or cubic interpolation with an accuracy of about 1/10th of the
sampling interval, which would be $\SI{10}{ps}$ in the case of 5 GSPS.

\subsubsection{STiC readout}

The STiC chip offers an alternative to the DRS5 readout. STiC is a mixed mode 16-channel ASIC chip in UMC $\SI{0.18}{\micro m}$ CMOS technology designed for SiPM readout with high time resolution. It is developed for Time-of-Flight measurements in high energy physics and medical imaging, in particular the EndoTOFPET-US project. The chip has a differential structure, however, it supports both differential and single-ended connection of SiPMs.
A 6 bit DAC allows to tune the voltage at each input terminal within $\approx \SI{1}{V}$. In this way the SiPM operating voltage can be adjusted and temperature and device-to-device fluctuations can be compensated.

\begin{figure}[tb!]
	\centering
		\includegraphics[width=0.48\textwidth]{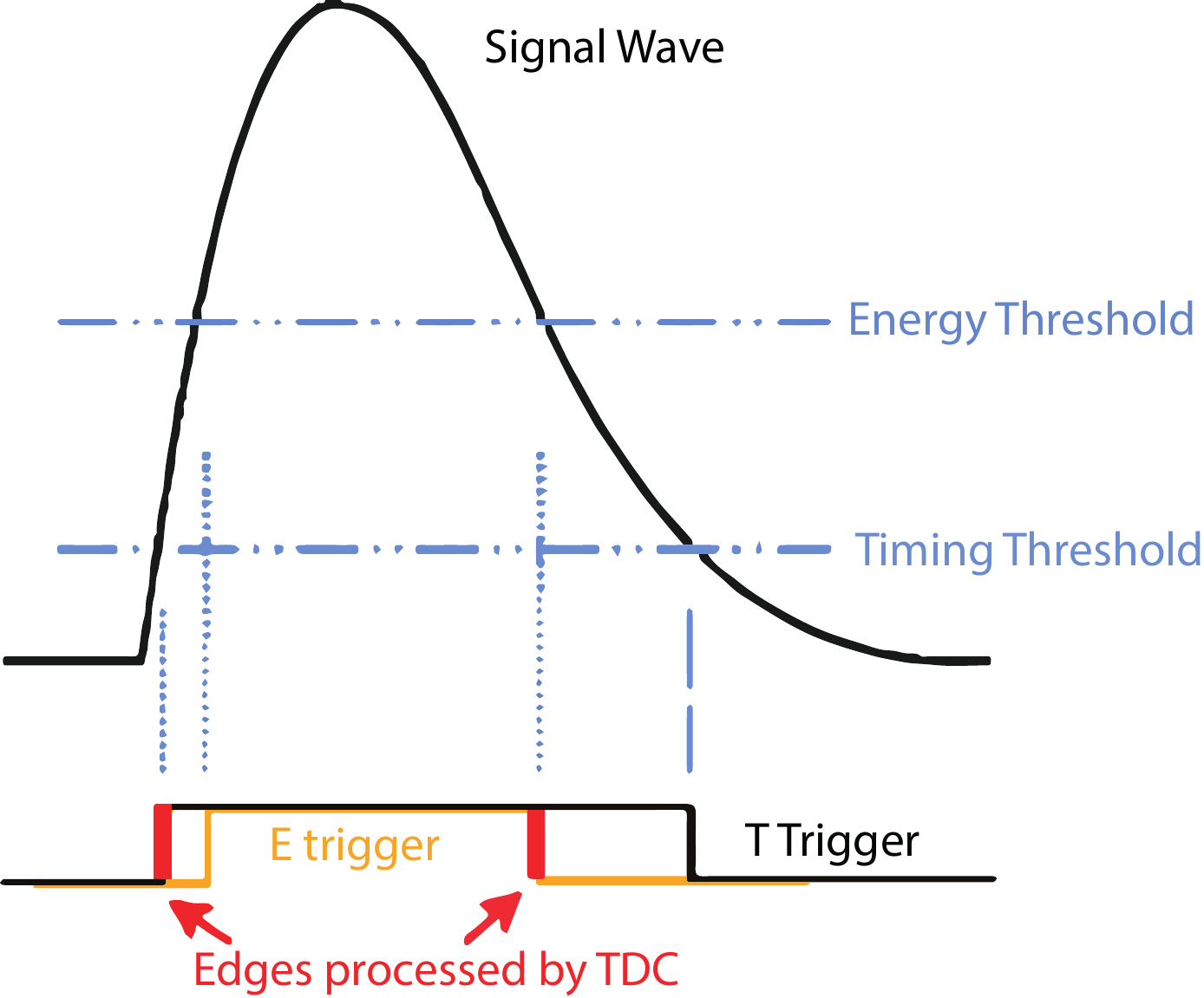}
	\caption{Dual threshold discrimination for energy and timing information.}
	\label{fig:stic}
\end{figure}

The time and charge information of the signal are encrypted into two time-stamps which are obtained by discriminating the signal with two different thresholds (see Figure~\ref{fig:stic}). The thresholds can be tuned in a range of $\approx 0.2 - 15$ pixel signals for the timing and up to 200 pixel signals for the charge. The time-stamps are then processed by an embedded TDC module with a resolution of $<\SI{20}{ps}$. A special linearization method is implemented to obtain a linear charge response in a very wide range. With the chip, a time-resolution of $\approx \SI{50}{ps}$ was measured for a 10 pixel signal of a MPPC S10362-33-50 without scintillator. For typical signal amplitudes of $\mathcal{O}(100)$ pixels, which are expected in the tile detector, the timing jitter of the chip is negligible.

The data rate of the current chip is limited to $\approx \SI{100}{kHz}$ per channel. However, the chip will be modified to allow for data rates of $\approx \SI{500}{kHz}$ per channel within 2013/2014. Until 2016, the data rate will be further increased to several MHz in order to match the requirements for phase II.

\subsubsection{FPGA based readout}

A further alternative for the fibre readout is the use of time to digital
converters implemented in FPGAs. Resolutions of $\mathcal O(\SI{1}{ns})$ can be
achieved fairly cheaply; much better performance requires the use of carry chain
techniques, which greatly reduces the number of channels per FPGA and makes
programming much more fickle. It has however been shown that resolutions of
$\mathcal O(\SI{10}{ps})$ can be achieved with this technology
\cite{Wu:2008zzo}. Whilst definitely not the optimal solution, FPGA based timing
could serve as a low-cost, low-risk solution for phase IB running.

\section{Read-out links}

In total there are three different types of read-out links in the Mu3e data acquisition system. The same type of links can be used for a small number of slow and fast control links in the opposite direction.  

The data from the MUPIX chips will be transmitted to the front-end FPGAs via LVDS links at \SI{800}{Mbit\per\s},
which is a quasi industry standard. The fast serializers and LVDS drivers for the MUPIX chip will be adopted from a similar chip design by a group from Bonn \cite{Putsch2011}. The link will be physically implemented as a matched differential pair of aluminum traces on the sensor flex-print, shielded by ground lines on both sides. It is foreseen to have a first set of LVDS repeaters just outside the acceptance of the detector. Very compact commercially available LVDS repeaters have already proven to be radiation tolerant. Outside the acceptance, the flex-print will then be of multilayer type, with differential signal inner layers, shielded by ground planes. The connection to the PCB housing the front-end FPGA is made by small form factor connectors, which support high bandwidth. Lastly the routing on the front-end FPGA PCB will be done with matched differential copper traces going to the pre-defined LVDS inputs of the FPGA itself.

There will be two types of optical high speed data links. The first one is going from the front-end FPGAs to the read-out boards, the second from the read-out boards to the FPGA PCIe boards in the event filter farm PCs.
The optical links from the front-end FPGAs to the read out boards have a bandwidth of \SI{3.125}{Gbit\per\s}, which fits well the FPGA specifications. Each FPGA has eight (or more) fast transceiver blocks. The optical interface can be implemented using radiation tolerant components developed by the Versatile Link group \cite{Vasey2012}. Especially the 12-way laser array with MTP/MPO standard connector would give a very compact form factor, the fall back solution would be a radiation hard dual laser transmitter system, the VTTx module developed at CERN \cite{Troska2012}. The data laser will have a wavelength of  \SI{850}{\nm} and the optical fibre is of 50/125 multi-mode type, since this is a standard both in industry and in particle physics detector readout. In the case of the 12-way emitter the fibres will go to a splitter and patch panel in order to combine fibres going to the same read-out board. In the case of dual transmitters the fibres can be combined in groups of 12 using break-out fibre cables. The fibres traveling from the experimental area to the filter farm counting house can be further combined in cables containing $8\times{12}$ fibres going to the same sub-farm. An optical patch panel for each sub-farm will allow to connect the long distance 96-fibre cables to 12 fibre patch cords going to the readout boards. The readout boards have 2 or 3 12-way optical inputs, which are populated with commercial receiver modules. The high-end FPGAs used on the read-out card have up to 66 high speed transceiver blocks, which will be used for the de-serialization of the data.

The last type of link connects the read-out cards to the FPGA PCIe boards in the event filter farm PCs. This optical link will be implemented as \SI{10}{Gbit\per\s} high speed link. Since each read-out board is connected to every sub-farm PC with one of the high speed links, they are point to point single fibre links. The fibres are of 50/125 multi-mode type operated at \SI{850}{\nm} to stay compatible with the links from the detector. If 12-way optical transmitters and receivers running at \SI{10}{Gbit\per\s} can be purchased, they will be used for this high speed link. Otherwise single fibre transmitters and receivers will be used both on the read-out board and the FPGA PCIe board side. In practice the high speed optical receivers for the FPGA PCIe boards could be on optical mezzanines such as the 8 channel St.~Luce card developed by TU Dortmund \cite{Swientek2012}.   

\section{Read-out cards}

The main tasks of the read-out cards is to act as switches between the front end
and the on-line reconstruction farm and to act as buffers between the
synchronous front end and the asynchronous back end. The board design and choice
of FPGAs is dominated by the number of fast links required. We plan to adapt an
existing development, e.g.~LHCb TELL1 cards \cite{Haefeli:2006cv} or PANDA
compute nodes \cite{Kuhn:2008zzb}, which would both fulfill our needs.

\section{Event filter interface}

The filter farm PCs will be equipped with FPGA cards in PCIe slots and optical
receiver daughter cards, as described in more detail in section
\ref{sec:HardwareImplementation}.

\section{Data collection}

The filter farm will output selected events at a data rate in the order of 50
MBytes/s in total. This data rate is low enough to be collected by a single PC
connected to the filter farm by common GBit Ethernet and written to local disks.
Then the data will be transferred to the central PSI computing center, where it
is stored and analyzed. For the central data acquisition the well established
MIDAS (Maximum Integrated Data Acquisition System) \cite{MIDAS:2001} 
software package will be used. This software is currently used in several 
major experiments such
as the T2K ND280 detector in Japan \cite{Abe:2011ks}, ALPHA at CERN and the MEG
experiment at PSI \cite{MEG}. It can easily handle the required data rate, and
contains all necessary tools such as event building, a slow control system
including a history database and an alarm system. A web interface allows
controlling and monitoring the experiment through the Internet. The farm PCs
will use MIDAS library calls to ship the data to the central DAQ PC. The
framework also offers facilities to send configuration parameters from a central
database (the ``Online DataBase'' or ODB) to all connected farm PCs and to
coordinate common starts and stops of acquisition (run control).

For the purpose of monitoring and data quality control of the experiment the
MIDAS system offers taps to the data stream for connections of analysis and
graphical display programs. The output of graphical user interface programs can
be fed back into the web interface of the MIDAS system so the experiment can be
monitored also remotely with just a Web browser.

\section{Slow control}

\begin{figure}
	\centering
		\includegraphics[width=0.48\textwidth]{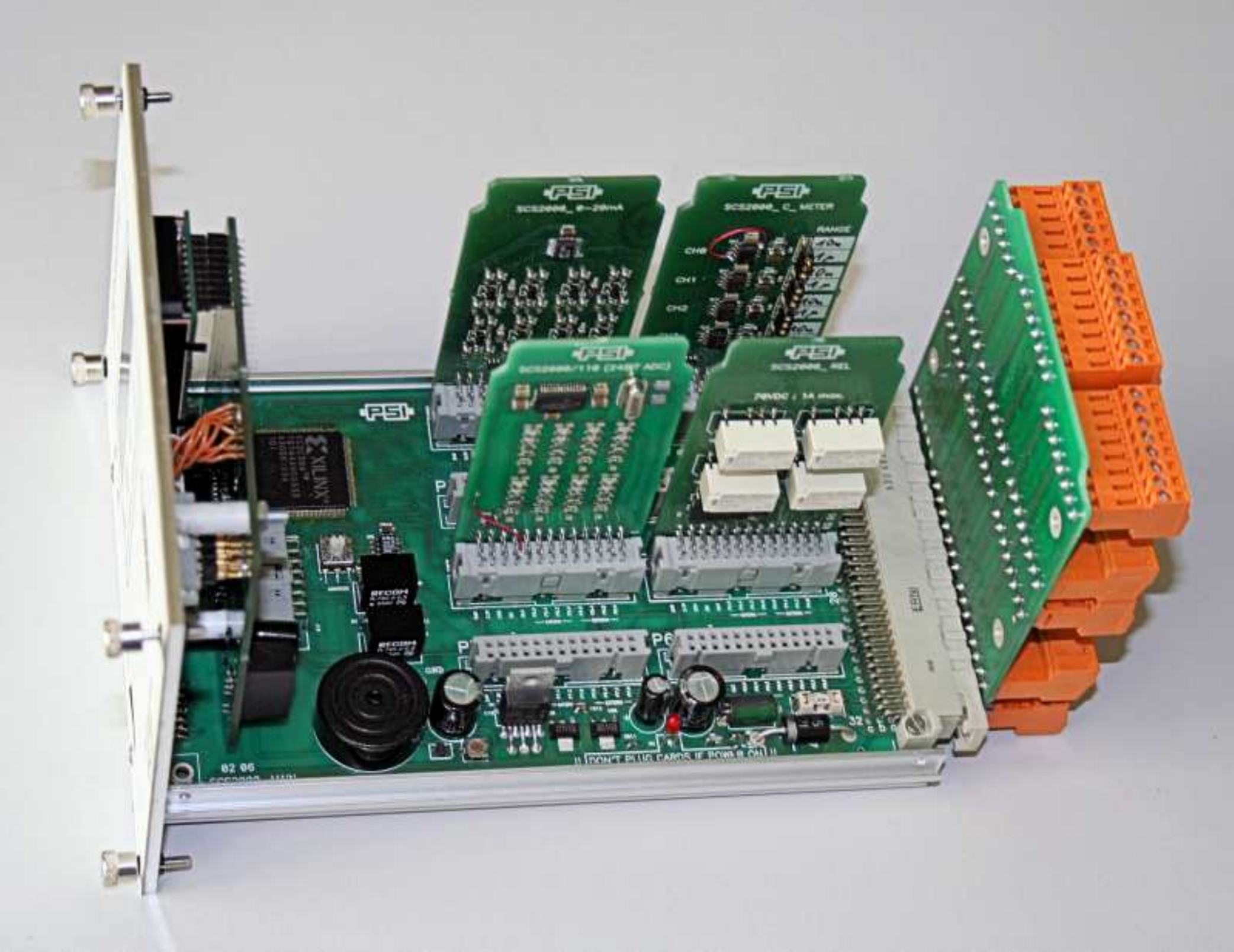}
	\caption{SCS-2001 unit as part of the MSCB slow control system. This unit
	has 64 input/output channels, which can be configured via plug-in boards as
	digital or analog channels. Many plug-in boards exist already such as PT100
	temperature sensor readout cards, analog high resolution inputs (24 bit
	resolution), valve control outputs and many more.}
	\label{fig:SCS2001}
\end{figure}

The slow control system deals with all ``slow'' data such as high voltages for
the SiPMs and silicon sensors, ambient temperatures and pressures. For the
configuration and control of the silicon pixel sensors the JTAG standard
\cite{JTAG} will be used. It is planned to use the MIDAS Slow Control Bus (MSCB)
system \cite{MSCB:2001} to link all distributed control and monitoring devices into a
single system. The MSCB system is also well established at several laboratories.
It uses a serial differential bus for communication, and simple micro controllers
in all control devices. The micro controllers perform local control loops such
as high voltage stabilization, and send measured values to the central DAQ
system for monitoring. Many devices already exist for this system, such as the
SCS-2001 unit shown in \ref{fig:SCS2001}. Since the system was developed at PSI,
it can be quickly adapted to new hardware. The high voltage control for the
SiPMs can for example be directly integrated into the carrier boards holding the
SiPMs, thus eliminating the need for high voltage cables. The optimized protocol
of the MSCB system allows the monitoring of many thousand channels with
repetition rates in the 100 ms range, which will be more than enough for this
experiment.

\balance
In addition to the MSCB system, the MIDAS slow control package contains
interfaces to the PSI beamline elements via the EPICS system \cite{EPICS}. This
allows monitoring and control of the beamline from the main DAQ system, which
has been proven very versatile in other experiments using this scheme.

All slow control data will be stored in the history system of the MIDAS system,
so that long term stabilities of the experiment can be effectively verified. The
slow control data is also fed into the main event data stream, so that any offline
analysis of the event data has this data available.

A special case is the configuration of the pixel detectors, which require many
million parameters, like the trim-DAC values for each pixel. Since the amount of
data here is considerably larger than for all other systems, an
extension of the slow control system is planned. A dedicated program manages,
visualizes and exchanges the pixel detector configuration parameters between an
optimized database and the pixel hardware. In this way the time required to
configure the pixel detectors can be minimized, while this program is still
connected to the main DAQ system. It can be synchronized with run starts and
stops, and can inject pixel monitoring data periodically into the event data
stream for offline analysis.

\chapter{Online Event Selection}
\label{sec:Farm}

\nobalance

\section{Selection Algorithms}
\label{sec:SelectionAlgorithms}

As in the final analysis, event selection in the filter farm can rely on the
coincidence of three tracks in time and vertex and on their kinematics.
Especially for high rate running, coincidence in time in the fibre detector is
not sufficient to reduce the data rate by three to four orders of magnitude.
Thus a track reconstruction will be required. The triplet based multiple
scattering fit described in chapter \ref{sec:Reconstruction} is well suited for online implementation
and current GPUs can perform $\num{e9}$ triplet fits per second\footnote{As
tested on a AMD Radeon 6990 using OpenCL under Linux.}, thus already fulfilling
the needs of Mu3e up to at least medium intensity (few  $\num{e8}$ muons/s)
running.

\begin{figure}
	\centering
		\includegraphics[width=0.48\textwidth]{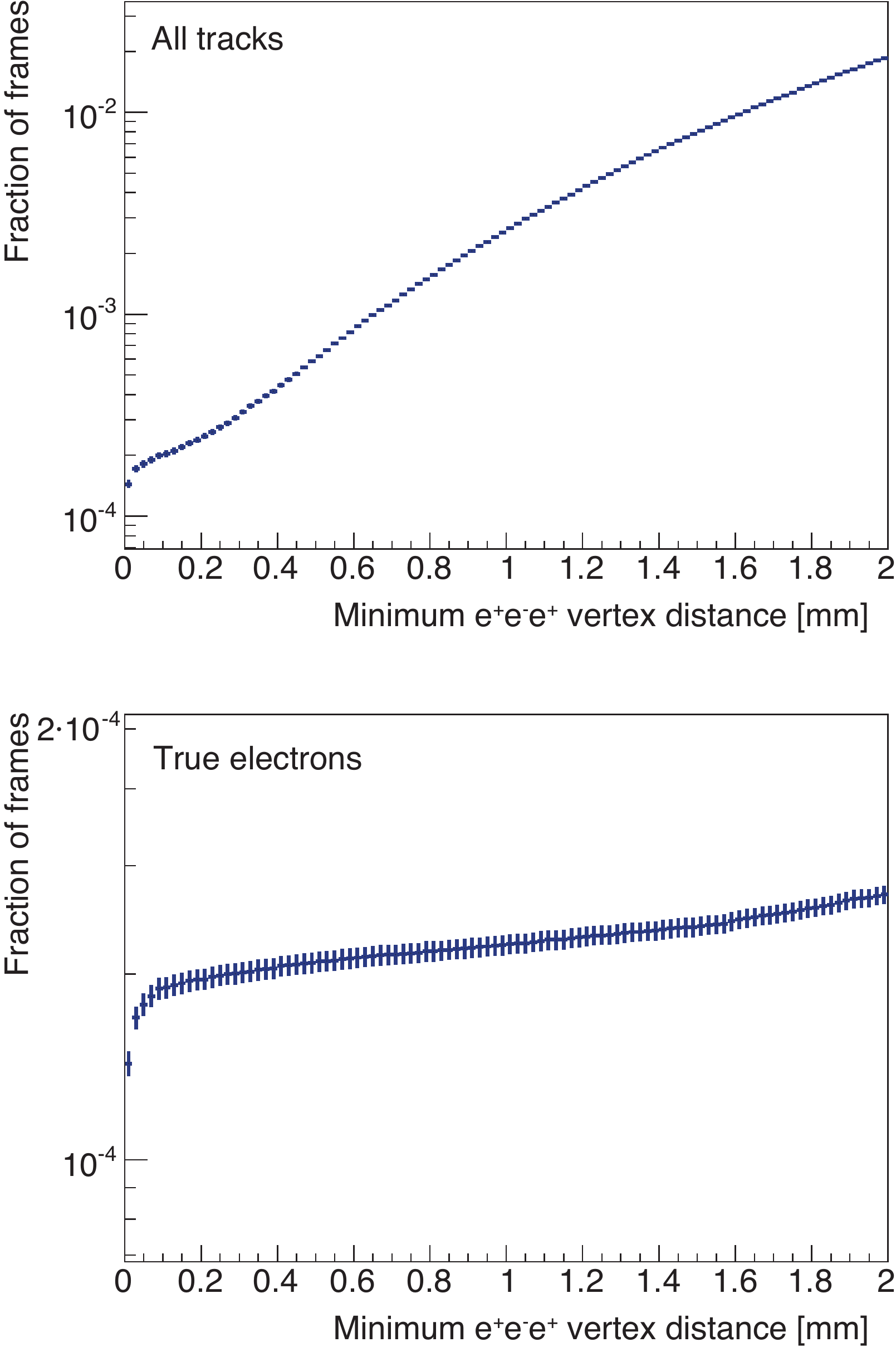}
	\caption{Fraction of $\SI{50}{ns}$ frames containing three vertices consistent with $e^+e^-e^+$ inside a given distance for a muon stop rate of $\SI{2e8}{Hz}$ for 3.15 million simulated frames. In the top plot, every crossing of a simulated electron/positron track is counted as a vertex; charge assignments are made purely on the apparent curvature, i.e.~recurling positrons are counted as electrons. In the bottom plot, only true electrons are counted.}
	\label{fig:fraction_phase1}
\end{figure}

\begin{figure}
	\centering
		\includegraphics[width=0.48\textwidth]{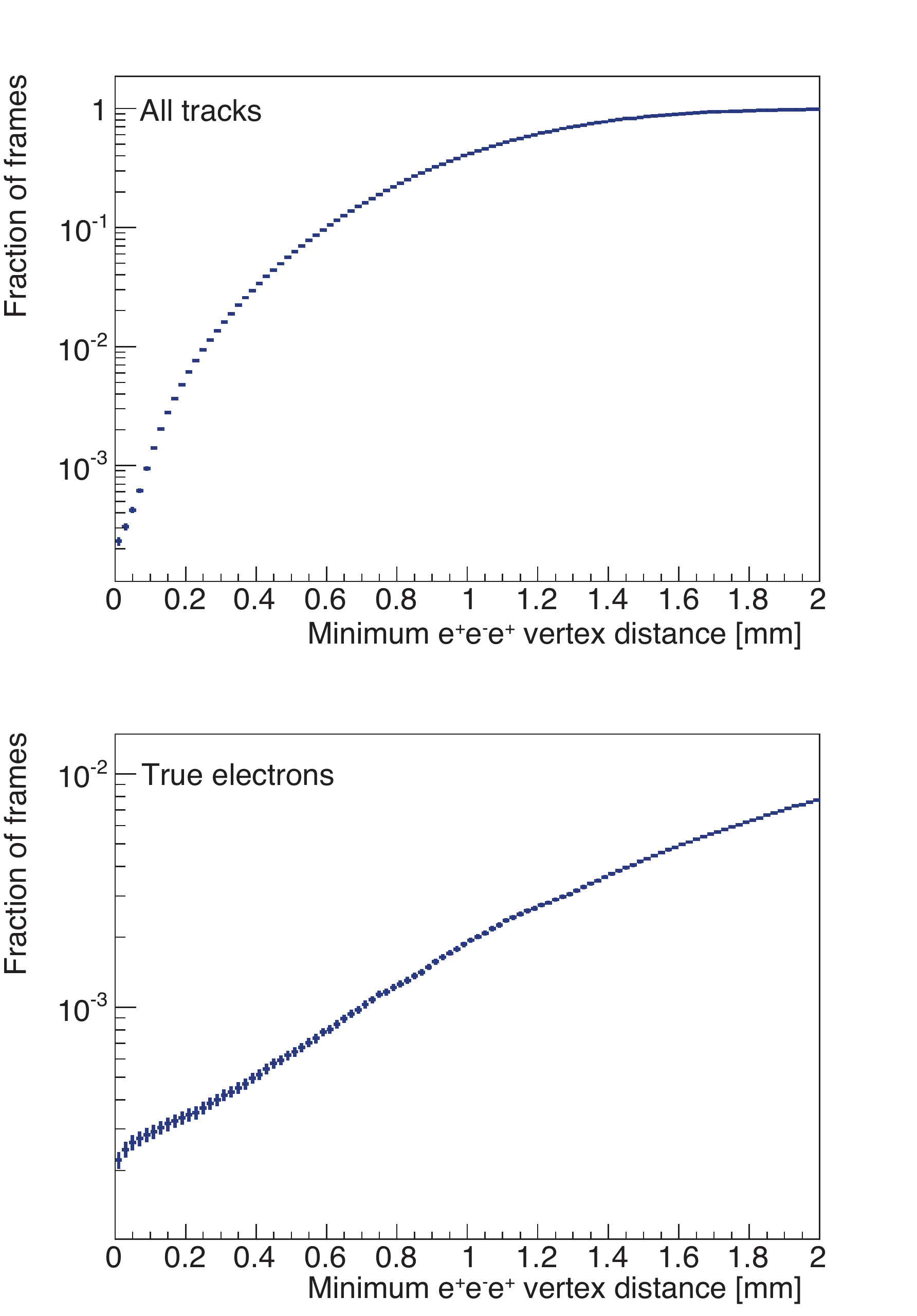}
	\caption{Fraction of $\SI{50}{ns}$ frames containing three vertices consistent with $e^+e^-e^+$ inside a given distance for a muon stop rate of $\SI{2e9}{Hz}$ for 680'000 simulated frames. In the top plot, every crossing of a simulated electron/positron track is counted as a vertex; charge assignments are made purely on the apparent curvature, i.e.~recurling positrons are counted as electrons. In the bottom plot, only true electrons are counted.}
	\label{fig:fraction_phase2}
\end{figure}

Triplets of the tracks thus reconstructed can then be fit to a common vertex.
Even loose vertex requirements can give a $\num{e3}$ reduction factor at
$\SI{2e9}{Hz}$ muon rate and $\num{e4}-\num{e5}$ for the phase I experiment
(see Figures~\ref{fig:fraction_phase1} and  \ref{fig:fraction_phase2}).
Combining the vertexing with modest kinematic requirements (e.g.~on the
three-particle invariant mass or the planarity) should produce the required data
reduction, leaving the timing information as a valuable offline cross-check (and
obviating the need for online timing reconstruction).

\section{Hardware Implementation}
\label{sec:HardwareImplementation}

\begin{figure}
	\centering
		\includegraphics[width=0.4\textwidth]{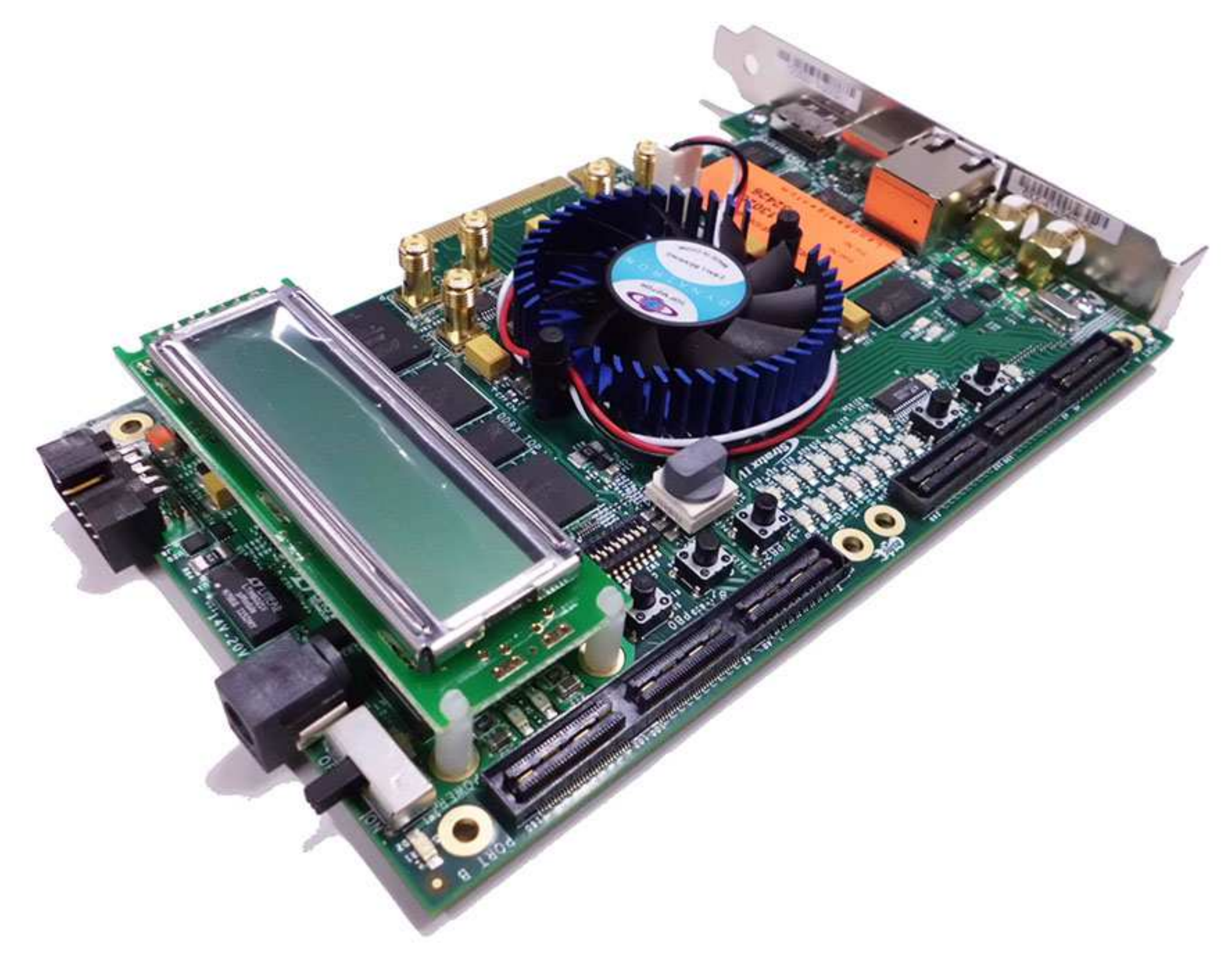}
	\caption{ALTERA Stratix IV PCIe development board.}
	\label{fig:alteradevboard}
\end{figure}

The data will arrive on the farm PCs via optical links on a PCIe FPGA board. The
FPGA will perform the event building and buffering and also allows to run simple
clustering and sorting algorithms. The event data are then transferred via DMA
over the PCIe 3 bus\footnote{Note that PCIe is actually not a bus protocol, but
offers switched point-to-point connections. The \emph{bus} designation is due to
the software-side backwards compatibility to the original PCI bus interface.} to
the memory of a graphics processing unit (GPU), where the selection algorithms
are run. The GPU then posts selected events and monitoring data to the main
memory of the PC, from where the CPU ships it via Ethernet to the central data
acquisition computer running the MIDAS software. At that computer, the data
streams from the farm PCs are combined into a single data stream, combined with
various slow control data, compressed and stored.

For the receiver FPGA cards, evaluation boards from either XILINX
\cite{XilinxDevKitUG}, or ALTERA (Figure~\ref{fig:alteradevboard})
\cite{AlteraDevKitUG,AlteraDevKitRM} or similar hardware built by the
collaboration could be used in conjunction with daughter boards with the optical
receivers (similar to e.g.~the optical receiver boards used in the LHCb readout
electronics \cite{Wiedner2004}). The maximum data rate over the PCIe 3.0 bus is
$\SI{16}{Gbyte/s}$, amply sufficient for phase I\footnote{For phase~I running,
the FPGA-GPU link can also be implemented on PCIe 2.0 (max. $\SI{8}{Gbyte/s}$),
which is better supported on currently available FPGAs.}. For the full phase~II
rate, the raw link speed is still sufficient, would however have to be fully and efficiently used. The
PCIe 4.0 standard, doubling this rate, should become commercially available
around 2017, compatible with phase II running; alternatively, the number of farm
PCs could be increased. \balance

The GPU boards will be obtained commercially as late as possible in order to
profit from the fast developments and sinking prices. As far as raw floating
point throughput is concerned, current high-end GPUs already pack enough power
for high rate running \cite{NVIDIAGTX680,AMDGCN}. Newer cards are however
expected to offer higher memory bandwidth and better caching. Also the
performance of the driver software (especially as far as the PCIe 3 bus is
concerned) and the GPU compilers is expected to improve. The two GPU vendors AMD
and NVIDIA offer fairly different architectures; which one performs better
depends a lot on the details of the algorithm to be implemented; we are
currently performing tests with both architectures and will choose a vendor once
we have a mature implementation.

We currently plan to host the farm PCs in individual tower casings, ensuring
enough space for the FPGA board and the high end GPU whilst allowing for air
cooling. At load, each tower will consume around $\SI{0.5}{KW}$, so adequate
cooling of the counting house is essential.

\chapter{Simulation}
\label{sec:Simulation}

\nobalance

This chapter describes the Geant4 \cite{Allison:2006ve, Agostinelli2003250} based 
simulation used to derive the figures and plots in this proposal.

\begin{figure}[b!]
	\centering
		\includegraphics[width=0.48\textwidth]{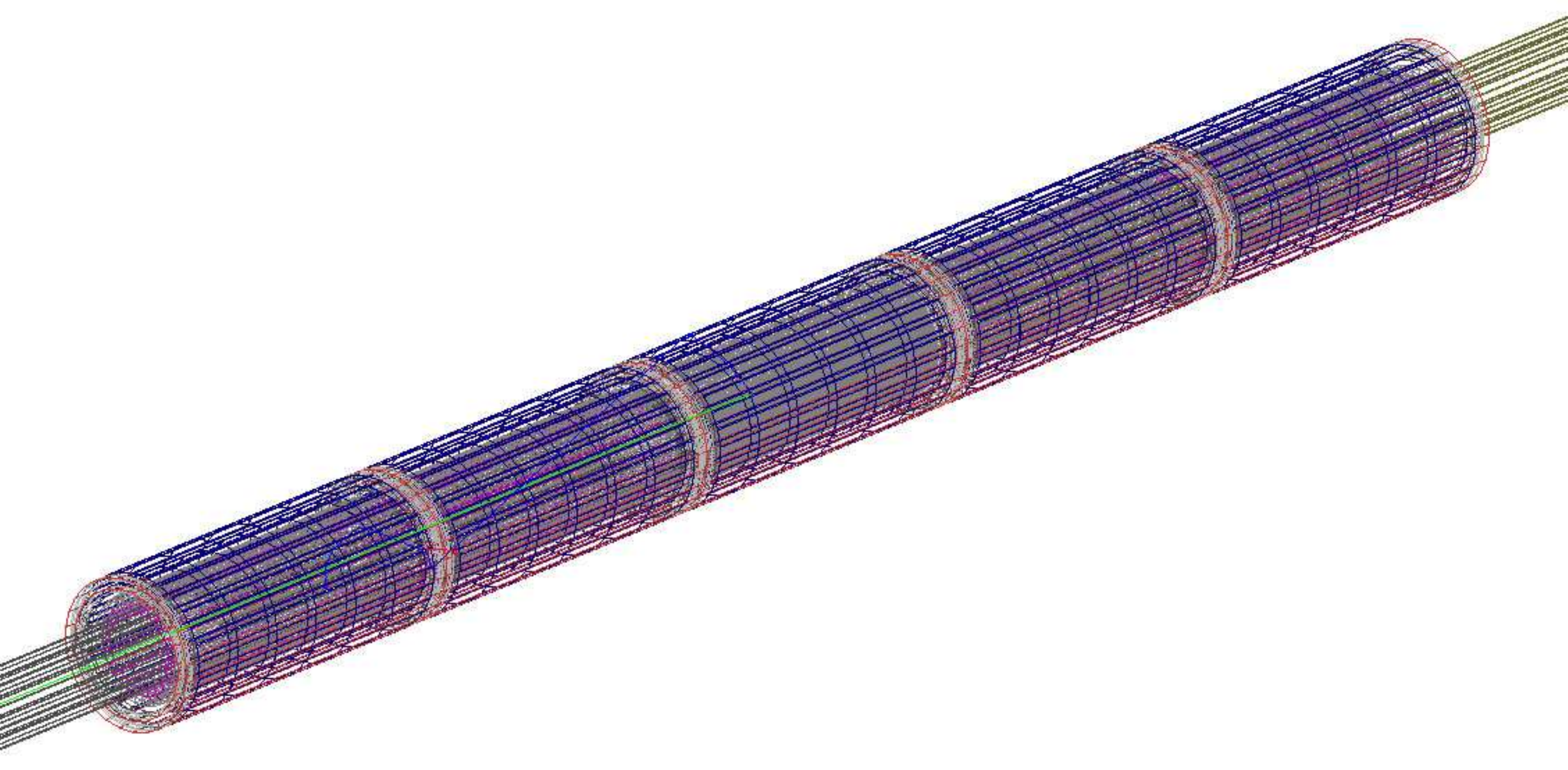}
	\caption{Wire frame view of the simulated detector.}
	\label{fig:longpipe}
\end{figure}

\section{Detector geometry}

\subsection{Beam delivery}
\label{sec:SimBeamDelivery}

\begin{figure*}
	\centering
		\includegraphics[width=0.98\textwidth]{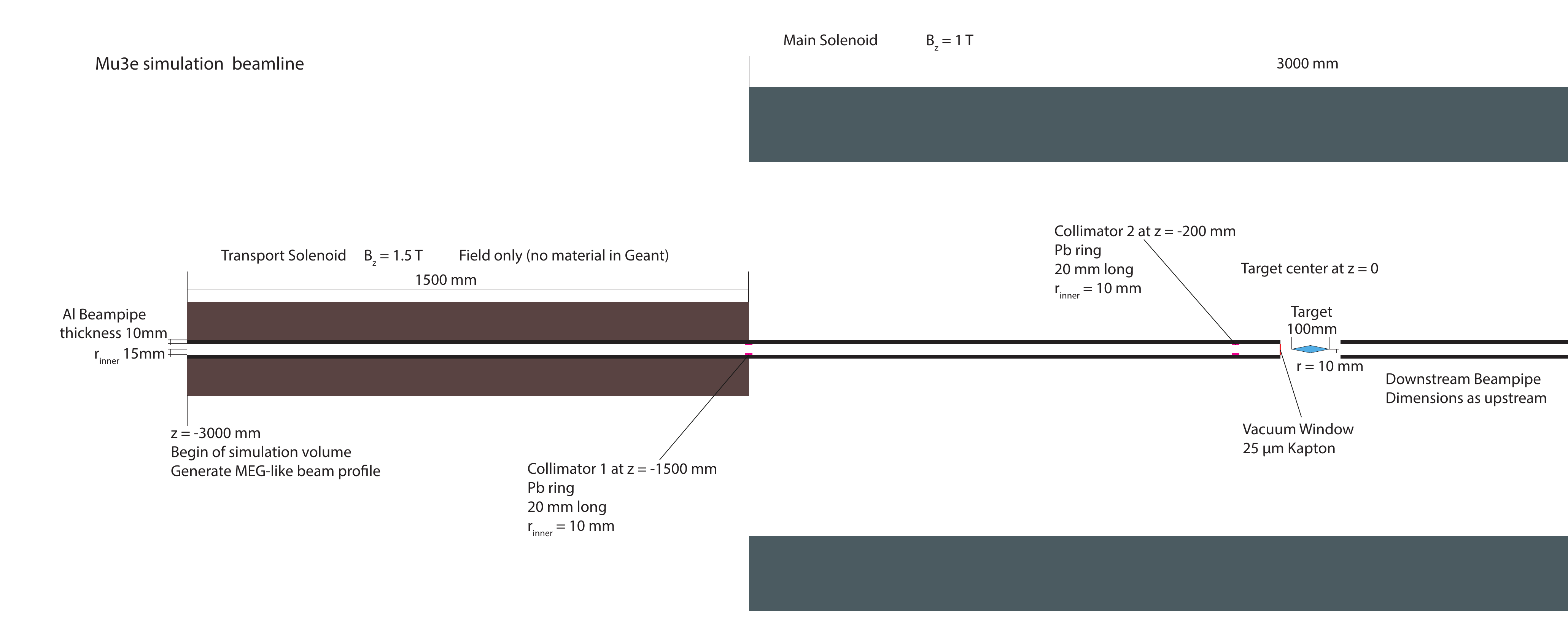}
	\caption{Beamline in the current simulation.}
	\label{fig:BeamlineSimulation}
\end{figure*}

In the simulation, the beam is started $\SI{3}{m}$ in front of the target inside
a beam transport solenoid. Beam particles are generated with a profile and
momentum spectrum like the one observed in MEG. $\SI{1.5}{m}$ before the target,
the beam enters the main solenoid and shortly before the target it exits the
beam vacuum through a thin window. Along the beamline, two thick lead
collimators reduce the beam to the target size. For an overview of the simulated
beamline elements, see Figure~\ref{fig:BeamlineSimulation}. In this simple setup,
about a third of the generated muons decay in the target, which, whilst not very efficient, gives a conservative estimate of beam-induced backgrounds.

\subsection{Target}

The target is simulated as a hollow aluminium double cone supported by three
nylon strings at each end and a nylon string along its axis, see also chapter
\ref{sec:Target}.

\begin{figure*}
	\centering
		\includegraphics[width=0.98\textwidth]{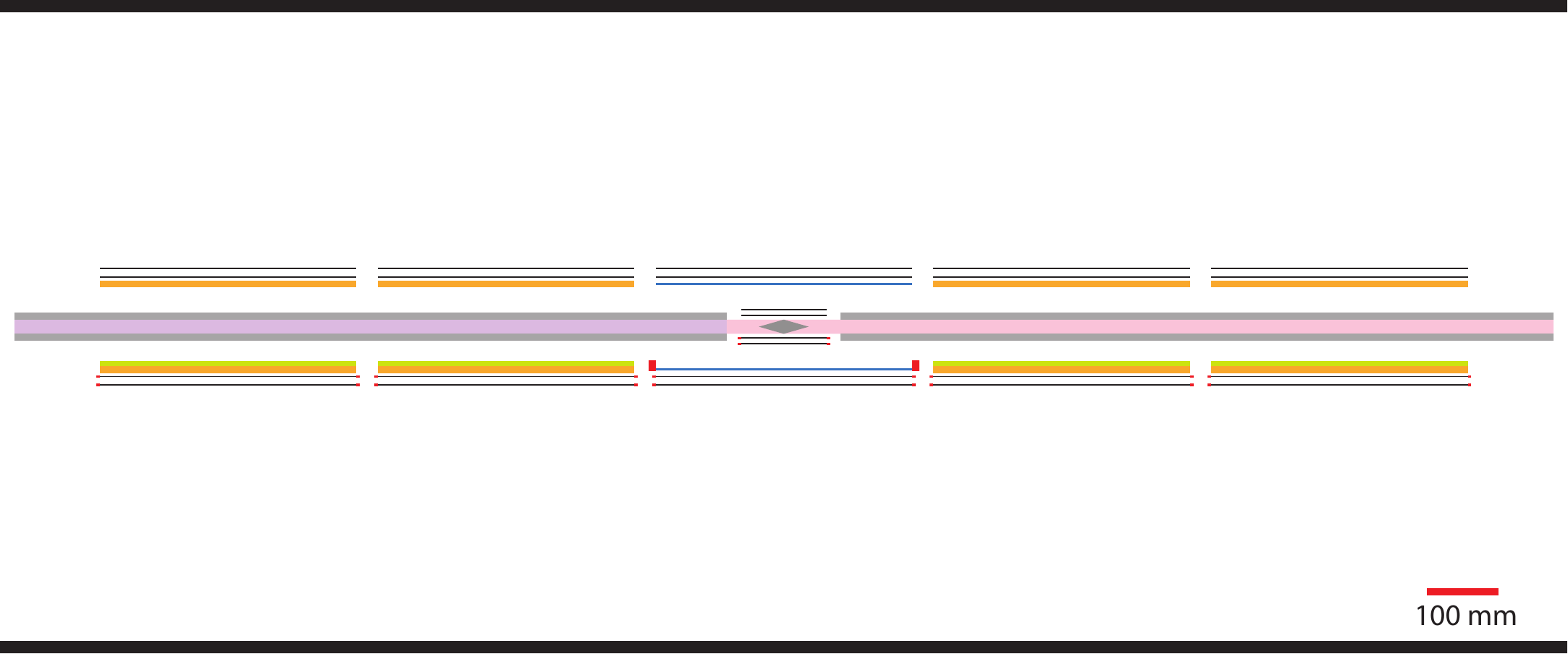}
	\caption{Geometry of the detector in the simulation. The top half only shows active (sensor) volumina, whereas the bottom half only shows support structures.}
	\label{fig:SimDet}
\end{figure*}

\subsection{Pixel detector}

\begin{figure}
	\centering
		\includegraphics[width=0.48\textwidth]{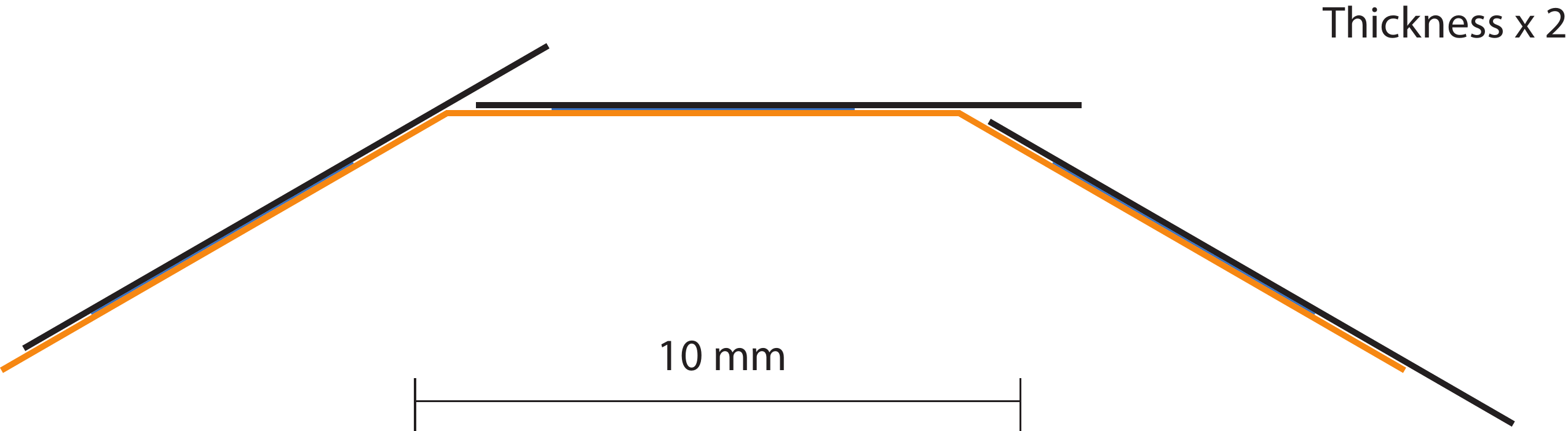}
	\caption{Pixel detector simulation geometry for the innermost layer. The sensor is shown in black, the aluminium traces in blue and the Kapton support in orange. Note that all thicknesses are stretched by a factor of 2.}
	\label{fig:Pixelsimulations}
\end{figure}

The pixel detector is simulated as $\SI{50}{\micro\meter}$ of silicon on top of
$\SI{15}{\micro\meter}$ of aluminium representing the traces on the flexprint
(covering half the available area) on top of $\SI{50}{\micro\meter}$ of Kapton,
with the silicon offset such that an overlap with next sensor is created, see
Figure~\ref{fig:Pixelsimulations}. Half a millimeter of the pixel sensor at the
edge is assumed to be inactive, the rest is divided into
$80\times\SI{80}{\micro\meter\squared}$ pixels. The simulated sensor layers are
supported at their ends by plastic and aluminium structures modeled on those in
the mechanical prototype shown in Figure~\ref{fig:InnerLayerPrototypes}.

\subsection{Scintillating fibres}

The fibre detector is simulated as three circular layers of
$\SI{250}{\micro\meter}$ scintillating fibres in the main simulation. A
detailed simulation including optical transport and the effect of fibre
cladding and coating also exists, see section \ref{sec:FibreSimulation}. The results of the detailed simulation
regarding light yield and propagation times will eventually be fed back into the
main simulation in a parameterized form. The response of the silicon
photomultipliers is simulated by the GosSiP package \cite{Eckert:2012yr}. In the
simulation, the fibres are supported at both ends by massive aluminium rings.

\subsection{Tile detector}

The simulated tile detector consists of plastic scintillator tiles mounted on an
aluminium tube. Also here, a separate detailed simulation including light
transport and silicon photomultiplier response is available and will have to be
fed back into the main simulation in a parameterized form.

\section{Magnetic field}

The simulated magnetic field can be read from arbitrary field maps or generated
in the code via integration over current loops. The propagation of muons in the
field includes spin tracking. For the simulations shown in this report, the
field was generated from 100 current loops spaced equally over $\SI{3}{\meter}$,
with currents normalized such that the longitudinal component of the field in
the center of the target is $\SI{1}{\tesla}$, supplemented by a
$\SI{1.5}{\tesla}$ field in the center of the beam transport solenoid, see
section \ref{sec:SimBeamDelivery} and Figure~\ref{fig:BeamlineSimulation}.

\section{Physics Processes}

\subsection{Multiple Coulomb scattering}

\begin{figure*}
	\centering
		\includegraphics[width=1.00\textwidth]{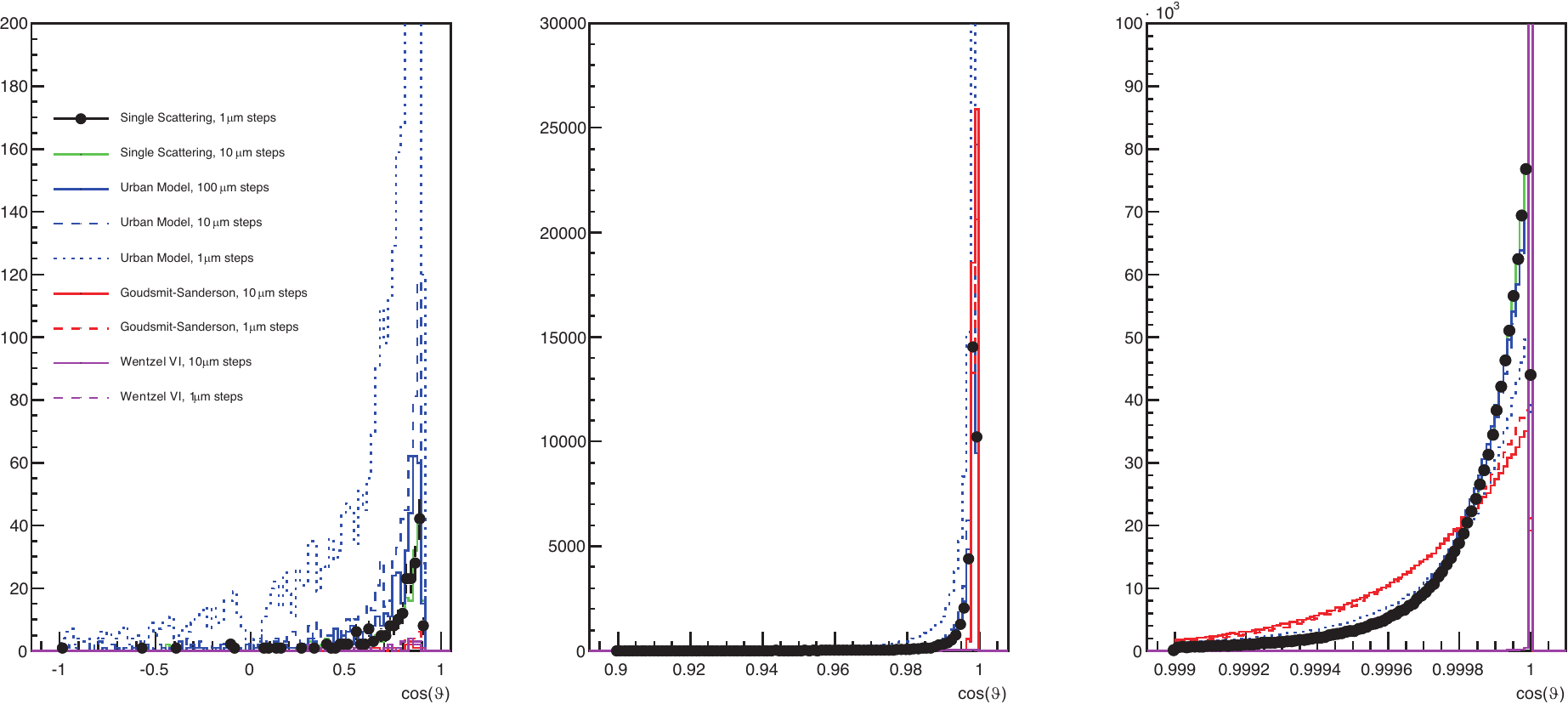}
	\caption{Comparison of multiple coulomb scattering models in different
	scattering angle ranges. The scatterer is a single silicon-Kapton assembly
	shot at at a right angle with $\SI{30}{MeV}$ positrons. The black dots and
	the green line show the single scattering model which serves as a reference;
	as expected, the single scattering model is not affected by the Geant step
	size. Of all the parameterizations, the Urb\'an model with a step size that
	treats each bit of material as a single volume performs best.}
	\label{fig:cosMS}
\end{figure*}

\begin{figure*}
	\centering
		\includegraphics[width=1.00\textwidth]{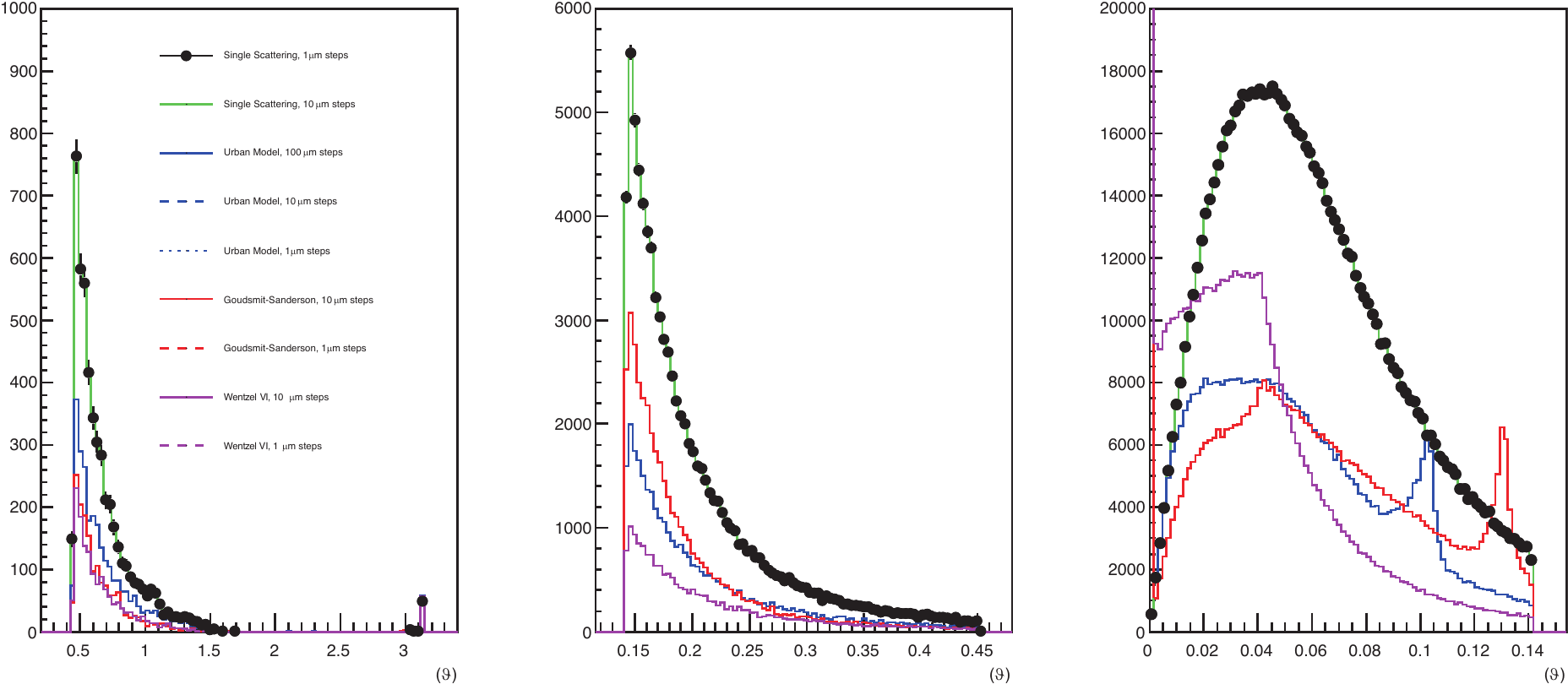}
	\caption{Comparison of multiple coulomb scattering models in
	$\SI{}{\meter}$ of helium gas for different scattering angle ranges. The
	test particles are $\SI{30}{MeV}$ positrons. The black dots and the green
	line show the single scattering model which serves as a reference. All the
	parameterizations are unfortunately inadequate.}
	\label{fig:HeMs}
\end{figure*}

Multiple coulomb scattering is the main limiting factor for the resolution of
the experiment; an accurate simulation is thus crucial. The best results are
obtained by simulating each individual scattering, which however results in
prohibitively large computing times. A large selection of multiple scattering
parameterizations are available in Geant4; in a test setup they were compared to
the single scattering model, see Figure~\ref{fig:cosMS}. The best overall
description is obtained from the Urb\'an-Model \cite{Urban:1004190} at large
step widths, which also has the shortest computation times. In the helium gas on
the other hand, none of the parameterizations performs adequately, see
Figure~\ref{fig:HeMs}.

We plan to verify the simulation results with beam telescope measurements in
2013, which should also lead to a usable parameterization of multiple scattering
in gases.

\subsection{Muon decays}

\subsubsection{Michel decay}

Geant4 implements the Michel decay including polarization of both the muon and
the positron based on \cite{Scheck:1977yg} and \cite{Fischer:1974zh}. The
spectra of the neutrinos do not follow the physical distribution, this does
however not affect the simulation for Mu3e. Somewhat more worrying is the fact
that the Michel matrix element contains radiative corrections but is not clearly
separated from the radiative decay matrix element.

\subsubsection{Radiative decay}

The radiative decay of the muon was implemented in Geant4 by the TWIST
collaboration \cite{Depommier2001} based on \cite{Fronsdal:1959zzb}. The code
does not describe the neutrino spectra and avoids the collinear and infrared
singularities by sampling the matrix element assuming a finite integral.

\subsubsection{Radiative decay with internal conversion}

The radiative decay with internal conversion is simulated using the hit and miss
technique on events generated evenly in phase space using the RAMBO code
\cite{Kleiss1986359} and applying the matrix element from
\cite{Djilkibaev:2008jy}. Unfortunately, there is currently no polarized version
of the matrix element available and thus the simulation is unpolarized. The hit
and miss technique is very expensive in terms of computation time, if the
complete phase space is to be simulated (as the matrix elements varies by more
than 16 orders of magnitude), this can however be overcome by restricting the
simulation to regions of particular interest, e.g.~high invariant masses of the
visible particles. 

\subsubsection{Signal}

The signal kinematics are highly model-dependent, see chapter
\ref{sec:DecayMu3e}. If not otherwise noted, we have used three particle phase
space distributions in the simulation, following the practice of SINDRUM and
earlier experiments.

\subsubsection{Special decays}

The simulation allows the simulation of overlap decays, where we force more than
one muon decay to happen at a single vertex. Thus we can simulate the accidental
backgrounds arising e.g.~from the overlap of an internal conversion decay and a
Michel decay without having to generate in excess of $\num{e16}$ frames.

\section{Time structure}

\begin{figure}
	\centering
		\includegraphics[width=0.48\textwidth]{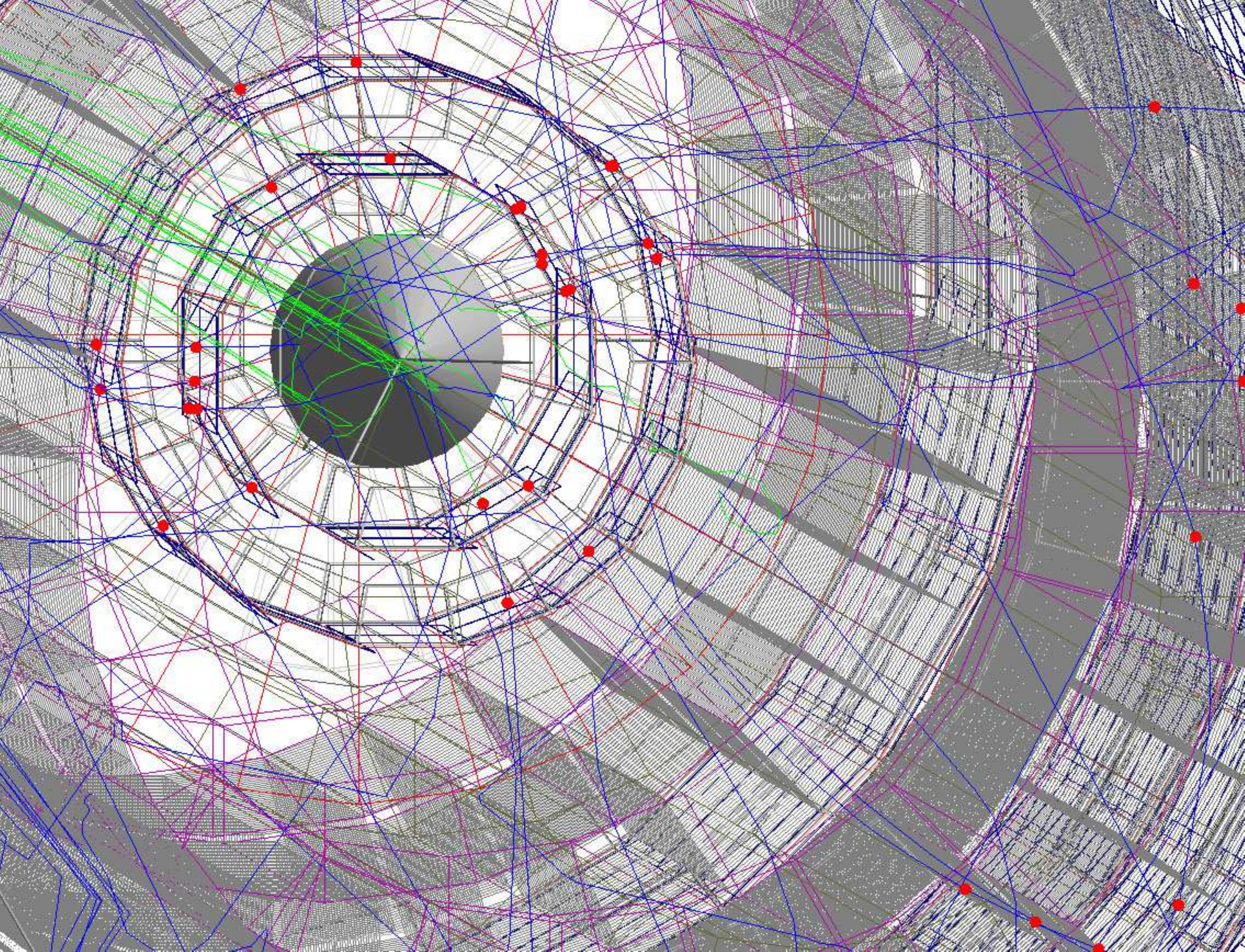}
	\caption{View of a simulated frame.}
	\label{fig:simulation}
\end{figure}

\balance

As the Mu3e experiment operates with a quasi continuous beam, the paradigms of
\emph{bunch crossing} and \emph{event} from collider experiments do not apply;
they have however informed the design of the Geant4 package. In our simulation,
particles are assigned a $\SI{64}{\bit}$ ID, which is unique over a run and thus
conserves identities and mother-daughter relationships across the boundaries of
read-out time frames. Before each step of tracking the particle through the
detector, it is checked whether the particle has just crossed into the next time
slice. If so, its information is stored, its time relative to the time slice
adjusted and tracking deferred to the next slice. Thus we ensure that we
correctly treat muons stuck in the target before decaying and decay products
crossing read-out frame boundaries while traversing the detector. In order to
simulate a steady state, where approximately the same number of muons enter the
target and decay, the first $\SI{5}{\milli\second}$ of simulation running,
during which the target is loaded, are usually thrown away and not used in
occupancy or efficiency studies.

Currently not simulated are effects of the $\SI{40}{\mega\hertz}$ structure of
the primary proton beam on the time structure seen in the detector; if this
would be needed, a measured structure could easily be superimposed on the
generation of muons in the simulation framework.

\chapter{Reconstruction}
\label{sec:Reconstruction}

\nobalance

\section{Track Reconstruction in the Pixel Tracker}

A precise track reconstruction of electrons is of highest importance for
the identification of the \mteee decay with a sensitivity of 1 out of
$10^{15}$ ($10^{16}$) ordinary Michel decays in phase I (II), which have to be suppressed by 16 orders of magnitude. 

Due to the high rate and the resulting high occupancy especially at phase II of the project with up to 100 tracks per readout frame, the reconstruction algorithm has to deal effectively with the combinatorial background in order to reduce the fake rate, i.e.~the rate of wrongly reconstructed tracks, to an acceptable level. 
The combinatorial problem is not only due to the high rate but also due to the
large bending of the low momentum electrons in the strong magnetic field of
$B=\SI{1}{T}$, which, depending on the position and flight direction can make several turns in the detector (recurlers).
Hit combinations can span over distances of more than half a meter. 
Hits of recurling tracks are found on opposite sides of the detector and still have to be correctly combined by the reconstruction program. 
This is of particular importance for the determination of the flight direction
and therefore charge of the particle. 
Only for a fully reconstructed track the time information provided by
the time of flight system can be correctly applied. 

As the full detector readout is triggerless, all muon decays have to be fully
reconstructed already on filter farm level, setting high demands on the speed of the online track reconstruction algorithm.
A further complication comes from the fact that the track resolution is
dominated by multiple scattering in the silicon pixel sensors and not by the
pixel size, in contrast to most other experiments.
Therefore, standard non-iterative circle fits of tracks
\cite{Kari91} as used in high energy experiments can not be used here.

In order to reduce multiple scattering, the number of sensor
layers are reduced to a minimum in the detector design which, unfortunately, 
also reduces redundancy for track reconstruction. 
Therefore, the track reconstruction also has to cope with a minimum of
information provided by only four sensor layers.

\section{Track Fitting and Linking}
For the track reconstruction two different approaches are followed in the Mu3e experiment, the broken line fit~\cite{Blobel:2011az,Kleinwort:2012yt} and the fast linear fit based on multiple scattering~\cite{AS2012a}. 
The broken line fit determines hit positions and scattering angles 
simultaneously and was implemented in 2D ~\cite{Blobel:2011az,Kiehn2012} and recently also
in 3D~\cite{Kleinwort:2012yt,Kiehn2015}. 
It is based on linearisation of a previous circle fit, works non-iteratively and provides the correlation between all fit
quantities. 
The broken line fit, however, requires knowledge of the assignments of hits to tracks from a previous linking step. 
Therefore, the broken line fit can only be used in the final step of the track
reconstruction, also because a previous track fit is required for the
linearisation procedure.

\begin{figure}
	\centering
		\includegraphics[width=0.48\textwidth]{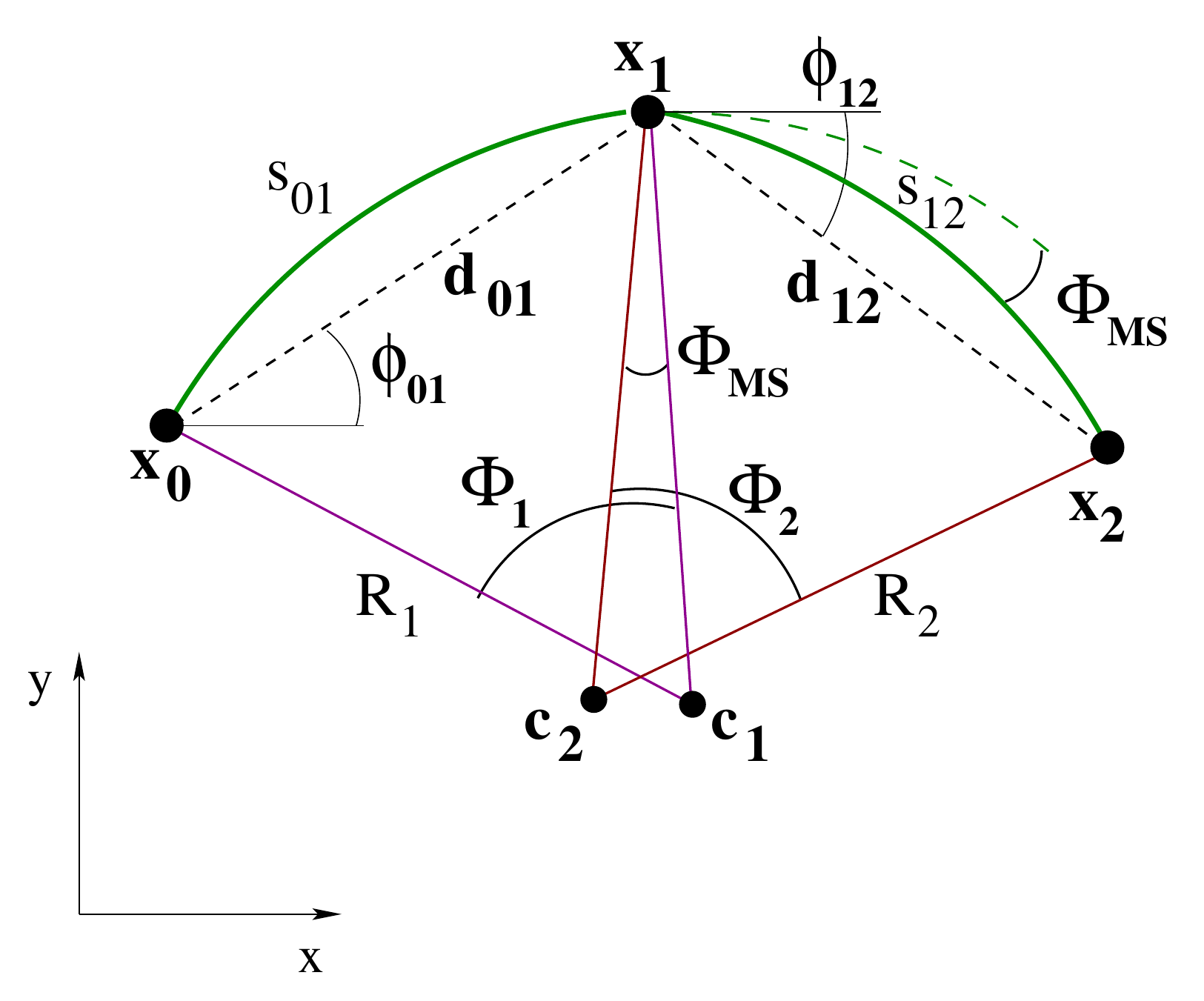}
	\caption{Sketch of the variables used in the multiple scattering fit.}
	\label{fig:sketch_MS}
\end{figure}

\begin{figure}[p]
	\centering	\includegraphics[width=0.48\textwidth]{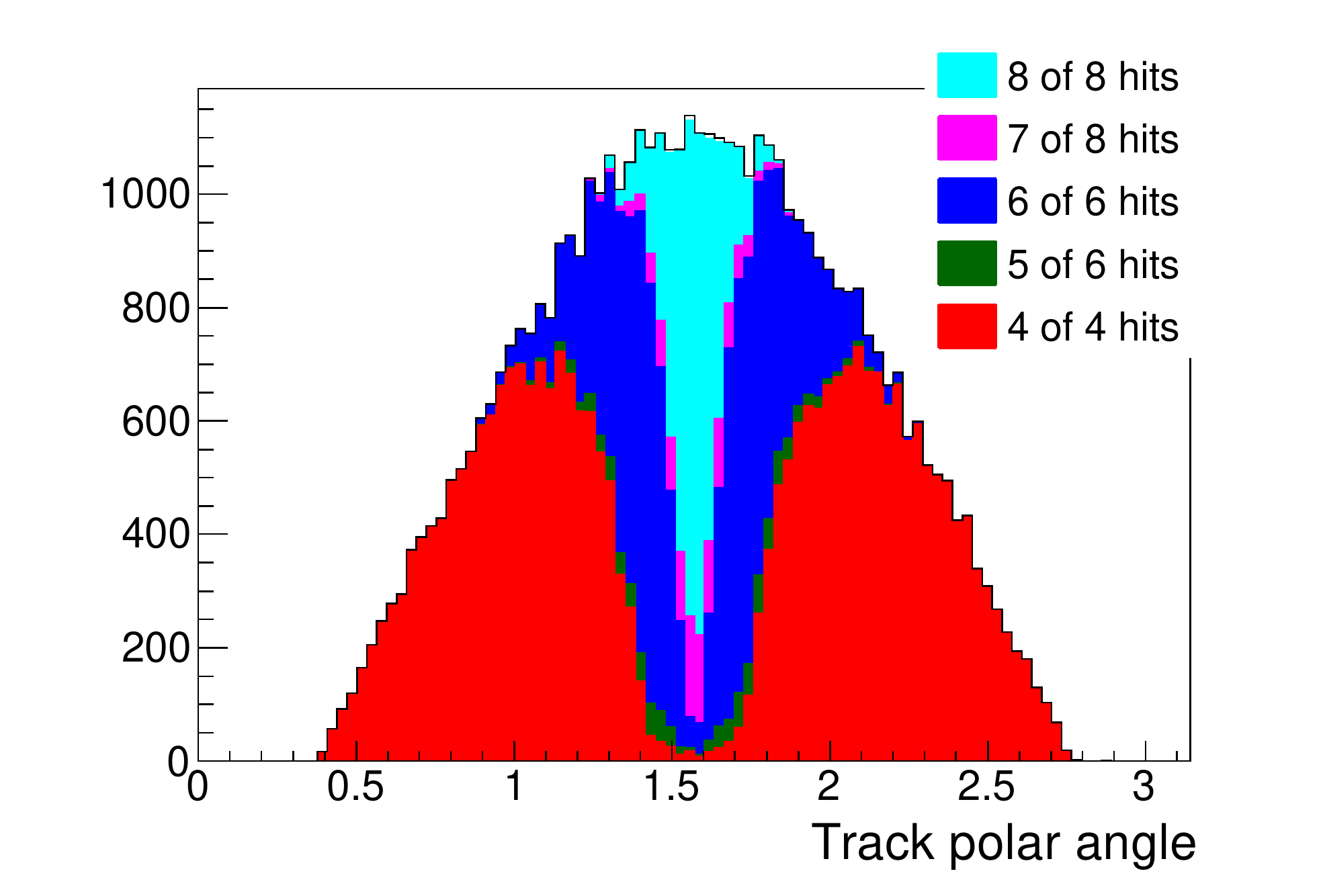}
	\caption{Track types versus track polar angle for Michel decays in phase IA.}
	\label{fig:tracks_ph1a}
\end{figure}

\begin{figure}[p]
	\centering
\includegraphics[width=0.48\textwidth]{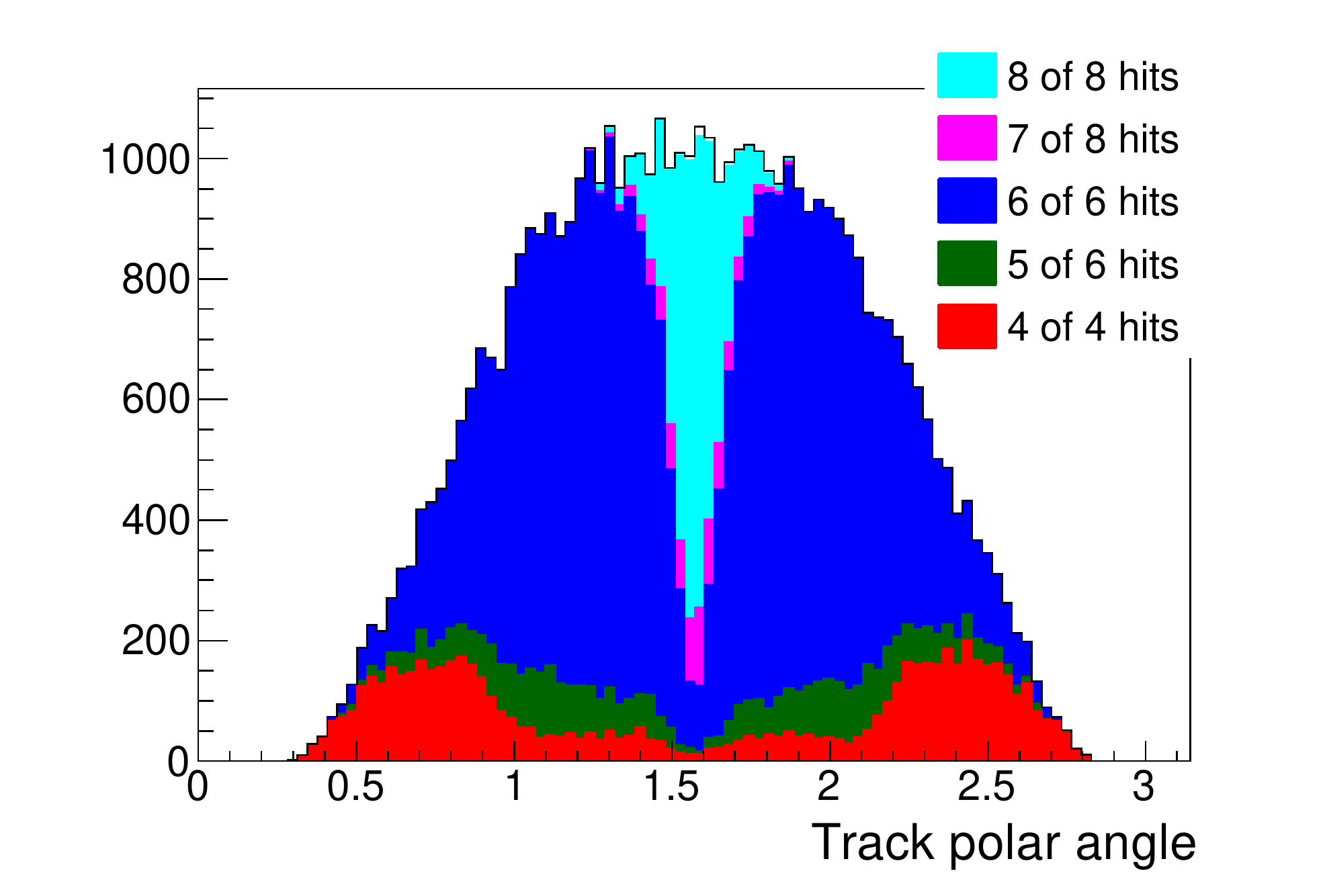}
	\caption{Track types versus track polar angle for Michel decays in phase IB.}
	\label{fig:tracks_ph1b}
\end{figure}

\begin{figure}[b!]
	\centering
\includegraphics[width=0.48\textwidth]{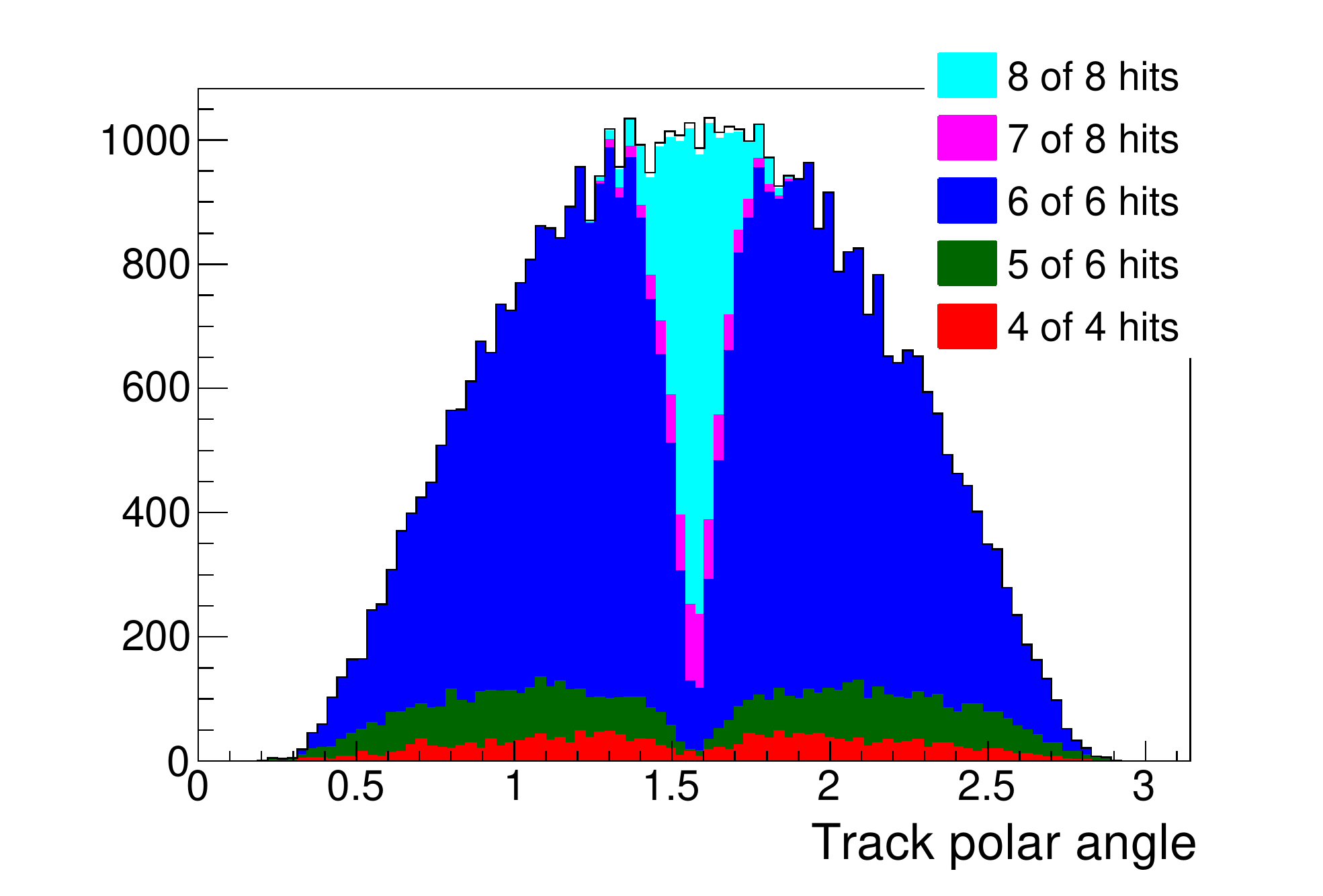}
	\caption{Track types versus track polar angle for Michel decays in phase II.}
	\label{fig:tracks_ph2}
\end{figure}

\begin{figure*}[t!]
	\centering	\includegraphics[width=1.00\textwidth]{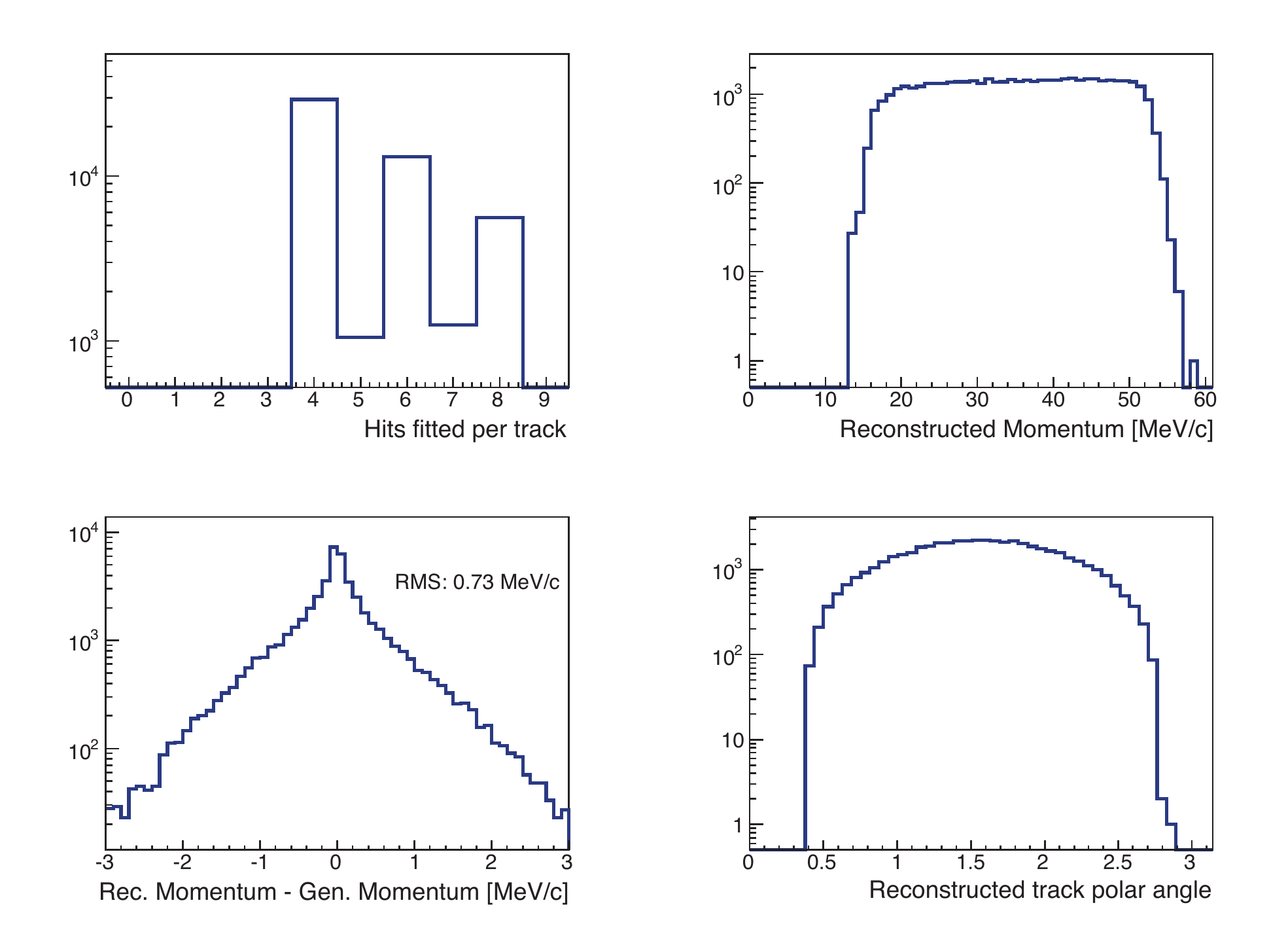}
	\caption{Tracking performance for Michel decays in phase IA.}
	\label{fig:recopix_ph1a_michel}
\end{figure*}

\begin{figure*}[t!]
	\centering	\includegraphics[width=1.00\textwidth]{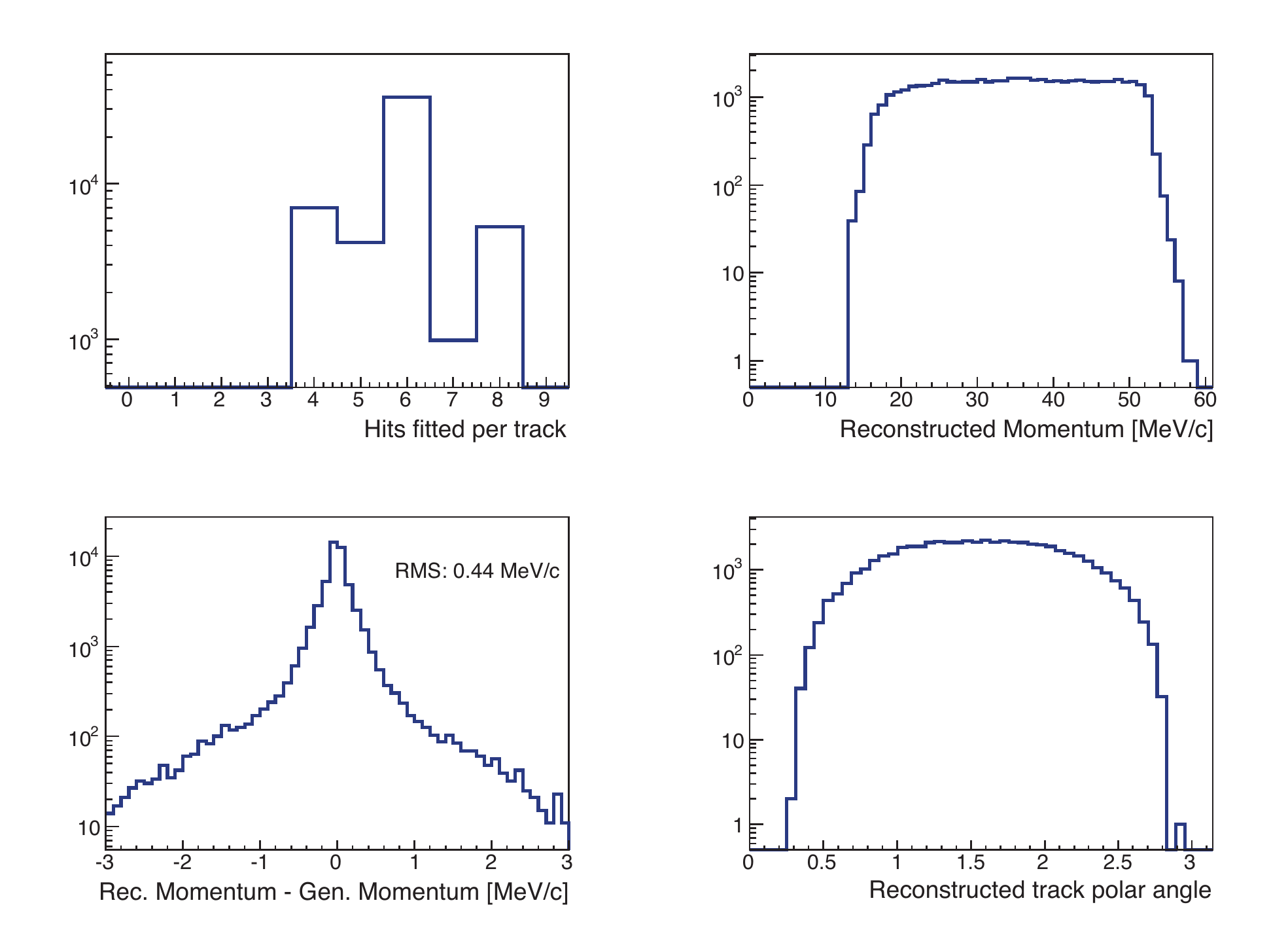}
	\caption{Tracking performance for Michel decays in phase IB.}
	\label{fig:recopix_ph1b_michel}
\end{figure*}

\begin{figure*}[t!]
	\centering	\includegraphics[width=1.00\textwidth]{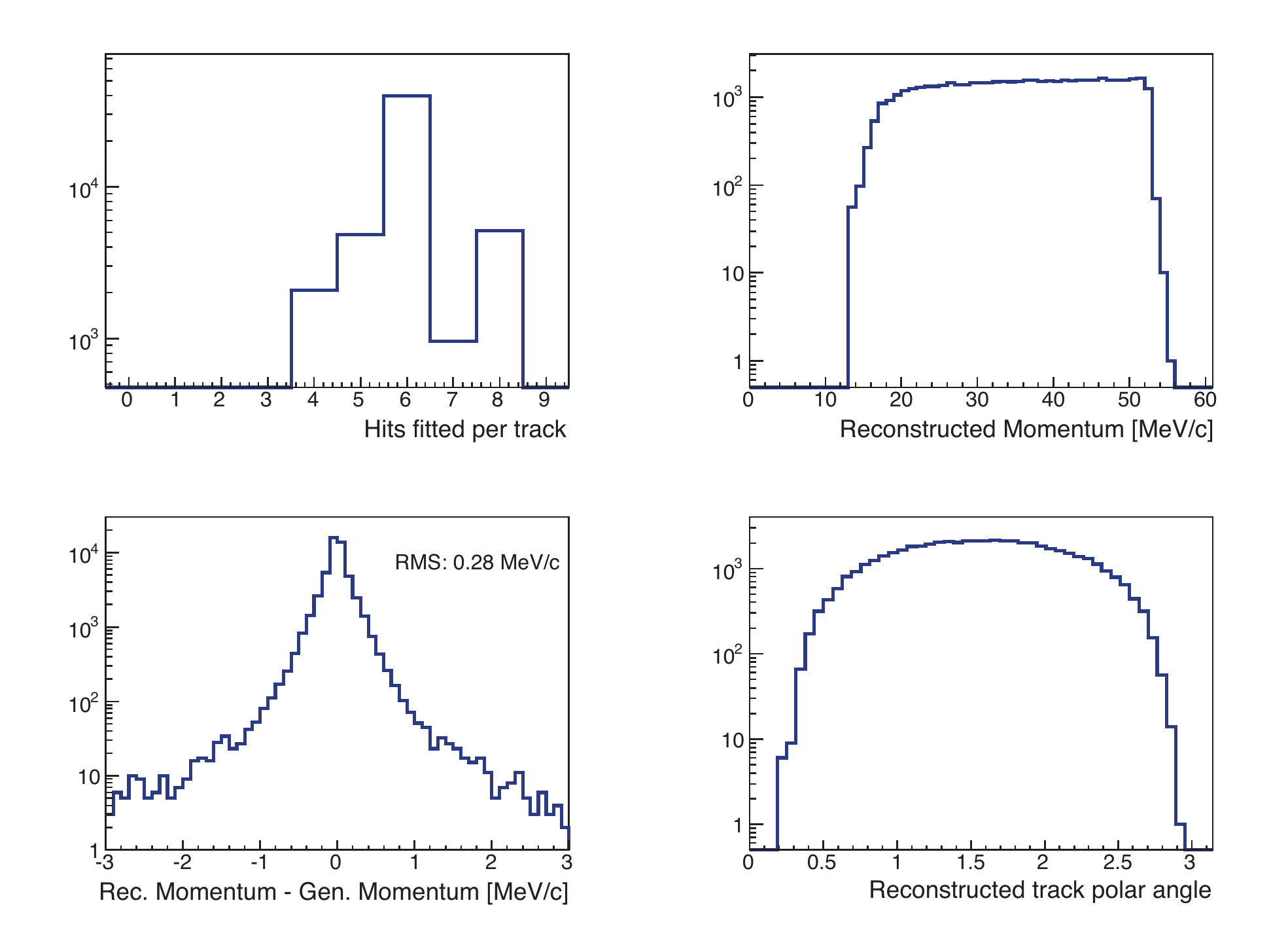}
	\caption{Tracking performance for Michel decays in phase II.}
	\label{fig:recopix_ph2_michel}
\end{figure*}

The fast three-dimensional multiple scattering (MS) fit \cite{AS2012a} is based on fitting the multiple
scattering angles at the middle hit position in a hit triplet combination, see
Figure~\ref{fig:sketch_MS}. 
In this fit, spatial uncertainties of the hit positions are ignored. 
This is a very good approximation for the Mu3e experiment as the pixel 
resolution uncertainty given by $\sigma_{\textnormal{pixel}} =  \SI[parse-numbers=false]{80 / \sqrt{12}}{\um}$ is much smaller than the uncertainty from multiple
scattering in the corresponding sensor layer. 
The MS-fit requires a detailed knowledge of the
material distribution in the detector for the calculation of the scattering
angle uncertainty. 
It minimises the azimuthal
and polar scattering angles at the sensor corresponding to the middle hit and exploits energy 
conservation\footnote{the energy loss in the Mu3e experiment is only about
  $80$~keV per sensor layer and can be neglected for track finding.}.
The hit triplet trajectory, represented by two connected helical curves, is described by the following two
equations:
\begin{eqnarray}
\sin^2{\frac{\Phi_1}{2}} & = & \frac{d_{01}^2}{4 R_{\rm{3D}}^2} \ + \ \frac{z_{01}^2}{R_{\rm{3D}}^2}
  \frac{\sin^2{(\Phi_1/2)}}{\Phi_1^2} \\
\sin^2{\frac{\Phi_2}{2}} & = & \frac{d_{12}^2}{4 R_{\rm{3D}}^2} \ + \ \frac{z_{12}^2}{R_{\rm{3D}}^2}
  \frac{\sin^2{(\Phi_2/2)}}{\Phi_2^2} \quad .
\label{eq:r3dphi1}
\end{eqnarray}
The quantities, also shown in the sketch of Figure~\ref{fig:sketch_MS}, are
the following: $R_{\rm{3D}}$ is the three dimensional track radius, which can
be directly related to the momentum of the particle for a given magnetic
field; $\Phi_{1}$ ($\Phi_{2}$) are the bending angles of the first (second) arc and
$d_{01}$ ($d_{12}$) and $z_{01}$ ($z_{12}$) are the distances between the hits
in the plane transverse and longitudinal to the solenoidal
magnetic field between the first and second (second and third) hit, respectively.
These equations can be linearised and solved in a fast non-iterative procedure~\cite{AS2012a}.

This linearized MS-fit is used as basis for the full reconstruction of tracks
in the pixel detector. 
Tracks with more than three hits are fitted by subsequently combining several hit triplets. In the current 
reconstruction program~\cite{AS2012b}, long tracks combining hits from several
pixel layers are reconstructed first, then shorter tracks with fewer hit
assignments are reconstructed. This procedure is repeated until no hits are
left or no further tracks are found.
Tracks with unresolved hit ambiguities are ignored in the following study to ensure high
quality tracks with low fake rate. Also tracks with less than four hits
combined are ignored.   

The number of hits linked to tracks depends on the single hit efficiency,
which is assumed to be 98\% in the following studies, on the track direction (polar
angle), the geometry of the pixel tracker and the geometrical acceptance of
the detector, which is largest in phase II with four recurl stations. 
The multiplicity of linked tracks as function of the polar angle is shown
in Figures~\ref{fig:tracks_ph1a}, \ref{fig:tracks_ph1b} and \ref{fig:tracks_ph2} for the three different detector configurations.
The multiplicity of linked hits is highest in the central region of the
detector, where recurling tracks can be fully reconstructed. 
In the current version of the reconstruction program, only single turns of tracks 
are reconstructed, corresponding to a maximum of eight hits - four hits
assigned to the outgoing trajectory and four hits assigned to the returning
trajectory. In cases where one hit is missing, only seven hits are linked.
For upstream or downstream going tracks the recurlers are not fully 
reconstructed, leading to tracks with mainly four linked hits or six linked
hits depending on the detector setup and the number of implemented recurl stations.
It should be noted here that there is a big qualitative difference between the
four-hit tracks and tracks with more than 4 hits (recurlers).
The bending radius of recurling tracks can be measured 
with much higher precision due to the larger bending in the magnetic field.
This leads to very different momentum resolutions for the
different detector setups. 

\balance

In phase IA without Recurl Stations most tracks are
reconstructed with only four linked hits. This yields a momentum resolution
for reconstructed Michel electrons of about $RMS(p) = \SI{0.73}{MeV\per c}$, see Figure~\ref{fig:recopix_ph1a_michel}. 
This resolution improves considerably to $RMS(p) = \SI{0.44}{MeV\per c}$ by adding two inner Recurl Stations
in phase IB (Figure~\ref{fig:recopix_ph1b_michel}) and it improves further to $RMS(p) = \SI{0.28}{MeV\per c}$ by adding two inner and two outer Recurl
Stations in phase II Figure~\ref{fig:recopix_ph2_michel}).
The reconstruction of the recurling tracks, which provide high resolution
momentum information, is crucial for the success of the experiment.

\section{Vertex Fitting}
A vertex fit will be used to precisely reconstruct the position of the muon
decay for signal events. 
The vertex fit intrinsically checks the consistency of the assumption that
all three signal candidate tracks originate from the same vertex. 
A common vertex fit of the three candidate electrons allows a suppression
of the combinatorial background of Michel decays by a factor of about $\num{3e8}$  due to the very good pointing resolution of the pixel detector,
which is mainly given by the multiple scattering at the first sensor layer.
Instead of using a common vertex fit, also the distance between target impact
points of reconstructed and extrapolated tracks can be used to discriminate signal against
background. For simplicity the latter approach is used here for the
following sensitivity study.

%!TEX root = ../RP.tex
\chapter{Sensitivity Study}
\label{sec:SensitivityStudies}

\nobalance

\begin{figure}[tb!]
	\centering
		\includegraphics[width=0.48\textwidth]{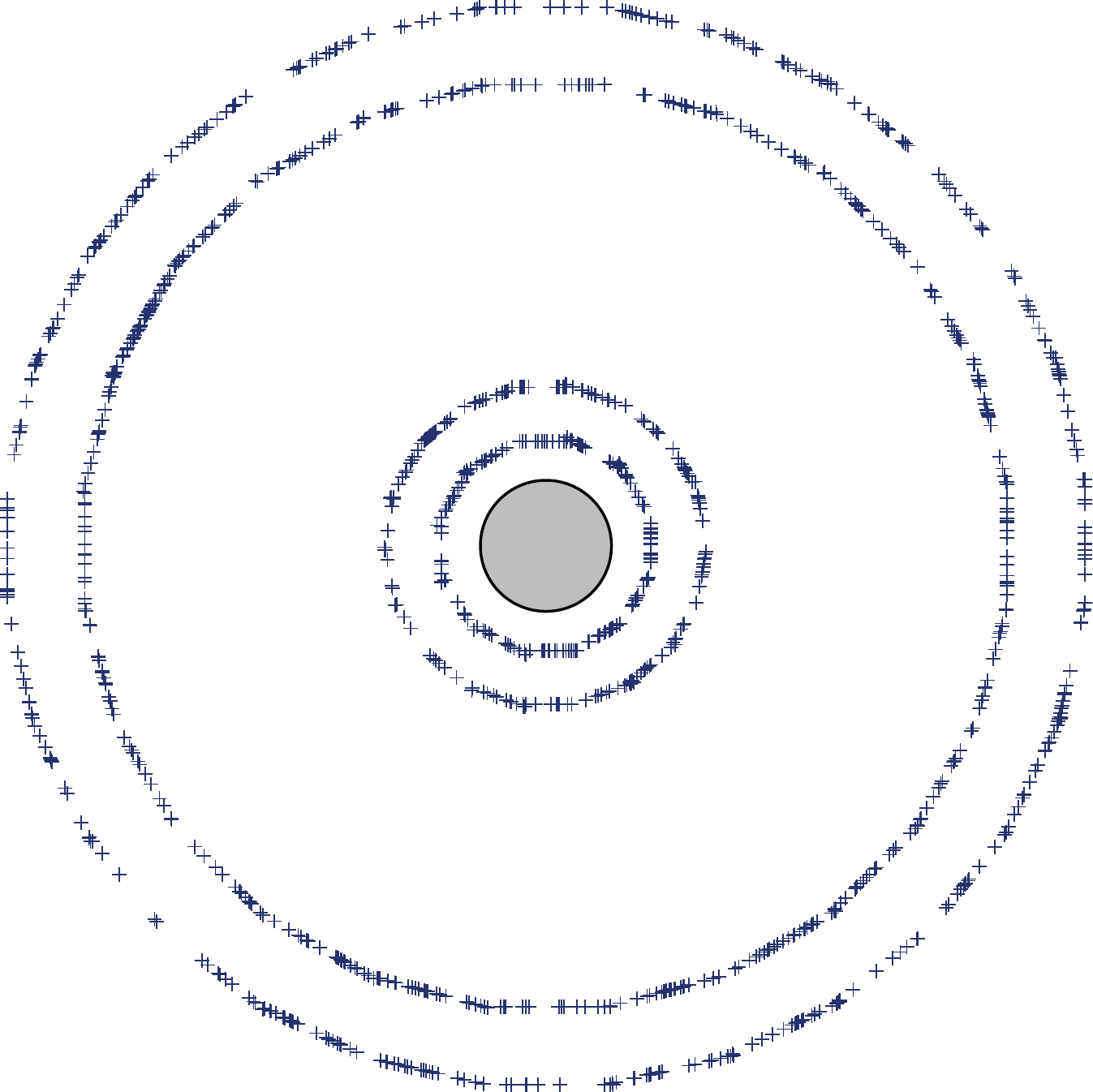}
	\caption{Hits in a simulated event at $\SI{2e9}{Hz}$ muon rate, viewed in the plane transverse to the beam.}
	\label{fig:display_xy}
\end{figure}

\begin{table*}
        \centering
                \begin{tabular}{lccc}
                        \toprule
                                        & Phase IA & Phase IB & Phase II\\
                        \midrule
                                Recurl stations  &      0 & 2 & 4\\
                                Scintillating fibres & No & Yes & Yes\\
                                Charge measurement & No & Yes & Yes\\
                                Target stopping rate$^1$ (Hz)& $\num{2e7}$& $\num{2e8}$ &$\num{2e9}$\\
                                \bottomrule
                \end{tabular}
        \caption{Mu3e configurations considered for the sensitivity
          studies. Note that the detector acceptance is about.~\SI{50}{\percent} for signal
          events, given by the geometry of the central pixel detector.\\
        $^1$ Highest possible rates are assumed in the simulation in order to be conservative.}
        \label{tab:Mu3eConfigurationsConsideredForSensitivtyStudies}
\end{table*}

A sensitivity study based on a full simulation of the pixel detector 
is performed to estimate the
background contribution at the different phases of the Mu3e experiment.
The simulation includes a first material description of the detector, which
varies between the different detector phases. 
The increasing level of detector instrumentation at later stages of the
experiment will provide more information for particle tracking on one hand.
On the other hand, the scintillating fibre tracker, which will be added at a later stage
of the experiment when running at high muon rates, will also add
multiple scattering and consequently increase confusion in the track linking step.
In addition, the factor 10 and 100 higher increased particle rates at the Phase IB and II,
respectively, will significantly increase the combinatorial background from
Michel decays and other accidentals. 
A detailed simulation is therefore required to estimated the sensitivity reach
of the different experimental phases. 
The simulated setups are summarized in Table~\ref{tab:Mu3eConfigurationsConsideredForSensitivtyStudies}.

\section{Simulation and Reconstruction Software}
In order to estimate the maximum achievable sensitivities at the different
stages of the project the track reconstruction program~\cite{AS2012b} is
interfaced to a) a full GEANT simulation of the detector and b) to a fast simulation
program, which is mainly used for the following sensitivity studies. 
Both simulations were cross checked to give comparable values for
the track parameter resolution. 

% Yes, we do have lots of plots, but there are sizeable differences between "analytic" and full simulation, which are mostly understood, but would need to be explained here...

In a first test it was proven that the track reconstruction program is relatively insensitive with regards to
changes of the single hit efficiency.
Using tight reconstruction criteria it is found that a \SI{5}{\percent} loss of signal hits corresponds to
an about \SI{5}{\percent} loss of tracks and an about \SI{10}{\percent} loss of signal events.

Noise hits also affect the track reconstruction. From noise studies
of the MUPIX2 prototype chip it is known that the noise rate is very
small (no noise hits were observed in an overnight run at a low threshold). 
To simulate the effect of noise 
100 additional hits are randomly generated in the pixel detectors,
corresponding to a noise rate of $10^{-6}$~Hz per pixel. 
This rate will be remeasured as soon as a first large scale detector
including a full readout chain becomes available.

\section{Signal Acceptance}
Signal events are generated by using a constant matrix
element in phase space. This choice is motivated by LFV models with
effective four-fermion contact interactions, which predict a constant matrix element. 
The so generated signal events are used for the following efficiency studies.

The resulting signal efficiency is
mainly given by the acceptance of the experiment, which is in total about \SI{50}{\percent} for the innermost
 detector layers being placed at radii of approximately \SI{2.0}{\centi\metre} and \SI{3.0}{\centi\metre}. 
The total acceptance consists of two contributions, one related to the
orientation of the decay plane with respect to the instrumented regions of the
experiment and one from the minimum bending radius (momentum), which can be
reconstructed in the experimental setup.

By moving the inner pixel layers closer to the target, the total acceptance 
can be slightly increased to about \SI{55}{\percent}.
Further improvements of the total acceptance to about \SI{60}{\percent} or more are only
possible by using a longer inner barrel design and by reconstructing particles
with a lower momentum threshold. 
The latter can be achieved in two ways - a) by including highly bent
trajectories in the track reconstruction or b) by decreasing the strength of the 
magnetic field, which however also compromises the momentum resolution.

In the following, a conservative design is chosen with inner sensor layers placed at
\SIlist[list-units=single]{2.0;3.0}{\cm} and outer sensor layers placed at
\SIlist[list-units=single]{7.0;8.0}{\cm}, providing a signal acceptance of about \SI{50}{\percent}.

\begin{figure}[p!]
        \centering      \includegraphics[width=0.48\textwidth]{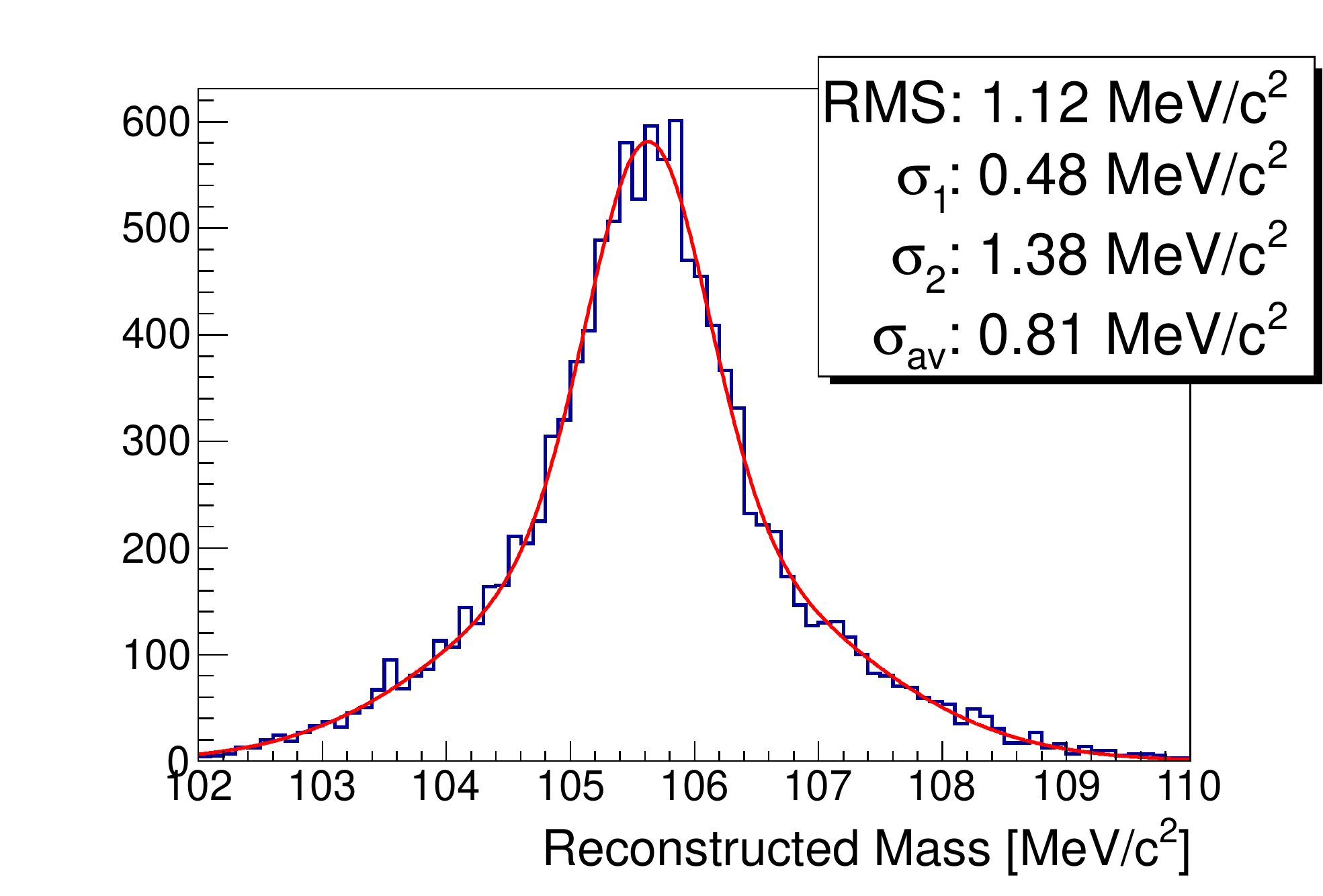}
        \caption{Reconstructed mass resolution for signal events in the phase IA configuration.}
        \label{fig:reso_ph1a_reco}
\end{figure}

\begin{figure}[p!]
        \centering      \includegraphics[width=0.48\textwidth]{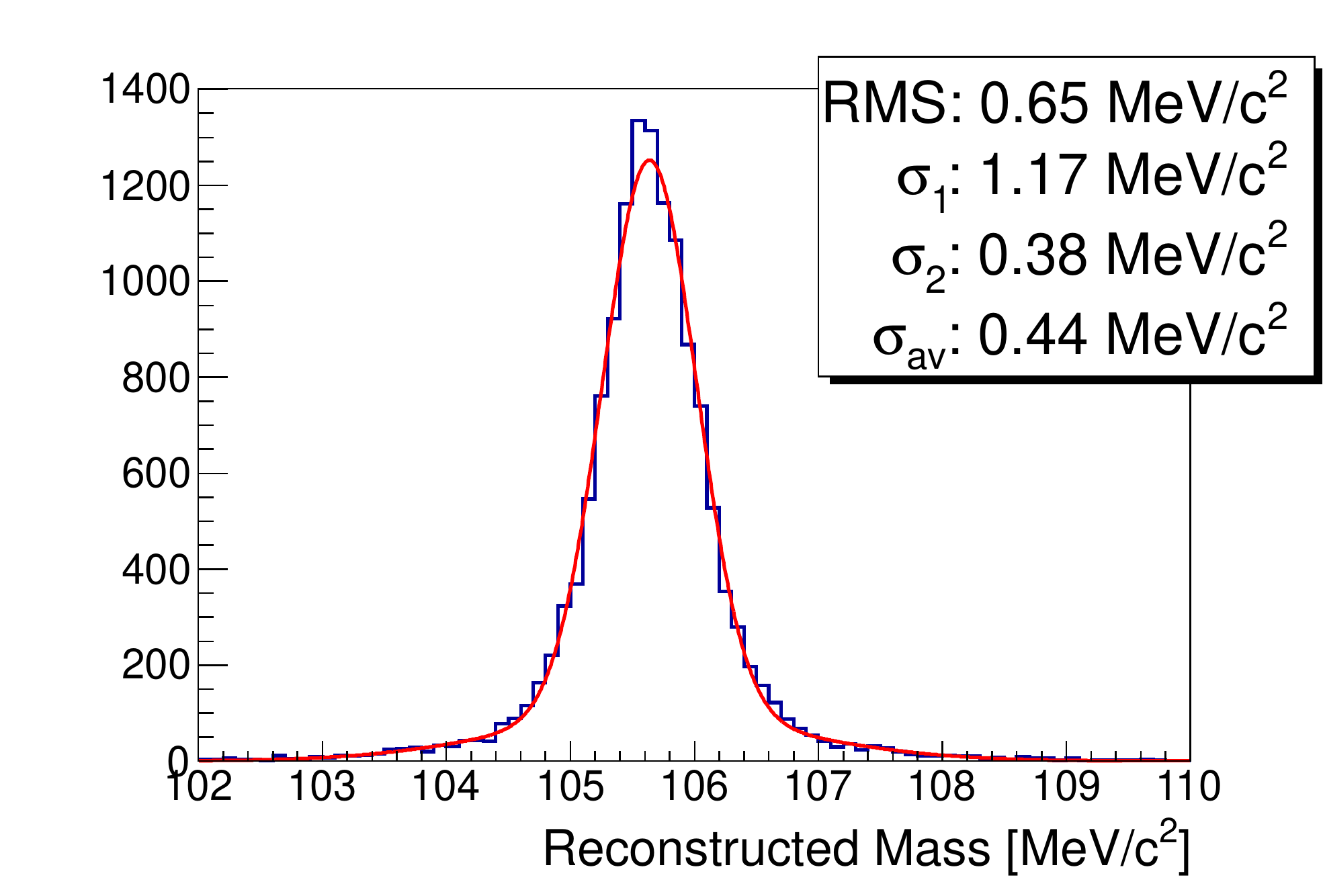}
        \caption{Reconstructed mass resolution for signal events in the phase IB configuration.}
        \label{fig:reso_ph1b_reco}
\end{figure}

\begin{figure}[p!]
        \centering      \includegraphics[width=0.48\textwidth]{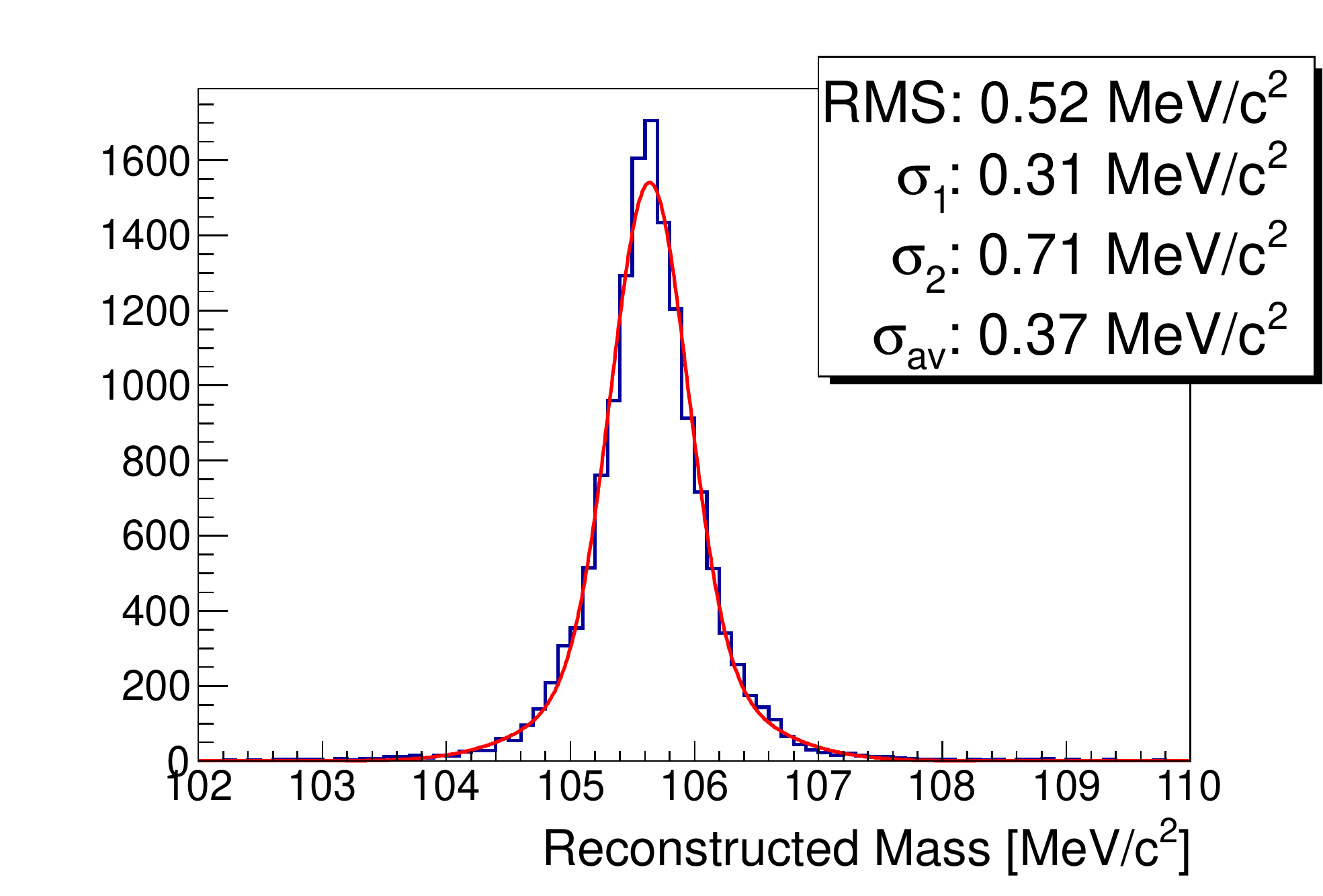}
        \caption{Reconstructed mass resolution for signal events in the phase II configuration.}
        \label{fig:reso_ph2_reco}
\end{figure}

The resulting mass resolutions calculated from the reconstructed track parameters 
 are shown in Figures~\ref{fig:reso_ph1a_reco}-\ref{fig:reso_ph2_reco} for the different phases. 
Compared to phase~IA, the mass resolutions improve considerably by adding the recurl stations at
phase~IB and phase~II and yield - despite the additional material from the
scintillating fibre tracker - resolutions well below \SI{1}{\mega\electronvolt}.
Of special importance for the background suppression is the
strong reduction of the tails in the phase II setup.

\section{Selection}
\subsection{Kinematic Selection of Signal Events}
\begin{figure}
        \centering      \includegraphics[width=0.48\textwidth]{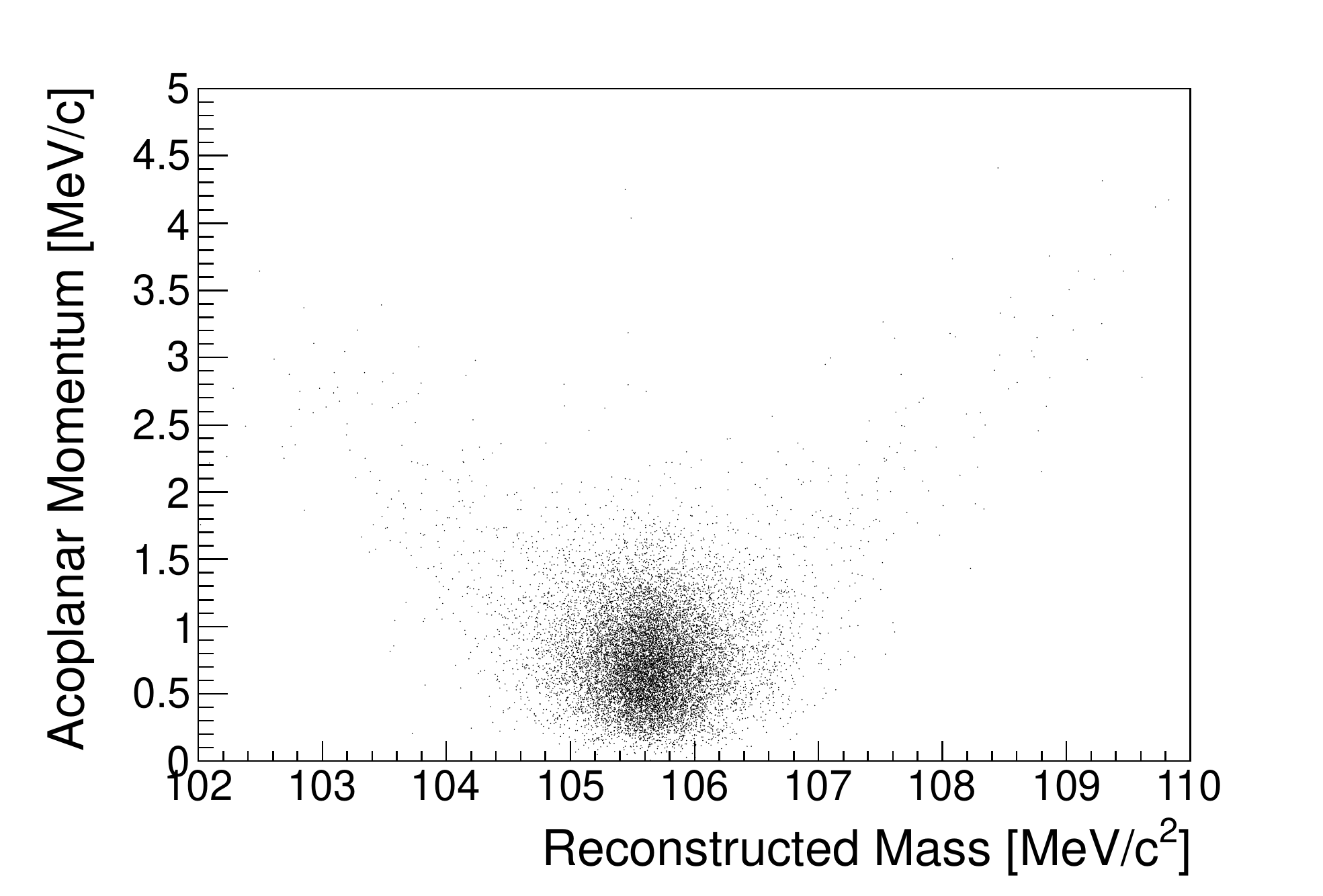}
       \caption{Reconstruced mass versus acoplanar momentum for the phase II detector.}
        \label{fig:reso_2d_ph2}
\end{figure}

The mass resolution can be further improved by rejecting badly reconstructed
signal events, where one or several tracks are either wrongly
reconstructed or suffer from large multiple scattering. A wrong measurement of
the track momentum or direction affects momentum balance and leads to a
measurable missing momentum. 
In the following we follow the strategy of the SINDRUM experiment
\cite{Bellgardt:1987du} and
define an acoplanar momentum vector $p_{\textnormal{acopl}}$, which is obtained by
projecting the vectorial sum of all track momenta 
into the decay plane defined by the three tracks. 
The correlation between the reconstructed invariant mass and the
variable $p_{\textnormal{acopl}}$ is shown in Figures~\ref{fig:reso_2d_ph2}. 
In addition to the main spot at small values of $p_{\textnormal{acopl}}$, which
originates from well measured signal events, two diagonal sidebands
originating from wrongly measured signal events are visible. 
By applying the cut $p_{\textnormal{acopl}}< \SI{1.4}{\mega\electronvolt\per\c}$, which will also be used to reject
background events in the following, most of the wrongly measured signal
events are rejected.

\begin{figure}[p!]
        \centering      \includegraphics[width=0.48\textwidth]{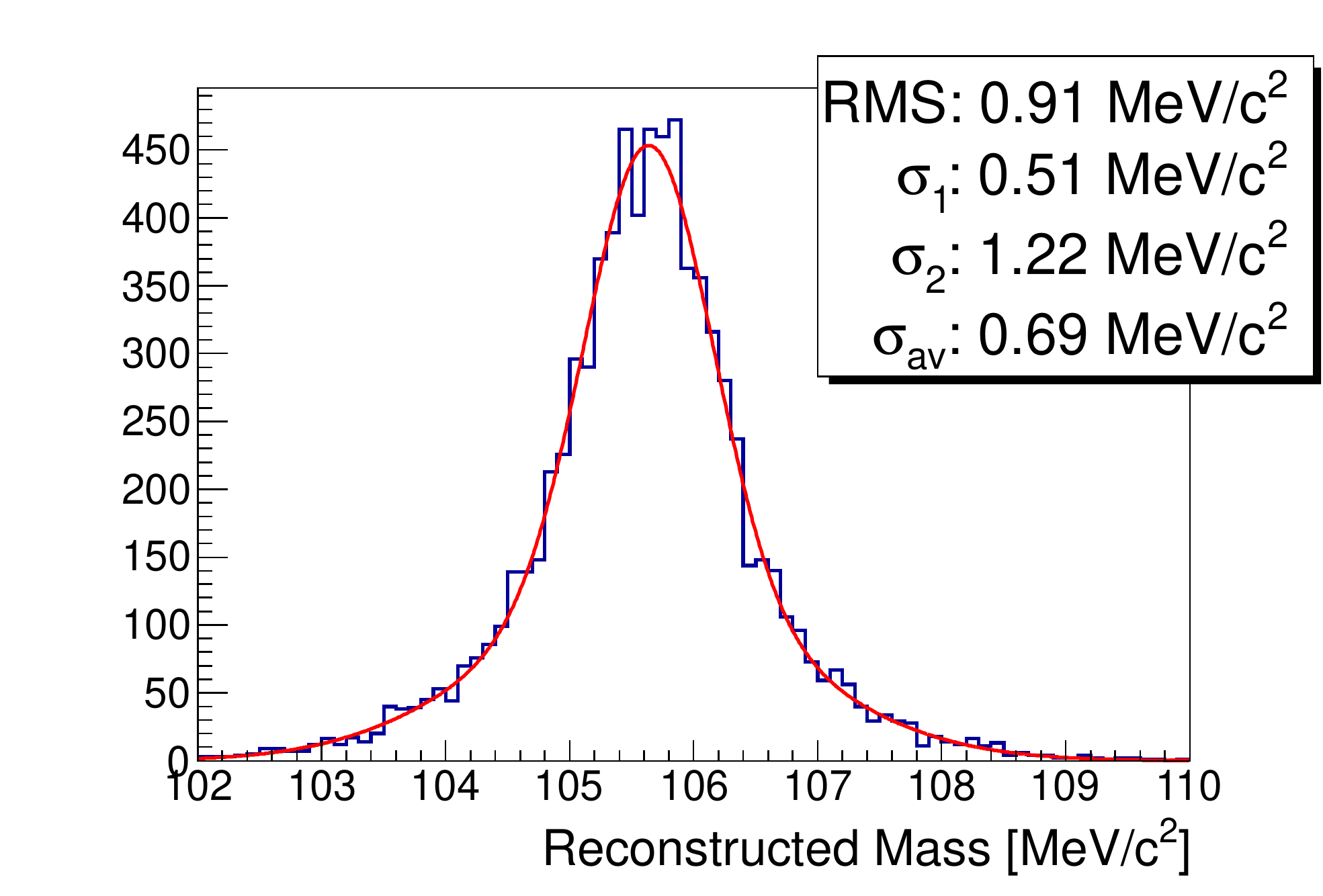}
        \caption{Reconstructed mass resolution for signal events after kinematic cuts in the phase IA configuration.}
        \label{fig:reso_ph1a_cut}
\end{figure}

\begin{figure}[p!]
        \centering      \includegraphics[width=0.48\textwidth]{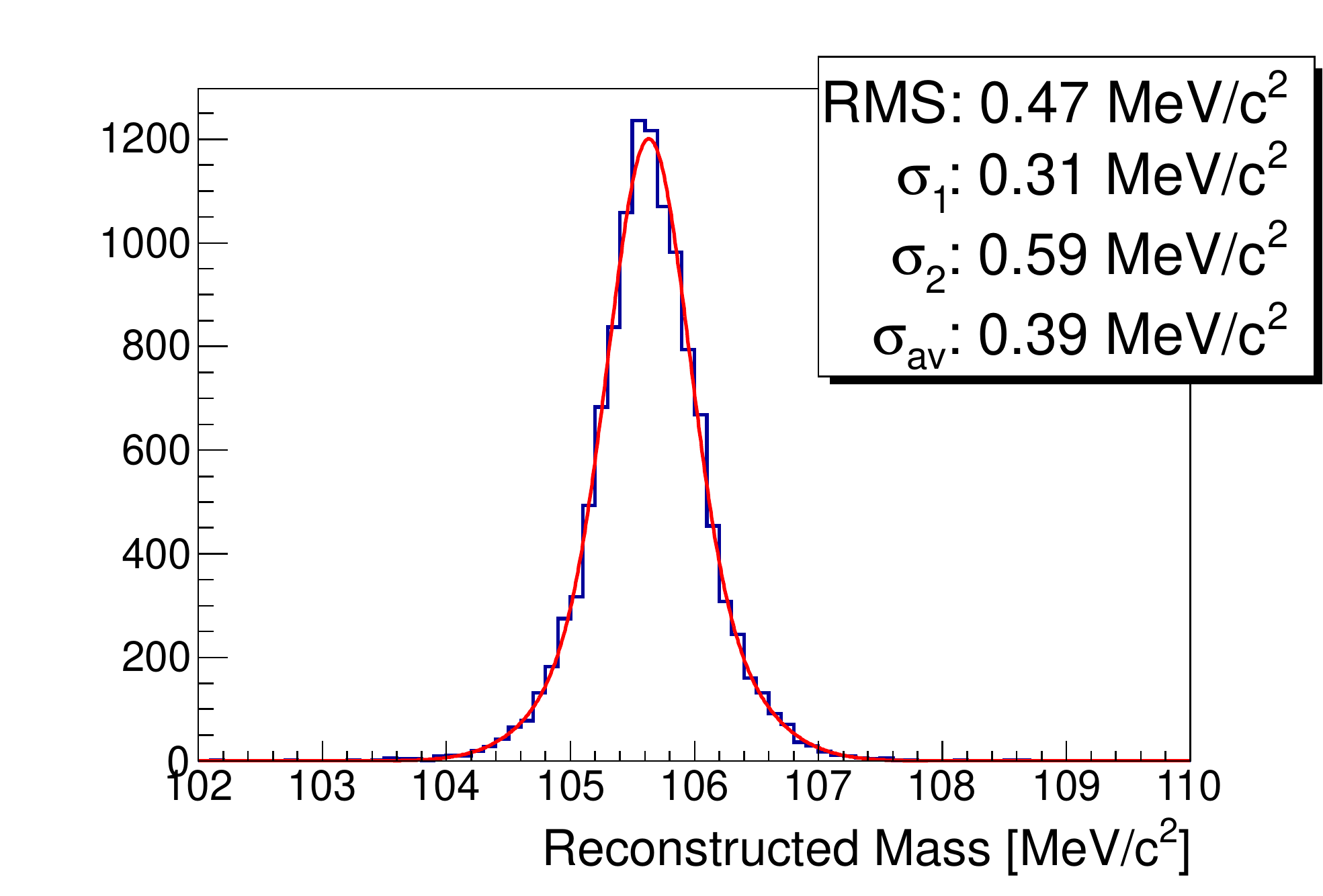}
        \caption{Reconstructed mass resolution for signal events after kinematic cuts in the phase IB configuration.}
        \label{fig:reso_ph1b_cut}
\end{figure}

\begin{figure}[p!]
        \centering      \includegraphics[width=0.48\textwidth]{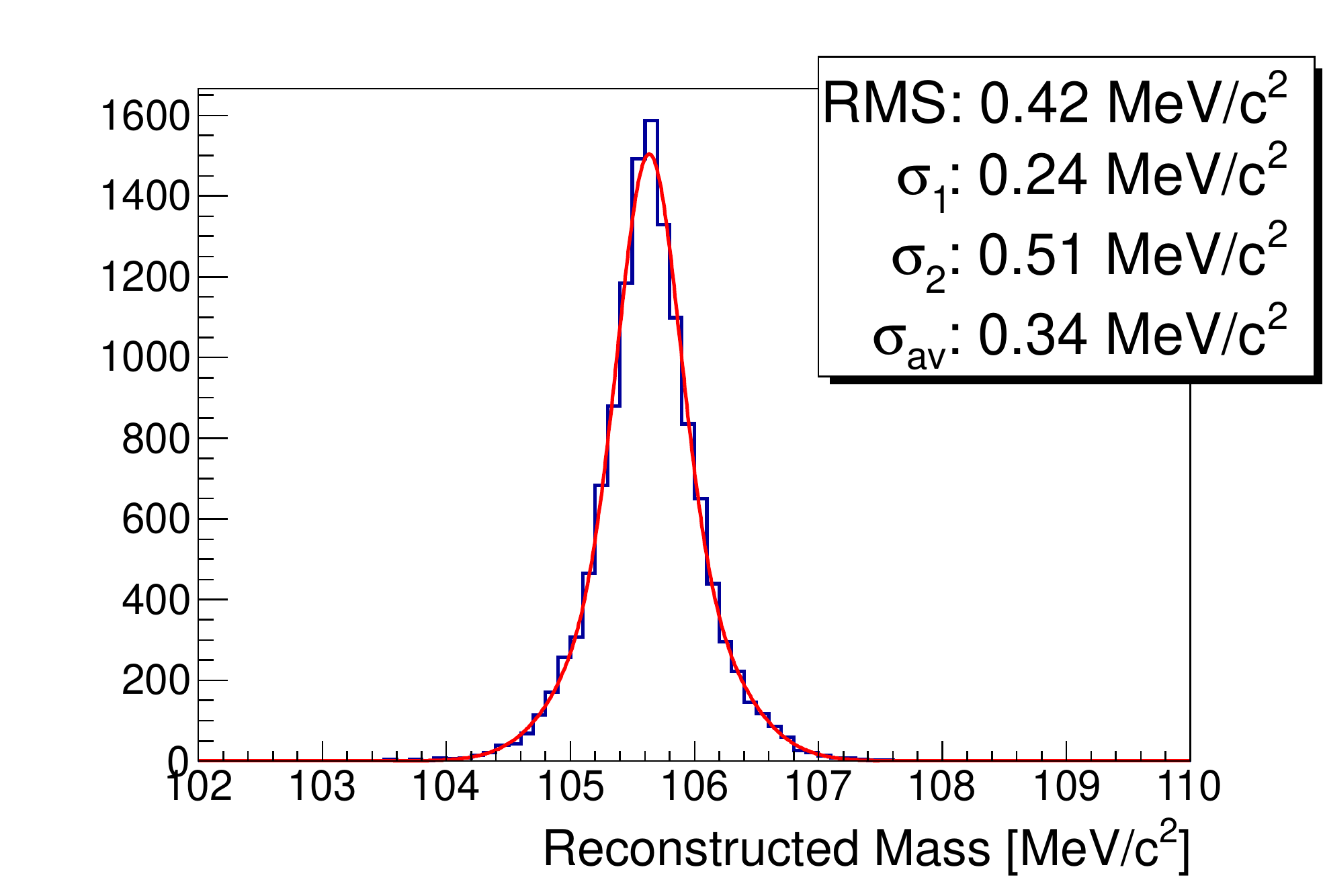}
        \caption{Reconstructed mass resolution for signal events after kinematic cuts in the phase II configuration.}
        \label{fig:reso_ph2_cut}
\end{figure}

The resulting mass resolution plots are shown in Figure~\ref{fig:reso_ph1a_cut}-\ref{fig:reso_ph2_cut}, which
show by about \SI{20}{\percent} improved resolutions and significantly reduced tails. 
The expected mass resolutions and signal efficiencies for the three stages of
the experiment are summarized in Table~\ref{tab:Efficiencies}.

\begin{table*}
        \centering
                \begin{tabular}{lccc}
                        \toprule
                                        & Phase IA & Phase IB & Phase II\\
                        \midrule
                                Michel decays:  &        &  & \\
                                \hspace{3mm}Efficiency (unpolarized) & \SI{50.0}{\percent} & \SI{53.4}{\percent} & \SI{52.4}{\percent}\\
                                \hspace{3mm}Momentum RMS & $\SI{0.73}{\mega\electronvolt\per\c}$ & $\SI{0.44}{\mega\electronvolt\per\c}$ & $\SI{0.28}{\mega\electronvolt\per\c}$\\
                                \hspace{3mm}Wrong charge fraction&  \SI{1.14}{\percent} & \SI{0.45}{\percent} & \SI{0.45}{\percent}\\
                                \midrule
                                Signal: & & & \\
                                \hspace{3mm}Reconstruction efficiency & \SI{39}{\percent} & \SI{46}{\percent} & \SI{48}{\percent} \\
                                \hspace{3mm}Energy sum RMS (reconstructed)& $\SI{1.12}{\mega\electronvolt\per\c\squared}$ & $\SI{0.65}{\mega\electronvolt\per\c\squared}$ & $\SI{0.52}{\mega\electronvolt\per\c\squared}$ \\
                                        \hspace{3mm}Efficiency after selection& \SI{26}{\percent} & \SI{39}{\percent} & \SI{38}{\percent} \\
                                \hspace{3mm}Energy sum RMS (selected)& $\SI{0.91}{\mega\electronvolt\per\c\squared}$ & $\SI{0.47}{\mega\electronvolt\per\c\squared}$ & $\SI{0.42}{\mega\electronvolt\per\c\squared}$ \\
                                \midrule
                                Track $d_{ca}$ resolution ($\sigma$) & $\SI{190}{\um}$ & $\SI{185}{\um}$ & $\SI{185}{\um}$ \\
                                \bottomrule
                \end{tabular}
        \caption{Efficiencies and resolutions used in the sensitivity study. $d_{ca}$ is the distance of closest approach of 
a track to the beam line. The drop in the efficiency after selection for phase II is due to the larger combinatorial background.}
        \label{tab:Efficiencies}
\end{table*}

The main criterion for the kinematic selection of signal events is the reconstructed invariant
mass. In a final analysis the number of signal and background events will be
determined by a fit of the invariant mass distribution, see also 
Figures~\ref{fig:smearplot_phase1a}-\ref{fig:smearplot_phase2} or by exploiting 
multivariate methods which include several estimators. 
For sake of simplicity the number of signal events is here determined from a
2-sigma mass window around the nominal muon mass.

\subsection{Reduction of the \mtenunu Background}
The acoplanar momentum cut is also very effective in reducing the dominating background from
radiative events with internal conversion \mtenunu. 
These events are characterized by missing energy carried away by the two
undetected neutrinos. 
Most of these background events have either a small value of the reconstructed
three electron mass $m_{eee}$ or show some momentum imbalance. 
In particular the class of dangerous background events, which have only little missing
energy, carried away by the two neutrinos, and are wrongly reconstructed such that
the reconstructed invariant mass matches the muon mass, show some significant momentum
imbalance. Most of these background events are rejected by the cut
$p_{\textnormal{acopl}} < \SI{1.4}{\mega\electronvolt\per\c}$.

\begin{figure}[p!]
        \centering      \includegraphics[width=0.48\textwidth]{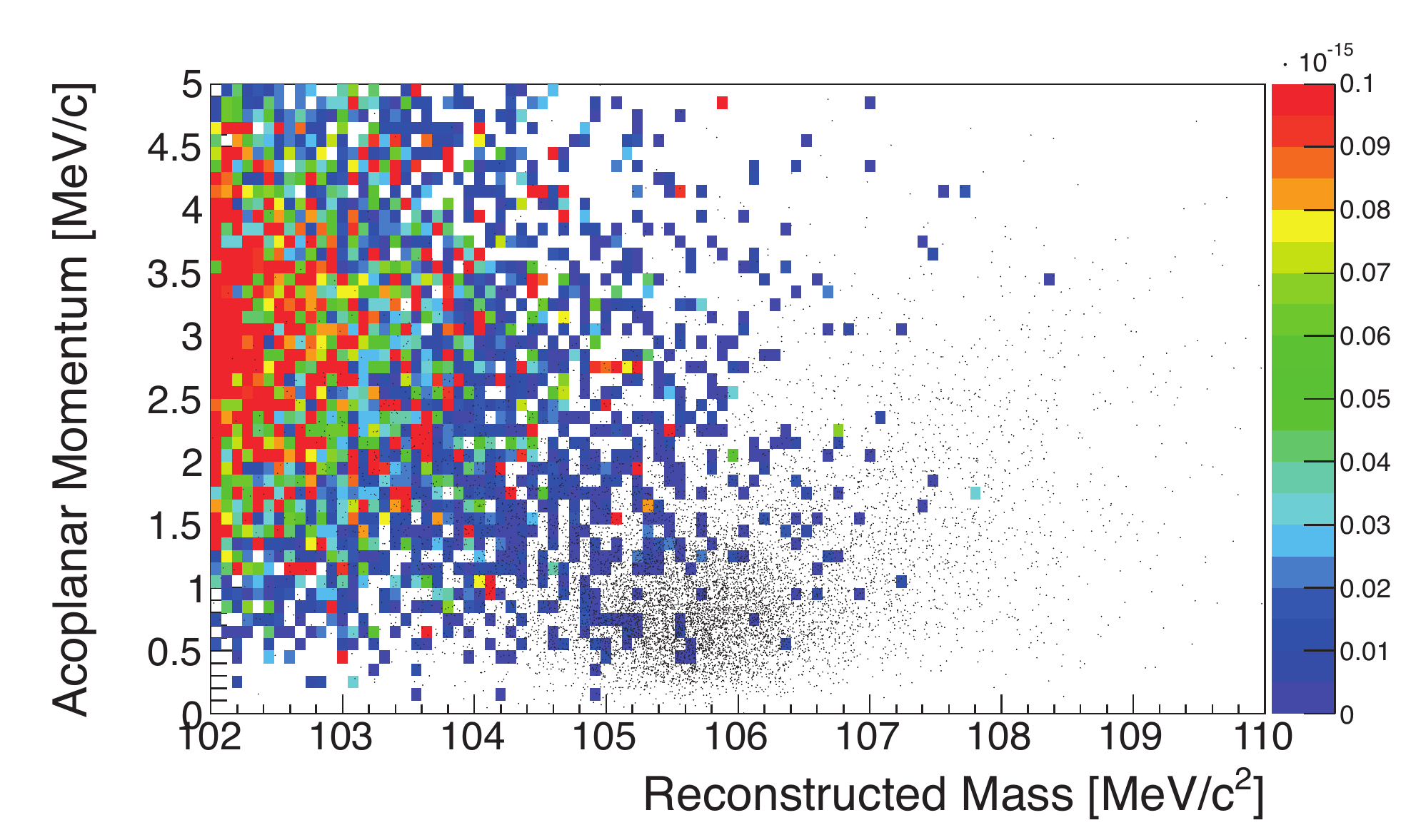}
        \caption{Internal conversion background (colours) and signal (black
          dots) in the acoplanar momentum - reconstructed mass plane for the phase IA detector configuration.}
        \label{fig:signalbg_phase1a}
\end{figure}

\begin{figure}[p!]
        \centering      \includegraphics[width=0.48\textwidth]{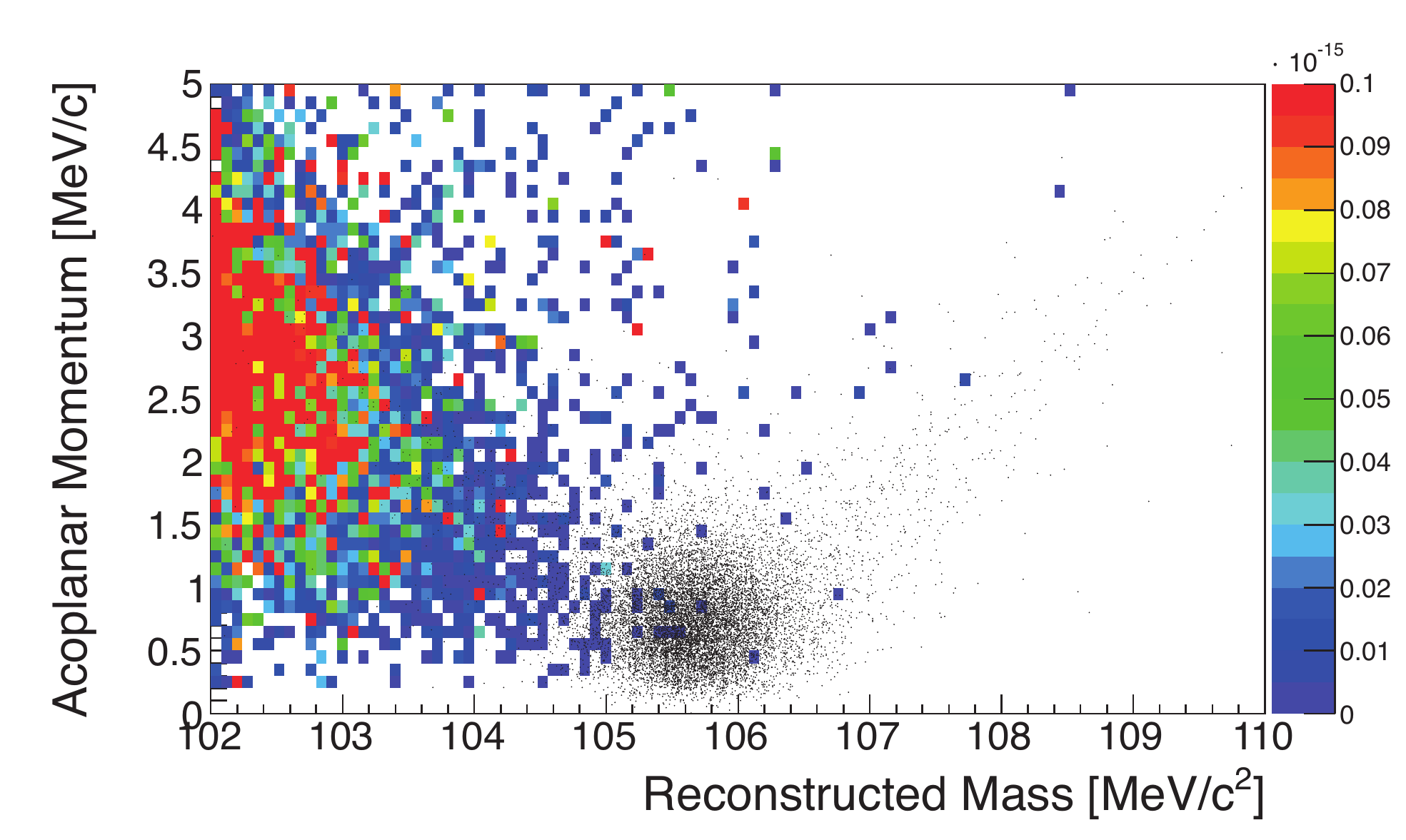}
        \caption{Internal conversion background (colours) and signal (black
          dots) in the acoplanar momentum - reconstructed mass plane for the phase IB detector configuration.}
        \label{fig:signalbg_phase1b}
\end{figure}

\begin{figure}[p!]
        \centering      \includegraphics[width=0.48\textwidth]{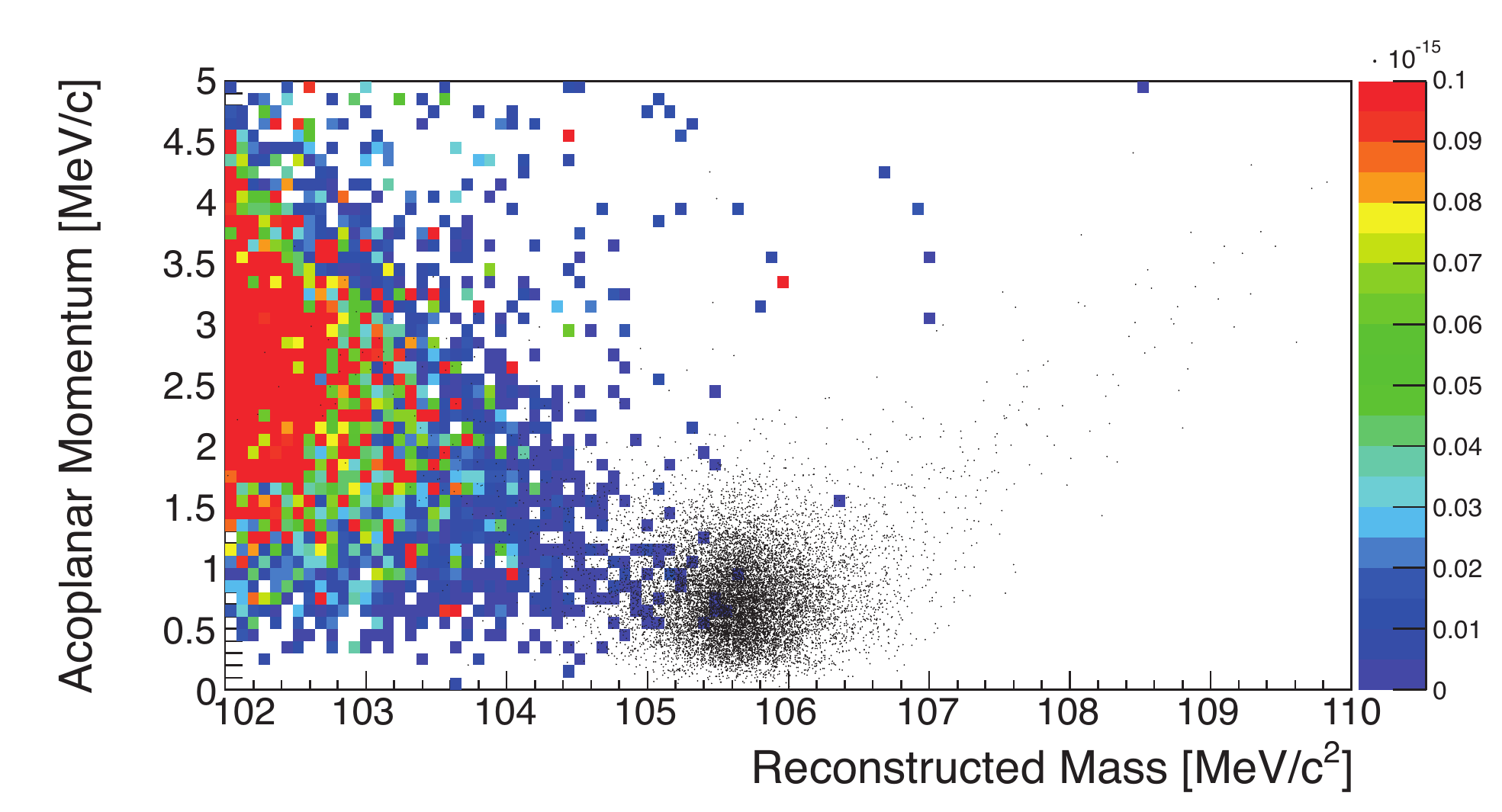}
        \caption{Internal conversion background (colours) and signal (black
          dots) in the acoplanar momentum - reconstructed mass plane for the phase II detector configuration.}
        \label{fig:signalbg_phase2}
\end{figure}

The separation of signal and background in the plane  $p_{\textnormal{acopl}}$ versus
$m_{eee}$ is shown for the three stages of the experiments in
Figures~\ref{fig:signalbg_phase1a}-\ref{fig:signalbg_phase2}. 
A clear separation of signal and background events at
a level of $10^{-17}-10^{-16}$ muon decays is visible. 
This separation improves with the upgraded experiment at phase~IB and~II.

\begin{figure}
        \centering      \includegraphics[width=0.48\textwidth]{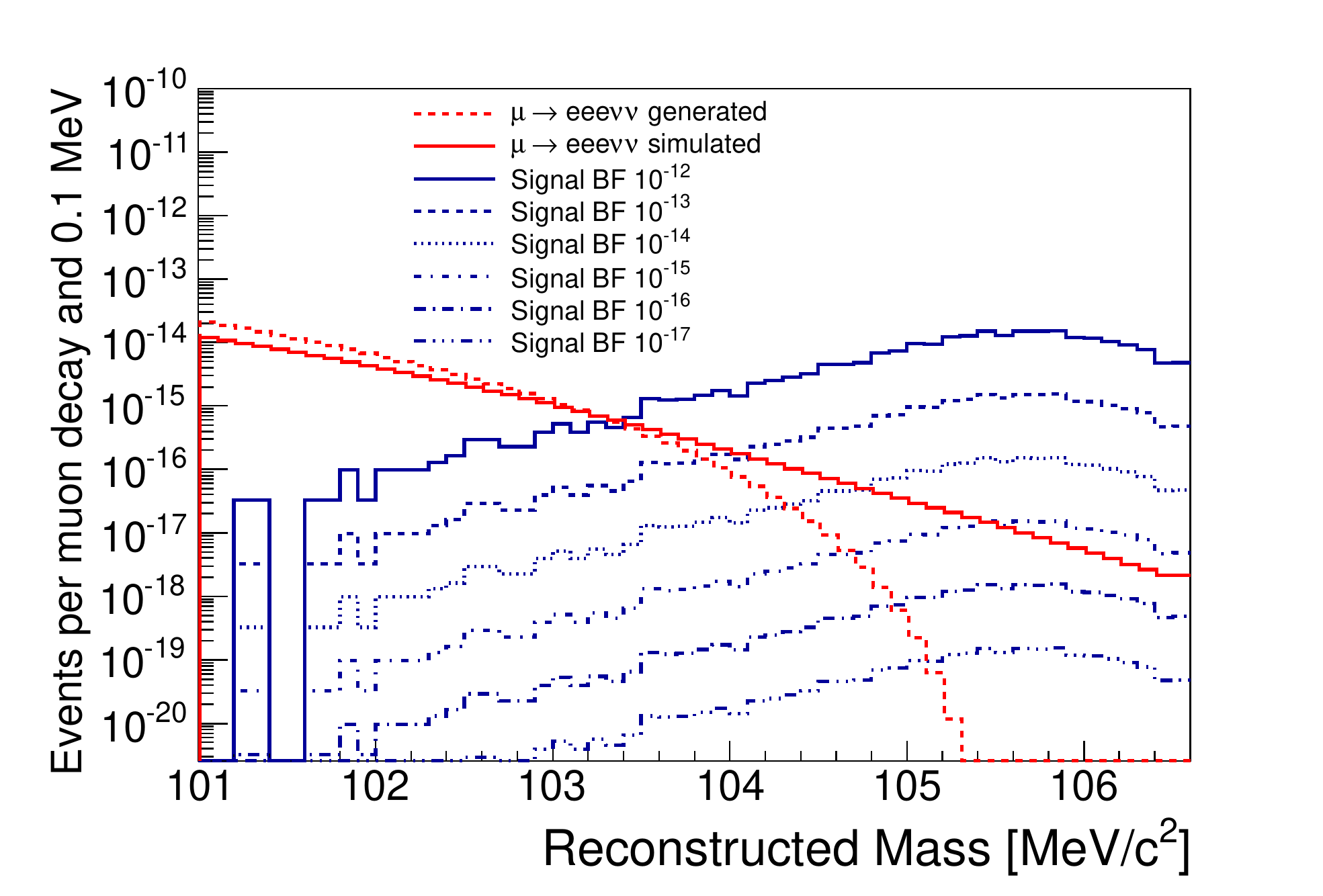}
        \caption{Tail of the internal conversion distribution overlaid with signal at different branching ratios for the phase IA detector. The resolution for the internal conversion decays was taken from \num{30000} simulated signal decays.}
        \label{fig:smearplot_phase1a}
\end{figure}

\begin{figure}
        \centering      \includegraphics[width=0.48\textwidth]{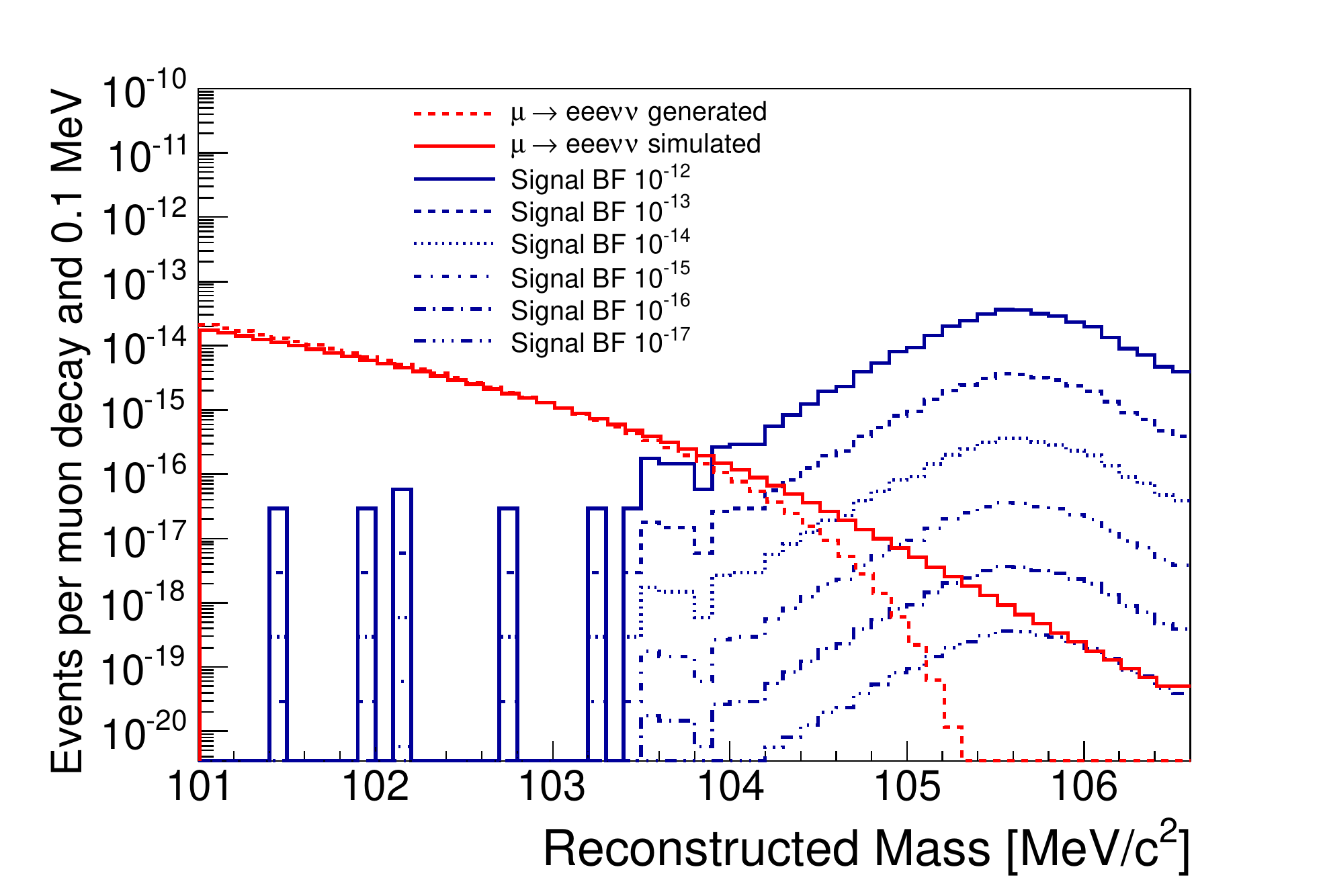}
        \caption{Tail of the internal conversion distribution overlaid with signal at different branching ratios for the phase IB detector. The resolution for the internal conversion decays was taken from \num{30000} simulated signal decays.}
        \label{fig:smearplot_phase1b}
\end{figure}

\begin{figure}
        \centering      \includegraphics[width=0.48\textwidth]{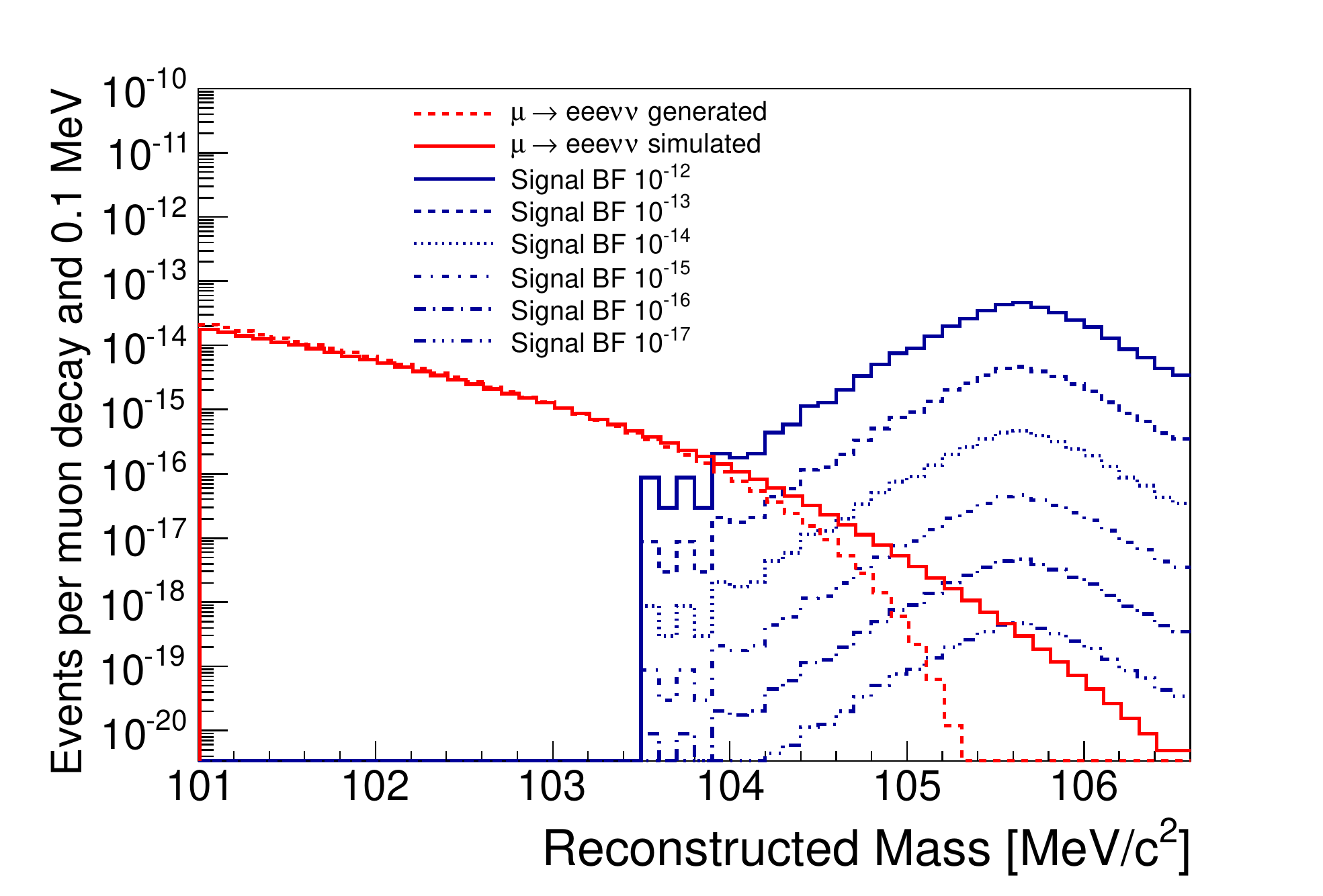}
        \caption{Tail of the internal conversion distribution overlaid with signal at different branching ratios for the phase II detector. The resolution for the internal conversion decays was taken from \num{30000} simulated signal decays.}
        \label{fig:smearplot_phase2}
\end{figure}

The projected invariant mass distribution of signal and background after applying the acoplanar
momentum cut is shown in the Figures~\ref{fig:smearplot_phase1a}-\ref{fig:smearplot_phase2}. 
Theses distributions are used as basis for the estimated sensitivity calculation.

\subsection{Reduction of Accidental Background}
The dominant contribution to the accidental background comes from
combinatorial background of three Michel decays $3 \times$(\Michel ) and 
from coincidences of a radiative muon decay with internal conversion \mteee
with a Michel decay \Michel.
Both types of backgrounds are significantly reduced by applying vertex and time
requirements. The combinatorial Michel background depends quadratically on the 
vertex and time resolution, whereas the coincidences  \mteee $\times$ \Michel
coincidence rate scale linearly with the vertex and the time resolution.

\balance
The vertex resolution in the transverse direction can be represented by the
distance of closest approach when the track is extrapolated to the muon decay
position. This values is about $\SI{185}{\um}$ in all phases of the experiment
and given in Table~\ref{tab:Efficiencies}. 
For a target size of $\SI{10}{\cm}$ in length and $\SI{2}{\cm}$ in diameter a reduction
factor of $\num{1e-4}$ can be derived per coincidence.

The estimation of the reduction factor from the timing cut, relevant for phases~Ib
and~II, is difficult as the design for the time of flight system has not been
finalised yet. 
Another difficulty comes from the fact that the timing resolutions of the
scintillating fibre detector the scintillating tiles detector are expected
to be different and that particles depending on their flight direction will 
be measured in both detectors or by the scintillating fibre detector only if staying
in the central region. In the latter case the particles will be
measured many times in several turns, what also leads to an
increase in precision. 
Preliminary studies indicate that timing resolutions of \SIrange[range-units=single]{200}{300}{\pico\second} in the scintillating fibre detector
and $<\SI{100}{\pico\second}$ in the Scintillating Tiles Detector can be achieved, see
sections~\ref{sec:Fibre} and \ref{sec:Tiles}.
For the sensitivity calculation it is assumed that all particles will be
measured with a time resolution of better than $\SI{250}{\pico\second}$ and that
reduction factors of \num{5e-3} per coincidence can be obtained at a
signal efficiency of \SI{90}{\percent}.

\section{Results}
\begin{figure*}
        \centering      \includegraphics[width=0.7\textwidth]{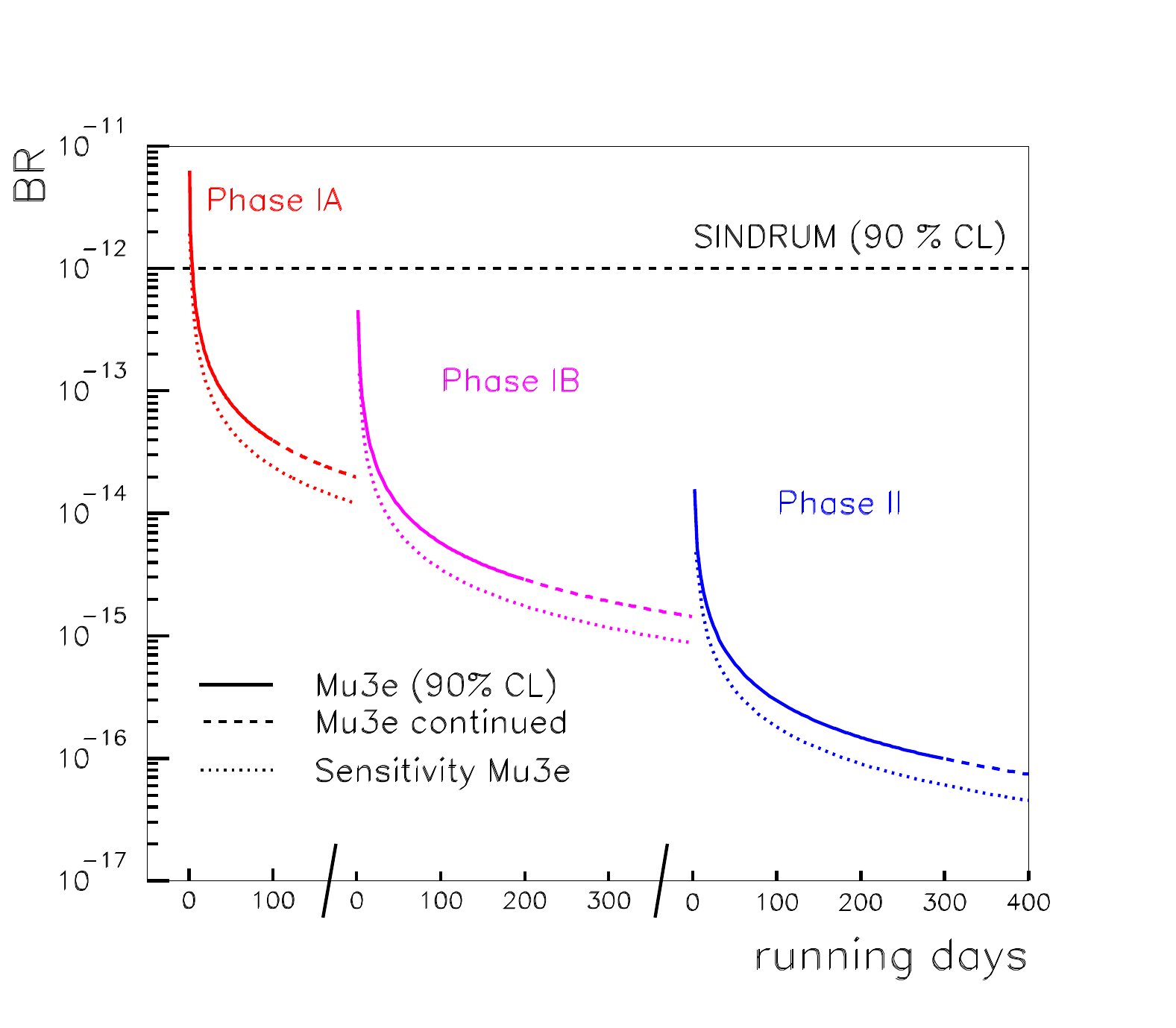}
        \caption{Projected sensitivity and projected limit  (90\% CL) of the \emph{Mu3e} experiment
        for the three construction phases~IA (red), IB (magenta) and~II (blue) as function of the running time. The
        curves were obtained from the corresponding numbers in Table~\protect{\ref{tab:Sensitivity}}. 
        For comparison also the (90\% CL) obtained from the
        SINDRUM experiment is indicated (black dashed line).}
        \label{fig:mu3e_projection}
\end{figure*}

The signal efficiencies and the expected background reduction factors of the
different selection cuts are summarized in Table~\ref{tab:Sensitivity}.
Single event sensitivities are calculated, which are defined
as the branching ratio at which one background event is expected. 
The sensitivity of the experiment scales with one over the number of muons on
target as long as the single event sensitivity is not reached and with one
over the square root of the number of muons on target after.
It can be seen that the single event sensitivity at phase~IB is higher than at
phase~II. This comes from the fact that the combinatorial background increases
at higher muon rates. However, the time to reach the aimed sensitivity of
$\num{1e-16}$ takes in phase~IB ten years of running whereas at phase~II this
sensitivity can be reached in 180 days because of the higher muon rate, see Figure~\ref{fig:mu3e_projection}.
Although the expected sensitivity at phase~IA, even without the time of flight
system, is  quite high with $\num{4e-16}$ it would take a long time of running to reach this sensitivity.
However, about one month of running is sufficient to reach a sensitivity of
$\num{1e-13}$, which is factor ten smaller than the current experimental bound.
\begin{table*}
        \centering
                \begin{tabular}{lccc}
                        \toprule
                                        & Phase IA & Phase IB & Phase II\\
                        \midrule
                        Backgrounds: & & &\\
                        \hspace{3mm}Michel & 0 & $<\num{2.5e-18}$ &$\num{5e-18}$ \\
                        \hspace{3mm}\mtenunu  & $ \num{1e-16}$ & $ \num{1e-17}$ & $ \num{1e-17}$ \\
                        \hspace{3mm}\mtenunu and accidental Michel      & 0 & $<\num{2.5e-21}$ &$\num{7.5e-18}$ \\
                \hspace{3mm}Total Background & $ \num{1e-16}$ & $ \num{1e-17}$ & $ \num{2.3e-17}$ \\
                \midrule
                        Signal: & & &\\
                        \hspace{3mm}Track reconstruction and selection efficiency & \SI{26}{\percent} & \SI{39}{\percent} & \SI{38}{\percent} \\
                        \hspace{3mm}Kinematic cut ($2\sigma$) & \SI{95}{\percent} & \SI{95}{\percent} & \SI{95}{\percent} \\
                        \hspace{3mm}Vertex efficiency $(2.5\sigma)^2$ & \SI{98}{\percent} & \SI{98}{\percent} & \SI{98}{\percent} \\
                        \hspace{3mm}Timing efficiency $(2\sigma)^2$ & - & \SI{90}{\percent} & \SI{90}{\percent} \\
                        \hspace{3mm}Total efficiency  & \SI{24}{\percent} & \SI{33}{\percent} & \SI{32}{\percent} \\
                        \midrule
                        Sensitivity: & & &\\
                        \hspace{3mm}Single event sensitivity & $ \num{4e-16}$ & $ \num{3e-17}$ & $ \num{7e-17}$ \\
                        \hspace{3mm}muons on target rate (Hz) & $ \num{2e7}$ & $ \num{1e8}$ & $ \num{2e9}$ \\
                        \hspace{3mm}running days to reach \num{1e-15} & \num{2600}
                        & 350  & 18 \\
                        \hspace{3mm}running days to reach \num{1e-16} & - & \num{3500} & 180 \\
                        \hspace{3mm}running days to reach single event sensitivity & \num{6500} & \num{11700} & 260 \\
                        \bottomrule
                \end{tabular}
        \caption{Signal efficiencies and estimated background reduction factors of the discussed
          selection cuts, which are used to determine the single event
          sensitivities for the three phases of the experiment. For the given muon
          on target rates running times are calculated to reach the given
          single event sensitivities.}
        \label{tab:Sensitivity}
\end{table*}

\part{The Mu3e Collaboration}

\chapter{The Institutes in Mu3e}
\label{sec:Institutes}

\nobalance

\section{Responsibilities}
\label{sec:Responsibilities}

The responsibilities of the institutes participating in Mu3e are outlined in Table~\ref{tab:responsibilities}. This list concerns the phase I experiment and does not imply funding responsibilities. 

\section{Collaborators}
\label{sec:Collaborators}

\subsection{University of Geneva}
\label{sec:UniversityOfGeneva}

\noindent Currently working on Mu3e:\\
\noindent {\bf Alain Blondel} \emph{Professor}\\
\noindent {\bf Alessandro Bravar} \emph{Professor}\\
\noindent {\bf Martin Pohl} \emph{Professor}\\
\noindent {\bf Antoaneta Damyanova} \emph{Master student}

\noindent Planned positions:\\
\noindent 1 \emph{Postdoc}\\
\noindent 1 \emph{Ph.D.~candidate}

\subsection{University of Z\"urich}
\label{sec:UniversityOfZUrich}

\noindent Currently working on Mu3e:\\
\noindent {\bf Ueli Straumann} \emph{Professor}\\
\noindent {\bf Peter Robmann} \emph{Senior scientist}\\
\noindent {\bf Roman Gredig} \emph{Ph.D.~candidate}

\subsection{Paul Scherrer Institut}
\label{sec:PaulScherrerInstitut}

\noindent Currently working on Mu3e:\\
\noindent{\bf Felix Berg} \emph{Ph.D.~candidate} \\
\noindent{\bf Malte Hildebrandt} \emph{Senior scientist} \\
\noindent{\bf Peter-Raymond Kettle} \emph{Senior scientist}\\
\noindent{\bf Angela Papa} \emph{Senior scientist}\\
\noindent{\bf Stefan Ritt} \emph{Senior scientist} \\
\noindent{\bf Alexey Stoykov} \emph{Senior scientist}

\noindent Funded open positions:\\
\noindent 1 \emph{Postdoc}\\
\noindent 1 \emph{Ph.D.~candidate}

\subsection{ETH Z\"urich}
\label{sec:ETHZUrich}

\noindent Currently working on Mu3e:\\
\noindent {\bf Christophorus Grab} \emph{Professor}\\
\noindent {\bf G\"unther Dissertori} \emph{Professor}\\
\noindent {\bf Rainer Wallny} \emph{Professor}

\noindent Planned positions:\\
\noindent 1 \emph{Ph.D.~candidate}

\subsection{University of Heidelberg}
\label{sec:UniversityOfHeidelberg}

\subsubsection{Physikalisches Institut (PI)}
\label{sec:PhysikalischesInstitut}

\noindent Currently working on Mu3e:\\
\noindent {\bf Andr\'e Sch\"oning} \emph{Professor}\\
\noindent {\bf Dirk Wiedner} \emph{Senior scientist}\\
\noindent {\bf Sebastian Bachmann} \emph{Senior scientist}\\
\noindent {\bf Niklaus Berger} \emph{Junior research group leader}\\
\noindent {\bf Bernd Windelband} \emph{Mechanical engineer}\\
\noindent {\bf Moritz Kiehn} \emph{Ph.D.~candidate}\\
\noindent {\bf Kevin Stumpf} \emph{Mechanical engineer}\\
\noindent {\bf Raphael Philipp} \emph{Master student}

\noindent Funded open positions:\\
\noindent 1 \emph{Postdoc}\\
\noindent 4 \emph{Ph.D.~candidates}

\noindent Planned positions:\\
\noindent 1 \emph{Postdoc}\\
\noindent 1 \emph{Ph.D.~candidate}

\newpage
\subsubsection{Kirchhoff Institut f\"ur Physik (KIP)}
\label{sec:KirchhoffInstitutFUrPhysik}

\noindent Currently working on Mu3e:\\
\noindent{\bf Hans-Christian Schultz-Coulon} \emph{Professor}\\ 
\noindent{\bf Wei Shen} \emph{Postdoc}\\
\noindent{\bf Patrick Eckert} \emph{Ph.D.~candidate}\\
\noindent{\bf Carlo Licciulli} \emph{Master student}\\
\noindent{\bf Tobias Hartwig} \emph{Bachelor student}

\noindent Planned positions:\\
\noindent 1 \emph{Postdoc}\\
\noindent 1 \emph{Ph.D.~candidate}

\subsubsection{Zentralinstitut f\"ur Technische Informatik}
\label{sec:ZentralinstitutFUrTechnischeInformatik}

\noindent Currently working on Mu3e:\\
\noindent{\bf Peter Fischer} \emph{Professor} \\
\noindent{\bf Ivan P\'eric} \emph{Senior scientist}

\begin{table}[p!]
	\centering
		\begin{tabular}{ll}
			\toprule
\sc Component & \sc Responsible \\ 
\toprule
Beam & PSI \\
\midrule
Magnet & UGS \\
       & Heidelberg PI \\
\midrule
Target & Heidelberg PI \\
       & PSI \\
\midrule
Pixel chip & ZITI Mannheim \\
					 & Heidelberg PI \\
					 & Heidelberg NB \\
\midrule
Pixel detector & Heidelberg PI \\
\midrule
Mechanics and cooling & Heidelberg PI \\
\midrule
Pixel on-detector electronics & Heidelberg NB \\
															 & Heidelberg PI \\
\midrule
Read-out electronics & Heidelberg NB \\
\midrule
Fibre detector & Geneva \\
							 & Z\"urich\\
							 & ETHZ\\
							 & PSI\\
\midrule
Tile detector & Heidelberg KIP\\
\midrule
Timing electronics & PSI \\
									 & Heidelberg KIP\\
									 & Geneva \\
							 		 & Z\"urich\\
							 		 & ETHZ\\
\midrule
Filter farm & Heidelberg NB \\
\midrule
Slow control & PSI \\
\midrule
Infrastructure & PSI\\

\bottomrule
		\end{tabular}
		\caption{Institutional interests and responsibilities in phase I Mu3e.   NB denotes the Emmy Noether junior reserch group at the \emph{Physikalisches Institut} led by N.~Berger. UGS is an external engineering consultancy tasked with magnet design and procurement. Note that especially for large common items such as the magnet, design and construction responisbility does not imply funding responsibility.}
		\label{tab:responsibilities}
\end{table}

\chapter{Schedule}
\label{sec:Schedule}

\balance

\section{Phase I Schedule}
\label{sec:PhaseISchedule}

Our goal is to take first data in 2015, with a pixel-only detector. Correspondingly, the timing critical items are the central pixel detectors and the magnet. Prototyping on the first is well underway and a first instrumented assembly corresponding to the vertex layers should be operational by late 2013. Magnet design has also started, however the tendering process will consume about half a year; we hope the actual magnet construction will be under way early in 2014.

For a second, longer run in 2016, we plan to integrate the fibre detector as well as the inner recurl stations complete with timing tiles. As the outer recurl stations are identical to the inner ones, we plan to maintain the production capacities and manufacture them during 2016.

Depending on the outcome of the 2016 run and the schedule for the high intensity beam line, 2017 could either be spend for phase II preparations or with an extended run at the existing beamline.

\section{Phase II Schedule}
\label{sec:PhaseIISchedule}

The phase II schedule is contingent on the schedule for the high intensity beamline. By 2017, we can have all required detector components ready. If the phase I experience shows that major parts of the detector will have to be rebuilt, earliest phase II startup would slip to 2018.

\chapter{Cost Estimates}
\label{sec:Cost}

\nobalance

\begin{table}[b!]
	\centering
		\begin{tabular}{lr}
			\toprule
			Item & Estimated cost  \\
			& (KCHF) \\
			\midrule
			Beam line & < 50 \\
			%Magnet    & 500\\
			Target    & < 10 \\
			Pixel detector (central) & 400\\
			Pixel detector (2 recurl stations) & 200\\
			Scintillating fibres & 200\\
			Tile detector (2 stations) & 110\\
			Mechanics and cooling & 200\\
			DAQ pixel detector & 50\\
			DAQ fibres & 200\\
			DAQ tiles & 190\\
			Central DAQ & 100\\
			Filter farm & 50\\
			Slow control & 50\\
			Infrastructure & 100\\
			\midrule
			Total & 1910\\
			\bottomrule
		\end{tabular}
	\caption{Estimated costs of the phase I experiment. The magnet (estimated at 500 KCHF) will be loaned to Mu3e by PI Heidelberg for the run time of the experiment.}
	\vspace{25.85mm}
	\label{tab:CostsPhaseI}
\end{table}

\begin{table}
	\centering
	
		\begin{tabular}{lr}
			\toprule
			Item & Estimated cost  \\
			& (KCHF) \\
			\midrule
			Beam line & u.a. \\
			Target    & < 10 \\
			Pixel detector (upgrade central) & 200\\
			Pixel detector (2 recurl stations) & 200\\
			Scintillating fibres & 200\\
			Tile detector (2 stations) & 110\\
			Mechanics and cooling & 100\\
			DAQ pixel detector & 50\\
			DAQ fibres & 200\\
			DAQ tiles & 190\\
			Central DAQ & 100\\
			Filter farm & 50\\
			Slow control & 50\\
			Infrastructure & u.a.\\
			\midrule
			Total & 1460\\
			\bottomrule
		\end{tabular}
	\caption{Estimated costs of the phase II experiment. Items marked u.a.~are under assesment; the new beam line required will be by far the most expensive item but also benefit many other users.}
	\label{tab:CostsPhaseII}
\end{table}

\appendix

\chapter{Appendix}
\label{sec:Appendix}

\section{Mu3e theses}
\label{sec:Students}

\balance

Several theses involving the Mu3e project have been completed:
\begin{itemize}
	\item M.~Kiehn, diploma thesis \emph{Track Fitting with Broken Lines 
	for the Mu3e Experiment}, Heidelberg University, 2012.
	\item M.~Zimmermann, bachelor thesis \emph{Cooling with Gaseous Helium 
	for the Mu3e Experiment}, Heidelberg University, 2012.
	\item H.~Augustin, bachelor thesis \emph{Charakterisierung von HV-MAPS}, 
	Heidelberg University, 2012.
	\item A.-K.~Perrevoort, master thesis \emph{Characterisation of High 
	Voltage Monolithic Active Pixel Sensors for the Mu3e Experiment}, 
	Heidelberg University, 2012.
\end{itemize}
These theses are available in portable document format (PDF) from 
\texttt{http://www.psi.ch/mu3e/documents}.
Several more theses are ongoing (thus titles are preliminary):
\begin{itemize}
	\item R.~Gredig, doctoral thesis \emph{Fibre Tracker for the Mu3e 
	Experiment}, Z\"urich University, started 2012.
	\item M.~Kiehn, doctoral thesis \emph{Track Reconstruction for the Mu3e 
	Experiment}, Heidelberg University, started 2012.
	\item A.~Damyanova, master thesis \emph{Fibre and SiPM characterization 
	for the Mu3e Experiment}, Geneva University, started 2012.
	\item R.~Philipp, master thesis \emph{Tests of High Voltage Monolithic 
	Active Pixel Sensors for the Mu3e Experiment}, Heidelberg University, started 2012.
\end{itemize}

\section{Acknowledgements}
\label{sec:IndAcknowledgements}

N.~Berger would like to thank the \emph{Deutsche Forschungsgemeinschaft} (DFG)
for funding his work on the Mu3e experiment through the Emmy Noether program
and thus supporting the experiment at a crucial early stage.

M.~Kiehn acknowledges support by the International Max Planck Research School
for Precision Tests of Fundamental Symmetries.

%\cite{*}
%\pagestyle{fancy}

\addcontentsline{toc}{chapter}{Bibliography}
\small
\bibliographystyle{common/literature/unsrt_collab_comma} %Style of Bibliography: plain / apalike / amsalpha / ...
\bibliography{common/literature/mu3e} 

\pagestyle{plain}
\end{document}